\documentclass[12pt]{iopart}

\usepackage{graphicx}
\usepackage{amssymb}
\newcommand{\dfrac}[2]{\frac{\displaystyle #1}{\displaystyle #2}}
\newcommand{\email}[1]{\ead{#1}}
\newcommand{\affiliation}[1]{\address{#1}}
\newcommand{\acknowledgments}{\ack}

\newcommand{\sss}[1]{{\scriptscriptstyle{#1}}}

\newcommand{\lta}{\lesssim}
\newcommand{\gta}{\gtrsim}
\newcommand{\dd}{\mathrm{d}}
\newcommand{\uPl}{\mathrm{Pl}}
\newcommand{\uin}{\mathrm{in}}
\newcommand{\uend}{\mathrm{end}}

\newcommand{\uinf}{\mathrm{inf}}
\newcommand{\ureh}{\mathrm{reh}}
\newcommand{\urad}{\mathrm{rad}}
\newcommand{\ucri}{\mathrm{cri}}
\newcommand{\unuc}{\mathrm{nuc}}
\newcommand{\uobs}{\mathrm{obs}}
\newcommand{\uphys}{\mathrm{phys}}
\newcommand{\uPgiP}{\mathrm{(gi)}}
\newcommand{\ue}{\mathrm{e}}
\newcommand{\uc}{\mathrm{c}}
\newcommand{\ub}{\mathrm{b}}

\newcommand{\uS}{\mathrm{S}}
\newcommand{\uM}{\mathrm{M}}
\newcommand{\uPSP}{(\uS)}
\newcommand{\uV}{\mathrm{V}}
\newcommand{\uT}{\mathrm{T}}
\newcommand{\uPTP}{(\uT)}
\newcommand{\uE}{\mathrm{E}}

\newcommand{\usssS}{\sss{\uS}}
\newcommand{\usssPSP}{\sss{\uPSP}}
\newcommand{\usssT}{\sss{\uT}}
\newcommand{\usssPTP}{\sss{\uPTP}}
\newcommand{\usssPMoneP}{\sss{(\uM=1)}}
\newcommand{\usssPl}{\sss{\uPl}}
\newcommand{\usssRH}{\ureh}
\newcommand{\usssV}{\sss{\uV}}
\newcommand{\usssE}{\sss{\uE}}
\newcommand{\nS}{n_\usssS}
\newcommand{\nT}{n_\usssT}
\newcommand{\alphaS}{\alpha_\usssS}
\newcommand{\alphaT}{\alpha_\usssT}
\newcommand{\zero}{{_0}}

\newcommand{\udm}{\mathrm{dm}}

\newcommand{\muS}{\mu_\usssS}
\newcommand{\muT}{\mu_\usssT}
\newcommand{\muST}{\mu_{\usssS,\usssT}}
\newcommand{\omegaT}{\omega_\usssT}
\newcommand{\omegaS}{\omega_\usssS}
\newcommand{\omegaST}{\omega_{\usssS, \usssT}}

\newcommand{\wstate}{w}
\newcommand{\calH}{\mathcal{H}}
\newcommand{\Mc}{M_\uc}
\newcommand{\OmegaL}{\Omega_\Lambda}
\newcommand{\OmegaCDM}{\Omega_\udm}
\newcommand{\OmegaB}{\Omega_\ub}
\newcommand{\CAMB}{\texttt{CAMB} }
\newcommand{\COSMOMC}{\texttt{COSMOMC} }
\newcommand{\mpl}{m_\usssPl}
\newcommand{\ie}{\textrm{i.e.}~}

\newcommand{\calO}{\mathcal{O}}

\newcommand{\ini}{\uin}

\newcommand{\Hzero}{H_0}
\newcommand{\Pstar}{P_*}
\newcommand{\Pnum}{P^{\usssPMoneP}}
\newcommand{\kstar}{k_*}
\newcommand{\kdiam}{k_\diamond}
\newcommand{\oscphase}{\psi}
\newcommand{\powosc}{p}
\newcommand{\freqosc}{\omega}
\newcommand{\amposc}{A_\freqosc}

\begin{document}

\title[Inflation after WMAP3]{Inflation after WMAP3: Confronting the
  Slow-Roll and Exact Power Spectra with CMB Data}

\author{J\'er\^ome Martin}
\email{jmartin@iap.fr}
\affiliation{Institut d'Astrophysique de Paris, UMR
7095-CNRS, Universit\'e Pierre et Marie Curie, 98bis boulevard Arago,
75014 Paris, France}

\author{Christophe Ringeval}
\email{c.ringeval@imperial.ac.uk}
\affiliation{Blackett Laboratory, Imperial College London, Prince
Consort Road, London SW7~2AZ, United Kingdom}

\date{today}

\begin{abstract}

The implications of the WMAP (Wilkinson Microwave Anisotropy Probe)
third year data for inflation are investigated using both the
slow-roll approximation and an exact numerical integration of the
inflationary power spectra including a phenomenological modelling of
the reheating era.  At slow-roll leading order, the constraints
$\epsilon _1<0.022$ and $-0.07<\epsilon _2<0.07$ are obtained at
$95\%$ CL (Confidence Level) implying a tensor-to-scalar ratio
$r_{10}<0.21$ and a Hubble parameter during inflation $H/\mpl <
1.3\times 10^{-5}$. At next-to-leading order, a tendency for $\epsilon
_3>0$ is observed. With regards to the exact numerical integration,
large field models, $V(\phi)\propto \phi^p$, with $p > 3.1$ are now
excluded at $95\%$ CL. Small field models, $V(\phi)\propto
1-(\phi/\mu)^p$, are still compatible with the data for all values of
$p$. However, if $\mu/\mpl<10$ is assumed, then the case $p=2$ is
slightly disfavoured. In addition, mild constraints on the reheating
temperature for an extreme equation of state $\wstate_\ureh\gta -1/3$
are found, namely $T_{\usssRH}>2\,\mbox{TeV}$ at $95\%$ CL. Hybrid
models are disfavoured by the data, the best fit model having $\Delta
\chi ^2\simeq +5$ with two extra parameters in comparison with large
field models. Running mass models remain compatible, but no prior
independent constraints can be obtained. Finally, superimposed
oscillations of trans-Planckian origin are studied. The vanilla
slow-roll model is still the most probable one. However, the overall
statistical weight in favour of superimposed oscillations has
increased in comparison with the WMAP first year data, the amplitude
of the oscillations satisfying $2\vert x\vert \sigma _0<0.76$ at
$95\%$ CL. The best fit model leads to an improvement of $\Delta
\chi^2\simeq -12$ for $3$ extra parameters. Moreover, compared to
other oscillatory patterns, the logarithmic shape is favoured.

\end{abstract}


%to separate title from main body
%\maketitle

%for JHEP only
%\abstract{The implications...}
%\begin{document}

\section{Introduction}
\label{sec:introduction}

The recent release of the three years WMAP data~\cite{Jarosik:2006ib,
  Spergel:2006hy, Hinshaw:2006ia,Page:2006hz} constitutes an important
step for the theory of inflation. One now has at our disposal high
accuracy data that can be used to probe the details of the
inflationary scenario and to learn about the physical conditions that
prevailed in the very early universe, at very high energies comparable
to the Grand Unified Theory (GUT) scale. There are many aspects that
would be interesting to study but, clearly, in a first approach, one
can restrict ourselves to (effective) single field models and see
whether it is already possible to constrain the shape of the inflaton
potential $V(\phi )$. In particular, it is important to known whether
it is necessary to go beyond a simple Harrison-Zeldovitch
(scale-invariant) power spectrum to correctly fit the data. If so, as
indicated by the results of reference~\cite{Spergel:2006hy} and from a
model building point of view, this means that the inflaton potential
is not completely flat and/or that inflation is not driven by a pure
cosmological constant. Equivalently, this also means that the
observations start feeling the non-trivial shape of the potential. In
this case, a non-vanishing second order derivative of the potential is
seen while the first order derivative of $V(\phi )$ presumably remains
unconstrained from below since, otherwise, primordial gravitational
waves would have been detected. Besides the previous issue, one would
also like to go further and to study which inflationary models remain
compatible with the data and which ones are ruled out. Addressing
these questions is the main goal of this article. Notice that we
restrict our considerations to WMAP3 data only although including
other data sets could allow us to obtain tighter constraints on the
inflationary models studied here. In a first step, we think it is more
reasonable to proceed this way in order not to mix the effects of
using different data sets with those originating from the new
numerical approach introduced in this article.

\par

To deal with this problem we proceed as follow. We first use the
slow-roll approximation and derive the constraints put by the third
year WMAP data on the first three slow-roll parameters (\ie second
order slow-roll approximation). Then, we compute exactly these
parameters for the large field, small field, hybrid and running mass
models. In particular, we show that some widely used approximate
expressions for $\epsilon_1$ and $\epsilon _2$ are no longer
sufficient to assess the likelihood of some models (small field
models) given the quality of the data. For this reason, in this
article, we always evaluate exactly the slow-roll parameters, using
simple numerical methods if necessary. The implications of the WMAP3
data using a slow-roll prior have also been investigated in
references~\cite{Alabidi:2006qa, Peiris:2006ug, Easther:2006tv,
Shafi:2006cs, deVega:2006hb}.

\par

In a second step, one frees ourselves from any approximation (except
the linear theory of cosmological perturbations) and calculate the
power spectra exactly by means of numerical computations. The models
that we study are the same than the ones already mentioned for the
slow-roll case, namely large field, small field, hybrid and running
mass models. These exact power spectra are computed mode by mode and
fed into a modified Cosmic Background Microwave (CMB) code, here a
modified \CAMB code~\cite{Lewis:1999bs}, which allows us to determine
the temperature and polarisation multipole moments. Finally, we
explore the corresponding parameters space by using Monte-Carlo
techniques as implemented in the \COSMOMC code~\cite{Lewis:2002ah}
together with the likelihood code developed by the WMAP
team~\cite{Spergel:2006hy}. This allows us to put constraints on the
free parameters characterising the models during the inflationary
phase but also, in principle, during the reheating phase although,
most of the time, the accuracy of the data is not sufficient to obtain
relevant limits on the reheating temperature.

\par

This article is organised as follows. In section~\ref{sec:inflation},
we briefly recall some basic facts about inflation, reheating and the
theory of inflationary cosmological perturbations of
quantum-mechanical origin. Section~\ref{sec:slowroll} is devoted to
the slow-roll approximation. In subsection~\ref{sec:basicsr}, the
derivation of the slow-roll scalar and tensor power spectra is
recalled. In subsection~\ref{sec:wmapsr}, we present the WMAP data
constraints on the slow-roll parameters at first order (\ie
$\epsilon_1$ and $\epsilon _2$) and also at second order (\ie the two
previous ones plus $\epsilon _3$). In subsections~\ref{sec:lfmodel} to
\ref{sec:rmmodel}, we calculate the slow-roll parameters for the large
field, small field, hybrid and running mass inflationary models.
These results are then compared to the constraints on $\epsilon_1$,
$\epsilon _2$ and $\epsilon _3$ obtained previously.
Section~\ref{sec:exact} is devoted to the the exact computations of
the inflationary power spectra for the four models mentioned above.
In subsection~\ref{sec:code}, we briefly describe the method and the
code used to perform the numerical calculations. In
subsections~\ref{sec:exactlf} to \ref{sec:exactrm}, the exact
numerical results are used to discuss the constraints put by the third
year WMAP data on the free parameters describing the models but also
(when possible) on the subsequent reheating phase. Finally, in
section~\ref{sec:tpl}, we investigate the presence of superimposed
oscillations in the CMB multipoles. In subsection~\ref{sec:basictpl},
we discuss a possible physical origin for those oscillations, namely
trans-Planckian effects during inflation. This allows us to use a
well-motivated and well-defined shape for the oscillatory power
spectra. Then, in subsection~\ref{sec:wmaptpl}, we compare these
spectra to the third year WMAP data and use them to put constraints on
the amplitude, the frequency and the phase of the superimposed
oscillations. In the last section~\ref{sec:end}, we recap our findings
and present our conclusions.

\section{Basics of inflation}
\label{sec:inflation}

\subsection{The background}
\label{sec:background}

\subsubsection{The accelerated phase}
\label{sec:acceleratedphase}

In this section, in order to describe the general setting and to fix
our notations, we recall some basic and well-known facts about
inflation at the background and perturbed levels. A phase of inflation
is supposed to make the universe homogeneous and isotropic on large
scales. It also drastically reduces its spatial curvature in agreement
with the observations which indicate that the spatial sections are
extremely flat. As a consequence, the metric tensor which describes
the geometry of the Universe can be taken of the
Friedman-Lema\^{\i}tre-Robertson-Walker (FLRW) form, namely
\begin{equation}
\label{metric0}
\dd s^2=-\dd t^2+a^2(t)\delta _{ij}\dd x^i \dd x^j
=a^2(\eta )\left(- \dd \eta ^2+\delta _{ij} \dd x^i \dd x^j
\right),
\end{equation}
where $\delta _{ij}$ is the Kr\"onecker symbol. The variable $t$ is
the cosmic time while $\eta$ is the conformal time and these two
quantities are related by $\dd t= a \dd \eta $. In the following, both
will be used, the choice of using one rather than the other being made
for convenience only and according to the problem at hand. Another
important time variable is the number of e-folds defined by the
following expression
\begin{equation}
N \equiv \ln \left(\frac{a}{a_\ini}\right),
\end{equation}
where $a_\ini$ is the value of the scale factor at some initial
time. As required to solve the flatness and homogeneity issue of the
FLRW model, the total number of e-folds during inflation must be
greater than $60$.

\par

The evolution of $a(\eta )$, the only free function in the above
metric element, is controlled by the Einstein equations. If matter is
assumed to be a perfect fluid, they read
\begin{equation}
\label{Einsteineqs}
\frac{3}{a^2} \left(\frac{a'}{a} \right)^2 = \frac{8\pi }{\mpl ^2}\rho \,
, \qquad -\frac{1}{a^2}\left[2\frac{a''}{a}
-\biggl(\frac{a'}{a}\biggr)^2\right]=\frac{8\pi }{\mpl^2}P \, ,
\end{equation}
where a prime denotes a derivative with respect to conformal time and
$\rho$, $P$ are respectively the energy density and the pressure of
the cosmological fluid driving the dynamics of the universe. In the
following, we will use the conformal Hubble parameter defined by
$\calH \equiv a'/a$. In addition, one also has the energy conservation
equation
\begin{equation}
\label{conservation}
\rho '+3{\cal H}(\rho + P)=0\, , 
\end{equation}
which can also be obtained from the Einstein equations by means of the
Bianchi identities.

\par

By definition, inflation is a phase of accelerated expansion for which
the scale factor
satisfies~\cite{Guth:1980zm, Linde:2005ht, Martin:2004um, Martin:2003bt}
\begin{equation}
\label{definf}
\frac{\dd^2a}{\dd t^2}>0 \, ,
\end{equation}
and this condition can also be re-written as 
\begin{equation}
\epsilon _1(\eta )\equiv
1-\left(\frac{a}{a'}\right)^2\left(\frac{a'}{a}\right)'<1\, .
\end{equation}
The quantity $\epsilon _1 $ is in fact nothing but the first
Hubble-flow (or slow-roll) parameter $-\dot{H}/H^2$ where $H =\calH/a$
denotes the physical Hubble parameter (see below). Inflation stops
when $\epsilon _1=1$. Another way to express the acceleration of the
scale factor is to combine equations~(\ref{Einsteineqs}). One gets
\begin{equation}
\label{accela}
\frac{\ddot{a}}{a}=-\frac{4\pi }{3\mpl ^2}\left(\rho +3 P \right) ,
\end{equation}
where a dot denotes a derivative with respect to the cosmic time
$t$. Clearly, inflation can be obtained if the fluid dominating the
universe has a negative pressure such that $P<-\rho /3$.

\par

A possible implementation of the inflationary scenario is to assume
that the matter content of the universe is described by a scalar field
$\phi(\eta )$ with a potential $V(\phi )$ since, when $V(\phi )$ is
sufficiently flat, the effective pressure of the scalar field can be
negative~\cite{Guth:1980zm, Linde:2005ht, Lyth:1998xn, Martin:2004um,
  Martin:2003bt}. In this article, we restrict ourselves to the case
of a single scalar field. In this case, the two Einstein equations
take the form
\begin{equation}
\label{EinsteinSF1}
\frac{3}{a^2}\calH^2  = \frac{8\pi }{\mpl^2}
\left[\frac{1}{2}\frac{(\phi ')^2}{a^2}+V(\phi )\right],
\end{equation}
and
\begin{equation}
\label{EinsteinSF2}
-\frac{1}{a^2}\left(2\calH' + \calH^2\right)=\frac{8\pi }{\mpl^2}
\left[ \frac{1}{2}\frac{(\phi ')^2}{a^2}-V(\phi )\right].
\end{equation}
Using the expressions of $\rho$ and $P$ for a scalar field, the
conservation equation~(\ref{conservation}) reduces to the Klein-Gordon
equation written in a FLRW background, namely
\begin{equation}
\label{KG}
\phi ''+2 \calH \phi '+a^2\frac{\dd V(\phi )}{\dd \phi }=0\, . 
\end{equation}
This equation can also be directly derived from (\ref{EinsteinSF1})
and (\ref{EinsteinSF2}). In fact, the physical interpretation is made
easier if the Klein-Gordon equation is written in terms of the number
of e-folds, namely~\cite{Ringeval:2005yn}
\begin{equation}
\label{Kleingordonefold}
\frac{1}{3-\epsilon _1}\frac{{\dd}^2 \phi}{{\dd}N^2} +\frac{{\rm
d}\phi }{{\dd}N}=-\frac{\mpl ^2}{8\pi }\frac{{\dd}\ln
V(\phi)}{{\dd}\phi }\, .
\end{equation}
We see that studying the evolution of the scalar field is in fact
equivalent to study the motion of particle in an effective potential
$\ln V(\phi )$ with a slightly variable mass and a constant friction
term. This is the reason why, in the following, when we consider
concrete models, we will also pay attention to the logarithm of the
potential in order to gain intuition about how inflation proceeds.

\par

If we are given a model, that is to say a concrete form for the
potential $V(\phi )$, then the equations of motion can be integrated
from the initial conditions $\phi _\ini $ and $\phi _\ini'$. In
general, the shape of the potentials that we consider does not allow
simple analytical solutions of the Einstein equations. In that case,
one has to rely on analytical approximations of numerical
calculations. However, the following potential~\cite{Lucchin:1984yf} 
\begin{equation}
\label{pot}
V(\phi)= M^4 \exp\left[-\frac{4\sqrt{\pi }}{\mpl}\sqrt{\frac{2+\beta
}{1+\beta }} \left(\phi - \phi_\ini \right)\right]\, ,
\end{equation}
is an example where the exact integration can be performed
analytically. One obtains power-law inflation for which the scale
factor and the scalar field are respectively given by
\begin{equation}
\label{defpl}
a(\eta )=\ell _0\vert \eta \vert ^{1+\beta }\,, \qquad \phi =\phi_\ini+
\frac{\mpl}{2}\sqrt{\frac{2+\beta }{\pi (1+\beta )}}(1+\beta )\ln
\vert \eta \vert\, .
\end{equation}
In this model the parameter $\epsilon _1$ is given by $\epsilon
_1=(2+\beta )/(1+\beta )$ and, therefore, inflation occurs if $\beta <
-2$ (we do not consider the case where $-2 < \beta < -1$ which cannot
be realised with a single scalar field). However, this model is not a
satisfactory model, since inflation never stops, at least if one does
not use another mechanism.

\subsubsection{The end of inflation}
\label{sec:endinflation}

When an exact integration is performed, it is very important to have a
description of the reheating phase. Without such a description, one
cannot relate the physical scales today to the physical scales during
inflation simply because one does not know the entire history of the
universe. As mentioned before, inflation ends for $\phi =\phi
_{\uend}$, or $\rho =\rho _{\uend}$ (or again for $H=H_{\uend}$), when
$\epsilon _1=1$. The difficulty is that the process of reheating can
be model dependent. For the large field inflation models, one may use
a phenomenological description based on the fact that the potential is
given by $V(\phi )\propto \phi ^p$. In this case, after the end of
inflation, the field starts oscillating around its
minimum~\cite{Turner:1983he,Kofman:1997yn}. In this regime, one can
show that the average energy density behaves as $\rho _{\uinf}\propto
a^{-6p/(p+2)}$ and, as a result, the scale factor is given by
$a(t)\propto t^{(p+2)/(3p)}$.  In other words, the equation of state
$P = \wstate \rho$ during reheating has a constant state parameter
given by
\begin{equation}
\label{omegareh}
\wstate_{\ureh}=\frac{p-2}{p+2}\, .
\end{equation}
For a massive scalar field, $p=2$, the energy density evolves as in a
matter-dominated epoch while for a quartic potential, $p=4$, the
energy density behaves as in radiation-dominated era. Nevertheless,
one always has
\begin{equation}
-\frac{1}{3} < \wstate _{\ureh} < 1 \,,
\end{equation}
for $p > 1$.

\par

The previous considerations and the description of the oscillatory
phase were just based on the Klein-Gordon equation~(\ref{KG}). It is
clear that, with this equation only, the decay of the inflaton field
into radiation cannot be described. In order to take into account this
effect, it is common to phenomenologically add a friction term $\Gamma
\dot{\phi }$ in the Klein-Gordon equation which now
reads~\cite{Turner:1983he}
\begin{equation}
\label{KGmodified}
\ddot{\phi }+3H\dot{\phi }+\Gamma \dot{\phi }+ \frac{{\rm
d}V(\phi )}{{\dd}\phi }=0 \, .
\end{equation}
In fact, this description cannot account for the complexity of the
reheating stage~\cite{Kofman:1997yn, Garcia-Bellido:1997wm,
  Felder:2000hj, Micha:2004bv, Senoguz:2004vu,
  Bassett:2005xm,Podolsky:2005bw, Desroche:2005yt, Allahverdi:2005mz,
  Allahverdi:2006wh, Allahverdi:2006iq}. But for our purpose, the
above treatment will be sufficient. Since we have modified the
Klein-Gordon equation one should also modify the conservation equation
for the radiation energy density in order to ensure the total energy
conservation:
\begin{equation}
\dot{\rho }_{\urad} =-4H\rho _{\urad} +\Gamma \rho _{\uinf}\,.
\end{equation}
In fact, the previous discussion can be generalized to other
inflationary models by simply assuming that the reheating era takes
place with a constant state parameter $\wstate_\ureh$ for which $a
\propto t^{2/(3+3\wstate_\ureh)}$ and $\rho \propto
a^{-3-3\wstate_\ureh}$. Note however that unlike for the large field
models, this parameter is not necessarily related to the inflationary
potential and equation~(\ref{omegareh}) is generally not satisfied.
However, the equations for $\rho _{\uinf}$ and $\rho _{\urad}$ can be
exactly integrated and the solutions read~\cite{Martin:2004um}
\begin{eqnarray}
\label{rhoinf}
\rho _{\uinf}(t) &=& \rho
_{\uend}\left(\frac{a}{a_{\uend}}\right)^{-3(1+\wstate_{\ureh})} \exp
\left[-\left(1+\wstate _{\ureh}\right) \Gamma
  \left(t-t_{\uend}\right)\right] ,\\ 
\label{rhorad}
\rho_{\urad}(t) &=& \Gamma t_{\uend}\rho _{\uend}
\left(\frac{a_{\uend}}{a}\right)^{4} \left[\left(1+\wstate
_{\ureh}\right) \Gamma t_{\uend}\right]^{(3\wstate
_{\ureh}-5)/(3+3\wstate _{\ureh})} \nonumber \\ & \times & \exp
\left[\left(1+\wstate _{\ureh}\right) \Gamma t_{\uend}\right] \left\{
\gamma \left[\frac{5-3\wstate _{\ureh}}{3+3\wstate
_{\ureh}},\left(1+\wstate _{\ureh}\right)\Gamma t\right]
\right. \nonumber \\ & - & \left. \gamma \left[\frac{5-3\wstate
_{\ureh}}{3+3\wstate _{\ureh}},\left(1+\wstate _{\ureh}\right) \Gamma
t_{\uend}\right] \right\} ,
\end{eqnarray}
where $t=t_{\uend}$ is the time at which the oscillations start (\ie
the time at which inflation ends). The function $\gamma(\alpha ,x)$
is the incomplete gamma function~\cite{Abramovitz:1970aa,
  Gradshteyn:1965aa} defined by $\gamma (\alpha ,x)\equiv \int
_0^x{\ue}^{-t} t^{\alpha -1}{\dd}t$. It was shown in
Ref.~\cite{Martin:2004um} that the expression of $\rho _{\urad}$ can
be simplified if one uses the expression of the incomplete gamma
function for small values of its argument. For times $t>t_{\uend }$ with
$t_\uend \ll t_\usssRH$ where $t_\usssRH \equiv \Gamma ^{-1}$, one
obtains the solution~\cite{Turner:1983he}
\begin{equation}
\label{rhoappro}
\rho _{\urad}(t) \simeq \Gamma t \rho _{\uend}
\left(\frac{t_{\uend}}{t}\right)^2 \frac{3+3\wstate _{\ureh}}
     {5-3\wstate _{\ureh}}
     \left[1-\left(\frac{t}{t_{\uend}}\right)^{-(5- 3\wstate
         _{\ureh})/(3+3\wstate _{\ureh})} \right] .
\end{equation}
Let us notice that the quantity $(5 - 3\wstate _{\ureh})/(3+3\wstate
_{\ureh})$ is always positive. Therefore, for $t \gg t_{\uend}$, the
second term in the above equation becomes negligible.
\begin{figure}
\begin{center}
\includegraphics[width=14cm]{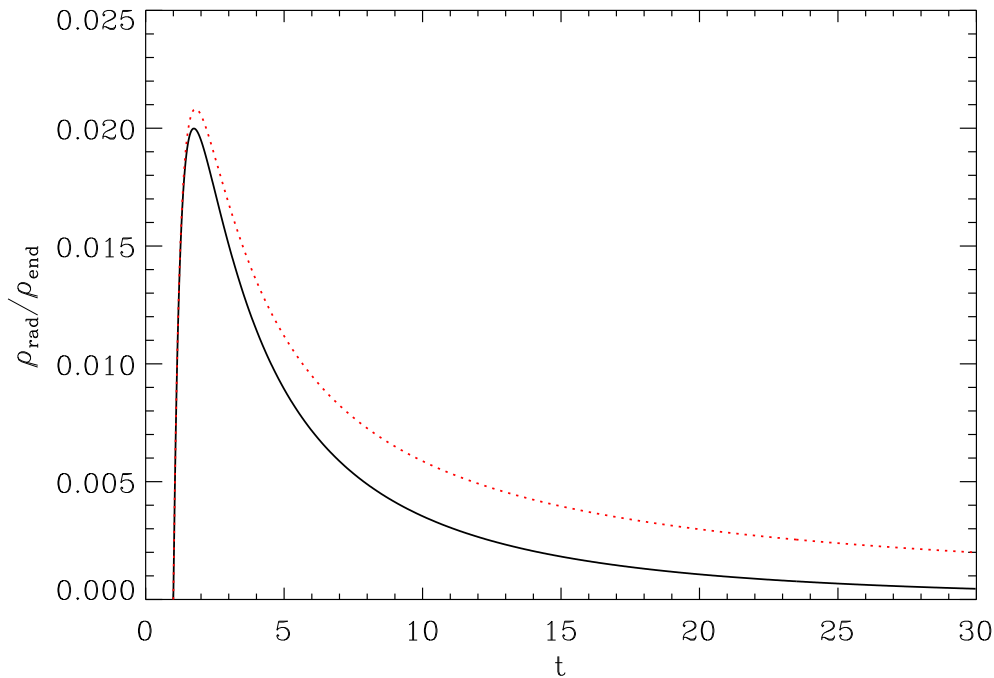} 
\caption{The quantity $\rho_{\urad}/\rho_{\uend}$ for
  $\wstate_{\ureh}=2$ or $p=2$, $t_{\uend}=1$ and $\Gamma =0.1$. The
  solid black line is the exact expression obtained from
  equation~(\ref{rhorad}) while the dotted red line is the
  approximation of (\ref{rhoappro}). At $t/t_{\uend}\gg 1$, the
  difference between the two curves is approximately a factor two.
  Here, the time $t$ and $\Gamma $ are measured in units of $\mpl$.}
\label{reheat}
\end{center}
\end{figure}
The exact~(\ref{rhorad}) and approximate~(\ref{rhoappro}) expressions
of $\rho _{\urad}(t)$ are presented in figure~\ref{reheat}. At
$t=t_\usssRH$, one can consider that the phase of reheating has been
completed. After thermalisation, the energy density of radiation takes
the form $\rho _{\urad}=g_*\pi ^2 T^4/30$ where $g_*$ is the number of
relativistic degrees of freedom. Expressed at $t=t_\usssRH$, this
quantity must be equal to $\rho _{\urad}$ in equation~(\ref{rhoappro})
and one arrives at
\begin{equation}
g_*^{1/4} T_\usssRH \simeq \left[\frac{30}{2 \pi^3
    (1+\wstate_\ureh)(5-3 \wstate_\ureh)}\right]^{1/4} \sqrt{\Gamma
\mpl} \,,
\end{equation}
where we have used $\rho _{\uend}=3H_{\uend }^2 \mpl^2/(8\pi)$ and
$t_{\uend}=2H_\uend^{-1}/[3 (1+\wstate_\ureh)]$. Moreover, The number
of e-foldings during reheating can be evaluated as
\begin{equation}
\label{efoldosci}
N_{\ureh}\simeq \frac{2}{3+3\wstate _{\ureh}}\ln
\left(\frac{3+3\wstate_\ureh}{2}\frac{H_{\uend}}{\Gamma }\right) .
\end{equation}
This result is important because it allows us to relate the physical
scales now to the physical scales during inflation. To do so, one must
take into account the fact that a large number of e-folds can exist
during reheating. This also means that the link between what happens
during inflation to what happens now depends on the details of the
reheating period. In our simplified model, this period is described by
a single parameter $\Gamma $ or, equivalently, $N_{\ureh}$. One can
also combine the two above equations in order to express the reheating
temperature only in terms of $N_{\ureh}$ and $\wstate _{\ureh}$. This
leads to
\begin{equation}
\label{reheatingT}
g_*^{1/4}T_\usssRH \simeq \frac{30^{1/4}}{\sqrt{\pi }}\rho
_{\uend}^{1/4} \left(\frac{3+3\wstate _{\ureh}}{5-3\wstate
  _{\ureh}}\right)^{1/4} {\ue}^{-3(1+\wstate _{\ureh})N_{\ureh}/4} .
\end{equation}
This expression also depends on $\rho _{\uend}$ but this quantity is
known once the background inflationary evolution is solved.

\subsection{The cosmological perturbations}
\label{sec:cosmopert}

Having integrated the background, we now turn to the cosmological
perturbations themselves. The power spectrum can be obtained by a mode
by mode integration. Before addressing this issue, let us recall some
basic definitions. The perturbed line element around a spatially flat
FLRW universe can be expressed
as~\cite{Bardeen:1980kt,Mukhanov:1990me}:
\begin{eqnarray}
\label{metricgi}
\dd s^2 & = & a^2(\eta )\left\{-\left(1-2\phi \right) \dd \eta
^2+2 \partial_i B  \dd x^i \dd \eta + \left[\left(1-2\psi
  \right) \delta _{ij} \right. \right. \nonumber \\ & + & \left.\left.
  2\partial_i \partial_j E
   +  h_{ij} \right] \dd
x^i \dd x^j \right \} .
\end{eqnarray}
The four functions $\phi $, $B$, $\psi $ and $E$ represent the scalar
sector whereas the transverse and traceless tensor $h_{ij}$, \ie
satisfying $h_i{}^i=\partial ^jh_{ij}=0$, represents the gravitational
waves. There are no vector perturbations because a single scalar field
cannot seed curly perturbations. At the linear level, the two types of
perturbations decouple and therefore can be treated separately. Since
the scalar sector suffers from a gauge dependence, it is more
convenient to work with the gauge-invariant Bardeen potentials $\Phi $
and the gauge-invariant perturbed scalar field $\delta\phi ^{\uPgiP}$
defined by~\cite{Bardeen:1980kt,Mukhanov:1990me}
\begin{eqnarray}
\label{defbardeen}
\Phi &=& \phi +\frac{1}{a}\left[a\left(B-E'\right)\right]'\, , \qquad
\delta \phi ^{\uPgiP} =\delta \phi +\phi '\left(B-E'\right) .
\end{eqnarray}
In the first of the two above equations, $\phi $ denotes the perturbed
time-time component of the metric element~(\ref{metricgi}) and not the
perturbed scalar field. In fact, it is clear that one has only one
degree of freedom because $\Phi $ and $\delta \phi ^{\uPgiP}$ are
related by the perturbed Einstein equations. One can therefore reduce
the study of the scalar sector to the study of a single variable,
namely the so-called Mukhanov-Sasaki variable defined
by~\cite{Mukhanov:1981xt}
\begin{equation}
v\left(\eta ,\mathbf{x}\right)\equiv a\left[\delta \phi^{\uPgiP}+ \phi
'\frac{\Phi }{\calH} \right].
\end{equation}
In order to set quantum initial conditions, it turns out to be more
convenient to work with the rescaled variable $\muS \equiv
-\sqrt{2\kappa }v$. On the other hand, the quantity of interest in the
primordial power spectrum is the comoving curvature perturbation
$\zeta \left(\eta ,\mathbf{x}\right)$ defined by
\begin{equation}
\zeta =-\frac{\muS}{2a \sqrt{\epsilon _1}} \,.
\end{equation}
Since no tensorial quantity can be used to generate an infinitesimal
coordinate transformation, the tensor sector is gauge invariant and
fully characterised by the quantity $\muT$ related to $h_{ij}$ through
the relation
\begin{equation}
h_{ij}=\frac{\muT}{a}Q_{ij} \,.
\end{equation}
The $Q_{ij}$ are the transverse and traceless eigentensors of the
Laplace operator on space-like hypersurfaces.

\par

The central result of the theory of inflationary cosmological
perturbations is that the quantities $\muST$ (or rather their
corresponding Fourier transform) both obey the equation of motion of a
parametric oscillator~\cite{Mukhanov:1981xt, Grishchuk:1974ny,
Grishchuk:1975uf, Martin:1997zd}
\begin{equation}
\label{paramoscillator}
\muST''+\omegaST^2(k,\eta ) \muST=0 \, ,
\end{equation}
where the time variation of the frequencies only depend on the behaviour 
of the background and is given by
\begin{equation}
\label{frequencies}
\omegaS^2\left(k,\eta \right)=k^2 -\frac{(a\sqrt{\epsilon
_1})''}{a\sqrt{\epsilon _1}}\, , \qquad \omegaT^2\left(k,\eta
\right)=k^2 -\frac{a''}{a} \, .
\end{equation}
In these expressions $k$ denotes the comoving wavenumber. In order to
solve the above equations, one postulates that the quantum fields are
initially placed in the vacuum state when the mode $k$ is well within
the Hubble radius, which amounts to assume that
\begin{equation}
\label{ini}
\lim _{k/(aH)\rightarrow +\infty }\muST(\eta )= \mp \frac{4\sqrt{\pi
}}{\mpl}\frac{{\rm e}^{-ik(\eta -\eta _{\rm i})}}{\sqrt{2k}},
\end{equation}
where $\eta _{\rm i}$ is the initial conformal time at the beginning
of inflation. Then, a mode by mode integration of
equation~(\ref{paramoscillator}) allows the determination of $\muST$
for a given $k$ at any time $\eta$.

\par

It is then straightforward to determine the resulting power
spectra. From a calculation of the two-point correlation functions,
one obtains
\begin{equation}
\label{spec}
k^3P_{\zeta }(k)=\frac{k^3}{8\pi ^2}\left \vert
\frac{\muS}{a\sqrt{\epsilon _1}}\right \vert ^2 , \qquad
k^3P_h(k) = \frac{2k^3}{\pi ^2}\left \vert \frac{\muT}{a}
\right \vert ^2 .
\end{equation}
These are the quantities that seed the subsequent CMB
anisotropies. Usually, the properties of these primordial power
spectra are characterised by the spectral indices and their
``running''. They are defined by the coefficients of Taylor expansions
of the power spectra with respect to $\ln k$, evaluated at an
arbitrary pivot scale $\kstar$, namely
\begin{equation}
\label{n}
\nS -1 \equiv \left. \frac{\dd \ln (k^3 P_\zeta)}{\dd \ln k}
\right \vert_{\kstar},
\qquad
\nT \equiv \left.\frac{\dd \ln (k^3P_h)}{\dd \ln k}
\right \vert_{\kstar}.
\end{equation}
For the runnings, one similarly has the two following expressions
\begin{equation}
\label{alpha}
\alphaS \equiv \left .\frac{\dd^2 \ln (k^3P_\zeta)}{\dd (\ln k)^2}
\right\vert_{\kstar}, \qquad \alphaT \equiv \left. \frac{\dd^2 \ln
  (k^3P_h)}{\dd ( \ln k)^2} \right \vert_{\kstar},
\end{equation}
and in principle, we could also define the running of the running and
so on.

\section{Testing the slow-roll models} 
\label{sec:slowroll} 

\subsection{Basics of slow-roll inflation}
\label{sec:basicsr}

It was shown in the previous section that the knowledge of the
background evolution is sufficient to calculate the time-dependent
frequencies $\omegaS\left(k,\eta \right)$ and $\omegaT(k, \eta )$
which are the only quantities needed to integrate (possibly
numerically) the equations of motion of the cosmological
perturbations. The slow-roll approximation allows to perturbatively
estimate $\omegaST\left(k,\eta \right)$ and, hence, to derive
approximated expression for the scalar and tensor power
spectra~\cite{Stewart:1993bc,Martin:1999wa,Martin:2000ak}. Although
there are several definitions of the slow-roll parameters, in this
article, we choose to work with the Hubble-flow parameters
$\{\epsilon_n\}$ defined by~\cite{Schwarz:2001vv,Schwarz:2004tz}
\begin{equation}
\label{flow}
\epsilon_{n+1} \equiv \frac{\dd \ln |\epsilon_n|}
{\dd N}\,, \qquad n\geq 0\, ,
\end{equation} 
where, as already mentioned, $N$ is the number of e-folds since some
initial time $\eta _\ini$. The above hierarchy starts from
$\epsilon_0=H_\ini/H$.  With this definition, all the $\epsilon_n$ are
typically of the same order of magnitude. One has \emph{slow-roll} inflation
as long as $|\epsilon_n| \ll 1$, for all $n>0$ while, as already
mentioned, inflation takes place if $\epsilon _1<1$.

\par

The physical interpretation of these parameters has been discussed in
Ref.~\cite{Schwarz:2004tz}. Let us briefly recall the results for
$\epsilon_1$ and $\epsilon _2$. Although the $\epsilon _n$ parameters
make no reference to the matter content of the Universe (they are only
defined in terms of the expansion rate), it is nevertheless
interesting, when we assume that a scalar field $\phi$ is responsible
for inflation, to express them in terms of $\phi $. One
obtains~\cite{Schwarz:2004tz}
\begin{equation}
\label{intersr}
\epsilon _1=3\frac{\dot{\phi }^2/2}{\dot{\phi }^2/2+V(\phi)}\, , \qquad
\frac{\dd}{\dd t}\left(\frac{\dot{\phi
}^2}{2}\right) = H\dot{\phi }^2\left(\frac{\epsilon _2}{2} -\epsilon
_1\right) .
\end{equation}
{}From the above expressions, one sees that $\epsilon _1/3$ measures the
ratio of the kinetic energy to the total energy (\ie kinetic plus
potential energy). Using the link between $\epsilon _1$ and $\epsilon
_2$, one has
\begin{equation}
\dot{\epsilon }_1=H\epsilon _1\epsilon _2 \,.
\end{equation}
Therefore, given the fact that $\epsilon _1$ is positive definite,
$\epsilon _2>0$ (respectively $\epsilon _2<0$) represents a model
where the kinetic energy itself increases (respectively decreases)
with respect to the total energy. From equation (\ref{intersr}),
$\epsilon _2=2\epsilon _1$ marks the frontier between models where the
kinetic energy increases ($\epsilon_2>2\epsilon _1$) and the models
where it decreases ($\epsilon_2<2\epsilon _1$).

\par

It was demonstrated in~\cite{Schwarz:2004tz,Leach:2002ar} that the
Hubble-flow parameters can be expressed in terms of the inflaton
potential and its derivatives. The exact expressions read
\begin{eqnarray}
V(\phi) &=& \frac{3H^2\mpl ^2}{8\pi }\left(1-\frac{\epsilon _1}{3}\right), \\
\frac{\dd V}{\dd\phi }& =& -3H^2\mpl
\sqrt{\frac{\epsilon_1}{4\pi }} \left(1-\frac{\epsilon
  _1}{3}+\frac{\epsilon _2}{6}\right),\\ \frac{\dd^2 V}{\dd\phi ^2} &
= & 3H^2\left(2\epsilon _1 -\frac{\epsilon
  _2}{2}-\frac23\epsilon _1^2+\frac56 \epsilon _1\epsilon _2
-\frac{1}{12}\epsilon _2^2-\frac16\epsilon _2\epsilon_3\right).
\end{eqnarray}
In order to have a meaningful model, the potential should be positive
and, hence, $\epsilon _1<3$. In fact, it is more interesting to
express the slow-roll parameters in terms of the potential and its
derivatives because, in practice, a model is defined by its
potential. Therefore, one has to invert the above expressions. At
leading order in these parameters, one obtains~\cite{Leach:2002ar}
\begin{eqnarray}
\label{epsV1}
H^2 & \simeq & \frac{8\pi}{3 \mpl ^2} V, \qquad \epsilon_1
\simeq\frac{\mpl ^2}{16 \pi} \left({V'\over V}\right)^2 ,\qquad
\epsilon_2 \simeq \frac{\mpl ^2}{4 \pi} \left[\left({V'\over
V}\right)^2 - {V''\over V}\right], \\ 
\label{epsV2}
\epsilon_2 \epsilon_3 &\simeq&
\frac{\mpl ^4}{32 \pi^2} \left[ {V''' V'\over V^2} - 3{V''\over
V}\left({V'\over V}\right)^2  + 2 \left({V'\over
V}\right)^4\right],
\end{eqnarray} 
where, in the present context, a prime denotes a derivative with
respect to the scalar field $\phi $. Since we have pushed the
calculation up to order two in the slow-roll parameters, one may be
worried about the fact that we only perform the inversion at leading
order. Following~\cite{Liddle:1994dx}, we can define two new slow-roll
parameters
\begin{equation}
\epsilon _\usssV \equiv \frac{\mpl ^2}{16 \pi}
\left( \frac{V'}{V}\right )^2, \qquad
\eta _\usssV \equiv \frac{\mpl ^2}{8\pi}\frac{V''}{V}\,.
\end{equation}
At the next-to-leading order, one obtains~\cite{Liddle:1994dx}
\begin{equation}
\label{inverting2}
\epsilon _1=\epsilon _\usssV -\frac{4}{3} \epsilon_\usssV^2 
+\frac{2}{3} \epsilon _\usssV \eta _\usssV + \dots \, .
\end{equation}
At leading order, $\epsilon _1=\epsilon _\usssV$ and one recovers the
expression~(\ref{epsV1}) for the first Hubble-flow parameter. As
expected, the corrections to this expression are quadratic in the
slow-roll parameters. Therefore, as long as the slow-roll parameters
are small, the predictions for a concrete model, \ie the corresponding
location in the $(\epsilon _1,\epsilon _2)$ plane, are almost
unchanged and, in any case, undetectable with the current data. We
will come back to this issue in the following and check this claim
explicitly for large fields models. Of course, if one wishes to
determine, say, the scalar spectral index at quadratic order, that is
to say to include the quadratic terms in the expression of $\nS-1$
(see below), then it would be mandatory to perform the inversion at
quadratic order as well.

\par 

We now turn to the perturbative expression of the power spectra. The
strategy consists in expanding the power spectra about the pivot scale
$\kstar$. The choice of this particular wavenumber is \emph{a priori}
arbitrary but must be chosen in a way minimising the uncertainties
coming from the perturbative expansion. Therefore, a good choice is
around the middle of the range of scales probed by the CMB (the size
of which is about three decades). Consequently, the usual choice is
$\kstar=0.05 \, \mbox{Mpc}^{-1}$. The expression of $P(k)$ reads
\begin{equation} 
\label{spectrumsr}
\frac{k^3P(k)}{k^3P_0(\kstar)} = a_0 + a_1 \ln \left(k\over \kstar\right) 
 + \frac{a_2}{2} \ln^2\left(k\over \kstar\right)
 + \dots \, ,
\end{equation}
where
\begin{equation}
k^3P_{\zeta \zero}(\kstar) =\frac{H^2}{\pi \epsilon_1 \mpl^2}\,,
\qquad k^3P_{h \zero}(\kstar)=16 \frac{H^2}{\pi \mpl^2}\,.
\end{equation}
The coefficients $a_i$ are then determined in terms of the slow-roll
parameters. These calculations are non trivial especially when one
goes beyond the leading order because, in that case, the usual
approximations in terms of Bessel functions are no longer valid. It is
therefore necessary to use more sophisticated methods either based on
the WKB approximation~\cite{Martin:2002vn, Casadio:2004ru,
  Casadio:2005xv, Casadio:2005em} and/or on the Green function
methods~\cite{Gong:2001he, Choe:2004zg}. The latter method has been
used in several works~\cite{Gong:2001he,Leach:2002ar} to derive the
coefficients $a_i$ and we only quote the results. For scalar
perturbations, one gets
\begin{eqnarray}
\label{eqn:as0}
a_{0}^{\usssPSP} &=& 1 - 2\left(C + 1\right)\epsilon_1 - C \epsilon_2
+ \left(2C^2 + 2C + \frac{\pi^2}{2} - 5\right) \epsilon_1^2 \nonumber
\\ & + & \left(C^2 - C + \frac{7\pi^2}{12} - 7\right)
\epsilon_1\epsilon_2 + \left(\frac12 C^2 + \frac{\pi^2}{8} -
1\right)\epsilon_2^2 \nonumber \\ & + & \left(-\frac12 C^2  +
\frac{\pi^2}{24}\right)  \epsilon_2\epsilon_3 \, , \\
a_{1}^{\usssPSP} & = & - 2\epsilon_1 - \epsilon_2 + 2(2C+1)\epsilon_1^2
+ (2C - 1)\epsilon_1\epsilon_2 + C\epsilon_2^2 - C\epsilon_2\epsilon_3
\, ,\\ a_{2}^{\usssPSP} &=& 4\epsilon_1^2 + 2\epsilon_1\epsilon_2 +
\epsilon_2^2 - \epsilon_2\epsilon_3 \, ,
\label{eqn:as2}
\end{eqnarray}
where $C \equiv \gamma_{\usssE} + \ln 2 - 2 \approx -0.7296$, $\gamma
_\usssE$ being the Euler constant. For the gravitational waves, the
coefficients $a_i$ read
\begin{eqnarray}
a_{0}^{\usssPTP} &=&
 1 - 2\left(C + 1\right)\epsilon_1 
 + \left(2C^2 + 2C + \frac{\pi^2}{2} - 5\right) 
 \epsilon_1^2 \nonumber \\ & + & \left(-C^2 - 2C + \frac{\pi^2}{12} - 2\right) 
 \epsilon_1\epsilon_2 \, ,\\
a_{1}^{\usssPTP} &=& 
 - 2\epsilon_1 + 2(2C + 1)\epsilon_1^2 
 - 2(C + 1)\epsilon_1\epsilon_2 \, ,
\label{eqn:at1} \\
a_{2}^{\usssPTP} &=& 4\epsilon_1^2 - 2\epsilon_1\epsilon_2 \, .
\label{eqn:at2}
\end{eqnarray}
We see that each coefficients $a_i$ starts at order $\epsilon
_n^i$. In fact, this can be proven analytically since the coefficients
$a_{i+1}$ can be obtained from the coefficient $a_i$ by
differentiation with respect to the number of e-folds. Equivalently,
if one is able to determine the coefficient $a_0$ at order $\epsilon
_n^i$, then one can derive a non-vanishing expression for the
coefficients $a_{j}$ up to $j=i$. The above considerations also lead
to the domain of validity of the
expression~(\ref{spectrumsr}). Clearly, one has a meaningful
expansion as long as
\begin{equation}
\epsilon _n \ln \left(\frac{k}{\kstar}\right) \ll 1 \, ,
\end{equation}
where we have again used the fact that $a_i$ is of order $\epsilon
_n^i$. In principle, this estimate is valid for any slow-roll
parameters. Given the fact that the CMB observations probe about three
decades in wavenumbers and assuming that a natural location for the
pivot scale $\kstar$ has indeed been chosen in the middle of this
range, one arrives at
\begin{equation}
\vert \epsilon _n \vert \ll 0.29\, .
\end{equation}
In practice, in the rest of this article, we will consider that the
equation~(\ref{spectrumsr}) remains valid as long as $\vert
\epsilon_n\vert <0.1$.

\par

The logarithm of the power spectra can also be Taylor expanded in
terms of the logarithm of the comoving wave number:
\begin{eqnarray} 
\label{plex}
\ln \frac{k^3P(k)}{k^3P_0(\kstar)} = b_0 + b_1 \ln \left(k\over \kstar\right)
+ \frac{b_2}{2} \ln^2\left(k\over \kstar\right) + \dots \, , 
\end{eqnarray}
where the coefficients $b_i$ are given by
\begin{eqnarray}
\label{bs0}
b_{0}^{\usssPSP} &=& 
 - 2\left(C + 1\right)\epsilon_1 - C \epsilon_2 
 + \left(- 2C + \frac{\pi^2}{2} - 7\right) 
 \epsilon_1^2 
 \nonumber \\ & + & \left(- C^2 - 3C + \frac{7\pi^2}{12} - 7\right) 
 \epsilon_1\epsilon_2  + \left(\frac{\pi^2}{8} - 1\right) 
 \epsilon_2^2  \nonumber \\ & + & \left(-\frac12C^2 + \frac{\pi^2}{24}\right) 
 \epsilon_2\epsilon_3 \, ,\\
b_{1}^{\usssPSP} &=& - 2 \epsilon_1 - \epsilon_2 - 2 \epsilon_1^2 
 - (2C + 3) \epsilon_1 \epsilon_2 
 - C \epsilon_2 \epsilon_3 \, ,\\
b_{2}^{\usssPSP} &=& - 2 \epsilon_1 \epsilon_2 - \epsilon_2 
\epsilon_3 \, ,
\label{bs2}
\end{eqnarray}
for the scalar power spectrum, while the $b_i$'s associated with the
tensor power spectrum read
\begin{eqnarray} 
 b_{0}^{\usssPTP} &=& 
 - 2\left(C + 1\right)\epsilon_1 
 + \left(- 2C + \frac{\pi^2}{2} - 7\right) 
 \epsilon_1^2 
 \nonumber \\ & + & \left(-C^2 - 2C + \frac{\pi^2}{12} - 2\right) 
 \epsilon_1\epsilon_2 \, , \\
b_{1}^{\usssPTP} &=& - 2\epsilon_1 - 2\epsilon_1^2 
 - 2(C + 1)\epsilon_1\epsilon_2 \, , \\
b_{2}^{\usssPTP} &=& - 2\epsilon_1\epsilon_2 \, .
\label{bt2}
\end{eqnarray}
The interest of the coefficients $b_i$ is that they are directly
related to observable quantities, namely to the spectral indices and
the runnings. Explicitly, one has 
\begin{equation}
\label{nalpha}
\nS -1 = b_{1}^{\usssPSP}\, ,\qquad \nT = b_{1}^{\usssPTP} \,
, \qquad \alphaS = b_{2}^{\usssPSP}\, , \qquad \alphaT=b_{2}^{\usssPTP} \, .
\end{equation}
In particular, one sees that, at first order in a consistent slow-roll
expansion, the scalar and tensor runnings vanish.

\par

Finally, the ratio of amplitudes of scalars and tensors at the pivot
reads
\begin{eqnarray}
r &\equiv &\frac{k^3P_h}{k^3P_{\zeta }}=16\epsilon_1 \left[1 + C
  \epsilon_2 + \left(C- \frac{\pi^2}{2} +5\right)\epsilon_1\epsilon_2
  + \left(\frac12C^2 - \frac{\pi^2}{8} +1\right) \epsilon_2^2
  \right. \nonumber \\ & + & \left. \left(\frac12 C^2 -
  \frac{\pi^2}{24}\right) \epsilon_2 \epsilon_3 \right]\, .
\label{eqn:Rsr}
\end{eqnarray}
At first order, this is nothing but the consistency condition of
inflation, \ie $r = - 8 \nT$. Therefore, any upper bound on the first
slow-roll parameter $\epsilon _1$ gives an upper bound on the
contribution of gravitational waves. In fact, the ratio $r$ itself is
not observable because one must take into account the evolution of the
transfer function when the modes re-enter the Hubble radius. If we
formulate the ratio tensor to scalar in terms of CMB multipole
moments, then one obtains
\begin{equation}
\label{r10}
r_{10}\equiv \frac{C_{10}^{\usssPTP}}{C_{10}^{\usssPSP}}=-f_{10}(h,
\Omega _{_{\Lambda}},\cdots)\nT\, .
\end{equation}
For the concordance model, one has $f_{10}\simeq 5$ which implies that
$r_{10}\simeq 10\epsilon _1$. We see that an upper bound on $\epsilon
_1$ puts an upper bound on $r_{10}$. This also implies an upper bound
on the energy scale of inflation. Indeed, on very large scales ($\ell
\ll 20$), the multipole moments can be written as
\begin{equation}
C_{\ell }^{\usssPSP}\simeq \frac{2H^2}{25 \mpl ^2\epsilon
_1}\frac{1}{\ell (\ell +1)}\, ,
\end{equation}
and the measurement of the quadrupole $C_2$ by WMAP
means that
\begin{equation}
\label{HQ}
\left(\frac{H}{\mpl }\right)^2=60\pi \epsilon
_1\left(\frac{Q_\mathrm{rms-PS}}{T}\right)^2 ,
\end{equation}
where~\cite{Hinshaw:2006ia}
\begin{equation}
\label{eq:Qrms}
\frac{Q_\mathrm{rms-PS}}{T} \equiv \sqrt{\frac{5C_2}{4\pi}} \simeq
6\times 10^{-6}\,.
\end{equation}
As announced, any upper bound on the first slow-roll parameter implies
an upper bound on $H/\mpl$. In the next section, the WMAP third year
data are used to constrain the slow-roll parameters and an upper limit
on $r_{10}$ and $H/\mpl $ is derived.

\subsection{WMAP data constraints on the slow-roll parameters}
\label{sec:wmapsr}

In this section, before turning to the specific models of inflation we
are interested in, we derive the bounds that the Hubble-flow
parameters have to satisfy given the third year WMAP data
(WMAP3). Following~\cite{Leach:2003us}, we used a modified version of
the \CAMB code~\cite{Lewis:1999bs} to compute the CMB temperature and
polarisation anisotropies seeded by the slow-roll scalar and tensor
primordial power spectra of equation (\ref{spectrumsr}). The parameter
space has been sampled by using Markov Chain Monte Carlo (MCMC)
methods implemented in the \COSMOMC code~\cite{Lewis:2002ah} and using
the likelihood estimator provided by the WMAP
team~\cite{Spergel:2006hy, Page:2006hz, Hinshaw:2006ia,
Jarosik:2006ib}. The likelihood code settings have been kept to their
default values which include the pixel based analysis at large scales,
a Gaussian likelihood for the beam, with diagonal covariance matrix,
and point source corrections~\cite{Hinshaw:2006ia, Lewis:2006ma,
Peiris:2006ug}. We have also checked the sensitivity of our results to
modifications of these options and no significant deviations has been
observed. The assumed cosmological model is a flat $\Lambda$CDM
universe involving a minimal set of four cosmological base parameters:
the number density of baryons $\OmegaB$, of cold dark matter
$\OmegaCDM$, the reionization optical depth $\tau$ and $\theta$ which
measures the ratio of the sound horizon to the angular diameter
distance (see Ref.~\cite{Lewis:2002ah}). Moreover, we have only
considered the WMAP3 data together with the Hubble Space Telescope
(HST) constraint ($\Hzero = 72 \pm 8\,
\mathrm{km/s/Mpc}$~\cite{Freedman:2000cf}) and a top hat prior on the
age of the universe between $10\,\mathrm{Gyrs}$ and $20
\,\mathrm{Gyrs}$. The convergence of the chains has been assessed by
using the Gelman and Rubin R--statistics implemented in \COSMOMC which
is a measure of the variance of the means divided by the mean value of
the variances between different chains~\cite{Gelman:1992,
Lewis:2002ah}. Unless otherwise specified, the iterations have been
stopped once $R-1 < 3\%$, which corresponds to a few hundred thousand
samples depending on the model explored.

\par

As described in the previous section, the set of primordial parameters
in the slow-roll approximation involves the value of the scalar power
spectrum $\Pstar = k^3 P_{\zeta \zero}(\kstar)$ and the Hubble-flow
parameters evaluated at the pivot scale $\kstar$.

\subsubsection{First order slow-roll expansion}
\label{sec:wmap1order}

\begin{figure}
\begin{center}
\includegraphics[width=10cm]{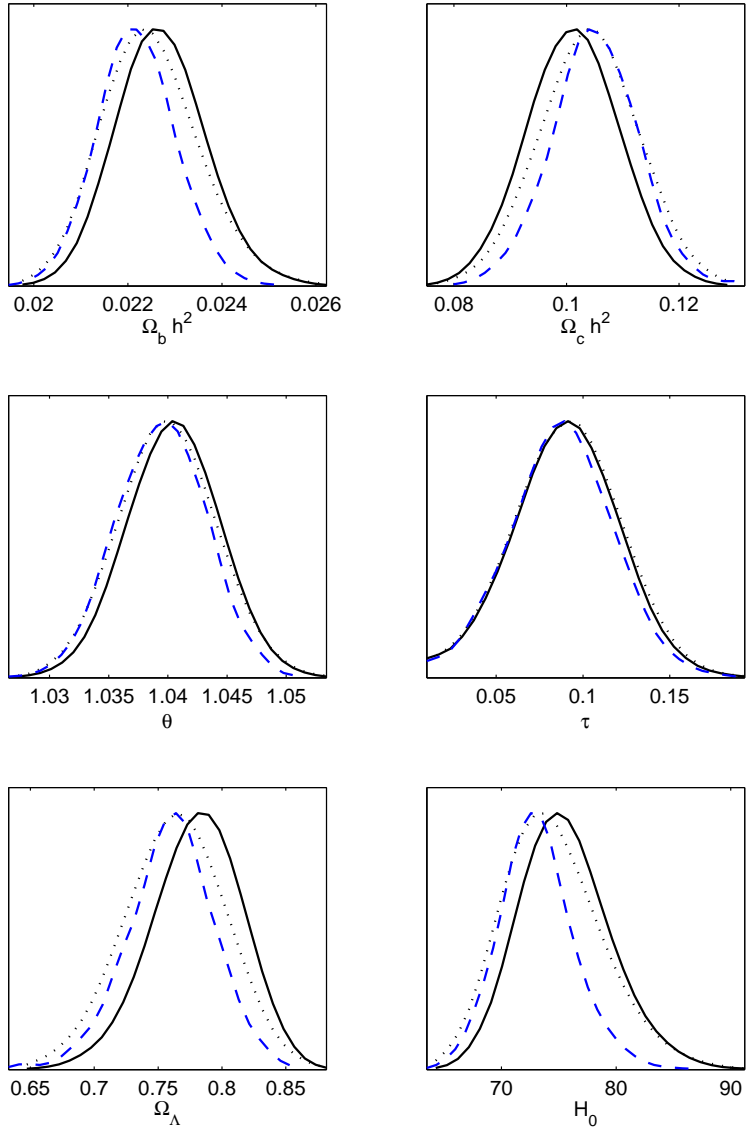}
\caption{Marginalised posterior probability distributions for the base
  $\Lambda$CDM cosmological parameters together with the cosmological
  constant and the Hubble parameter, obtained at first order in
  slow-roll expansion. The solid black lines correspond to an uniform
  prior choice on $\epsilon_1$ while the dashed blue ones to an
  uniform prior on $\log(\epsilon_1)$. The dotted black lines is the
  mean likelihood for the former prior.}
\label{sr1st_cosmo_1D}
\end{center}
\end{figure}

At first order in the slow-roll expansion, the relevant parameters are
$\epsilon_1$ and $\epsilon_2$ while the running of the spectral index
vanishes. This accounts for three primordial parameters including the
amplitude of the scalar power spectrum leading to seven model
parameters in total. The best fit model has $\chi^2=11252.2$ (for
comparison, the standard power law parametrisation almost leads to the
same best fit value $\chi^2=11252.4$).

\par

In addition to the usual uniform top hat priors on the cosmological
parameters~\cite{Lewis:2002ah}, we have chosen a uniform prior on
$\ln(10^{10} \Pstar)$ in the range $[2.7,4.0]$, as well as a uniform
prior on $\epsilon_2$ in $[-0.2,0.2]$. As discussed
in~\cite{Parkinson:2006ku, Pahud:2006kv}, a prior choice on
$\epsilon_1$ is not innocuous. This parameter encodes the amount of
gravitational waves contributing to the CMB anisotropies and its order
of magnitude is not known. Moreover, as shown in the next sections,
even in the simplest single field inflationary models under scrutiny
in this work, the orders of magnitude of $\epsilon_1$ may vary
considerably. As a result, we have performed two MCMC analysis: one
where $\epsilon_1$ has an uniform prior in the range
$[10^{-5},10^{-1}]$ and the other with a Jeffreys' prior by choosing
an uniform prior on $\log(\epsilon_1)$ in $[-5,0]$.  We have plotted
in figure~\ref{sr1st_cosmo_1D} the marginalised posterior probability
distributions for the base cosmological parameters together with
$\Hzero$ and $\OmegaL$ for convenience. The solid and dashed line
correspond to the uniform and Jeffreys' prior on $\epsilon_1$,
respectively.

\begin{figure}
\begin{center}
\includegraphics[width=12cm]{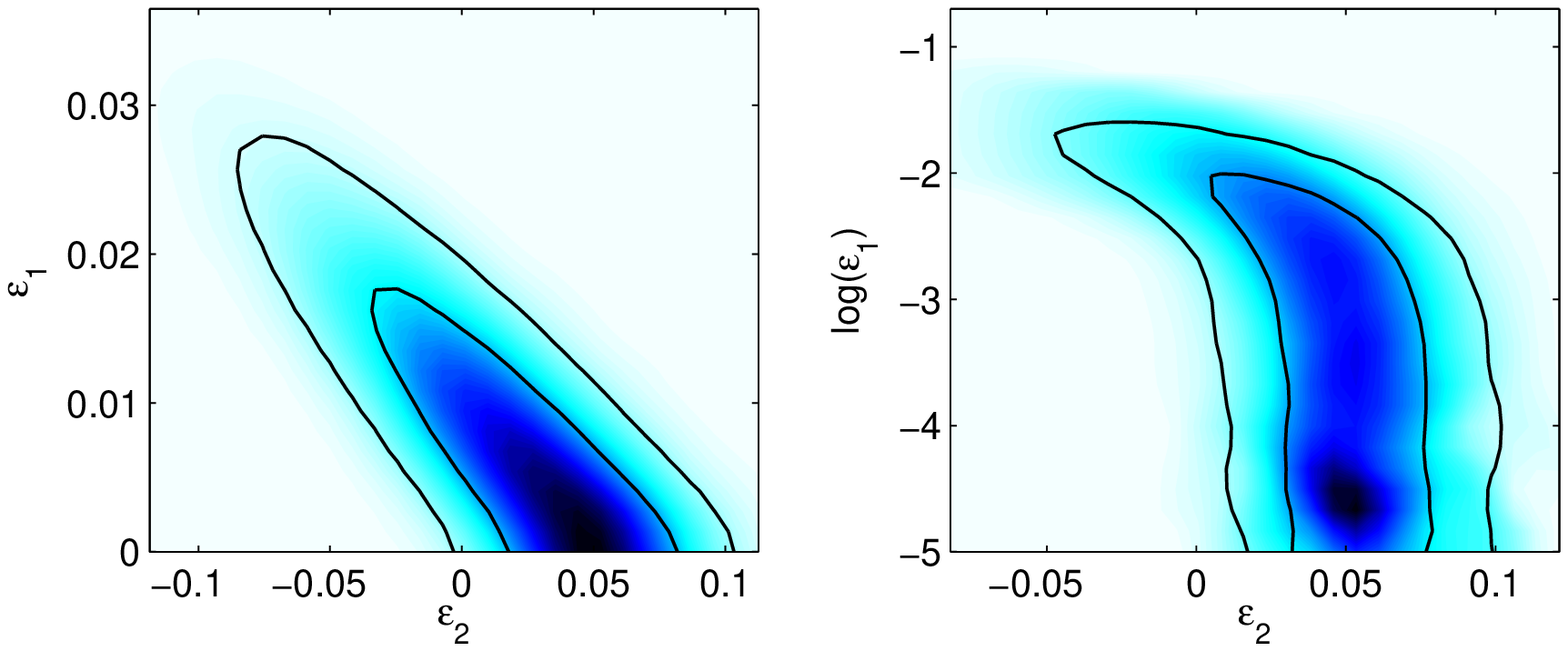}
\caption{$68\%$ and $95\%$ confidence intervals of the two-dimensional
  marginalised posteriors in the slow-roll parameters plane, obtained
  at leading order in slow-roll expansion. The shading is the mean
  likelihood and the left plot is derived under an uniform prior on
  $\epsilon_1$ while the right panel corresponds to an uniform prior
  on $\log(\epsilon_1)$.}
\label{sr1st_eps12_2D}
\end{center}
\end{figure}

The deviations induced by the prior choice on $\epsilon_1$ are the
consequences of the degeneracy between the two slow-roll parameters
(see figure~\ref{sr1st_eps12_2D}). As discussed in~\cite{Leach:2003us}
this is the result of both their influence on the spectral index
$\nS=1-2 \epsilon_1 - \epsilon_2$ and the tensor mode contribution to
the CMB anisotropies encoded in $\epsilon_1$. Since a Jeffreys' prior
on $\epsilon_1$ gives more statistical weight to its small values, the
accessible volume in the parameter space is enlarged in a region where
$\epsilon_1$ is small and the tensor modes have not observable
effects. In some sense, this prior choice tends to favour pure scalar
mode models. On the other hand, in the region where $\epsilon_1$ has a
significant effect, $\epsilon_2$ is pushed to lower values in such a
way that the scalar spectral index remains compatible with the data
while the cosmological parameters react to a significant tensor mode
contribution.

\begin{figure}
\begin{center}
\includegraphics[width=11cm]{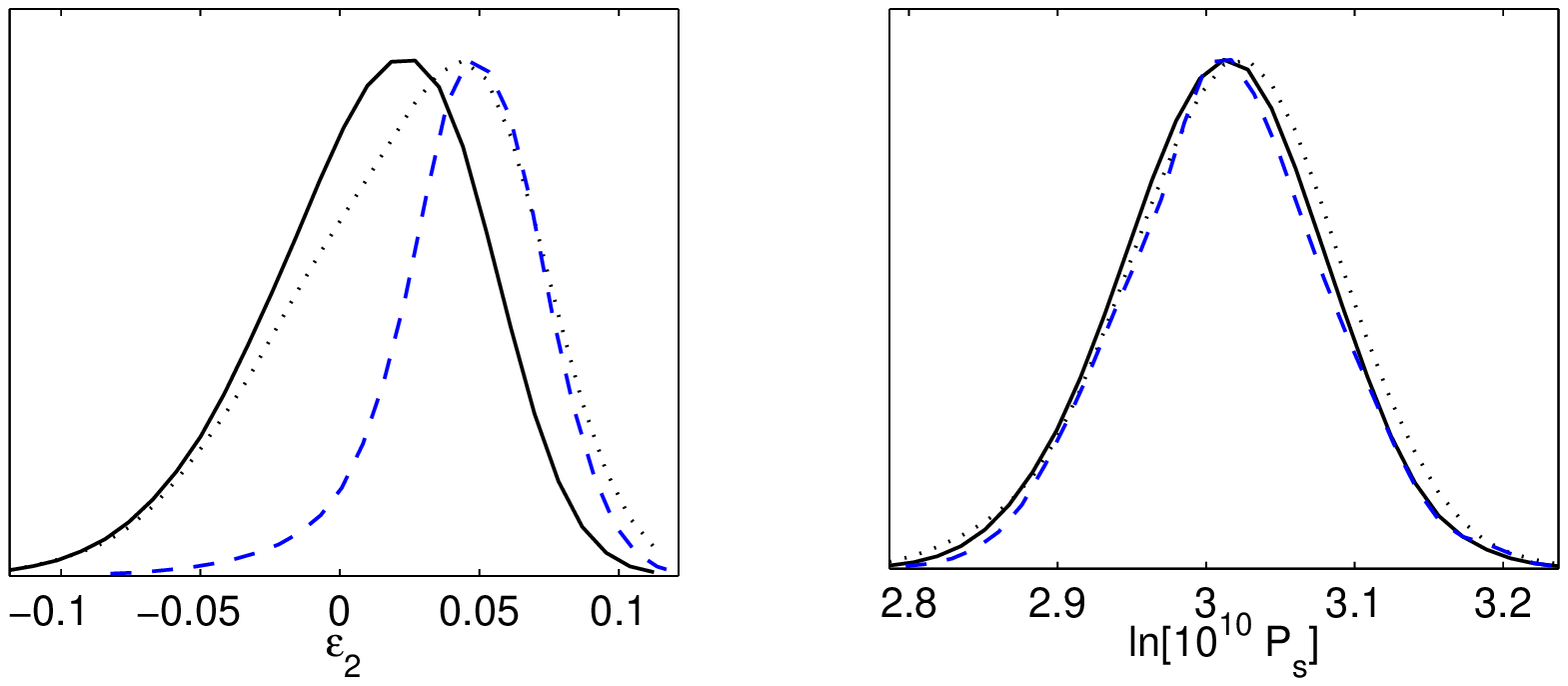}
\includegraphics[width=11cm]{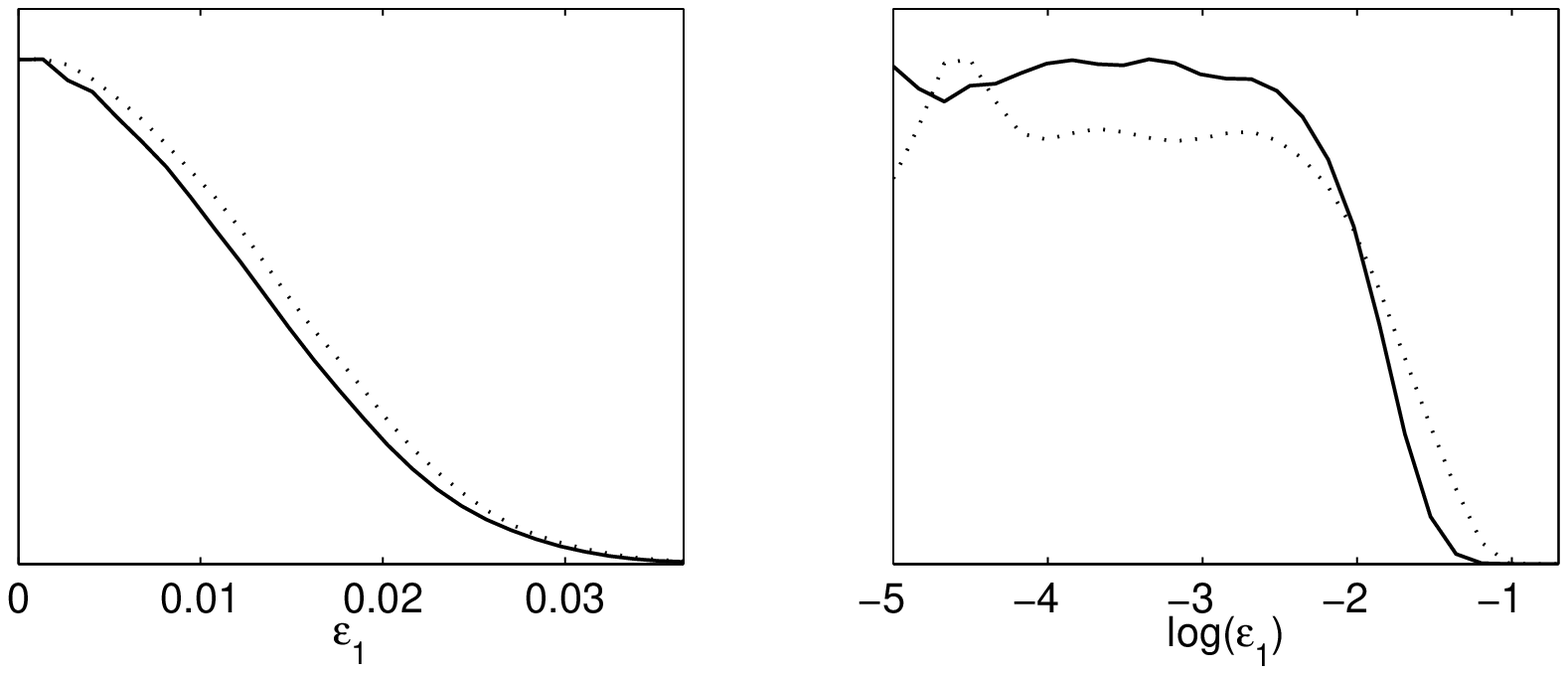}
\caption{Marginalised posterior probability distributions for the
  primordial parameters in the first order slow-roll expansion. As in
  figure~\ref{sr1st_cosmo_1D}, the black solid curves are derived
  under an uniform prior on $\epsilon_1$ whereas the dashed blue
  curves correspond to a flat prior on $\log(\epsilon_1)$. At
  two-sigma level of confidence one has the marginalised upper bound
  $\epsilon_1 < 0.022$.}
\label{sr1st_prim_1D}
\end{center}
\end{figure}

In figure~\ref{sr1st_prim_1D}, we have plotted the marginalised
posteriors for the primordial parameters. Similarly, the probability
distribution on $\epsilon_2$ ends up being affected by the
$\epsilon_1$ prior choice. Since for small values of $\epsilon_1$ the
spectral index approaches $\nS \simeq 1- \epsilon_2$, under the
Jeffreys' prior on $\epsilon_1$ one finds the positive values of
$\epsilon_2$ favoured: $-0.02 <\epsilon_2< 0.09$ at $2\sigma$ level
while $-0.07 < \epsilon_2 < 0.07$ under the uniform $\epsilon_1$
prior. We also obtain the marginalised upper bound $\epsilon _1<0.022$
at $95\%$ of confidence, or in terms of the observed tensor to scalar
ratio and scale of inflation:
\begin{equation}
r_{10}<0.21\, ,\qquad \frac{H}{\mpl}<1.3 \times 10^{-5},
\end{equation}
again at $2\sigma$.

\par

Notice that without marginalising over $\epsilon_2$ one obtains
slightly weaker $2 \sigma$ limits $\epsilon _1<0.028$ and
$r_{10}<0.28$. These can be compared to the unmarginalised bounds
coming from the WMAP first year data and given in
reference~\cite{Leach:2003us}, namely $\epsilon _1<0.032$ and
$r_{10}<0.32$.

\subsubsection{Second order slow-roll expansion}
\label{sec:wmap2order}

As discussed in section~\ref{sec:slowroll}, at second order in the
slow-roll approximation, one has to consider the third Hubble-flow
parameter $\epsilon_3$. In the following, we reiterate the previous
MCMC analysis on the parameters space enlarged by $\epsilon_3$ under
an uniform prior choice in $[-0.1,0.1]$. All the other priors have
been kept as in the first order analysis, as well as the comparison
between the uniform and Jeffreys' prior on $\epsilon_1$. We are
therefore dealing with a model involving eight parameters.

\begin{figure}
\begin{center}
\includegraphics[width=12cm]{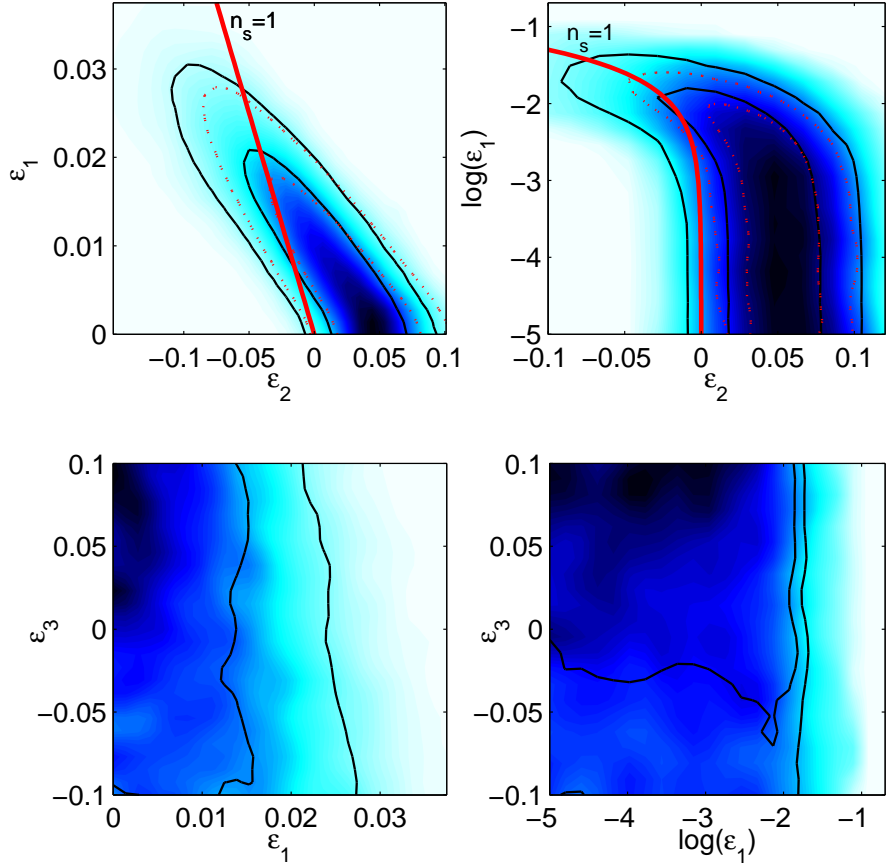}
\caption{One and two-sigma confidence levels of the two-dimensional
marginalised posteriors in the Hubble-flow parameter planes (solid
black contours). The dotted red contours correspond to the same
confidence intervals derived at first order in slow-roll expansion
(see figure~\ref{sr1st_eps12_2D}). The left panels are associated with
an uniform prior on $\epsilon_1$ whereas the right ones correspond to
an uniform prior on $\log(\epsilon_1)$. As can be seen in the bottom
right panel, positive values of $\epsilon_3$ are slightly favoured, but
in a prior dependent way while $\epsilon_3=0$ remains within the
one-sigma contour.}
\label{sr2nd_eps123_2D}
\end{center}
\end{figure}

As expected for well-constrained parameters, we find no significant
deviation on the base cosmological parameters and scalar power
spectrum amplitude between the first and second order slow-roll
models. The two-dimensional marginalised posteriors for the
Hubble-flow parameters are represented in
figure~\ref{sr2nd_eps123_2D}. Similarly, the one and two-sigma
confidence intervals in the plane $(\epsilon_1,\epsilon_2)$ are found
to be slightly enlarged compared to their first order equivalents, but
not more than what one may expect from the inclusion of a new
parameter.

\begin{figure}
\begin{center}
\includegraphics[width=11cm]{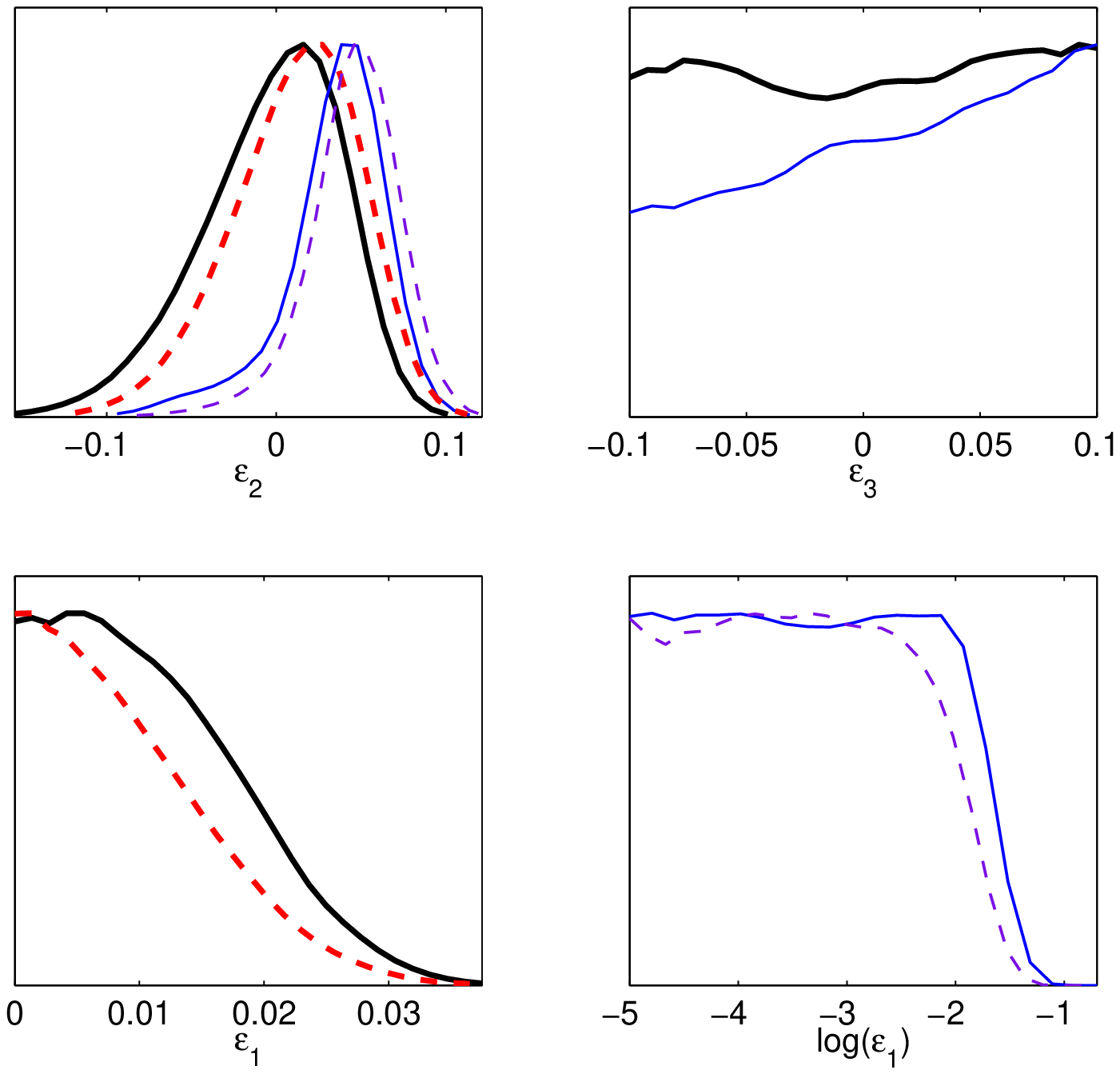}
\caption{Marginalised posterior probability distributions for the
  Hubble-flow parameters up to second order in the slow-roll
  expansion. The solid thick black curves corresponds to a uniform
  prior on $\epsilon_1$ whereas the solid thin blue curves are derived
  under an uniform prior on $\log(\epsilon_1)$. The dashed curves are
  the corresponding first order slow-roll posteriors of
  figure~\ref{sr1st_prim_1D}.}
\label{sr2nd_prim_1D}
\end{center}
\end{figure}

The marginalised posteriors associated with the Hubble-flow parameters
are plotted in Fig.~\ref{sr2nd_prim_1D}. Apart slightly weaker
constraints on $\epsilon_1$ and $\epsilon_2$, one may notice a weak,
but non-significant, tendency of running associated with positive
values of $\epsilon_3$. However, this effect appears only when a
Jeffrey's prior is chosen on $\epsilon_1$ and disappears under the
uniform prior choice~\cite{Leach:2003us, Easther:2006tv,
Huang:2006um}.

\subsection{Large field models}
\label{sec:lfmodel}

\begin{figure}
\begin{center}
\includegraphics[width=7.5cm]{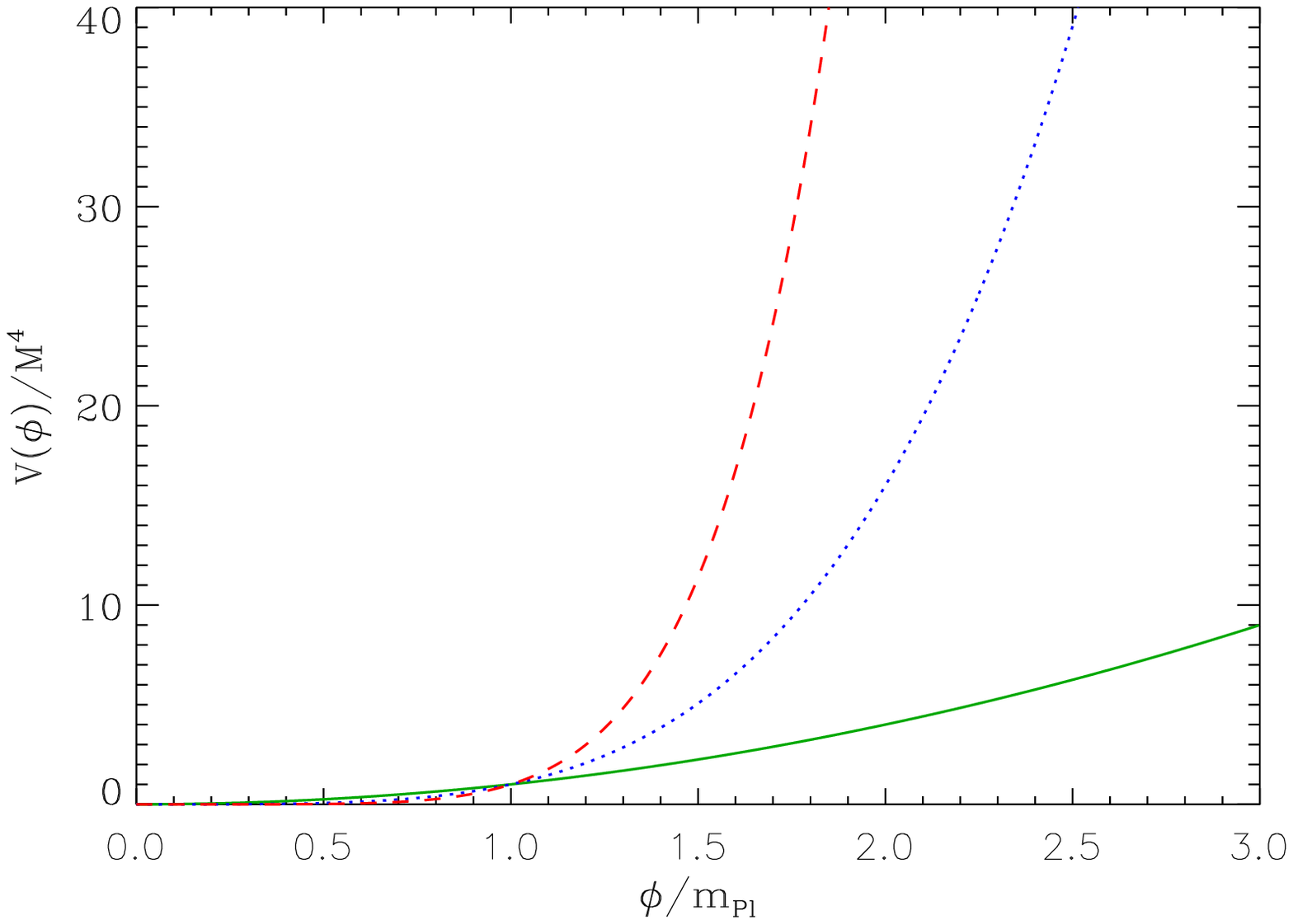}
\includegraphics[width=7.5cm]{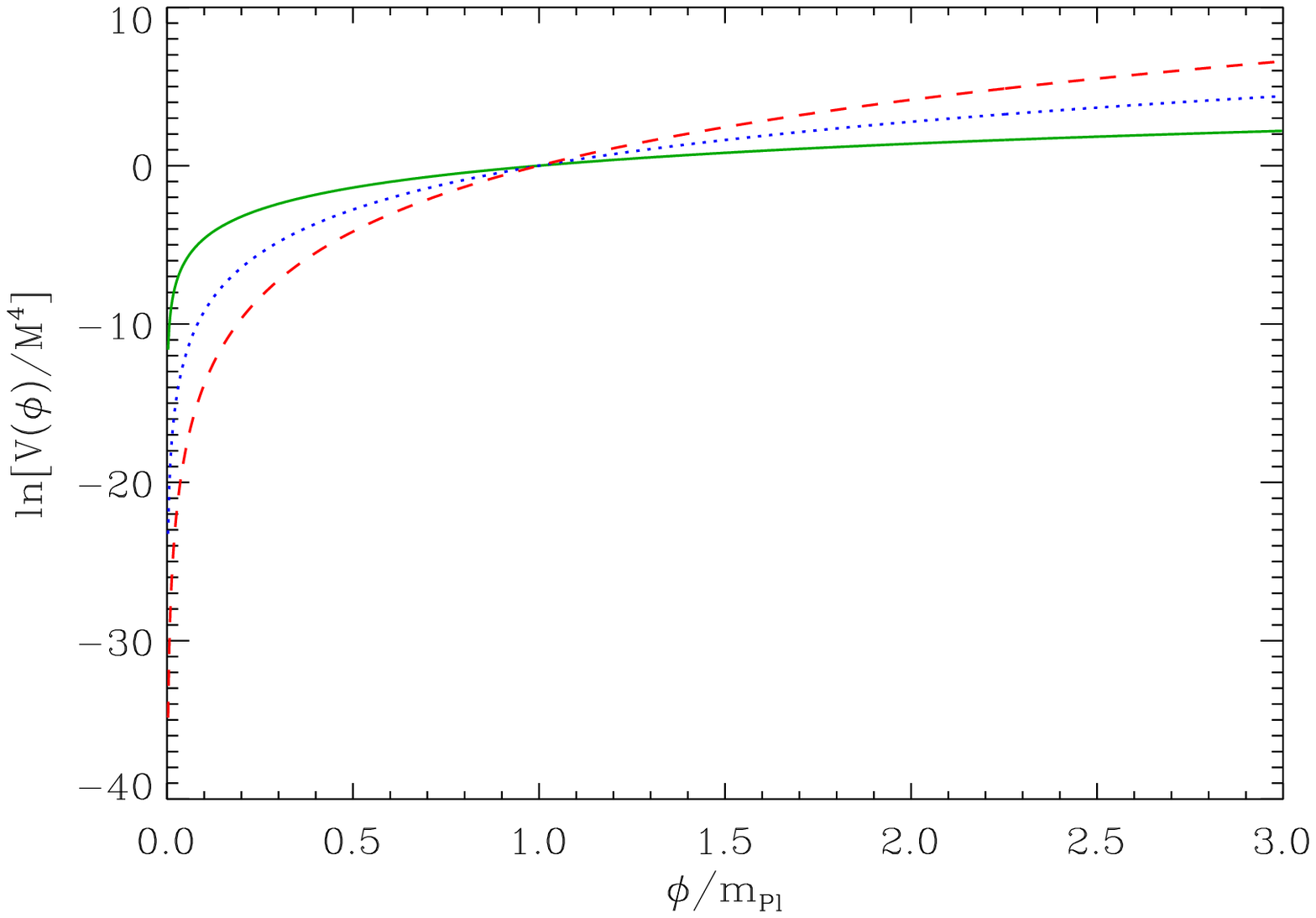}
\includegraphics[width=7.5cm]{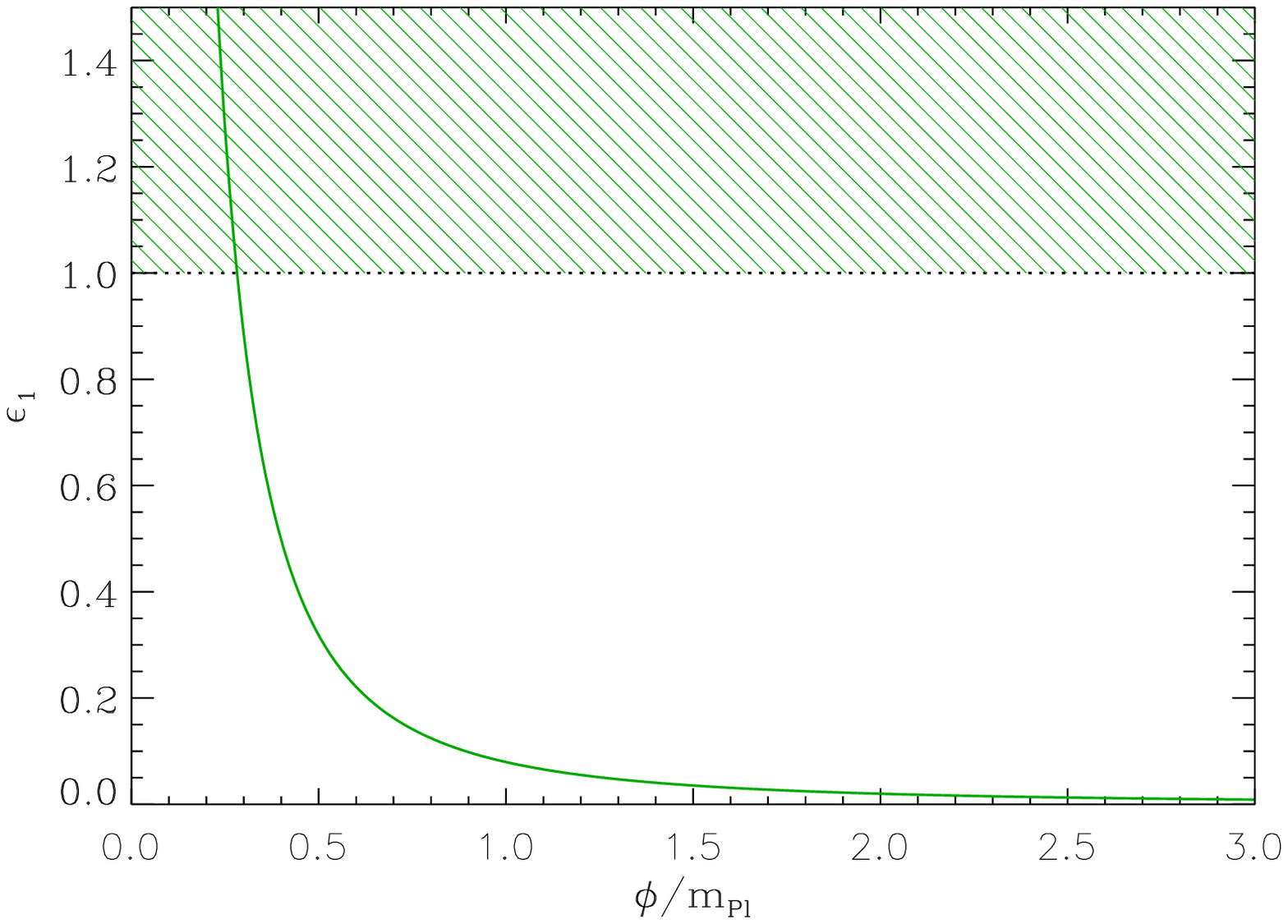}
\includegraphics[width=7.5cm]{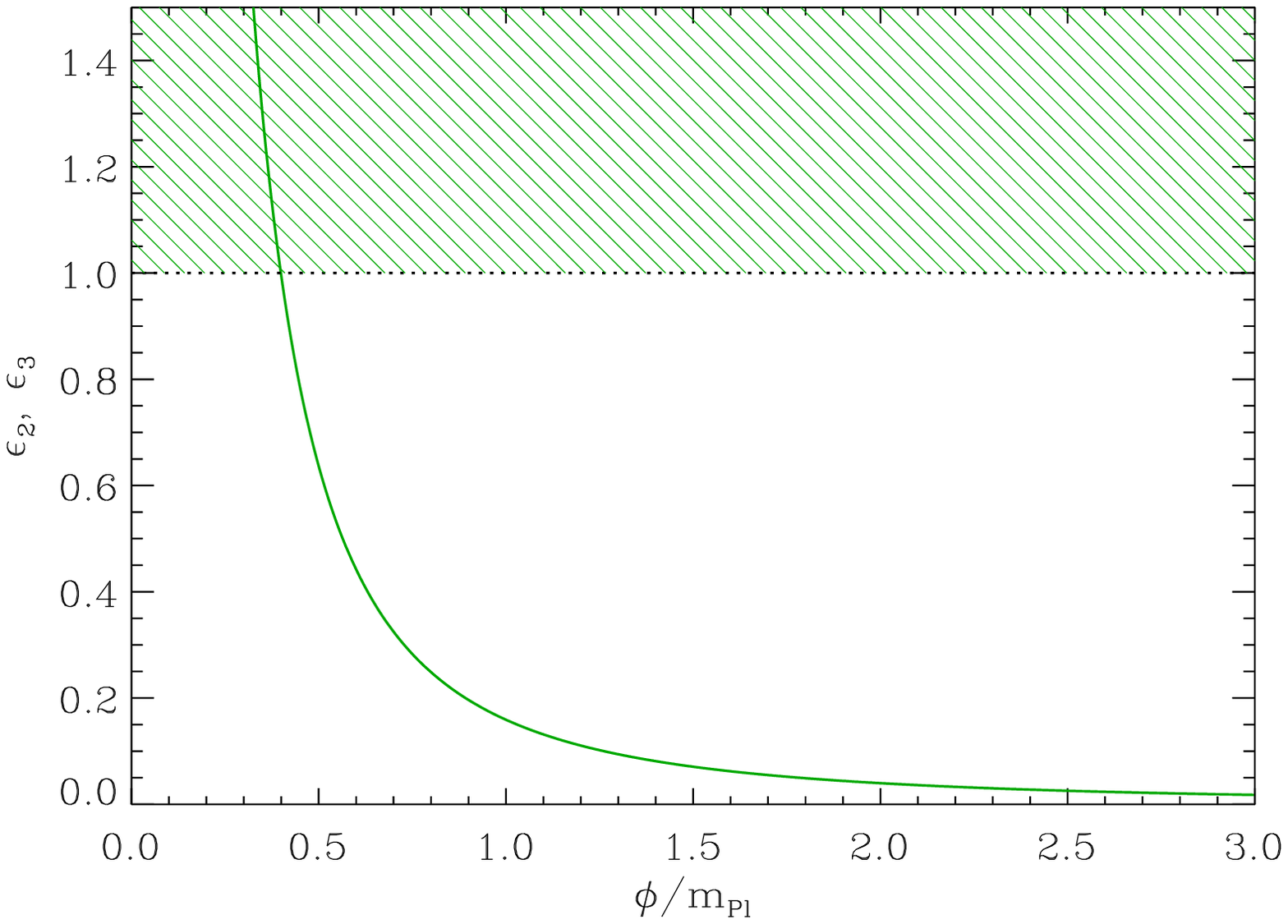}
\caption{Top left panel: large field potential for $p=2$ (solid green
  line), $p=4$ (dotted blue line) and $p=6$ (dashed red line). Top
  right panel: logarithm of the potentials for the same values of $p$
  (and the same line style code). The required flatness of the
  potential becomes obvious on this plot. Bottom left panel: slow-roll
  parameter $\epsilon _1$ for a large field potential with $p=2$. The
  shaded area indicates the breakdown of the slow-roll inflation
  (strictly speaking when the acceleration stops). Bottom right panel:
  slow-roll parameters $\epsilon _2$ and $\epsilon _3$ for a large
  field potential with $p=2$. Only one curve appears because $\epsilon
  _2=\epsilon_3$. On this plot, the shaded region signals the
  breakdown of the slow-roll approximation but not necessarily the end
  of the accelerated phase.}
\label{potlf}
\end{center}
\end{figure}

Large fields models (or chaotic inflation models although this is not
very appropriate~\cite{Vilenkin:2004vx}) are characterised by the
monomial potential given by~\cite{Linde:1984st}
\begin{equation}
  V(\phi) = M^4\left(\frac{\phi}{\mpl}\right)^p.
\end{equation}
There are only two free parameters, $M$ and $p$ and the energy scale
$M$ is uniquely determined by the amplitude of the CMB anisotropies,
and thus the WMAP normalisation. This is probably the simplest
inflationary scenario since, in the slow-roll approximation,
everything can be integrated analytically. This family of potentials
is represented in figure~\ref{potlf}.
{}From the expression of the potential, the three slow-roll parameters
can be easily computed and reads
\begin{equation}
\epsilon _1 =\frac{p^2}{16\pi }\frac{\mpl ^2}{\phi ^2}\, ,\qquad
\epsilon_2=\frac{p}{4\pi }\frac{\mpl ^2}{\phi ^2}\,, \qquad \epsilon
_3=\epsilon_2\, .
\end{equation}
These slow roll parameters are represented in the two bottom panels in
figure~\ref{potlf}. There are monotonic functions of $\phi $ and they
are decreasing as $\phi$ is increasing. One can immediately deduce
that, for a given $p$, the model in the plane $(\epsilon _1,\epsilon
_2)$ is represented by the trajectory $\epsilon _1=(p/4)\epsilon
_2$. We can also estimate the value of the inflaton at the end of
inflation, defined to be the time at which $\epsilon _1=1$ (see
before). This leads to
\begin{equation}
\label{phiendlf}
\frac{\phi _{\uend}}{\mpl}=\frac{p}{4\sqrt{\pi }}\, .
\end{equation}
Moreover, the slow-roll equation of motion leads to a solution which
is completely explicit. Integrating the following quadrature
\begin{equation}
N=-\frac{8\pi }{\mpl^2}\int _{\phi_\ini}^{\phi} 
\frac{V(\chi )}{V'(\chi )} \dd \chi = -\frac{8\pi }{p}
\int_{\phi_\ini/\mpl}^{\phi/\mpl} x \dd x \, ,
\end{equation}
leads to an explicit expression $N=N(\phi )$ which can be inverted and
reads
\begin{equation}
\label{solLarge}
  \frac{\phi}{\mpl} = \sqrt{\left(\frac{\phi_\ini}{\mpl}\right)^2
  -\frac{p}{4\pi}N}\, .
\end{equation}
This expression also allows us to obtain the total number of
e-folds. It is sufficient to evaluate the above expression for $\phi
=\phi _{\uend}$, $\phi _{\uend}$ being given by~(\ref{phiendlf}). One
arrives at
\begin{equation}
N_\usssT=\frac{4\pi}{p}
\left(\frac{\phi_\ini}{\mpl}\right)^2-\frac{p}{4}\, ,
\end{equation} 
which can be very large if the initial energy density of the inflaton
field is close to the Planck scale $\mpl^4$. However, the model
remains under control only if the initial energy density is smaller
than $\mpl^4$ and this imposes a constraint on the initial value of
the field, namely 
\begin{equation}
\frac{\phi_\uin}{\mpl} \lesssim \left(\frac{\mpl}{M}\right)^{4/p}.
\end{equation}
Let us notice that, when the inflaton energy density approaches the
Planck energy density, quantum effects become important. In this case,
the formalism of stochastic inflation must be
used~\cite{Vilenkin:1983xp, Vilenkin:1983xq, Goncharov:1987ir,
  Linde:1993xx, Starobinsky:1986fx, Liguori:2004fa, Martin:2005ir,
  Martin:2005hb}

\par

We now turn to the explicit determination of the slow-roll
parameters. We have seen that the model is represented by the
trajectory $\epsilon _1=(p/4)\epsilon _2$ but observable models only
lie in a limited portion of this straight line. Indeed, the
Hubble-flow parameters should be evaluated when the scales of
astrophysical interest today left the Hubble radius during
inflation. Let us call the value of the inflaton at that time $\phi
_*$. Then $\phi _*$ can be expressed in terms of $N_*$, the number of
e-folds between the time of Hubble exit and the end of inflation:
\begin{equation}
  N_*=-\frac{8\pi }{p} \int_{\phi_*/\mpl}^{\phi_{\uend}/\mpl}
   x \dd x \, ,
\end{equation}
from which one deduces
\begin{equation}
\frac{\phi _*^2}{\mpl^2}=\frac{p}{4\pi}\left(N_*+\frac{p}{4}\right).
\end{equation}
Therefore, the slow roll parameters can be expressed as
\begin{equation}
\epsilon _1 =\frac{p}{4\left(N_*+p/4\right)}\, ,\qquad 
\epsilon_2=\frac{1}{N_*+p/4}\,, \qquad \epsilon
_3=\epsilon_2\, .
\end{equation}
The number of e-folds $N_*$ between the end of inflation and the
Hubble length exit can be thought as the arc length along the straight
line representing the large field models. The value of $N_*$ can be
calculated once one knows the entire history of the Universe,
including the reheating phase~\cite{Liddle:2003as}. In the following,
we will consider that
\begin{equation}
\label{limitN*}
40<N_*<60\,,
\end{equation}
when dealing with the slow-roll models and in fact, as explained
in~\cite{Liddle:2003as}, one could consider an even more restricted
range for the quartic model $p=4$.

\par

\begin{figure}
\begin{center}
\includegraphics[width=14cm]{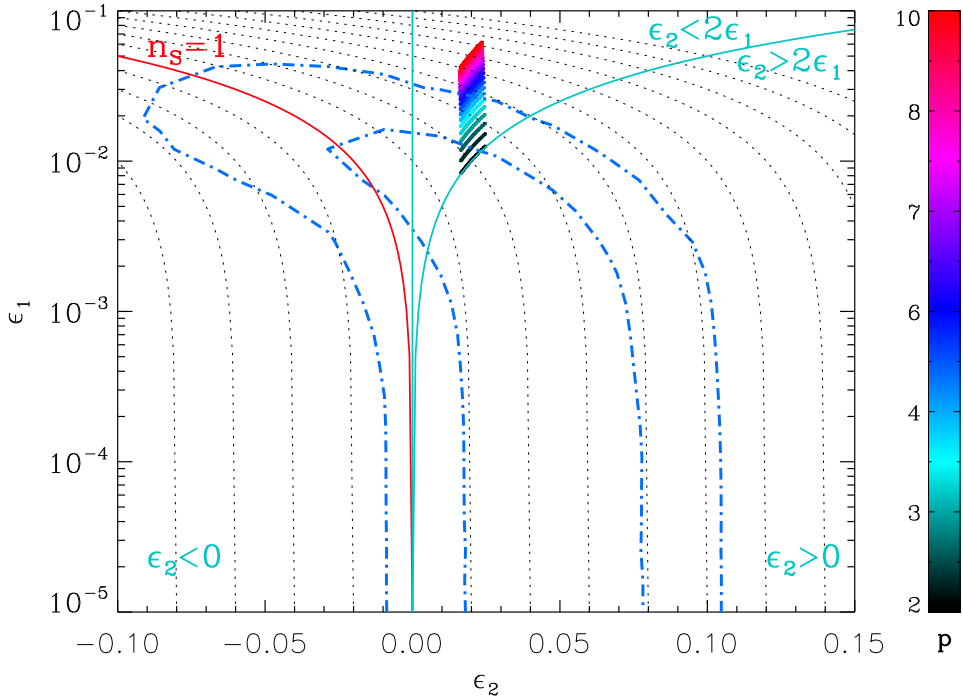}
\caption{Slow-roll predictions for large field models in the plane
  $(\epsilon _1, \epsilon _2)$. The dotted black lines indicate the
  lines of constant spectral indices from $\nS=0.8$ (on the right) to
  $\nS=1.1$ (on the left), the red line corresponding to a scale
  invariant spectrum $\nS=1$. The solid blue lines separate the
  regions $\epsilon _2<0$, $\epsilon _2>0$, $\epsilon _2<2\epsilon _1$
  and $\epsilon _2>2\epsilon _1$ associated with different energetic
  evolutions (see text). As expected, large field models lie in the
  region $\epsilon _2<2\epsilon _1$. The dotted-dashed blue curves
  indicate the $1\sigma $ and $2\sigma $ confidence intervals given
  the WMAP3 data (see section~\ref{sec:wmapsr}). Each coloured segment
  represents the prediction of a model given $p$ (see colour bar) and
  for a number of e-folds $N_*$ between the end of inflation and the
  Hubble exit varying in $[40,60]$.}
\label{lf}
\end{center}
\end{figure}

The slow-roll predictions for the large field models are represented
in figure~\ref{lf}.
As expected, the whole family lies in the region $\epsilon _2>0$ and
$\epsilon _2<2\epsilon _1$. According to the previous discussion, this
means that, as inflation proceeds, the kinetic energy increases with
respect to the total energy density and, at the same time, the
absolute value of this kinetic energy density decreases. From
figure~\ref{lf}, all the models with $p\gta 4$ lie outside the
$2\sigma $ contour. The quadratic (or massive) model remains
compatible with the data and predicts quite a high contribution of
gravitational waves, up to $r_{10}\sim 10\%$ level. Having found the
compatible values of the parameter $p$, our next move is to estimate
the numerical value of the parameter $M$. This can be done from the
measurement of the CMB quadrupole~(\ref{eq:Qrms}) made by the WMAP
satellite
\begin{equation}
\label{multi}
  \frac{Q_\mathrm{rms-PS}^2}{T^2}
  =\frac{1}{60\pi\epsilon_*}\frac{H_*^2}{\mpl^2}
  =\frac{2}{45\epsilon_*}\frac{V_*}{\mpl^4}\,.
\end{equation}
In the case of large fields model, this implies
\begin{equation}
\label{scaleMlf}
\left(\frac{M}{\mpl}\right)^4=\frac{(45/2)p}{\left(4N_*+p\right)^{p/2+1}}
\left(\frac{16 \pi }{p}\right)^{p/2}\frac{Q_\mathrm{rms-PS}^2}{T^2}\,,
\end{equation} 
and given the constraints on $p$ and $N_*$, this leads to
\begin{equation}
\label{consMlf}
4\times 10^{-4} \lta \frac{M}{\mpl }\lta 1.1 \times 10^{-3} .
\end{equation}
We recover the conclusion that, for large field models, inflation take
place close to the Grand Unified Theory (GUT) scale.

\par

To conclude this section, let us come back to the question raised
after equation (\ref{epsV2}), namely the error caused by expressing
the slow-roll parameters $\epsilon _n$ in terms of the potential $V$
and its derivatives. If we take into account the more accurate
equation~(\ref{inverting2}), then the expression of the parameter
$\epsilon _1$ for the quadratic model becomes
\begin{equation}
\epsilon _1=\frac{1}{4\pi }\frac{\mpl ^2}{\phi ^2}
\left(1-\frac{1}{6\pi }\frac{\mpl ^2}{\phi ^2} + \dots \right).
\end{equation}
{}From this expression, if one tries to calculate a new value of
$\phi_{\uend}$, one immediately faces the problem that the above
expression leads to a second order algebraic equation with a negative
discriminant, \ie an equation which does not admit real
solutions. This is not surprising since, near the end of inflation,
the correction is \emph{a priori} of order one. In order to have an
order of magnitude estimate of the corrections, one can still work
with the former value of $\phi_{\uend}$ in (\ref{phiendlf}). Then, as
expected, one finds that the correction is of order $1/N_*^2\simeq
4\times 10^{-3}$ and, hence, does not in any way modify the
conclusions established before about the compatibility of the large
field models with the WMAP data.

\subsection{Small field models}
\label{sec:sfmodel}

We now turn to another class of models, namely small field models. In
this case, the inflaton potential can be written
as~\cite{Linde:1981mu, Albrecht:1982wi, Kinney:1995cc}
\begin{equation}
\label{potentialsf}
V(\phi) = M^4 \left[1 -\left(\frac{\phi}{\mu}\right)^{p}\right].
\end{equation}
The potential is characterised by three parameters, the energy scale
$M$, $\mu $ and the index $p$, \ie one more parameter with respect to
the large field models.

\begin{figure}
\begin{center}
\includegraphics[width=7.5cm]{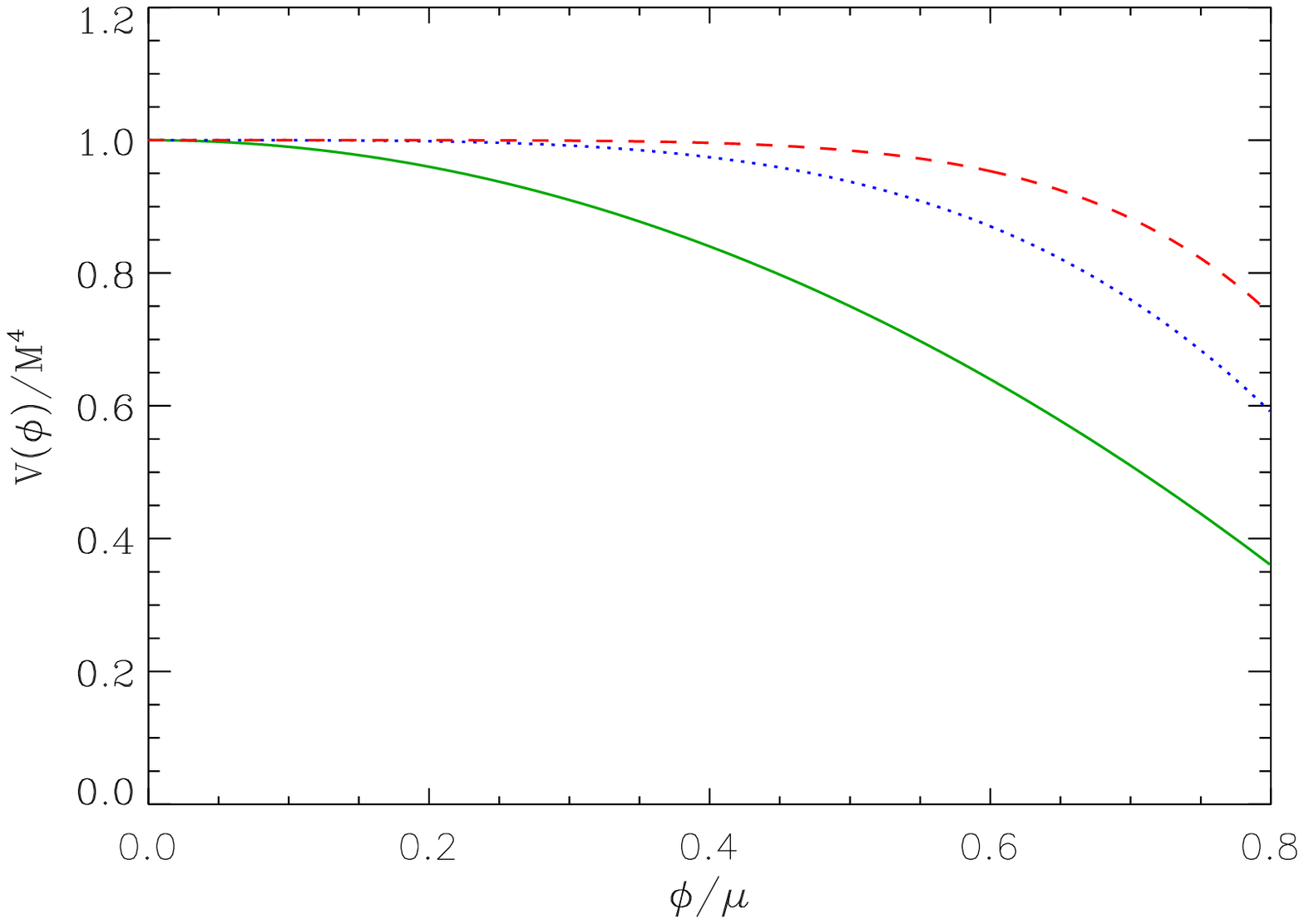}
\includegraphics[width=7.5cm]{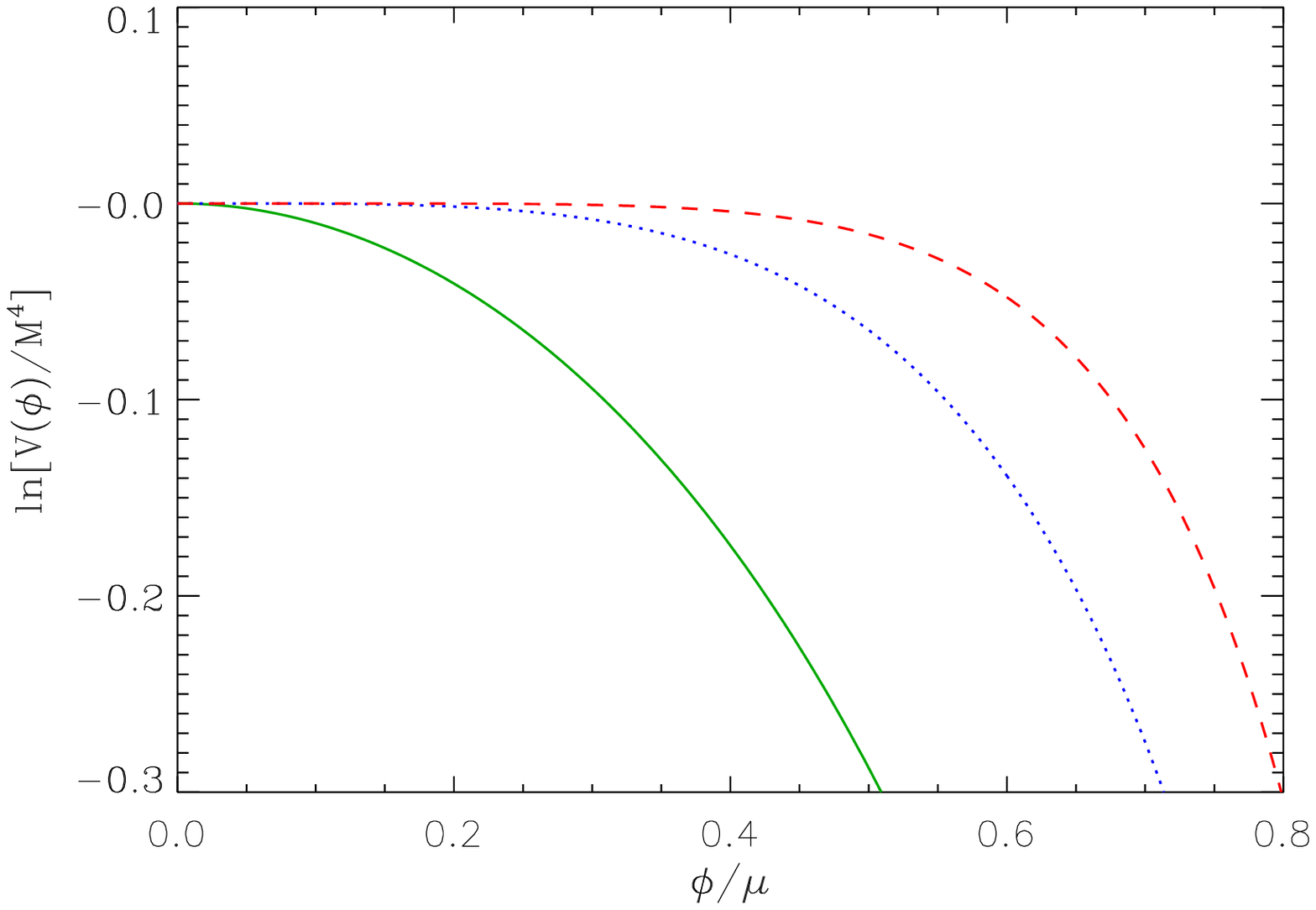}
\includegraphics[width=7.5cm]{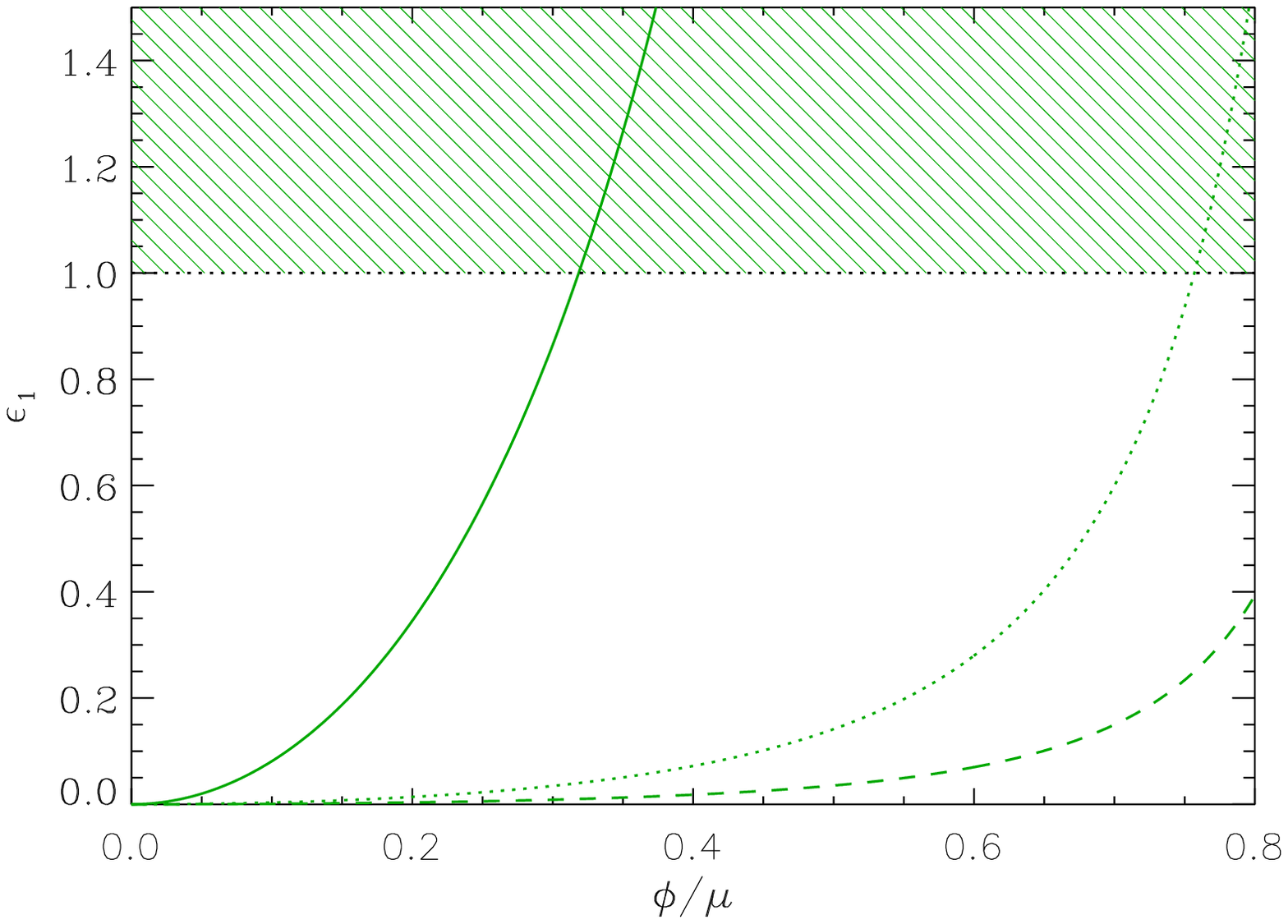}
\includegraphics[width=7.5cm]{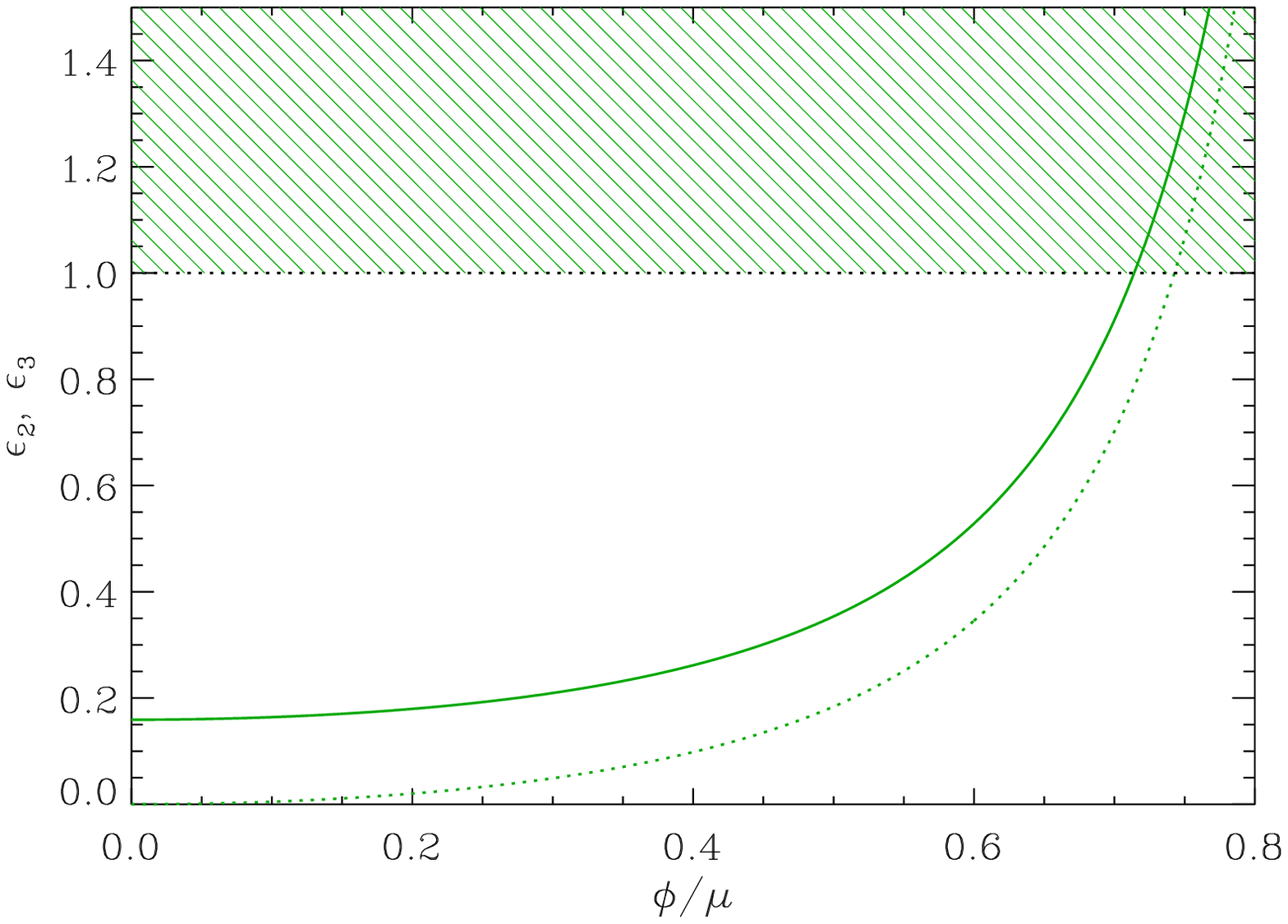}
\caption{Top left panel: small field potentials for $p=2$ (solid green
  line), $p=4$ (dotted blue line) and $p=6$ (dashed red line). Top
  right panel: logarithm of the same potentials than in the top left
  panel. Bottom left panel: slow-roll parameter $\epsilon _1$ for the
  model $p=2$ with $\mu /\mpl=0.1$ (solid line), $\mu/\mpl =0.5$
  (dotted line) and $\mu/\mpl=1$ (dashed line). The dashed area
  indicates where the slow-roll breaks down. Bottom right panel:
  slow-roll parameters $\epsilon _2$ (solid line) and $\epsilon _3$
  (dotted line) for $\mu /\mpl=1$. Note that the slow-roll parameter
  $\epsilon _2$ is non-vanishing in the limit $\phi/\mu \rightarrow 0$
  where its value reads $(\mpl/\mu)^2 p/(4\pi)$.}
\label{potsf}
\end{center}
\end{figure}

The potential, as well as its logarithm, are represented in
figure~\ref{potsf}. For these models, it is straightforward to calculate
the first three slow-roll parameters. They read
\begin{eqnarray}
\label{srsf}
\epsilon_1 &=& \frac{p^2}{16\pi}\left(\frac{\mpl}{\mu }\right)^2
\frac{(\phi/\mu)^{2p-2}}{\left[1-(\phi/\mu)^p\right]^2}\, ,
\\ \epsilon_2 & = & \frac{p}{4\pi}\left(\frac{\mpl}{\mu }\right)^2
\left(\frac{\phi }{\mu }\right)^{p-2}
\frac{(\phi/\mu)^{p}+p-1}{\left[1-(\phi/\mu)^p\right]^2}\, ,
\\ \epsilon_3 &=& \frac{p}{8\pi}\left(\frac{\mpl}{\mu }\right)^2
\frac{\left(\phi/\mu\right)^{p-2}}{\left[1-\left(\phi/\mu\right)^p\right]^2
  \left[\left(\phi/\mu\right)^p+p-1\right]} \left[2\left(\frac{\phi
  }{\mu }\right)^{2p}\right. \nonumber \\ & + &
  \left. (p-1)(p+4)\left(\frac{\phi }{\mu }\right)^p
  +(p-1)(p-2)\right] ,
\label{srsf3}
\end{eqnarray}
and they are represented in the two bottom panels of
figure~\ref{potsf}. As for large field models, they are monotonic
functions of the field. However, these parameters are now increasing
as the vacuum expectation value (vev) of the inflaton is increasing
during small field inflation.

\par

Then, we follow the same steps as in the previous subsection. Our
first goal is therefore to determine when inflation stops. Requiring
$\epsilon_1(\phi_\uend)=1$ leads to an algebraic equation, namely
\begin{equation}
\label{eqphiend}
\left(\frac{\phi_\uend}{\mu}\right)^{2p}-2
\left(\frac{\phi_\uend}{\mu}\right)^p-\frac{p^2}{16\pi }
\left(\frac{\mpl}{\mu }\right)^2
\left(\frac{\phi_\uend}{\mu}\right)^{2p-2}+1=0\, .
\end{equation}
This equation can be solved explicitly for $p=2$ and the solution
reads
\begin{equation}
\label{end}
\frac{\phi_\uend}{\mu}= \frac{1}{4\sqrt{\pi }}\frac{\mpl }{\mu
} \left(-1+\sqrt{1+\frac{16\pi\mu^2}{\mpl^2}}\right)\, .
\end{equation}
In fact, among the two solutions (\ref{eqphiend}) admits, we have
chosen the one with the minus sign. This is because, since the vev of
the inflaton field is increasing as inflation proceeds, the
accelerated phase stops for the ``smallest'' value of $\phi
_\uend$. Let us notice that, in the limit, $\mu /\mpl \rightarrow
+\infty $ one has $\phi _\uend\rightarrow \mu $ but we always have
$\phi _\uend<\mu $, in other words the limit is approached by lower
values\footnote{Of course, this limit is physically questionable
  because the expression~(\ref{potentialsf}) could viewed as a Taylor
  expansion in $\phi/\mu $. As long as $\phi \simeq \mu$, the accurate
  form of the potential should be given and, consequently, other power
  of $\phi/\mu $ considered~\cite{German:2001tz}}. On the other hand,
if we now assume $\mu /\mpl \ll 1$, then the previous solution can be
approximated by
\begin{equation}
\label{approx2}
\frac{\phi_\uend}{\mu}\simeq 2\sqrt{\pi }\frac{\mu }{\mpl}\, ,
\end{equation}
and, therefore, under the assumption $\mu /\mpl \ll 1$ one has
$\phi_\uend/\mu \ll 1$.

\par

The above considerations were established for the case $p=2$. If
$p\neq 2$ then, as already mentioned, equation~(\ref{eqphiend}) can no
longer be explicitly solved but we can still approximate its solutions
in the limits considered before. Indeed, for $\mu /\mpl \gg 1$,
equation~(\ref{eqphiend}) reduces to $y^2-2y+1=0$, where $y=(\phi
_\mathrm{end}/\mu)^p$, the only solution of which is $\phi _\uend=\mu
$. Consequently, as in the case $p=2$, $\phi_{\uend}\rightarrow \mu $
when $\mu/\mpl \rightarrow +\infty $. On the other hand, since we
always require $\phi _\uend/\mu<1$, the two terms $\left(\phi
_\mathrm{end}/\mu \right)^{2p}$ and $\left(\phi _\uend/\mu
\right)^{p}$ in (\ref{eqphiend}) can be neglected in comparison to
$1$, provided $\mu /\mpl \ll 1$. Then, keeping the third term on the
left-hand side leads to
\begin{equation}
\label{phiendgeneral}
\frac{\phi_\uend}{\mu}\simeq \left[\frac{16
    \pi}{p^2}\left(\frac{\mu }{\mpl}\right)^2\right]^{1/(2p-2)} ,
\end{equation}
and one can check that, for $p=2$, this reproduces
equation~(\ref{approx2}). If, on the contrary, one does not have $\mu
/\mpl \ll 1$ (or $\mu/\mpl\gg 1 $), then equation~(\ref{eqphiend}) can
only be solved numerically.

\par

The next step is to obtain the classical field trajectory for these
models. This can be done if the slow-roll approximation is satisfied
but, even in this case, the classical trajectory can only be found
implicitly. In terms of total number of e-folds $N$, we have for the
small field potential (\ref{potentialsf})
\begin{equation}
  N=\frac{8\pi }{p}\frac{\mu ^2}{\mpl ^2}
  \int_{\phi_\ini/\mu}^{\phi /\mu}   x^{1-p} \left(1
  -x^p\right)\dd x \, .
\end{equation}
giving
\begin{equation}
\label{trajecsf}
N=\frac{4\pi }{p}\frac{\mu ^2}{\mpl ^2}\,\left\{
\left(\frac{\phi_\ini}{\mu}\right)^2 - \left(\frac{\phi }{\mu
}\right)^2 + \frac{2}{p-2}\left[\left(\frac{\phi_\ini}{\mu
}\right)^{2-p} - \left(\frac{\phi }{\mu }\right)^{2-p}\right]\right\},
\end{equation}
for $p\neq2$, while for $p=2$ one has
\begin{equation}
\label{trajecsfp2}
  N = 2\pi \frac{\mu
  ^2}{\mpl^2}\left[\left(\frac{\phi_\ini}{\mu}\right)^2 -
  \left(\frac{\phi }{\mu }\right)^2 +2\ln \left(\frac{\phi }{\phi
  _\ini}\right)\right].
\end{equation}
For $p=2$, one can invert the above relation and express $\phi $ in
terms of the number of e-folds to get
\begin{equation}
\label{trajecsfquadratic}
  \frac{\phi }{\mu} = \sqrt{-W_0
  \left\{-\left(\frac{\phi_\ini}{\mu}\right)^{2} \exp
  \left[-\left(\frac{\phi_\ini}{\mu }\right)^2+\frac{N}{2\pi}
  \left(\frac{\mpl}{\mu}\right)^2\right] \right\}}\,,
\end{equation}
where $W_0(x)$ is the principal branch of the Lambert
function~\cite{Valluri:2000aa}. This special function\footnote{It is
  also called \texttt{ProductLog[ ]} in Mathematica.} is the solution
of the equation
\begin{equation}
W(x)\e ^{W(x)}=x\,.
\end{equation}
In our case, we have to choose the principal branch since
$\phi/\mu<1$.

\par

Finally, the last step consists in determining the link between $N_*$
and $\phi _*$. For $p\neq 2$, using the classical trajectory obtained
before, one has to solve the following equation
\begin{eqnarray}
\label{eqphistarsf}
\left(\frac{\phi _*}{\mu }\right)^2 +\frac{2}{p-2} \left(\frac{\phi
_*}{\mu }\right)^{2-p} &=& \frac{pN_*}{4\pi
}\left(\frac{\mpl}{\mu}\right)^2 + \left(\frac{\phi _\mathrm{end}}{\mu
}\right)^2 \nonumber \\ & & +\frac{2}{p-2} \left(\frac{\phi
_\mathrm{end}}{\mu }\right)^{2-p}\, ,
\end{eqnarray}
where $\phi_\uend$ and $N_*$ are known from the previous discussion.

\par

For $p=2$, this equation can be explicitly solved and one gets
\begin{equation}
\label{phistarquadratic}
  \frac{\phi_*}{\mu} = \sqrt{-W_0
  \left\{-\left(\frac{\phi_\uend}{\mu}\right)^{2}
  \exp\left[-\left(\frac{\phi_\uend}{\mu
  ^2}\right)^2-\frac{N_*}{2\pi}\left(\frac{\mpl}{\mu
  }\right)^2\right]\right\}}\, ,
\end{equation}
In the limit $\mu /\mpl \ll 1$, the argument of the Lambert function
is small and $W_0(x)\simeq x$ leading to
\begin{equation}
\label{phistarlfmusmall}
\left(\frac{\phi_*}{\mu}\right)^2 \simeq
\left(\frac{\phi_\uend}{\mu}\right)^{2} \exp
\left[-\left(\frac{\phi_\uend}{\mu}\right)^2-\frac{N_*}{2\pi}
\left(\frac{\mpl}{\mu}\right)^2\right].
\end{equation}
On the other hand, in the limit $\mu /\mpl \rightarrow +\infty $ then,
as established before, $\phi_{\uend}\rightarrow \mu $ and the
argument of the Lambert function goes to $-1/e$. Since $W_0(-1/e)=-1$,
one obtains $\phi _* \rightarrow \mu $.

\par

If $p\neq 2$, then equation~(\ref{eqphistarsf}) cannot be solved
explicitly. However, in the limit $\mu /\mpl \ll 1$, one has
$\phi_*\ll \mu $ and $\phi _\uend/\mu \ll 1$ allowing
(\ref{eqphistarsf}) to be approximated as
\begin{equation}
\frac{2}{2-p} \left(\frac{\phi _*}{\mu} \right)^{2-p}\simeq \frac{pN_*}{4\pi}
\left(\frac{\mpl}{\mu} \right)^2 +\frac{2}{2-p} \left(\frac{\phi _\uend}{\mu}
  \right)^{2-p}.
\end{equation}
Using the expression of $\phi _\uend$ obtained in
(\ref{phiendgeneral}) renders the second term in the right-hand side
of the previous equation proportional to $(\mpl/\mu)^{(p-2)/(p-1)}$.
Under the considered limit $\mu/\mpl \ll 1$, this term can be
neglected in comparison to the the first one which is proportional to
$(\mpl/\mu)^2$. Putting everything together, one gets
\begin{equation}
\frac{\phi_*}{\mu} \simeq \left[\frac{p(p-2)N_*}{8\pi }
\left(\frac{\mpl}{\mu }\right)^2\right]^{1/(2-p)} .
\end{equation}
Let us stress again that the above result is valid only for $\mu /\mpl
\ll 1$. If this is not the case, one has to rely on numerical
calculations to find the correct value of $\phi _*$.

\par

The values of the slow-roll parameters directly stem from the previous
considerations. If $p=2$, it is sufficient to use the value of $\phi
_*$ found in (\ref{phistarquadratic}) into the
expressions~(\ref{srsf}) to (\ref{srsf3}). The only underlying
approximation being in that case the inversion mentioned in
section~\ref{sec:cosmopert}. Besides, assuming the limit $\mu /\mpl\ll
1$, one arrives at
\begin{eqnarray}
\label{srapproxsf}
\epsilon _1 &\simeq& \exp \left[-\frac{N_*}{2\pi }\left(\frac{\mpl }{\mu
  }\right)^2\right] ,\\
\label{srapproxsf2}
\epsilon _2 & \simeq & \frac{1}{2\pi
}\left(\frac{\mpl }{\mu }\right)^2 , \\ \epsilon _3 & \simeq
&6 \exp \left[-\frac{N_*}{2\pi }\left(\frac{\mpl }{\mu
  }\right)^2\right] .
\label{srapproxsf3}
\end{eqnarray}
These expressions coincide with the formulas already used in the
literature~\cite{Kinney:1995cc, Martin:2003bt, Alabidi:2006qa} and it
is crucial to keep in mind that they are valid for $\mu /\mpl \ll 1$
only.

\begin{figure}
\begin{center}
\includegraphics[width=14cm]{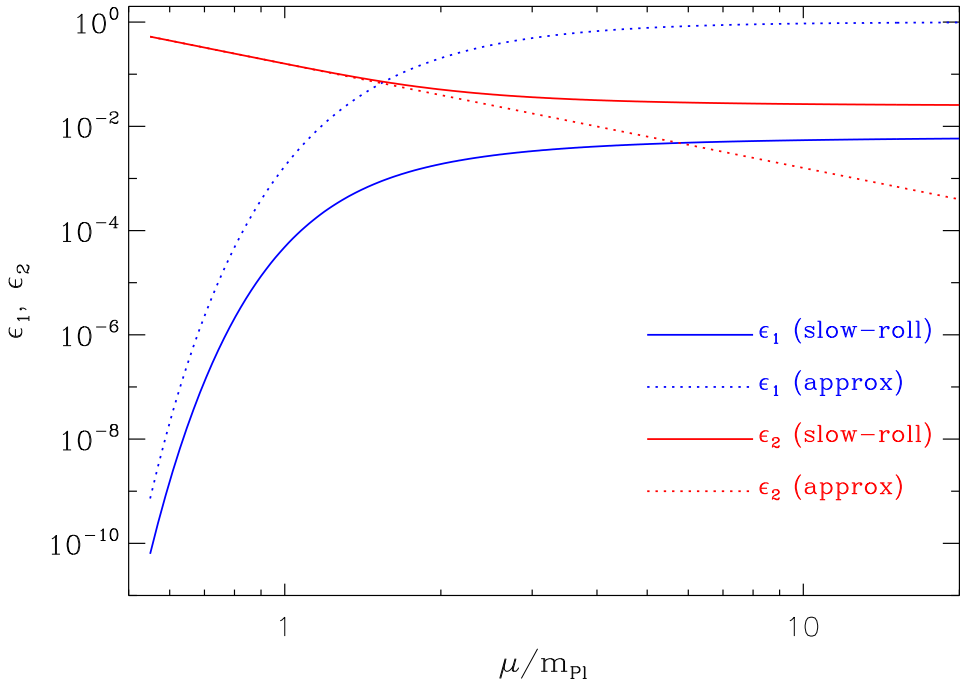} 
\caption{Slow-roll parameters $\epsilon_1$ (blue lines) and
  $\epsilon_2$ (red lines) for the small field model characterised by
  $p=2$ and $N_*=40$, plotted for different values of the scale $\mu
  /\mpl$. The dotted lines are the approximations established in
  equations (\ref{srapproxsf}) to (\ref{srapproxsf3}) whereas the
  solid lines are the slow-roll parameters expressed in terms of the
  Lambert function: the only approximation involved in that case being
  the inversion discussed in section~\ref{sec:cosmopert}.}
\label{compeps}
\end{center}
\end{figure}

In figure~\ref{compeps} the slow-roll parameters $\epsilon _1$ and
$\epsilon _2$ stemming from the approximate
equations~(\ref{srapproxsf}) and (\ref{srapproxsf2}) are represented
together with the ones obtained from the exact expression of $\phi_*$
in (\ref{phistarquadratic}). The plot is made for $N_*=40$ (and
$p=2$). One sees that as long as $\mu/\mpl \ll 1$, the two expressions
of $\epsilon _1$ and $\epsilon _2$ are in good agreement but when $\mu
\simeq \mpl $ the difference is no longer negligible. It turns out
that this difference is of utmost importance in view of the current
data because using only the equations (\ref{srapproxsf}) to
(\ref{srapproxsf3}) leads to the incorrect conclusion that the small
field model $p=2$ is ruled out. In fact, using directly the slow-roll
equations shows that this model is still compatible with the
observations. Indeed, very roughly speaking, the WMAP3 data are
compatible with $\epsilon _1< 0.03$ and $\epsilon _2\simeq 0.05$. If
one decides to use (\ref{srapproxsf2}) for $\epsilon _2$, then $\mu
\simeq 1.8 \mpl $ and inserting back this value into
(\ref{srapproxsf}), one finds $\epsilon _1\simeq 0.13$, \ie a value in
tension with the observations. The true story is quite different. As
can be seen in figure~\ref{compeps}, the slow-roll value $\epsilon
_2\simeq 0.05$ is perfectly compatible with a value for $\epsilon _1$
satisfying the WMAP3 bound. The model is thus still compatible with
the observational constraints. This conclusion is indeed confirmed by
the exact numerical integration performed in the next section. In
addition, we will demonstrate below that $\mu /\mpl \ll 1$ implies a
very low energy scale during inflation that can be, in some cases to
be discussed in the following, below the nucleosynthesis scale, \ie
already ruled out. In this situation, the small field models make
sense only if $\mu \gta \mpl $, that is to say precisely the situation
where it is necessary to carefully evaluate the values of the
slow-roll parameters.

\par

\begin{figure}
\begin{center}
\includegraphics[width=14cm]{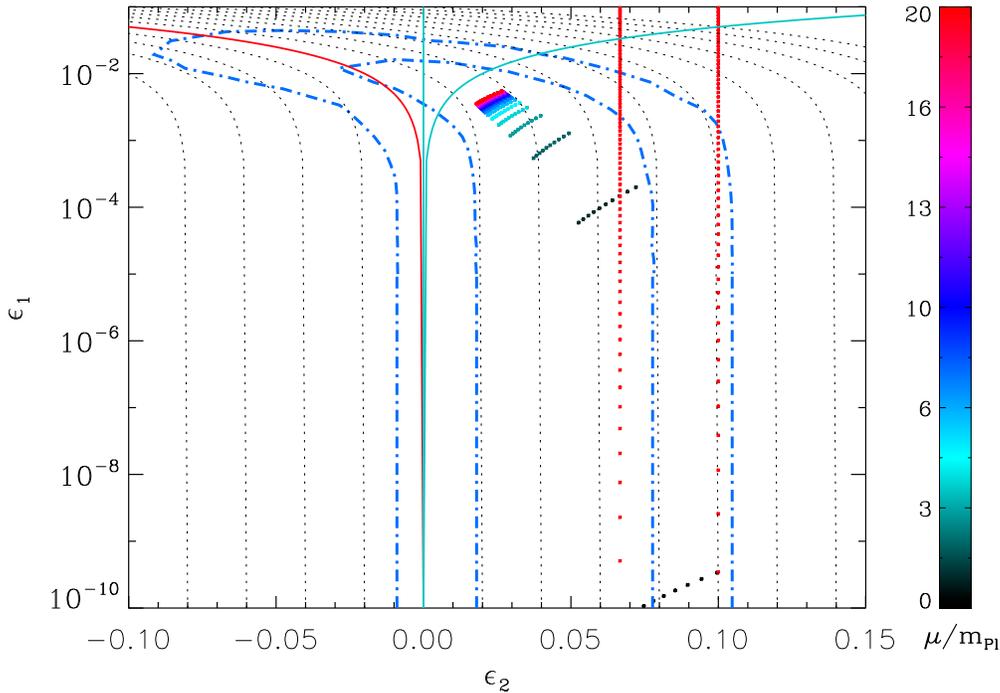}
\caption{Comparison of the slow-roll predictions for small field
  models with $p=3$ and $0.1<\mu /\mpl<20$ in the plane $(\epsilon _1,
  \epsilon _2)$. The solid red and blues lines, as well as the blue
  dotted-dashed contours, have the same meaning as in
  figure~\ref{lf}. The approximated slow-roll values are obtained
  using the equations (\ref{srapproxsfpneq2}) and
  (\ref{srapproxsfpneq2_2}) and are represented by the two red dotted
  vertical lines. The red dotted line on the left corresponds to
  models with different values of $\mu/\mpl $ at fixed $N_*=60$ while
  the red dotted line on the right is obtained for $N_*=40$. The
  correct slow-roll parameter values are obtained by numerical
  integrations and are represented by the line segments. Each line
  segment corresponds to a model with a fixed value of $\mu/\mpl $,
  indicated by the colour bar, along which the quantity $N_*$ varies
  from $40$ to $60$.}
\label{sfcomp}
\end{center}
\end{figure}

Let us now turn to the expressions of the slow-roll parameters in 
the case $p \neq 2$. If $\mu/\mpl \ll 1$, then one has
\begin{eqnarray}
\label{srapproxsfpneq2}
\epsilon _1 & \simeq & \frac{p^2}{16 \pi }\left(\frac{\mpl }{\mu }\right)^2
\left[N_*\frac{p(p-2)}{8\pi }\left(\frac{\mpl }{\mu
}\right)^2\right]^{-\frac{2(p-1)}{p-2}},\\
\label{srapproxsfpneq2_2}
\epsilon _2 & \simeq &
\frac{2}{N_*}\frac{p-1}{p-2}\,, \qquad \epsilon _3 \simeq \frac{1}{N_*}\, ,
\end{eqnarray}
which matches to the expressions usually used in the literature. On
the other hand, if the limit $\mu/\mpl \ll 1$ is not satisfied, the
slow-roll parameters can only be estimated from numerical
calculations. In figure~\ref{sfcomp}, we have precisely compared the
approximations given by the equations (\ref{srapproxsfpneq2}) and
(\ref{srapproxsfpneq2_2}) to the directly computed slow-roll
parameters in a model with $p=3$. The approximated values are
represented by the two red dotted curves, the left one corresponding
to the choice $N_*=60$ and the right one to $N_*=40$. These two curves
appear as vertical lines because the expression of $\epsilon_2$ in
(\ref{srapproxsfpneq2}) does not depend on $\mu/\mpl$. The correct
values for the slow-roll parameters obtained by numerical
determination of $\phi _*$ are represented by the line segments, $N_*$
varying from $40$ to $60$ along each segment. Although both methods
are in good agreement for $\mu/\mpl \ll 1$, they differ when $\mu \gta
\mpl $. As it was the case for $p=2$, we see that using only
(\ref{srapproxsfpneq2}) would lead us to the erroneous conclusion that
the model $p=3$ is compatible with the data only if $\mu/\mpl \ll
1$. On the contrary, the correct values of the slow-roll parameters
indicate that models with $\mu \gta \mpl $ are simply in perfect
agreement with WMAP3 data. Therefore, given the accuracy of the
current CMB data, it becomes mandatory to carefully estimate the
slow-roll parameters in the case of small field models. This will be
confirmed by the full numerical integration of the power spectrum
performed in the next section.

\begin{figure}
\begin{center}
\includegraphics[width=7.5cm]{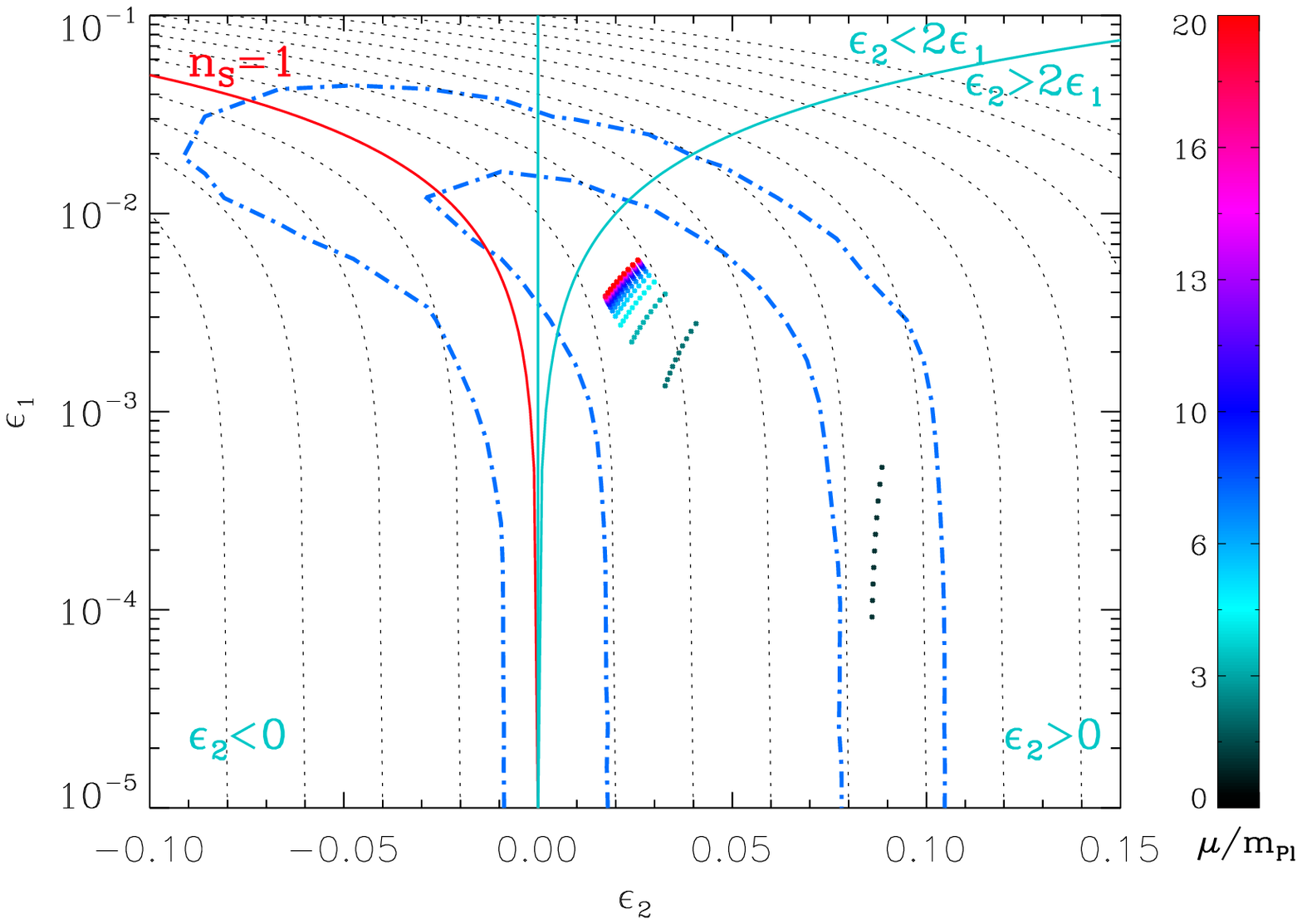}
\includegraphics[width=7.5cm]{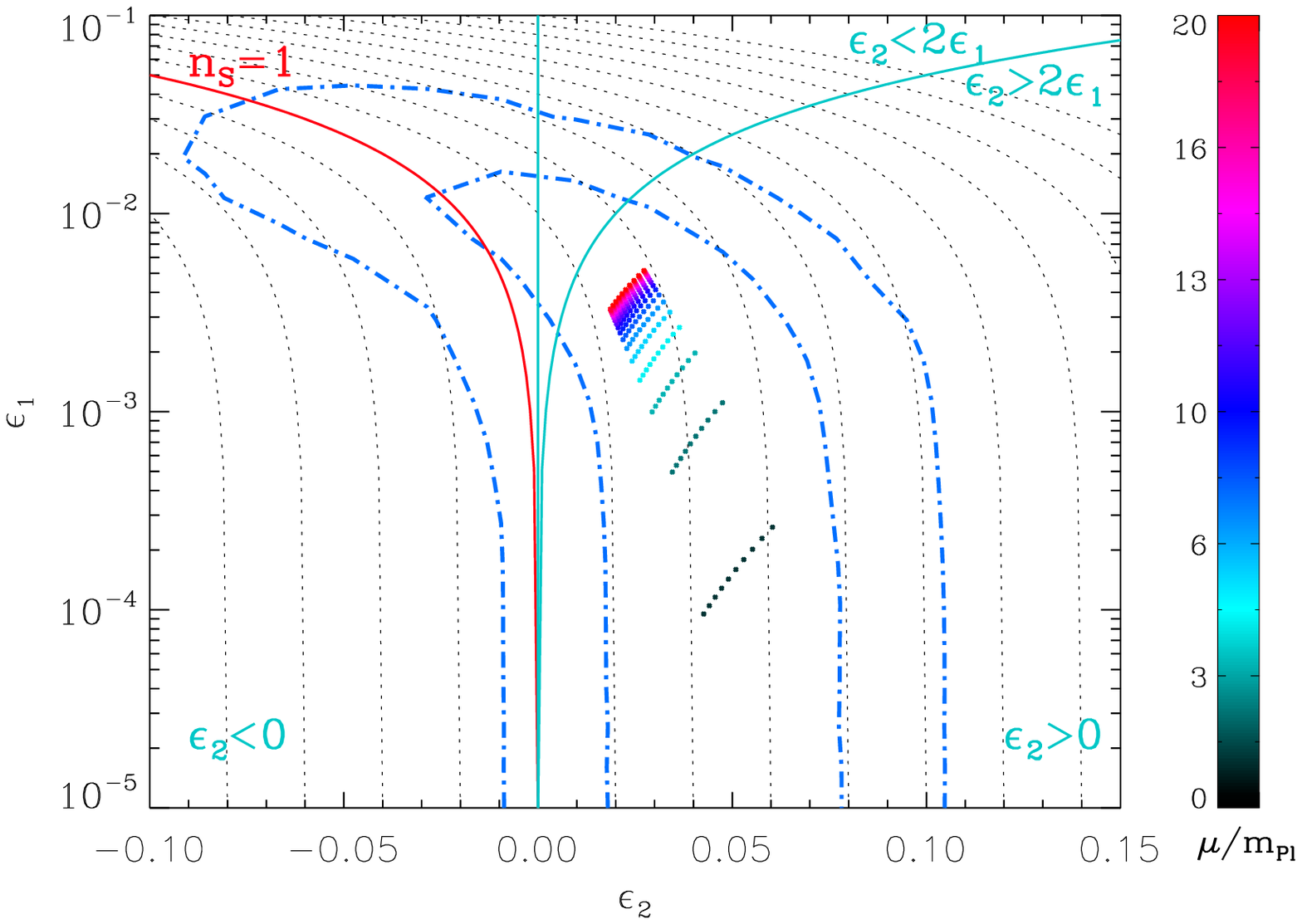}
\includegraphics[width=7.5cm]{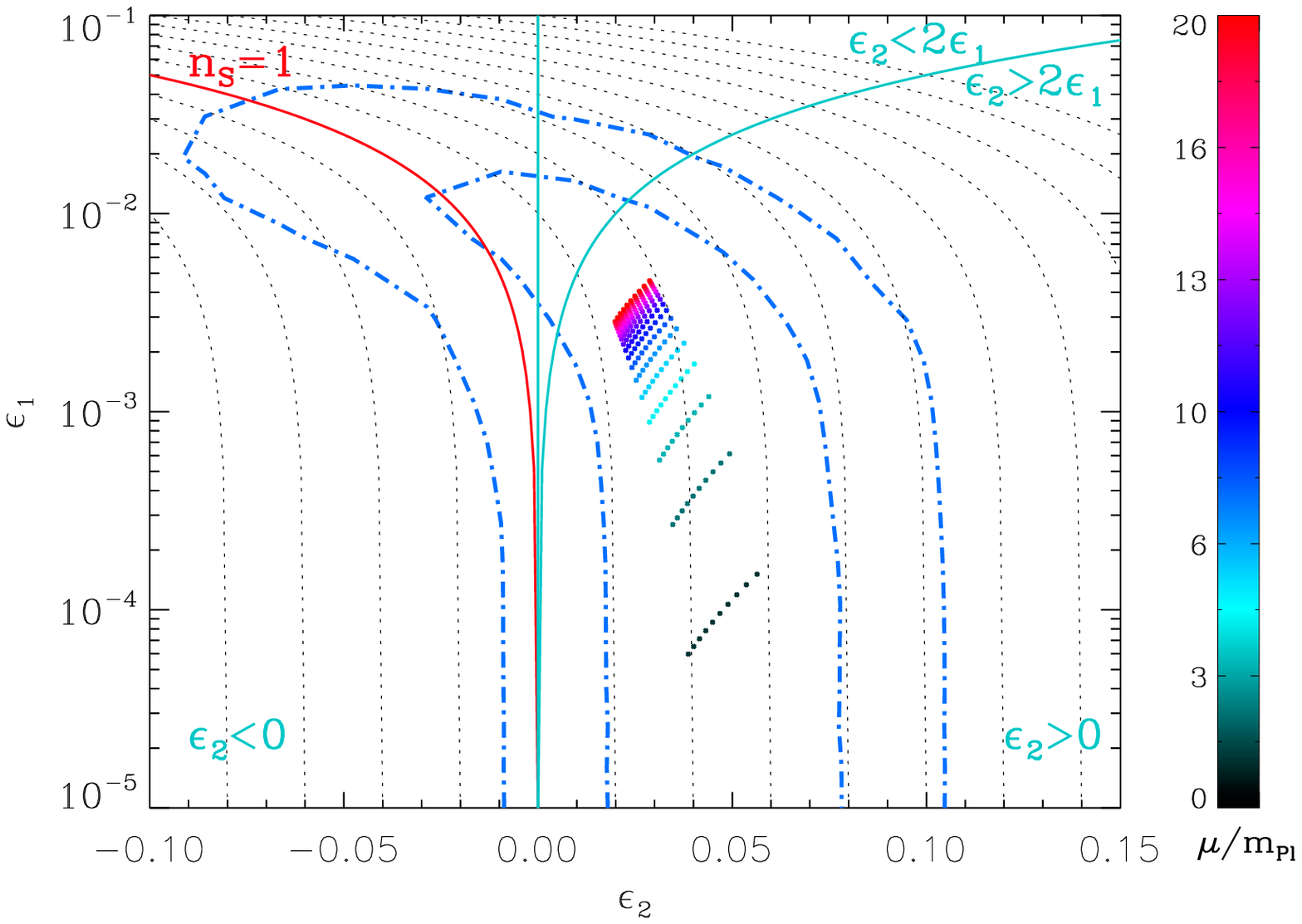}
\includegraphics[width=7.5cm]{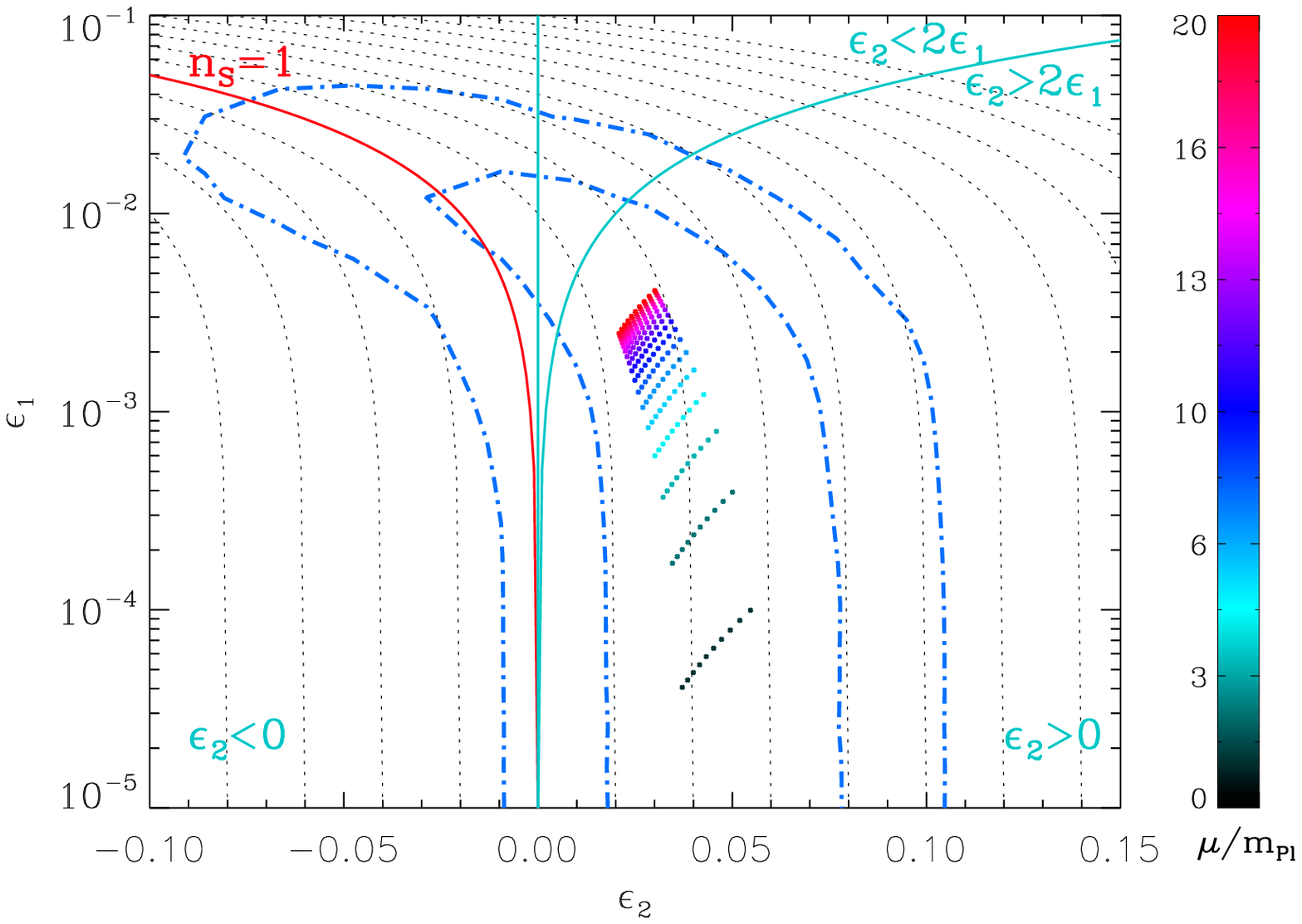}
\caption{Small field models in the plane $(\epsilon_1,\epsilon_2)$ for
  $0.2<\mu/\mpl <20$, $p=2$ (top left panel), $p=4$ (top right panel),
  $p=6$ (bottom left) and $p=8$ (bottom right). The solid red and
  blues lines, as well as the blue dotted-dashed contours have the
  same meaning as in figure~\ref{lf}. Each curved segment represents
  the slow-roll predictions along which $N_*$ varies from $40$ to
  $60$.}
\label{sf}
\end{center}
\end{figure}

Let us now more precisely consider the predictions associated with the
small field models. In figure~\ref{sf}, the slow-roll parameters
values are represented for $0.2<\mu/\mpl <20$ and for four models,
namely $p=2,4,6$ and $p=8$. As expected, the models are located in the
region where $\epsilon _2>2\epsilon _1>0$, \ie in the region where the
kinetic energy grows with time during inflation. One can see in
figure~\ref{sf} that, for any value of $p$ (including $p=2$), the
models are in good agreement with the data if $\mu \gta \mpl$. This
conclusion is confirmed by the full numerical integration done in the
next section. In fact, only the very small values of $\mu/\mpl $ are
problematic since they end up being associated with a too large
slow-roll parameter $\epsilon _2$. This can be seen in the top left
panel ($p=2$) where the line segments representing the $\mu /\mpl\lta
1$ models are only marginally compatible with the confidence intervals
or even ruled out. In fact, in the top left panel, this is not so
apparent at first sight since the models corresponding to the segment
line which lies between the one and two sigma contours are still
acceptable fits to the data. However, one has to realize that, for
even smaller values of $\mu /\mpl$, the corresponding line segments
are actually outside the figure, hence the above claim. Moreover, one
notices that the segment line has the tendency to become vertical
which indicates that the value of $\epsilon _2$ is independent from
$N_*$. This is in full agreement with equation~(\ref{srapproxsf2})
giving $\epsilon _2$.

\par

Finally, one word is in order on the CMB normalisation in those models
and the values of $\epsilon _1$. It is known that the very small
values of $\epsilon _1$ imply a very small contribution of
gravitational waves. But they also imply a quite low energy scale
during inflation. This allows us to derive a lower bound on $\epsilon
_1$. Indeed, the normalisation of the spectrum is given by
\begin{equation}
\label{normasf}
\frac{V_*}{\mpl ^4}\simeq \frac{45 \epsilon
_1}{2}\frac{Q_\mathrm{rms-PS}^2}{T^2}\, .
\end{equation}
Now, it is physically motivated that inflation must take place at an
energy scale at least greater than, say, the TeV scale. This means
$V_*/\mpl ^4\gta 10^{-64}$ and for $p=2$ (and $N_*=50$), one gets $\mu
\gta 0.25 \mpl$. Such a strong lower bound on $\mu$ is essentially due
to the exponential behaviour of the slow-roll parameter with respect
to the model parameters. In fact, this condition can equally be worked
out for $p\ne2$. For instance, with $p=2.1$ one obtains $\mu \gta 0.03
\mpl$, $p=2.5$ leads to $\mu \gta 7\times 10^{-6} \mpl$, $p=3$ to $\mu
\gta 3\times 10^{-9} \mpl$ and $p=10$ corresponds to $\mu \gta 8\times
10^{-21} \mpl$. These results sustain the remark made before
motivating a careful determination of the slow-roll parameters in the
regime $\mu\gta \mpl$.

\subsection{Hybrid inflation}
\label{sec:hybmodel}

\begin{figure}
\begin{center}
\includegraphics[width=7.5cm]{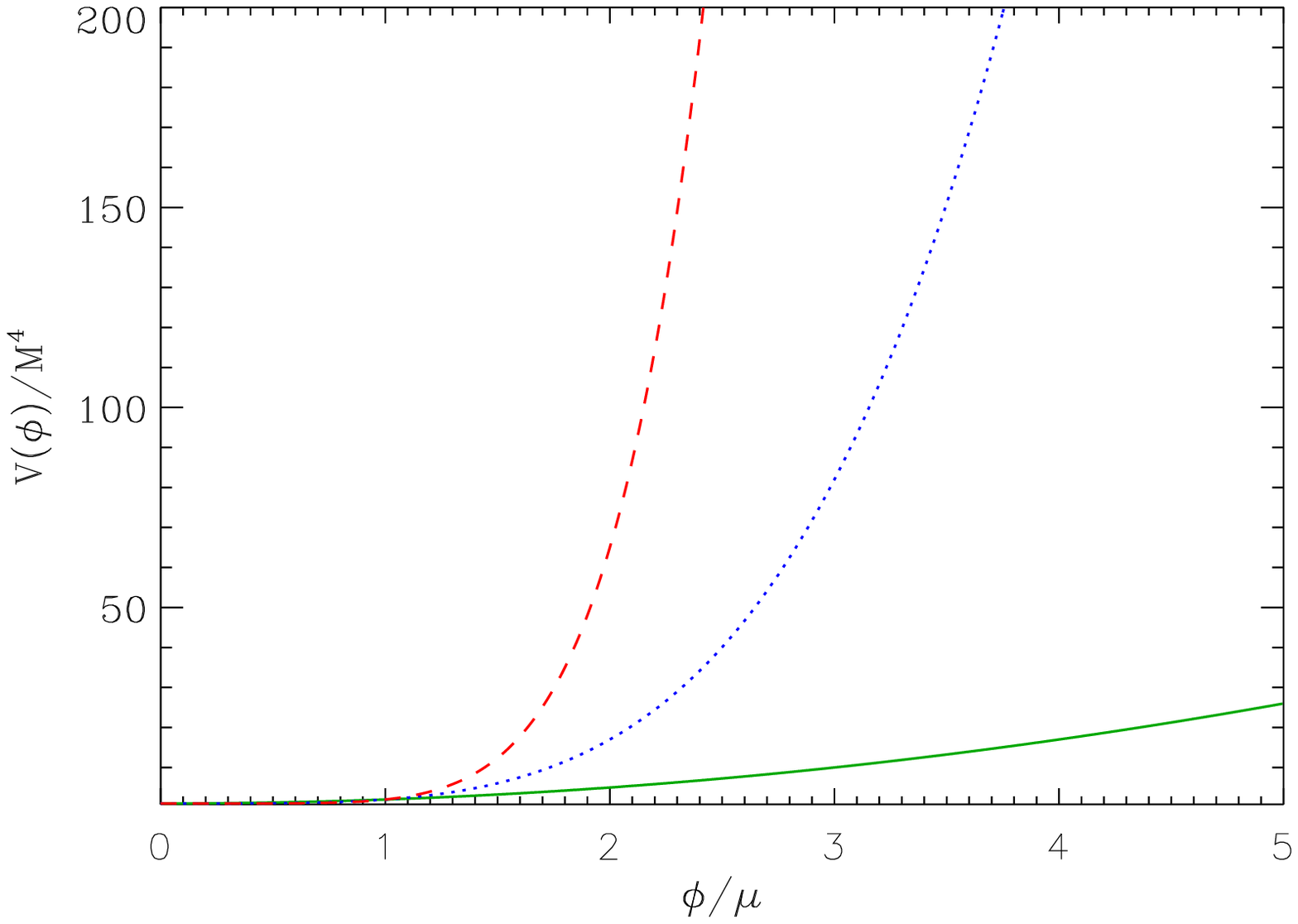}
\includegraphics[width=7.5cm]{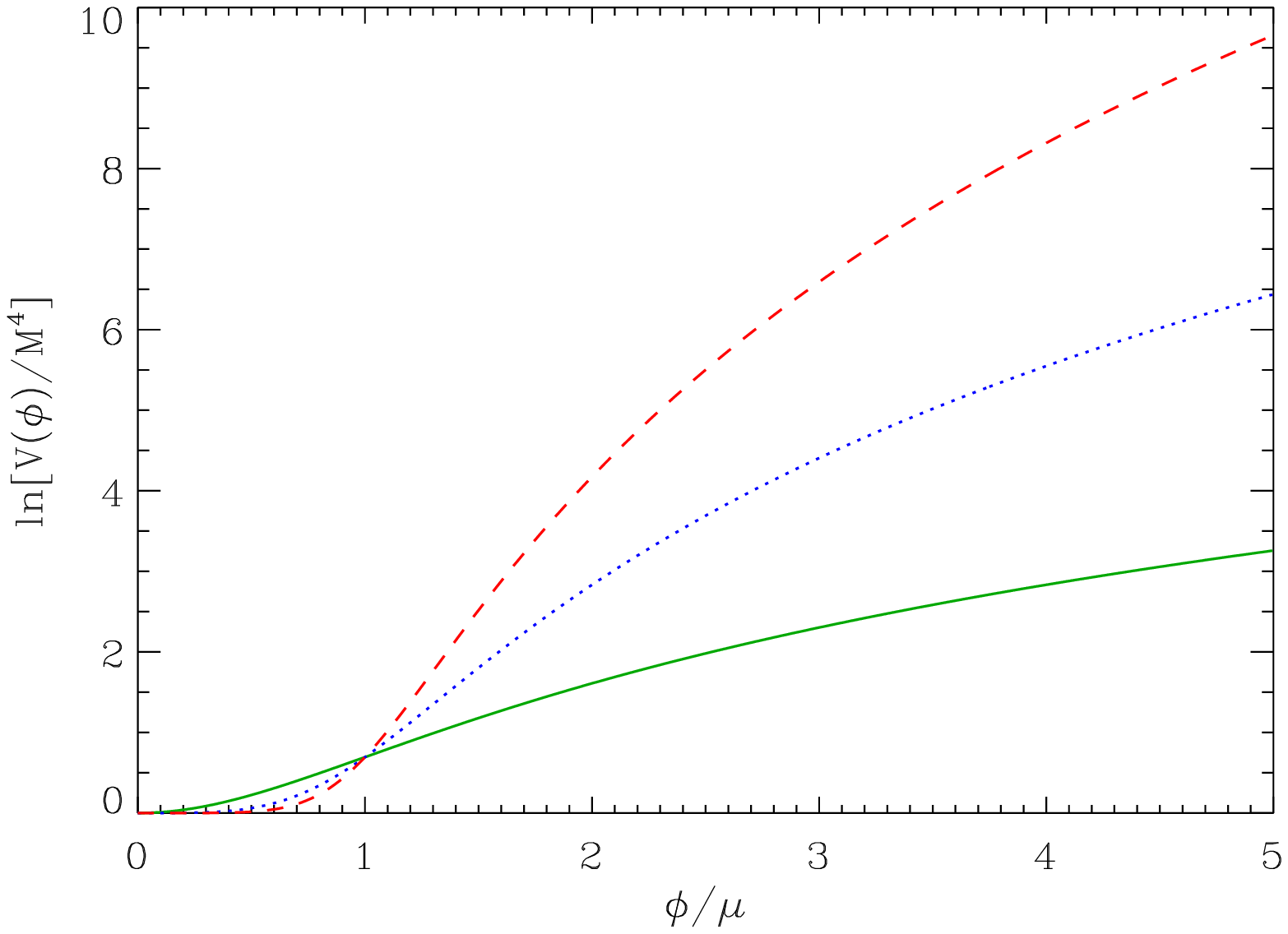}
\includegraphics[width=7.5cm]{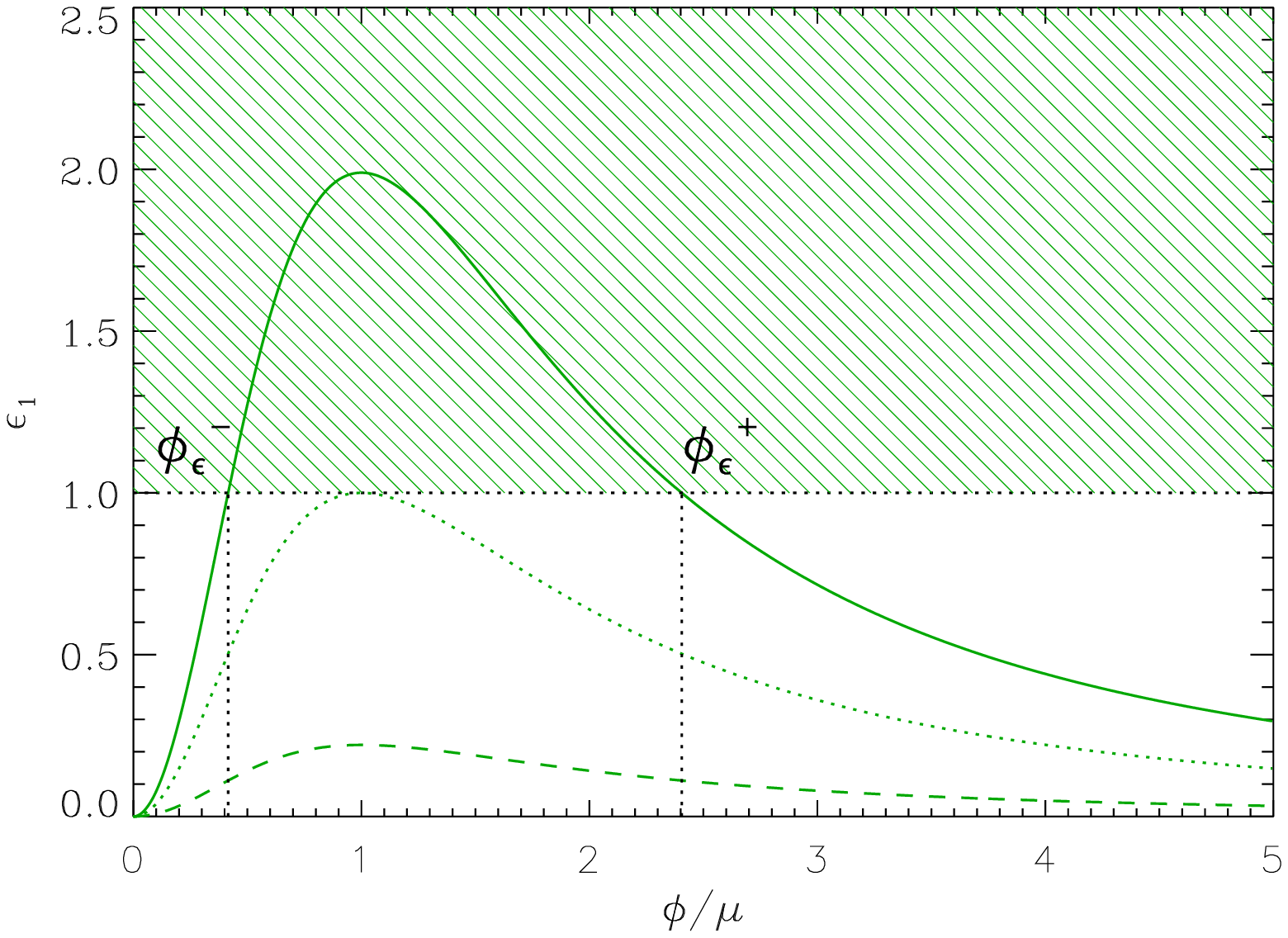}
\includegraphics[width=7.5cm]{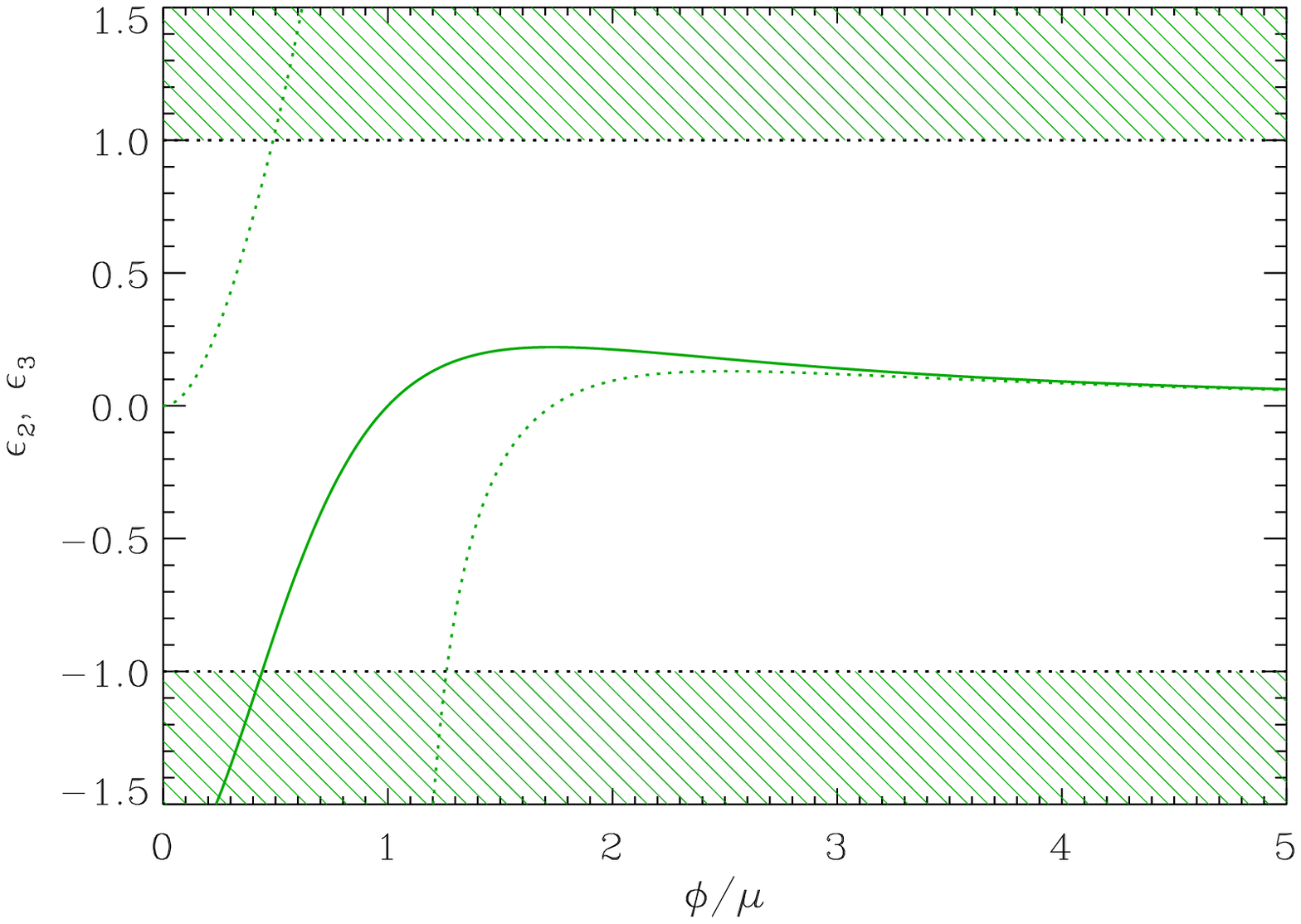}
\caption{Top left panel: hybrid field potential for $p=2$ (solid green
  line), $p=4$ (blue dotted line) and $p=6$ (red dashed line). Top
  right panel: logarithm of the potential for $p=2,4,6$ (same colour
  code as in the top left panel). Bottom left panel: slow-roll
  parameter $\epsilon _1$ for $p=2$ and $\mu/\mpl=0.1$ (solid line),
  $\mu /\mpl =1/(4\sqrt{\pi })\simeq 0.14$ (dotted line) and $\mu
  /\mpl=0.3$ (dashed line). Bottom right panel: slow-roll parameter
  $\epsilon _2$ (solid line) and $\epsilon _3$ (dotted line) both for
  $\mu/\mpl =0.3$.}
\label{pothyb}
\end{center}
\end{figure}

Let us now turn to hybrid inflation. This case is slightly different
since hybrid inflation is in fact a two-field model with the
potential~\cite{Linde:1991km,Copeland:1994vg,Lyth:1998xn}
\begin{equation}
\label{twofieldpot}
V\left(\phi ,\psi \right)=\frac12 m^2\phi ^2+\frac{\lambda '
}{4}\left(\psi ^2-\Delta ^2\right)^2+\frac{\lambda }{2}\phi ^2\psi
^2\, ,
\end{equation}
where $\phi $ is the inflaton, $\psi $ the waterfall field and
$\lambda ' $ and $\lambda $ are two coupling constants. Inflation
proceeds along the valley given by $\psi =0$ and, in this case, the
potential reduces to an effective single field potential that can be
written as
\begin{equation}
\label{potentialhyb}
V(\phi)=M^4\left[1 +\left(\frac{\phi}{\mu}\right)^{p}\right] ,
\end{equation}
with $p=2$ and where one has used the following redefinitions
\begin{equation}
\label{paramhybrid}
M=\frac{\lambda '{}^{1/4}\Delta }{\sqrt{2}}\, ,\qquad \mu
    =\sqrt{\frac{\lambda ' }{2}}\frac{\Delta ^2}{m}\, .
\end{equation}
As for small field models, the effective
potential~(\ref{potentialhyb}) depends on three parameters, namely
$M$, $\mu $ and $p$. In fact, as mentioned before, $p=2$ for the two
field model given in (\ref{twofieldpot}) but, to consider the most
general situation we leave $p$ unspecified in the following
equations. Indeed, one could consider a model where the inflationary
valley is described by, say, a quartic potential since the instability
mechanism is independent from the shape of the valley. Moreover, in
the second part of this article, when we perform the full numerical
analysis, we will consider $p$ as a free parameter and obtain its
corresponding posterior probability distribution. Finally, let us
emphasize that since we use a single field modelisation of hybrid
inflation, such a representation cannot account for multifield effects
as the generation of isocurvature modes or cosmic
strings~\cite{Ringeval:2001xd, Rocher:2004et, Ringeval:2005yn,
  Bastero-Gil:2006cm}. The potential~(\ref{potentialhyb}) and its
logarithm are represented in figure~\ref{pothyb} for different values
of the power $p$.

\par

The calculation of the slow-roll parameters proceeds as usual. The
explicit expressions are given by
\begin{eqnarray}
\label{epsilonhyb}
  \epsilon_1 &=& \frac{p^2}{16\pi}\left(\frac{\mpl}{\mu }\right)^2
  \frac{(\phi/\mu)^{2p-2}}{\left[1+(\phi/\mu)^p\right]^2}\, , \\
\label{epsilonhyb2}
  \epsilon_2 & = & \frac{p}{4\pi}\left(\frac{\mpl}{\mu }\right)^2
  \left(\frac{\phi }{\mu }\right)^{p-2}
  \frac{(\phi/\mu)^{p}-p+1}{\left[1+(\phi/\mu)^p\right]^2}\, , \\
\label{epsilonhyb3}
  \epsilon_3 &=& \frac{p}{8\pi}\left(\frac{\mpl}{\mu }\right)^2
  \frac{\left(\phi/\mu\right)^{p-2}}{\left[1+\left(\phi/\mu\right)^p\right]^2
    \left[\left(\phi/\mu\right)^p-p+1\right]} \left[2\left(\frac{\phi
    }{\mu }\right)^{2p}\nonumber \right. \\ & - &
    \left. (p-1)(p+4)\left(\frac{\phi }{\mu }\right)^p
    +(p-1)(p-2)\right]\, .
\end{eqnarray}
The three parameters are plotted in figure~\ref{pothyb} in the case
$p=2$. We see that the parameter $\epsilon _1$ has a maximum at
$\phi/\mu =1$ which corresponds to the inflexion point of $\ln V$. If
$\mu <1/(4\sqrt{\pi })$ then $\epsilon _1<1 $ for all values of $\phi
$. We come back to this point below when we discuss how to stop
inflation. Another specific feature of hybrid inflation in comparison
to large and small field models is that, as can be seen on the bottom
right panel, the parameters $\epsilon _2$ and $\epsilon _3$ can be
negative. In particular
\begin{equation}
\lim_{\phi/\mu \rightarrow 0} \epsilon _2 = - \frac{p(p-1)}{4\pi} \left(
\frac{\mpl}{\mu}\right)^2 \left(\frac{\phi}{\mu}\right)^{p-2},
\end{equation}
and $\epsilon _3$ blows up in the limit $(\phi/\mu )^p \rightarrow
p-1$.

\par

We now discuss how inflation ends and how to calculate $\phi
_{\uend}$. In the hybrid scenario there are \emph{a priori} two
mechanisms for ending inflation. Either inflation stops by instability
when the inflaton reaches a value
\begin{equation}
\label{cri}
\phi _{\ucri}=\frac{\lambda ' }{\lambda }\Delta\, ,
\end{equation}
for which the mass in the direction perpendicular to the inflationary
valley becomes negative, or the slow-roll conditions are violated and
$\epsilon_1=1$. The latter condition happens for a field value $\phi
_\epsilon$ solution of
\begin{equation}
\label{eqphiendhybrid}
\left(\frac{\phi_\epsilon}{\mu}\right)^{2p}+2
\left(\frac{\phi_\epsilon}{\mu}\right)^p-\frac{p^2}{16\pi }
\left(\frac{\mpl}{\mu }\right)^2
\left(\frac{\phi_\epsilon}{\mu}\right)^{2p-2}+1=0\, .
\end{equation}
The difference between this equation and (\ref{eqphiend}) resides only
in the sign of the $\left(\phi_\epsilon/\mu \right)^p$ term. This
equation admits a solution only if the following condition is
fulfilled: $\mu/\mpl<1/(4\sqrt{\pi })$. In figure~\ref{pothyb}, such
situations correspond to the case where $\epsilon _1$ can be greater
than one. If $\mu/\mpl>1/(4\sqrt{\pi })$ then $\epsilon _1$ is always
smaller than one and inflation can only stop by instability. The above
equation (\ref{eqphiendhybrid}) cannot be solved explicitly unless
$p=2$. In this case, one obtains
\begin{equation}
\label{epshybrid}
\frac{\phi_\epsilon}{\mu} = \frac{1}{4\sqrt{\pi }}\frac{\mpl }{\mu }
\left(1\pm \sqrt{1-\frac{16\pi\mu^2}{\mpl^2}}\right).
\end{equation}
Of course, one recovers the fact that the solutions exist only if
$\mu/\mpl<1/(4\sqrt{\pi })$. The positive sign corresponds to the
largest root, $\phi _\epsilon ^+$, while the minus sign corresponds to
the smallest one, $\phi _\epsilon ^-$ (see figure~\ref{pothyb}). In
the limit $\mu /\mpl \ll 1$ the previous equation takes the form
\begin{equation}
\label{approx2hybrid}
\frac{\phi_\epsilon}{\mu}\simeq \frac{\mpl}{2\mu \sqrt{\pi }}\, ,
\end{equation}
and one recovers the large field value of $\phi_\uend$ [see for
instance equation~(\ref{phiendlf})]. More generally, in the limit of
small $\mu/\mpl$, one can even approximately solve
(\ref{eqphiendhybrid}) for $p\ne2$. However, contrary to the small
field models case and besides the term proportional to $p^2$, one
should keep the term $\left(\phi_\epsilon/\mu\right)^{2p}$ rather than
$1$. This leads to
\begin{equation}
\frac{\phi_\epsilon }{\mu}\simeq \frac{p\mpl }{4\mu \sqrt{\pi }}\, ,
\end{equation} 
which is the large field model expression of $\phi _\uend$ for a
potential with an arbitrary power $p$ in the field. Therefore, the
final value $\phi _\uend$ of the inflaton is the maximum of $\phi
_{\ucri}$ and $\phi _{\epsilon}$. To decide which mechanism is
realised in practice requires the knowledge of the model
parameters. However, one crucial interest of hybrid inflation is that
inflation can proceed for small values of the inflaton vev. As we have
seen before, if the inflaton vev is large, then the model is
equivalent to a large field model which was already considered in a
previous subsection. Therefore, in the following, we will be focused
on hybrid inflation taking place for $\phi <\phi _\epsilon ^-$ only:
the so-called vacuum dominated regime. Since $\epsilon_1$ is always
lower than unity in such cases (see figure~\ref{pothyb}), hybrid
inflation must stop by instability. As a result, $\phi _{\ucri}<\phi
_\epsilon ^-$ will be considered in the following as a free parameter
accounting for a total of four inflationary parameters.

\par

The next step consists in calculating the classical
trajectory. Straightforward manipulations similar to the ones
performed for small field models, lead to 
\begin{equation}
N= \kappa^2 \frac{\mu ^2}{2p}\,\left\{
\left(\frac{\phi_\ini}{\mu}\right)^2 - \left(\frac{\phi }{\mu
}\right)^2 - \frac{2}{p-2}\left[\left(\frac{\phi_\ini}{\mu
  }\right)^{2-p} - \left(\frac{\phi }{\mu
  }\right)^{2-p}\right]\right\}\, ,
\end{equation}
for $p\neq2$, where $\kappa \equiv \sqrt{8\pi }/\mpl $ is the reduced
Planck mass. For $p=2$ one has
\begin{equation}
N=\kappa^2 \frac{\mu
  ^2}{4}\left[\left(\frac{\phi_\ini}{\mu}\right)^2 - \left(\frac{\phi
  }{\mu }\right)^2 -2\ln \left(\frac{\phi }{\phi _\ini}\right)\right],
\end{equation}
which can be inverted to express $\phi$ in terms of the number of
e-folds:
\begin{equation}
  \frac{\phi }{\mu} = \sqrt{W_0
  \left\{\left(\frac{\phi_\ini}{\mu}\right)^{2} \exp
  \left[\left(\frac{\phi_\ini}{\mu }\right)^2-\frac{N}{2\pi}
  \left(\frac{\mpl }{\mu }\right)^2\right] \right\}}\,.
\end{equation}
Once again, $W_0(x)$ denotes the principal branch of the Lambert
function~\cite{Valluri:2000aa}. The above expression is very similar
to equation~(\ref{trajecsfquadratic}), except for the signs.

As for the other models, the last step consists in determining the
link between $N_*$ and $\phi _*$. For $p\neq 2$, using the classical
trajectory obtained before, one has to solve the following equation
\begin{eqnarray}
\left(\frac{\phi _*}{\mu }\right)^2 -\frac{2}{p-2} \left(\frac{\phi
  _*}{\mu }\right)^{2-p} & = &\frac{pN_*}{4\pi
}\left(\frac{\mpl}{\mu}\right)^2 + \left(\frac{\phi _\ucri}{\mu
}\right)^2 \nonumber \\ & - & \frac{2}{p-2} \left(\frac{\phi
  _\ucri}{\mu }\right)^{2-p},
\end{eqnarray}
where $\phi_\ucri$ and $N_*$ are known from the previous steps. In
general, this equation can only be solved numerically. However, if
$p=2$, one gets
\begin{equation}
\label{phistarhyb}
  \frac{\phi_*}{\mu} = \sqrt{ W_0
  \left\{\left(\frac{\phi_\ucri}{\mu}\right)^{2}
  \exp\left[\left(\frac{\phi_\ucri}{\mu
  }\right)^2+\frac{N_*}{2\pi}
  \left(\frac{\mpl}{\mu}\right)^2\right]\right\}} .
\end{equation}

The slow-roll parameters for $p=2$ are then obtained by inserting the
above equation into the expressions~(\ref{epsilonhyb}) to
(\ref{epsilonhyb3}). We therefore have an explicit form for the three
parameters $\epsilon_1$, $\epsilon _2$ and $\epsilon _3$ which become
explicit functions of $\mu $, $\phi _{\ucri}$ and $N_*$. In the
so-called vacuum dominated regime we are interested in, the term
$(\phi/\mu)^2$ in the expression of the potential~(\ref{potentialhyb})
ends up being a small correction only. Note that in the opposite
situation, \ie the inflaton dominated regime, the model is equivalent
to chaotic inflation which was already treated in the section devoted
to large fields models. This question is discussed in more details
below.

\par

\begin{figure}
\begin{center}
\includegraphics[width=7.5cm]{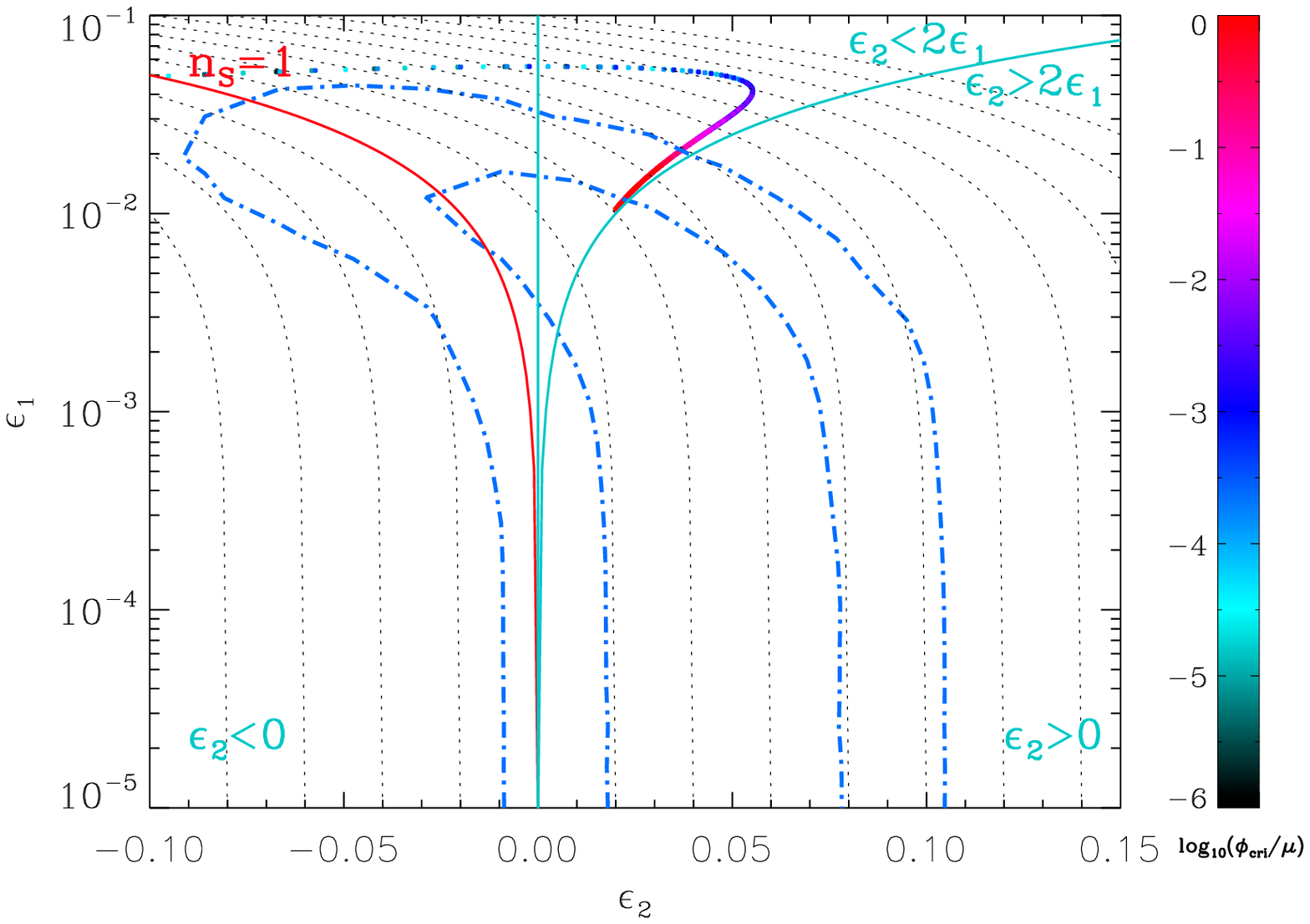}
\includegraphics[width=7.5cm]{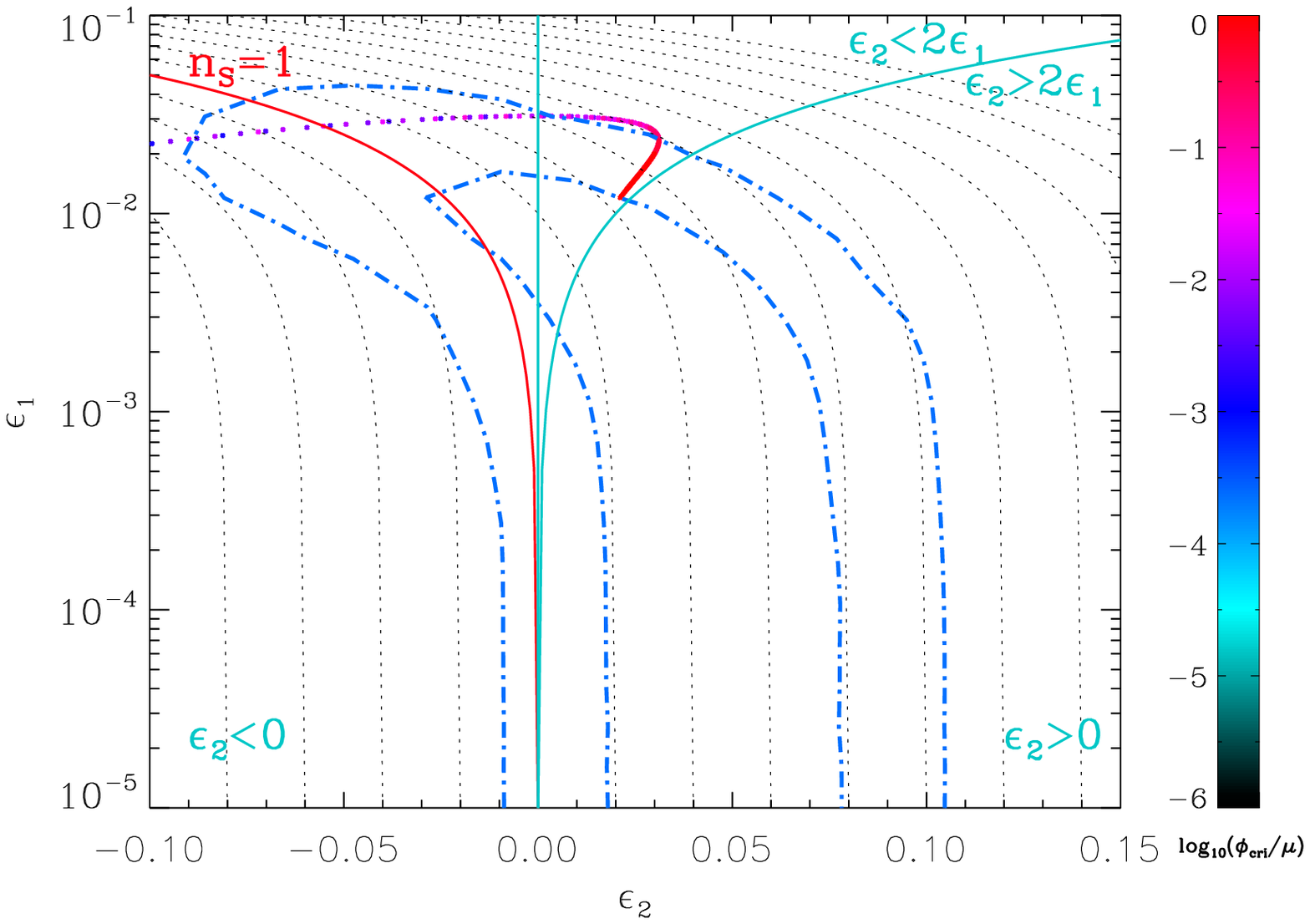}
\includegraphics[width=7.5cm]{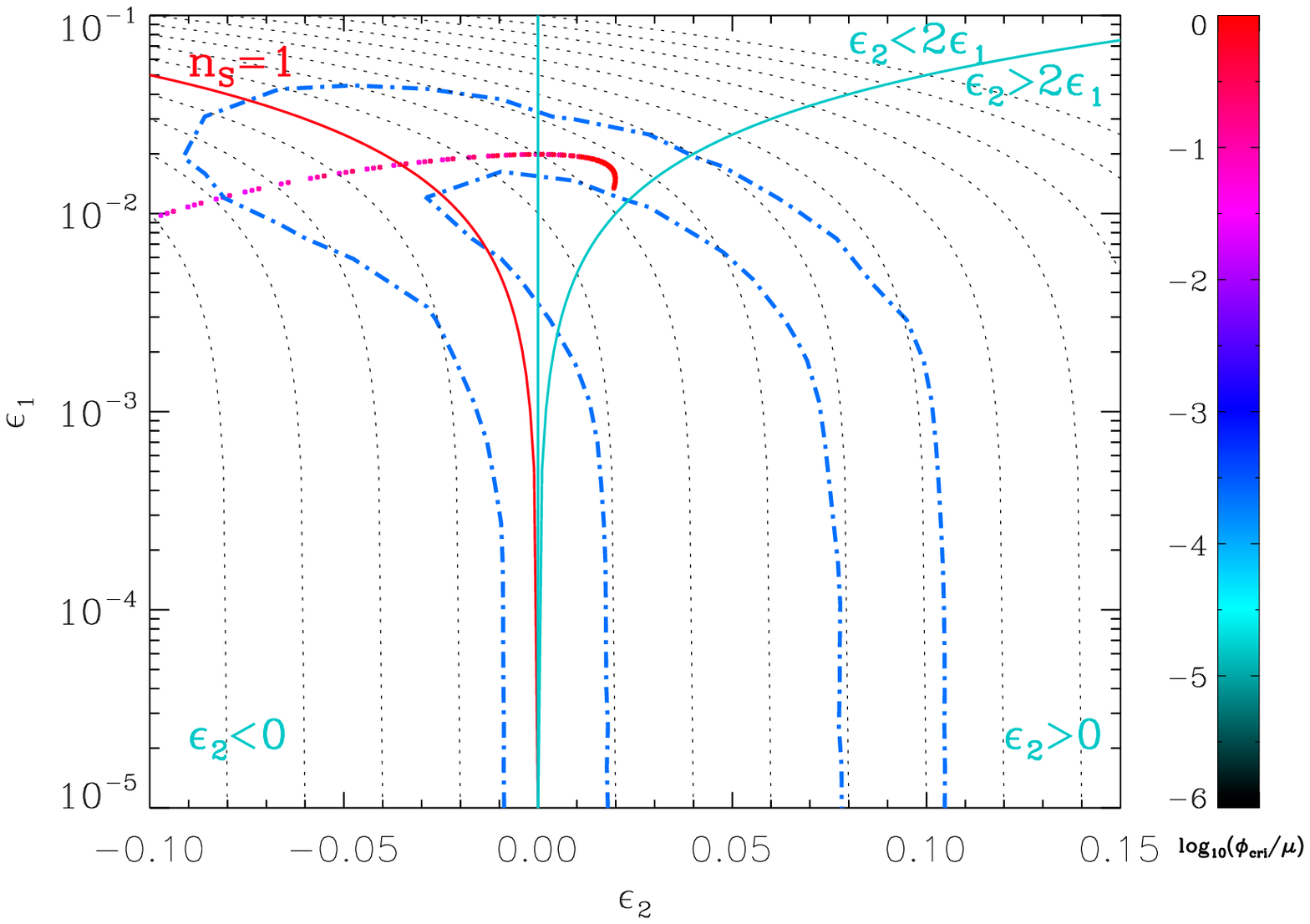}
\includegraphics[width=7.5cm]{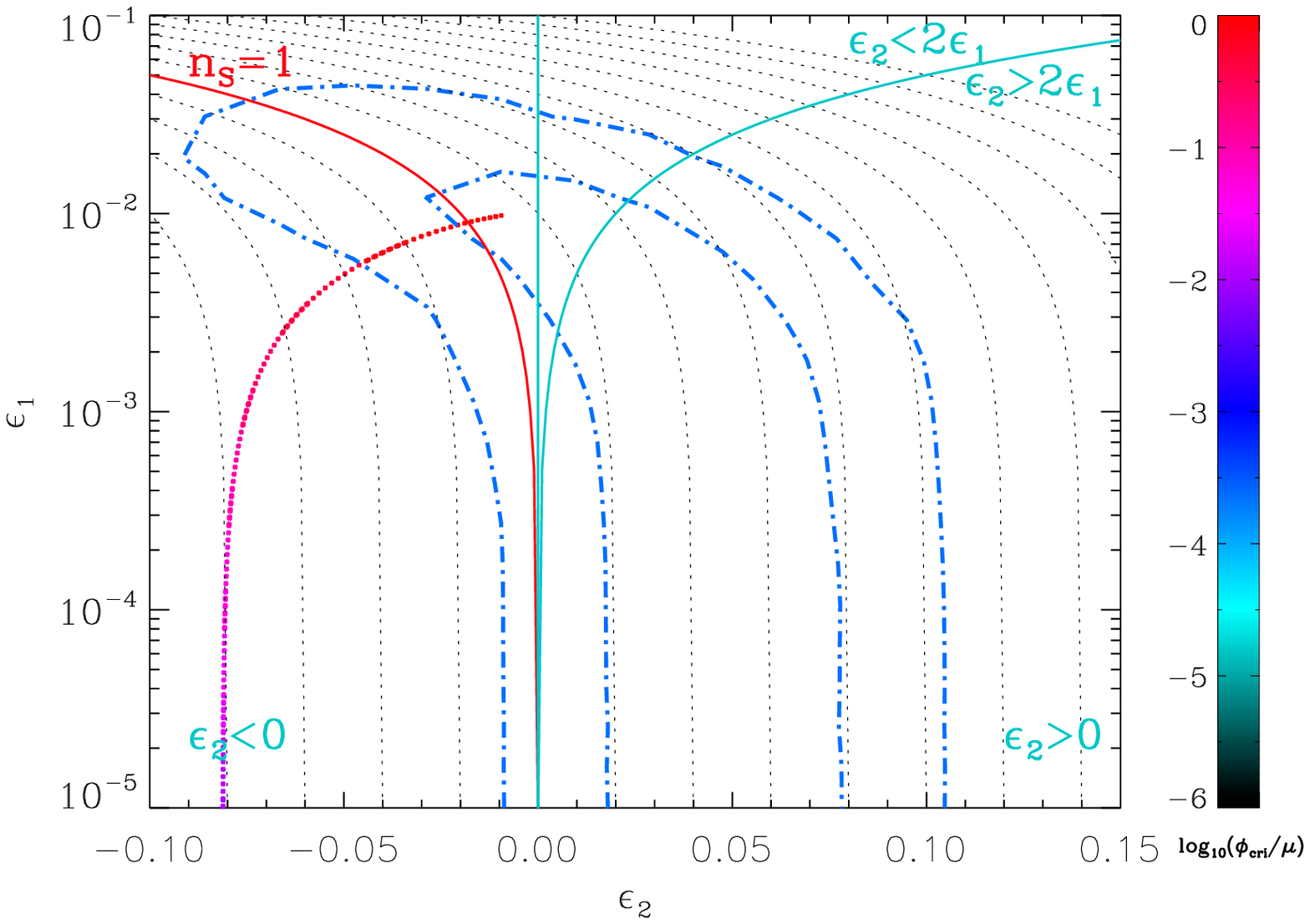}
\includegraphics[width=7.5cm]{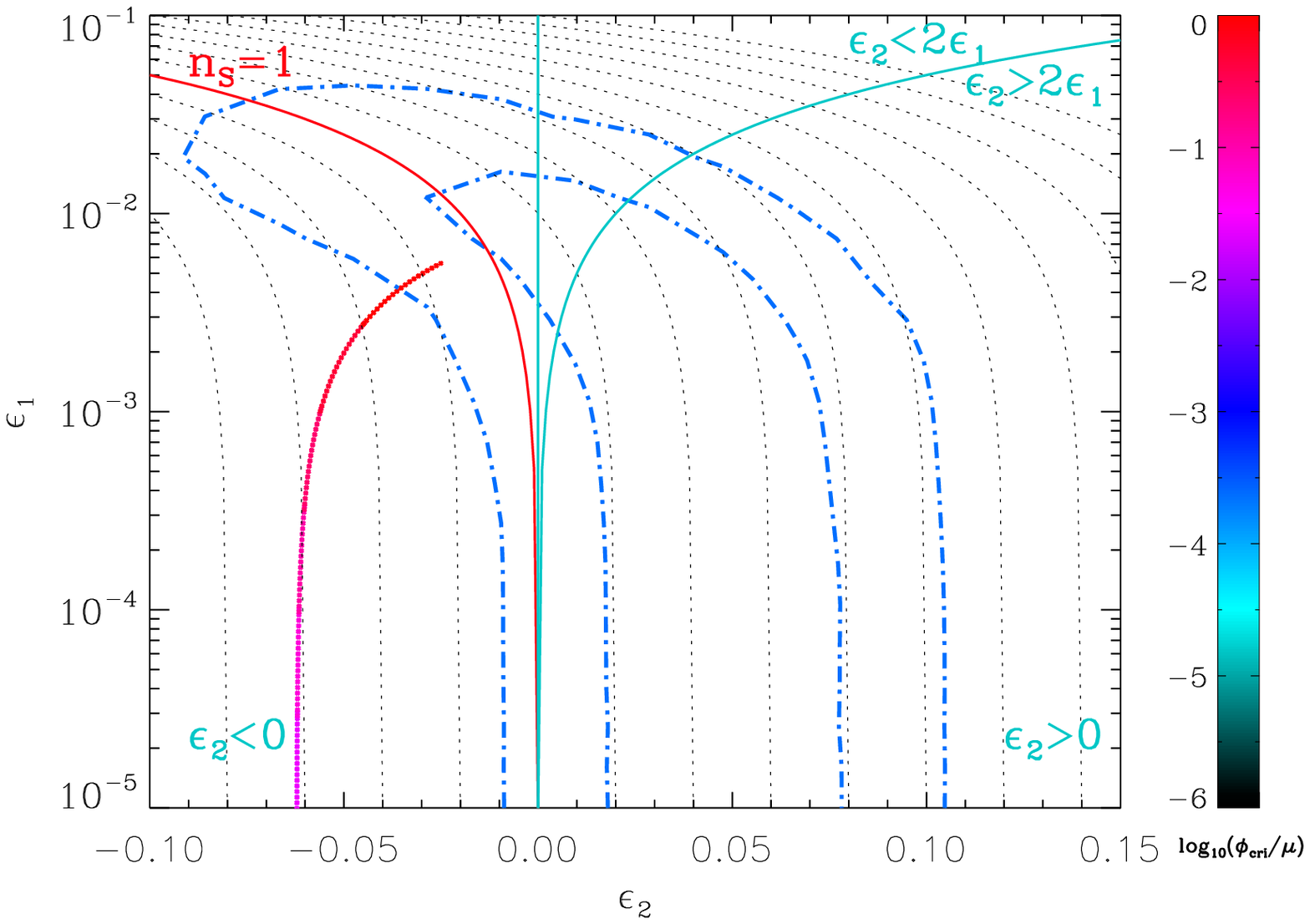}
\includegraphics[width=7.5cm]{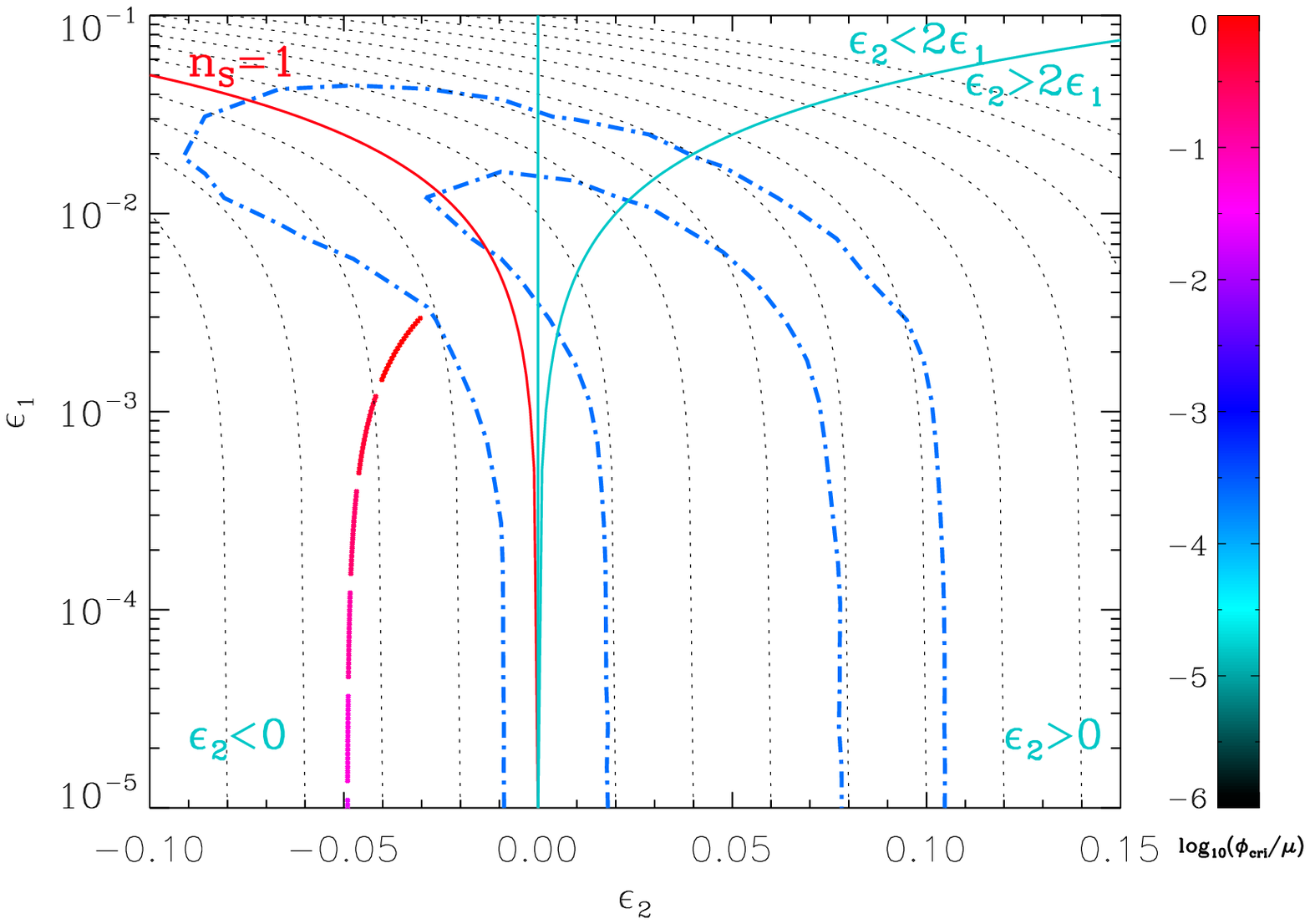}
\caption{Top left panel: slow-roll predictions for hybrid inflation
  with $p=2$, $\mu /\mpl=0.6$, $10^{-6}<\phi
  _{\ucri}/\mu<10^{-1}$. The colour bar indicates the value of
  $\log(\phi_{\ucri}/\mu )$ and the number of e-folds $N_*$ is between
  $40$ and $60$. The other panels display different values of $\mu
  /\mpl $, the other parameters being unchanged: $\mu /\mpl=0.8$ (top
  right), $\mu /\mpl=1$ (middle left), $\mu /\mpl=1.4$ (middle right),
  $\mu /\mpl=1.6$ (bottom left) and $\mu /\mpl=1.8$ (bottom
  right). The solid red and blues lines, as well as the blue
  dotted-dashed contours, have the same meaning as in
  figure~\ref{lf}.}
\label{hyb}
\end{center}
\end{figure}

In Fig.~\ref{hyb}, we have plotted the slow-roll predictions in the
plane $(\epsilon _1,\epsilon _2)$ for different values of $\mu /\mpl $
and $10^{-6}<\phi _{\ucri}/\mu <10^{-1}$. The values of $\mu /\mpl $
that we consider go from $\mu /\mpl =0.6$ (top left panel) to $\mu
/\mpl =1.8$ (bottom right panel). Let us now discuss in more details
these plots.

For $\mu /\mpl =0.6$, some models lie in the region $\epsilon _2>0$
and others in the region $\epsilon _2<0$. The first ones corresponds
to $\phi_{\ucri}/\mu $ close to the upper limit chosen for the plot,
namely $10^{-1}$ while the second ones are valid for small values as
one can check on the colour bar. The interpretation is as
follows. When the value of $\phi_{\ucri}$ is close to its upper bound,
inflation proceeds in a regime where the term $(\phi /\mu)^2$ in the
potential~(\ref{potentialhyb}) is dominant. In this case, the model
behaves as a quadratic large field model and this explains why the
models are concentrated along the line $\epsilon _2=2\epsilon _1$.
When the value of $\phi _{\ucri}/\mu $ becomes smaller, inflation can
proceeds for very small values of the inflaton vev and one enters the
vacuum dominated regime which is the main characteristic of hybrid
inflation. Consequently, as can be seen in figure~\ref{hyb}, the
points representing the corresponding models leave the large field
region $\epsilon _2<2\epsilon _1$ and penetrate in the region
$\epsilon _2<0$. In this region, the kinetic energy decreases
absolutely but also relatively with the total energy density. With
$\mu /\mpl =0.6$, these models are now disfavored by the WMAP3 data at
more than $2\sigma $.

\par

However, the situation can be different when one starts to increase
the value of $\mu /\mpl $. As we see in the top right panel of
figure~\ref{hyb}, and even more clearly in the remaining panels, the
models are no longer equivalent to quadratic large field models and
the line $\epsilon _2=2\epsilon _1$ is never reached. This effect is
amplified as $\mu /\mpl$ increases since then, the relative importance
of the term $(\phi /\mu )^2$ is diminished. These remarks are
nevertheless valid provided the range of $\phi _{\ucri}$ is kept
fixed. Clearly, as $\mu/\mpl $ increases, one could always increase
the upper limit on $\phi _{\ucri}/\mu $ to counter-balance the
above-mentioned effect. As can be seen in the left and right middle
panels, all the models are almost concentrated in the vacuum dominated
region $\epsilon _2<0$ and have the tendency to produce a blue tilted
power spectrum, a standard characteristic of hybrid inflation. Some
models with $\epsilon _2<0$ are still perfectly compatible with the
data as can be seen in the middle right panel. Then, as $\mu/\mpl $ is
further increased, the models become excluded due to their too high
blue spectral index (bottom left and bottom right panel). Again, in
this case, to re-obtain models in agreement with the data, it would be
necessary to modify the upper bound on $\phi _{\ucri}/\mu$. The
parameter space of hybrid inflation and its compatibility with the CMB
data is directly explored in the following section where the power
spectra are numerically integrated.

\subsection{Running-mass inflation}
\label{sec:rmmodel}

\begin{figure}
\begin{center}
\includegraphics[width=7.5cm]{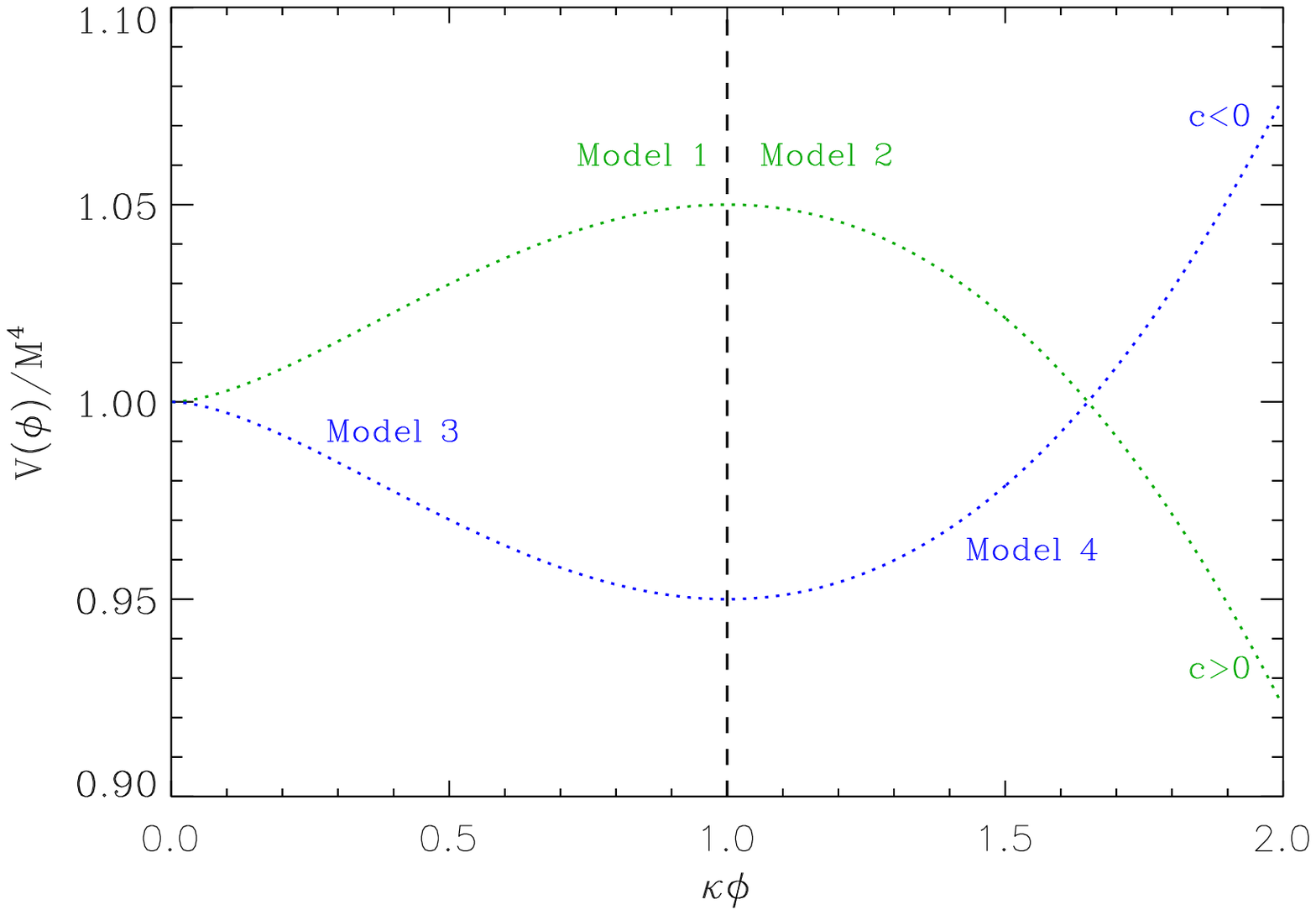}
\includegraphics[width=7.5cm]{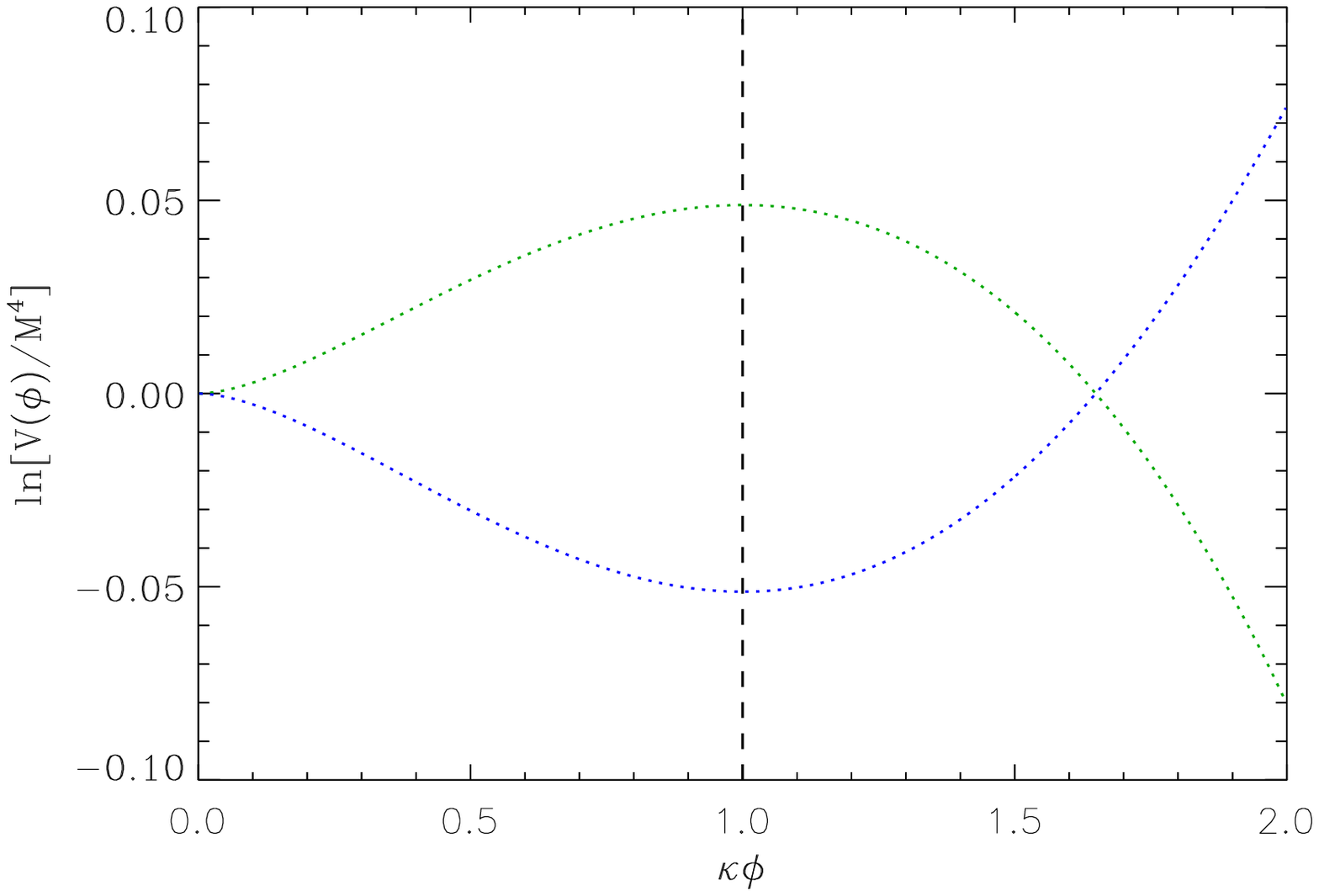}
\includegraphics[width=7.5cm]{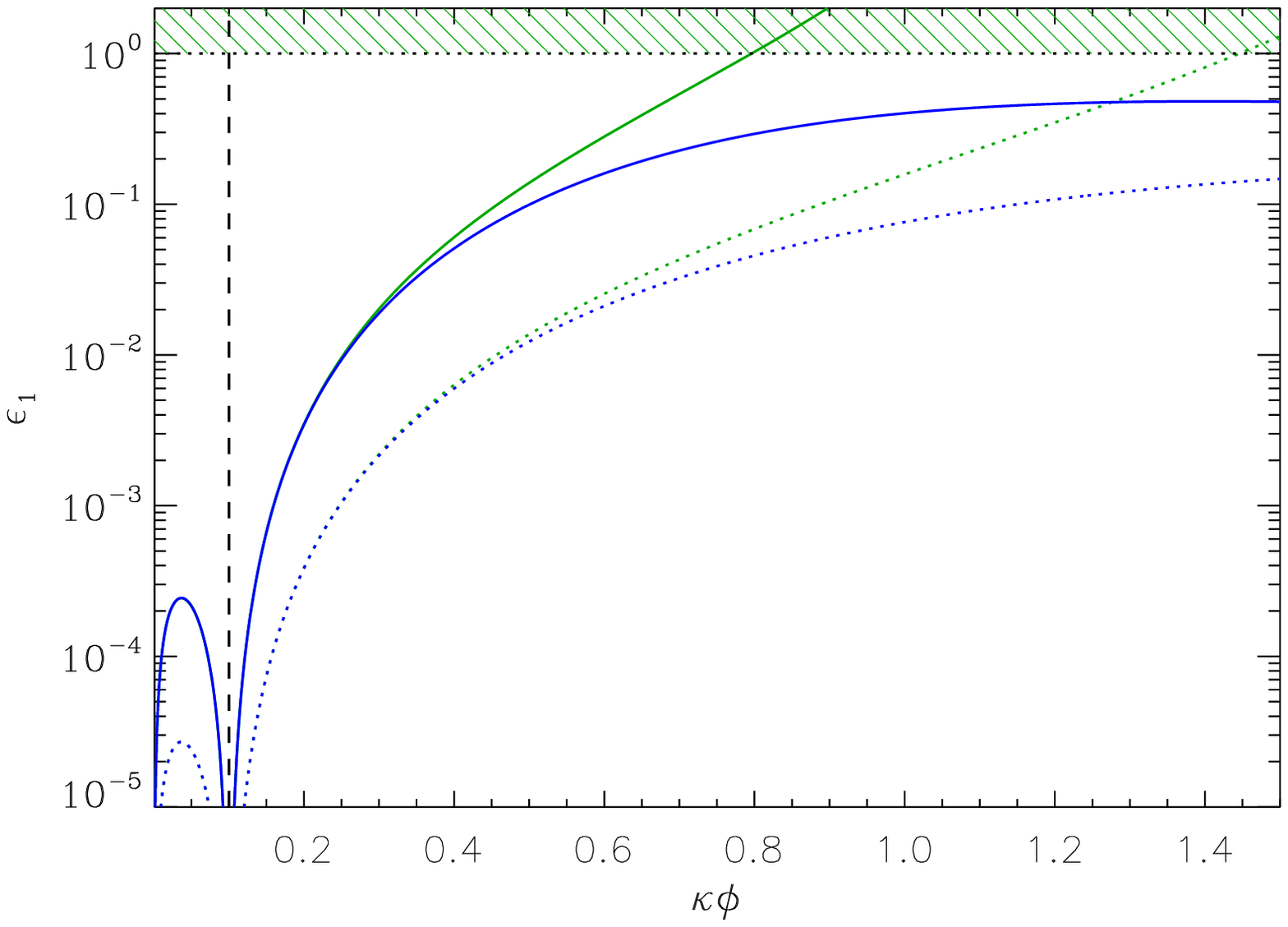}
\includegraphics[width=7.5cm]{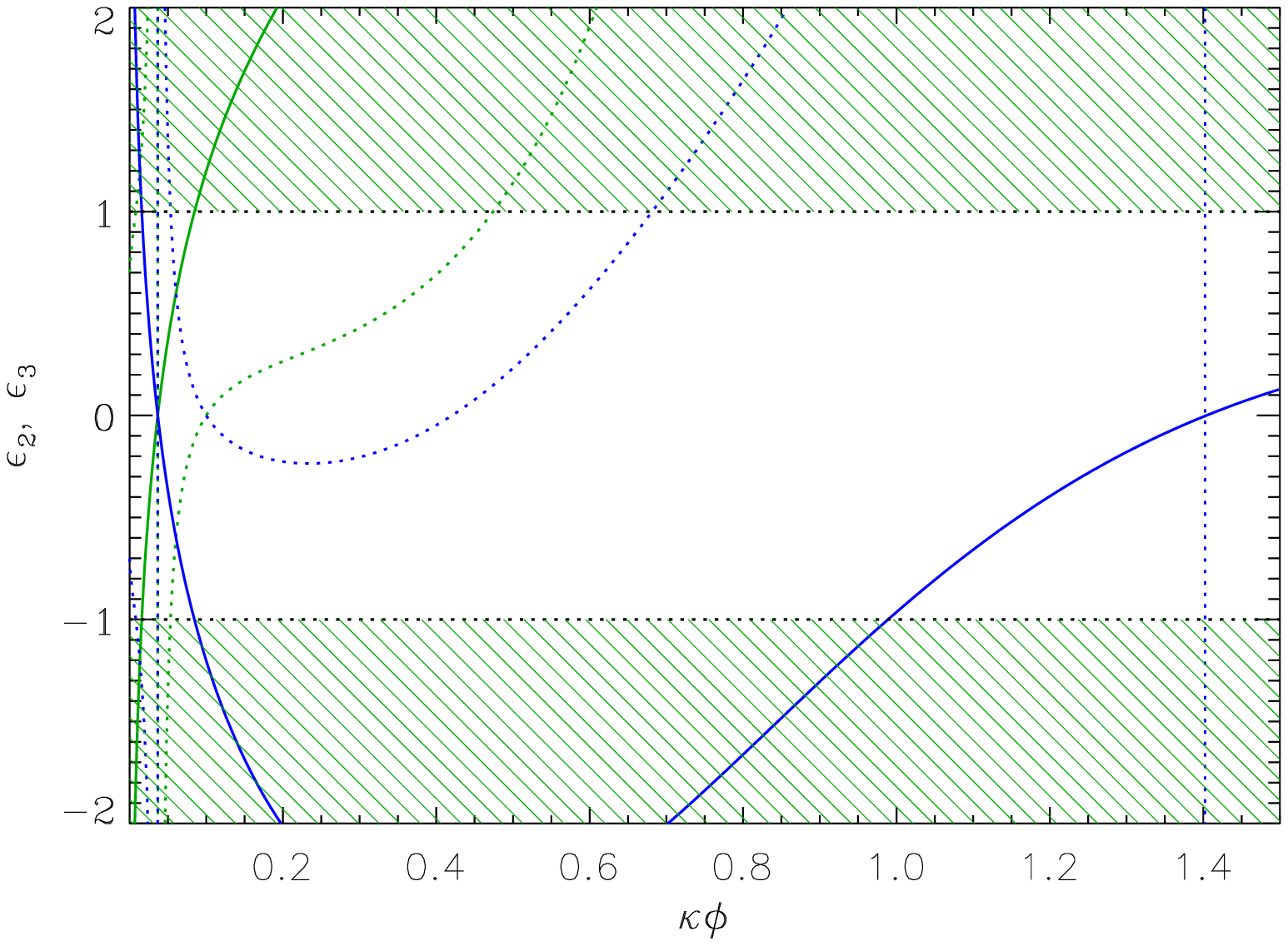}
\caption{Top left panel: running mass potentials for $\kappa \phi_0=1$
  and $c=0.2$ (dotted green line) and $c=-0.2$ (dotted blue line). The
  vertical dashed line indicates the position of $\phi _0$. Top right
  panel: logarithm of the potential (same colour code).  Bottom left
  panel: slow-roll parameter $\epsilon_1$ for $\kappa \phi_0=0.1$ and
  $c=0.6$ (solid green line), $c=0.2$ (dotted green line), $c=-0.6$
  (solid blue line) and $c=-0.2$ (dotted blue line) with $\kappa \phi
  _0=0.1$. Bottom right panel: slow-roll parameters $\epsilon_2$ and
  $\epsilon _3$ for $c=0.6$ (blue lines, solid for $\epsilon _2$,
  dotted for $\epsilon _3$) and for $c=-0.6$ (green lines, solid for
  $\epsilon _2$, dotted for $\epsilon _3$) with $\kappa \phi _0=0.1$.}
\label{potrunning}
\end{center}
\end{figure}

The last type of model that we consider is the running-mass model
(RM)~\cite{Covi:1998mb,Covi:2002th,Covi:2004tp}. Using the same
parametrisation as before, the potential reads
\begin{equation}
V(\phi) = M^4\left[1-\lambda \left(-\frac{1}{2} +\ln \frac{\phi
  }{\mu}\right)\frac{\phi ^2}{\mu ^2} \right],
\end{equation}
which is a function of $\phi /\mu $ only. This potential is
characterised by three free parameters, $M$, $\mu$ and $\lambda
$. However it turns out to be more convenient to use a slightly
different parametrisation (essentially to facilitate the comparison
with the existing literature). For this reason, we re-write $\lambda $
as $\lambda /\mu ^2\equiv \kappa ^2c/2$. Then, the potential takes the
form
\begin{equation}
\label{potentialrunning}
  V(\phi) = M^4\left[1-\frac{c}{2}\left(-\frac{1}{2} +\ln
\frac{\phi }{\phi _0}\right)\kappa ^2\phi ^2\right],
\end{equation}
where $\phi_0 \equiv \mu$. In this expression, $M$, $c $ and $\phi _0$
are free parameters. Let us recall that $c$ can be positive or
negative~\cite{Covi:2002th} while $\phi=\phi _0$ is an extremum of
$V(\phi )$, a maximum if $c>0$ and a minimum if $c<0$. The potential
and its logarithm are represented in figure~\ref{potrunning}.

\par

Running mass inflation can be realised in four different
ways~\cite{Covi:1998mb}, denoted as RM1 to RM4, according to where the
vev of the inflaton field is located along the potential, see
figure~\ref{potrunning} (top panels). RM1 corresponds to the case
where $c>0$ and $\phi <\phi_0$. In this case, $\phi $ decreases during
inflation. RM2 also corresponds to $c>0$ but with $\phi >\phi_0$ and
$\phi $ increases during inflation. RM3 refers to the situation where
$c<0$ and $\phi <\phi _0$ all the time. In this case, $\phi $
increases during inflation. Finally, RM4 has $c<0$ and $\phi >\phi_0$
decreases as inflation proceeds. Using the
potential~(\ref{potentialrunning}), one can calculate the three
slow-roll parameters $\epsilon _1$, $\epsilon _2$ and $\epsilon
_3$. Their explicit expression read
\begin{eqnarray}
\label{eps1running}
\epsilon _1 &=& \frac{c^2}{2}\kappa ^2\phi ^2 \ln ^2 \frac{\phi }{\phi
  _0}\left[1-\frac{c}{2}\left(-\frac12+\ln \frac{\phi }{\phi_0}\right)
  \kappa ^2\phi ^2 \right]^{-2}\, ,\\
\label{eps2running}
\epsilon _2 &=& 2c \left[1+\frac{c}{4}\kappa ^2\phi ^2
  +\left(1-\frac{c}{4}\kappa ^2\phi ^2\right)\ln \frac{\phi }{\phi _0}
  +\frac{c}{2}\kappa ^2\phi ^2\ln ^2\frac{\phi }{\phi _0} \right]
\nonumber \\ & \times & \left[1-\frac{c}{2}\left(-\frac12+\ln
  \frac{\phi }{\phi _0}\right)\kappa ^2\phi ^2\right]^{-2},\\
\label{eps3running}
\epsilon _3 &=& c\ln \frac{\phi }{\phi _0} \left[1+\frac{c}{4}\kappa
  ^2\phi ^2 +\left(1-\frac{c}{4}\kappa ^2\phi ^2\right)\ln \frac{\phi
  }{\phi _0} +\frac{c}{2}\kappa ^2\phi ^2 \ln ^2\frac{\phi }{\phi _0}
  \right]^{-1}\nonumber \\ & \times & \left[1+\frac{c}{2}\kappa ^2\phi
  ^2+\frac{c^2}{16} \kappa ^4\phi ^4+c\left(2\kappa ^2\phi
  ^2+\frac{c}{2}\kappa ^4\phi ^4\right)\ln \frac{\phi }{\phi _0}
  \right. \nonumber \\ & + & \left. c \left(3\kappa ^2\phi ^2 -
  \frac{c}{2}\kappa ^4\phi ^4\right) \ln ^2 \frac{\phi }{\phi _0} +
  \frac{c^2}{2}\kappa ^4\phi ^4\ln ^3 \frac{\phi }{\phi _0}\right]
\nonumber \\ & \times & \left[1-\frac{c}{2}\left(-\frac12+\ln
  \frac{\phi }{\phi _0}\right)\kappa ^2\phi ^2\right]^{-2} .
\end{eqnarray}
The slow-roll parameters are represented in the bottom panels in
figure~\ref{potrunning}.

\par

Let us now study how inflation stops in these models. \emph{A priori},
the end of inflation is found from the condition $\epsilon _1=1$ and
for this reason, it is interesting to look at the behaviour of
$\epsilon _1$ in more details (see the bottom left panel in
figure~\ref{potrunning}). First of all, one notices that $\epsilon
_1=0$ for $\phi =\phi _0$ (at $0.1/\kappa$ in this plot) which marks
the limit between RM1 and RM3 on one side and RM2 and RM4 on the other
hand. Secondly, we also remark that the curves corresponding to a
model with the same value of $\vert c\vert $ are almost identical when
$\kappa \phi \ll 1$. This comes from equation~(\ref{eps1running})
where, in this limit, the denominator approaches unity while the
numerator depends on $c^2$ only. If we approximate the denominator by
one, then the maximum of $\epsilon _1$ in the range $0<\phi <\phi_0$
is located at $\phi \simeq \phi _0/e$ for which $\epsilon _1 \simeq
c^2\kappa ^2 \phi _0^2/(2e^2) \ll 1$, assuming the physical values of
the parameters $c$ and $\phi _0$ we are interested in. This means
that, for RM1 and RM3, one always has $\epsilon _1\ll 1$ and inflation
cannot stop by violation of the slow-roll conditions. Therefore, one
must use another mechanism and, naturally, we will consider that
inflation ends by instability at some critical value $\phi _{\ucri}$.

\par

In the regime where $\phi >\phi _0$, corresponding to RM2 and RM4, the
curves representing $\epsilon _1$ for different $c$ but the same
$\vert c\vert $ separate. From (\ref{eps1running}), one sees that this
is due to the influence of the denominator. In the bottom left panel
in figure~\ref{potrunning}, one notices that, for RM4, the situation
is very similar to what was discussed before, \ie inflation cannot
stop due to lower than unity $\epsilon _1$ values. Hence, one must
also use the instability mechanism for this model.

\par

Finally, it remains RM2. For this model, the inequality $\epsilon
_1<1$ can be violated and inflation could stop normally. In practice,
this happens for large values of $\kappa \phi $ and meanwhile the
other slow-roll parameters have already reached values greater than
one meaning that the slow-roll approximation has already broken
down. However, as long as $\epsilon _1<1$, inflation is still
proceeding. In the following, we will also assume that inflation stops
by instability (inverted hybrid mechanism).

\par

In summary, for the four running mass models, we always consider that
inflation ends at some value $\phi _{\ucri}$ which is therefore viewed
as an additional free parameter. Let us notice that, even if $\epsilon
_1\ll 1$ while inflation is proceeding, one could have $\epsilon
_2\gta 1$ at some point~\cite{Covi:1998mb}. In such a case, inflation
would not stop but the slow-roll approximation would break down when
$\epsilon _2=1$. As a result, the running mass inflation models under
scrutiny account for a total of four primordial parameters.

\par

We now turn to the calculation of the slow-roll parameters. Our first
step is to obtain the classical trajectory, that is to say the number
of e-folds in terms of the vev of the inflaton field. In the case of
the running-mass model~(\ref{potentialrunning}), it reads
\begin{eqnarray}
\label{eq:rmefold}
N &=& \frac{1}{c}\left(\ln \left\vert\ln \frac{\phi }{\phi
  _0}\right\vert-\ln \left\vert\ln \frac{\phi _\ini}{\phi
  _0}\right\vert\right) -\frac{1}{4}\left(\kappa ^2 \phi ^2-\kappa
^2\phi _\ini^2\right) \nonumber \\ & + &\frac{1}{4}\left(\kappa \phi
_0\right)^2\left[ {\rm Ei}\left(2\ln \frac{\phi }{\phi _0}\right)
  -{\rm Ei}\left(2\ln \frac{\phi _\ini}{\phi _0}\right)\right].
\end{eqnarray}
where the exponential integral
function~\cite{Abramovitz:1970aa,Gradshteyn:1965aa} is defined by
${\rm Ei}(x)\equiv -\int _{-x}^{+\infty} {\dd}t\e ^{-t}/t$.  This
expression cannot be inverted explicitly. However, in the limit
$\kappa \phi \ll 1$ (which is necessary to theoretically justify the
shape of the running mass potential), the above expression can be
approximated by
\begin{equation}
\label{Napproxrun}
N\simeq \frac{1}{c}\left(\ln \left\vert \ln \frac{\phi }{\phi
    _0}\right\vert -\ln \left\vert \ln \frac{\phi _\ini}{\phi
    _0}\right\vert \right). 
\end{equation}
This form allows an explicit expression of the inflaton vev as a
function of the number of e-folds, namely
\begin{equation}
\label{trajecrunning}
\phi \left(N\right) =\phi _0\exp\left({\rm e}^{cN}\ln
\frac{\phi _\ini}{\phi _0}\right)\, .
\end{equation}
{}From this expression, it is straightforward to calculate $\phi
_*$. Remembering that inflation is supposed to stop by instability,
one arrives at
\begin{equation}
\phi _*=\phi _0\exp\left({\rm e}^{-cN_*}\ln \frac{\phi
_\ucri}{\phi _0}\right)\, .
\end{equation}
Equipped with this value, it is then sufficient to evaluate the
slow-roll parameters for this value of the inflaton vev using the
equations (\ref{eps1running}), (\ref{eps2running}) and
(\ref{eps3running}).

\par

\begin{figure}
\begin{center}
\includegraphics[width=7.5cm]{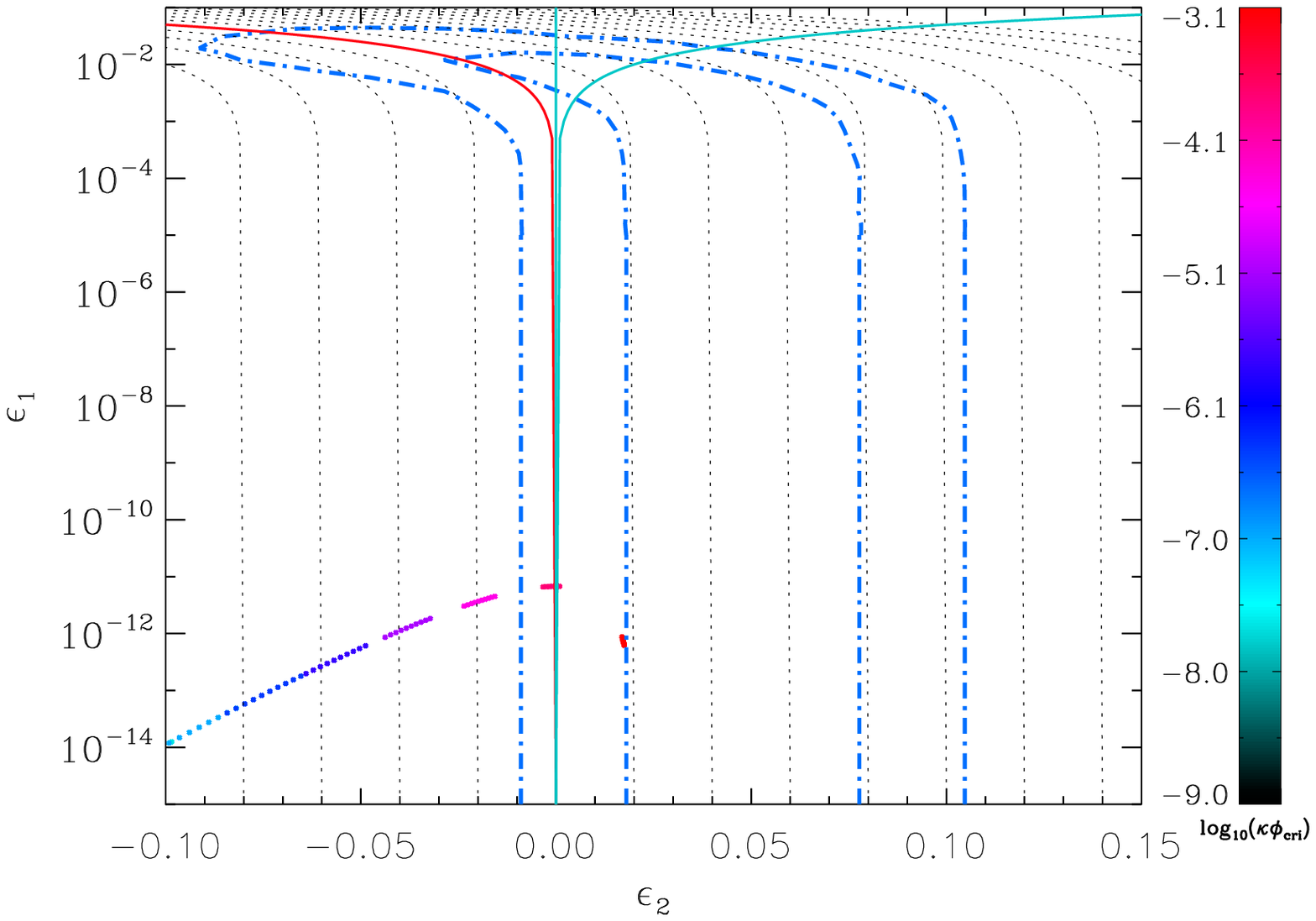}
\includegraphics[width=7.5cm]{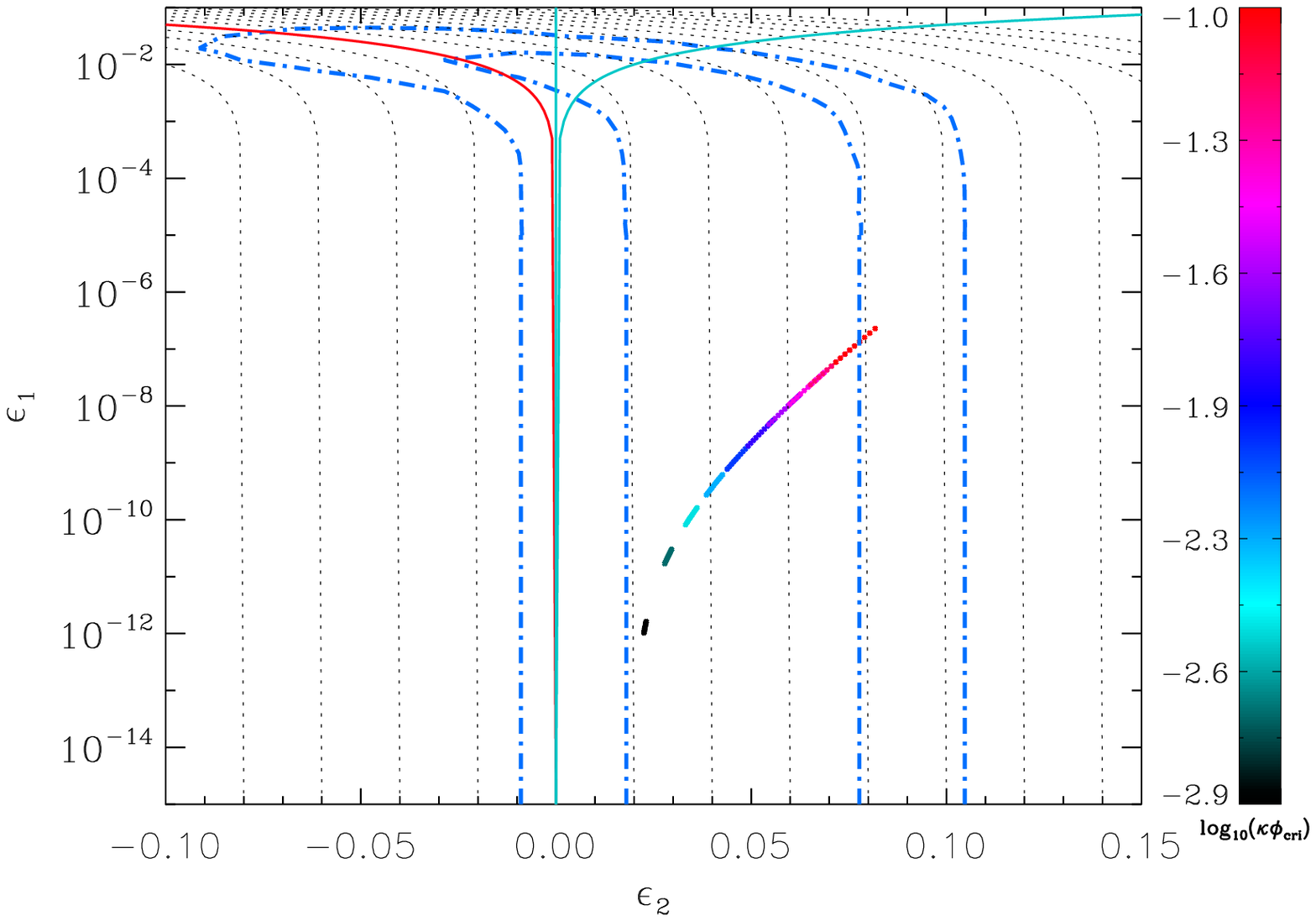}
\includegraphics[width=7.5cm]{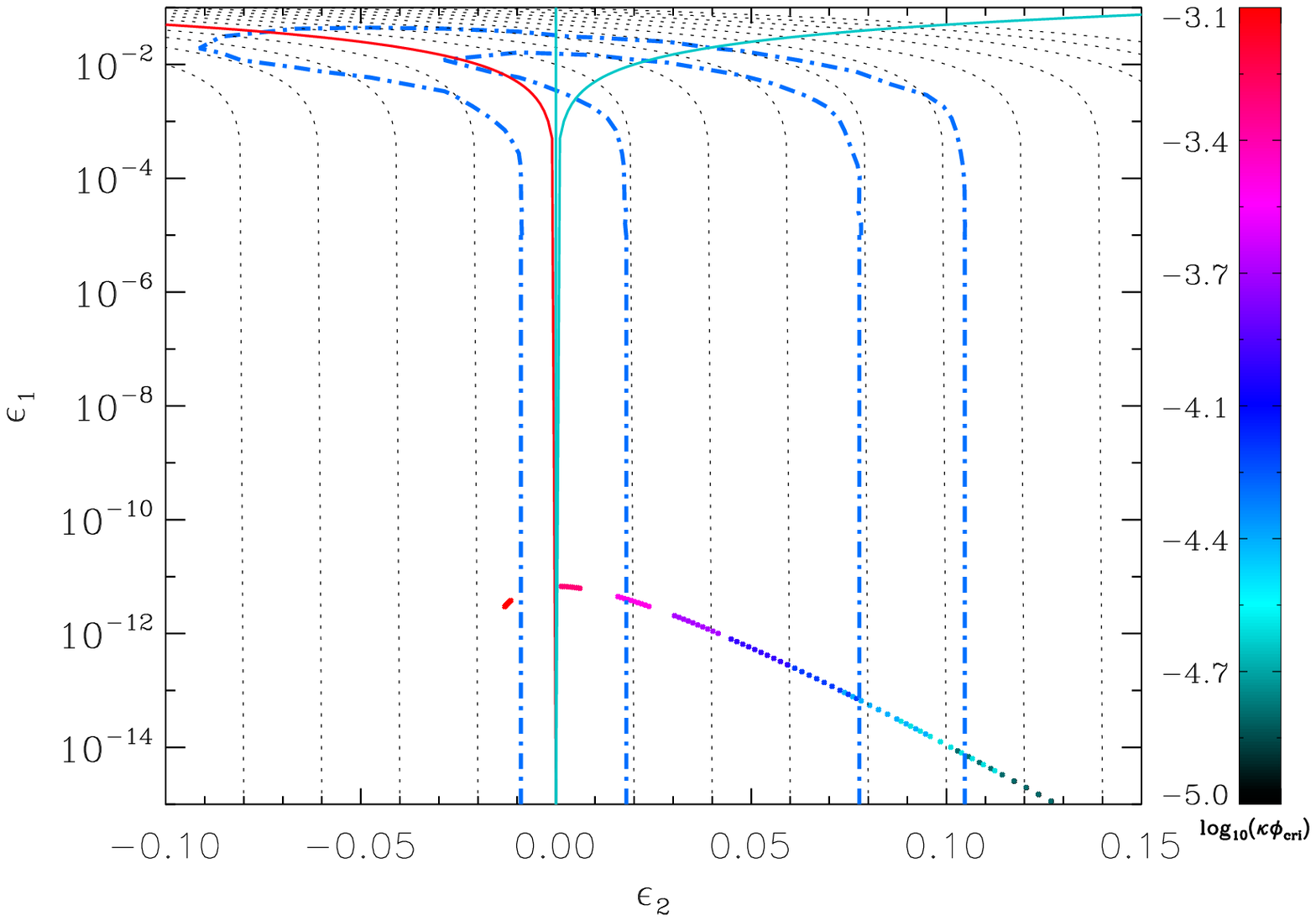}
\includegraphics[width=7.5cm]{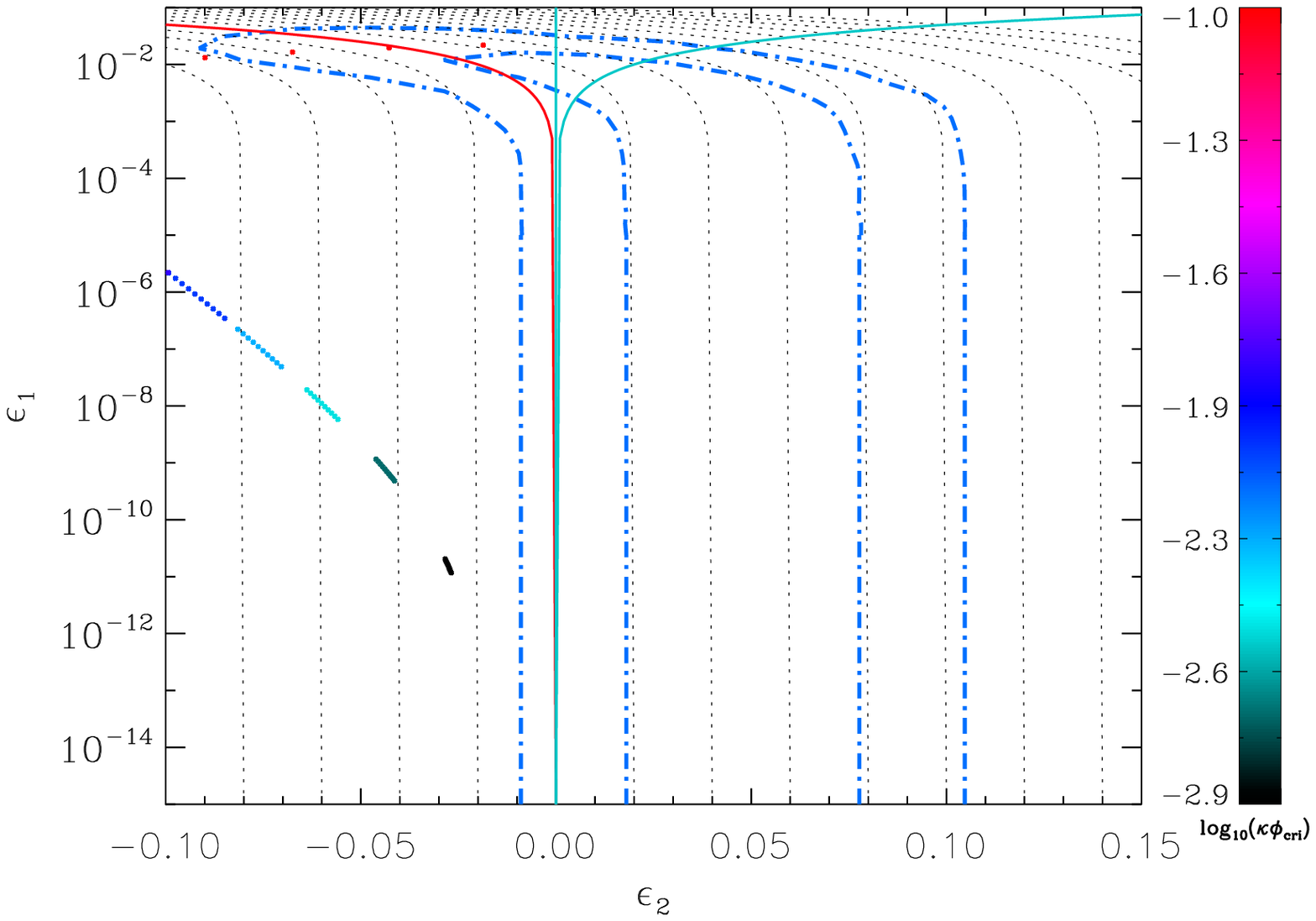}
\caption{Top left panel: RM1 model with $c=0.01$, $\kappa \phi
  _0=0.001$ and $10^{-9}<\kappa \phi _{\ucri}<10^{-3.1}$. Top right
  panel: RM2 model with $c=0.01$, $\kappa \phi _0=0.001$ and
  $10^{-2.9}<\kappa \phi _{\ucri}<10^{-1}$. Bottom left panel: RM3
  model with $c=-0.01$, $\kappa \phi _0=0.001$ and $10^{-5}<\kappa
  \phi _{\ucri}<10^{-3.1}$. Bottom right panel: RM4 model with
  $c=-0.01$, $\kappa \phi _0=0.001$ and $10^{-2.9}<\kappa \phi
  _{\ucri}<10^{-1}$. The solid red and blues lines, as well as the
  blue dotted-dashed contours, have the same meaning as in
  figure~\ref{lf}.}
\label{rm}
\end{center}
\end{figure}

It may be interesting to present some analytical estimates of the
slow-roll parameters. Instead of using the rather long
equations~(\ref{eps1running}), (\ref{eps2running})
and~(\ref{eps3running}), one can approximate them by assuming that the
denominator is just given by $M^4$, as done
in~\cite{Covi:2002th,Covi:2004tp}. In that case, the two first order
slow-roll parameters read
\begin{eqnarray}
\epsilon _1 &\simeq & \frac{c^2}{2}\left(\kappa \phi _0\right)^2
\exp\left[2{\ue}^{-cN_*}\ln \left(\frac{\phi _{\ucri}}{\phi
    _0}\right)\right] {\ue}^{-2cN_*}\ln ^2 \left(\frac{\phi _{\rm
    cri}}{\phi _0}\right)\, , \\ \epsilon _2 &\simeq & 2c \left\{1+{\rm
  e}^{-cN_*}\ln \left(\frac{\phi _{\ucri}}{\phi
  _0}\right)+c\exp\left[2{\ue}^{-cN_*}\ln \left(\frac{\phi _{\rm
      cri}}{\phi _0}\right)\right] \right. \nonumber \\ & \times &
\left. {\ue}^{-2cN_*}\ln ^2 \left(\frac{\phi _{\ucri}}{\phi
  _0}\right)\right\}.
\end{eqnarray}
In fact, it turns out to be convenient to define the following
quantity
\begin{eqnarray}
s\equiv c \ln \left(\frac{\phi _0}{\phi _*}\right),
\end{eqnarray}
which can also be written as 
\begin{eqnarray}
\label{eqs}
s=-c\, {\ue}^{cN_*}\ln \left(\frac{\phi _{\ucri}}{\phi _0}\right)\, .
\end{eqnarray}
For RM1 and RM4, $s>0$ while for RM2 and RM3 one has $s<0$. This
quantity can also be used to estimate $\phi_{\ucri}$ since $\phi
_{\ucri}\simeq \phi _0\exp[-s\exp(-cN_*)/c]$. Then, the approximate
equations giving the first two slow-roll parameters can be re-written
as
\begin{eqnarray}
\label{eps1eps2rm}
\epsilon _1 &\simeq & \frac{s^2}{2}\left(\kappa \phi _0\right)^2{\rm
e}^{-2s/c}\, , \qquad \epsilon _2 \simeq 2c
\left(1-\frac{s}{c}+\frac{s^2}{c}{\ue}^{-2s/c}\right).
\end{eqnarray}
The last equations means that the trajectory in the plane $(\epsilon
_1,\epsilon _2)$ can be expressed as $\epsilon_2\simeq 2(c-s)+4\epsilon
_1/(\kappa \phi _0)^2$. If we neglect $\epsilon _1$ (see below), then
one recovers the formula already derived in~\cite{Covi:1998mb,
  Covi:2002th, Covi:2004tp}, namely $\nS-1\simeq 2(s-c)$. The same route
for the third slow-roll parameter gives
\begin{eqnarray}
\epsilon _2\epsilon _3 & \simeq & 2c^2\left[-\frac{s}{c}+3c\left(\kappa
  \phi _0\right)^2 {\ue}^{-s/c}-3\frac{s^3}{c^2}\left(\kappa \phi
  _0\right)^2 {\ue}^{-2s/c} \right. \nonumber \\ & + &
  \left. 2\frac{s^4}{c^2}\left(\kappa \phi _0\right)^4
       {\ue}^{-4s/c}\right] ,
\end{eqnarray}
and if, again, one neglects $\epsilon _1$, making use of
equations~(\ref{bs2}) and (\ref{nalpha}) gives the the scalar running
$\alphaS \simeq 2sc$.

\par

In figure~\ref{rm}, we have represented the slow-roll predictions for
the four versions of the running mass model in the plane $(\epsilon
_1,\epsilon _2)$, together with the $1\sigma $ and $2\sigma$ WMAP3
confidence intervals. The top left panel corresponds to the RM1 model
with $c=0.01$, $\kappa \phi_0=0.001$ and $10^{-9}<\kappa \phi
_{\ucri}<10^{-3.1}$. The $\epsilon _1 $ parameter appears to be
extremely small and, hence, the spectral index is approximately
$\nS-1\simeq \epsilon _2$. According to the value of $\kappa \phi
_{\ucri}$, the spectral index can either be red for ``large'' values
of $\kappa \phi _{\ucri}$, \ie $\kappa \phi _{\ucri}$ relatively close
to $\kappa \phi_0$, or blue for ``small'' values of $\kappa \phi
_{\ucri}$. The relation $\nS-1\simeq 2(s-c)$ also reads
\begin{equation}
\label{nsrmapprox}
\nS-1\simeq -2c\left[{\ue}^{cN_*}\ln \left(\frac{\phi _{\ucri}}{\phi
_0}\right)+1\right],
\end{equation}
and allows us to understand the behaviour of the spectral index.

\par

For the RM1 model, $\phi _{\ucri}<\phi _0$ and the logarithmic term in
the above equation is negative. When $\phi _{\ucri}\lta \phi _0$, this
term is small, the constant term dominates in the squared bracket and,
since $c>0$, the spectral index is red. When $\phi _{\ucri}\ll \phi
_0$, the logarithm dominates, the squared bracket is negative and the
spectral index becomes blue.

\par

The top right panel corresponds to the RM2 model with $c=0.01$,
$\kappa \phi_0=0.001$ and $10^{-2.9}<\kappa \phi _{\ucri}<10^{-1}$:
the spectral index is always red. This can be interpreted by means of
equation~(\ref{nsrmapprox}). Indeed, we now have $\phi _{\ucri}>\phi
_0$ and hence the squared bracket is always positive ensuring that the
spectral index remains lower than unity. Moreover, the larger $\phi
_{\ucri}$, the redder $\nS$ in agreement with what is observed.

\par

The bottom left panel represents the RM3 model with $c=-0.01$, $\kappa
\phi _0=0.001$ and $10^{-5}<\kappa \phi _{\ucri}<10^{-3.1}$. The
spectral index can be red or blue depending on the value of $\kappa
\phi _{\ucri}$. Since $\phi _{\ucri}<\phi _0$ the logarithm term is
negative. If it dominates ($\phi _{\ucri}\ll \phi _0$) then the
bracket in equation~(\ref{nsrmapprox}) is negative and therefore the
spectrum is red since $-c>0$. On the contrary, if the constant term
dominates ($\phi _{\ucri}\lta \phi _0$), then the spectrum can be
blue.

\par

Finally, the bottom right panel corresponds to the RM4 model with
$c=-0.01$, $\kappa \phi _0=0.001$ and $10^{-2.9}<\kappa \phi
_{\ucri}<10^{-1}$. Since $\phi _{\ucri}>\phi _0$ and $-c>0$, the
spectral index is always greater than one. This model can be
compatible with the data only for ``small'' values of $\phi _{\ucri}$
as can be seen in figure~\ref{rm}.

\par

The previous observations agree with the existing literature, in
particular with reference~\cite{Covi:1998mb}.

\section{Testing exactly the inflationary models}
\label{sec:exact}

In this section, we do not use the slow-roll approximation but
integrate numerically both the background evolution and the
cosmological perturbations. This approach allows an exact
determination of the power spectra of scalar and tensor modes assuming
only the linear perturbation theory in General
Relativity~\cite{Salopek:1988qh, Grivell:1999wc, Adams:2001vc,
  Makarov:2005uh, Ringeval:2005yn}. For a given model, these power
spectra depend on the parameters characterising the potential and
introduced in the previous sections. By coupling this mode by mode
integration during inflation to a modified version of \CAMB, we can
use the MCMC techniques implemented in \COSMOMC to derive the
constraints these parameters have to satisfy given the third year WMAP
data. Let us stress that this method allows us to get marginalised
posterior distributions directly on the potential parameters, out of
any intermediate assumption. This approach is therefore different to
the ones used so far and includes by construction a marginalisation on
the reheating. As a result, the constraints we obtain include the
effects coming from varying the number of efolds at which the
cosmological perturbations can be generated~\cite{Liddle:2003as,
  Leach:2003us, Sanchez:2005pi, Alabidi:2006qa, Peiris:2006ug,
  Easther:2006tv}.

\par

The base cosmological parameters and their priors, as well as the HST
data and the top age prior, are already described in
section~\ref{sec:wmapsr}. Therefore, in the next section, we only
discuss the method used to sample the inflationary parameters.

\subsection{Method}
\label{sec:code}

Let us sketch how the numerical integration is performed for a
potential of the form $V(\phi )=M^4 U(\phi)$.

The first step is to integrate the background evolution and this is
done using the number of e-folds $N$ as time variable. The energy
scale $M$ is initially set to an arbitrary non-physical value $M=1$
(see below). The initial conditions $\phi _\ini$ and $\dd\phi
_\ini/{\dd}N$ are chosen such that there are at least $60$ e-folds of
inflation, the end of inflation being defined to be the time at which
the exact Hubble flow parameter $\epsilon _1=1$. Provided this
condition is fulfilled, the initial conditions are in fact irrelevant
thanks to the presence of the inflationary
attractor~\cite{Ringeval:2005yn}. Once the background integration
performed, the function $\phi(N)$ is numerically known in the range
$N\in [0,N_\usssT]$ (where $N_\usssT>60$ is the total number of
e-folds during inflation) as well as all the other background
functions, like the Hubble parameter $H(N)$.

\par

In a second step, the equation~(\ref{paramoscillator}) controlling the
evolution of the perturbations is numerically solved. This equation is
fully determined only once the time-dependent frequencies
$\omegaST(k,N)$, given in (\ref{frequencies}), are known. This
requires the knowledge of $H(N)$ and its derivatives (up to third
order), which simply comes from the background integration discussed
above. In addition one needs the choice of some comoving wavenumbers
$k$ that will be the ones of astrophysical interest today. As a
result, it is compulsory to be able to relate a comoving scale $k$
during inflation to a physical scale $k_{\uphys}$ defined at the
present time. For that purpose, the complete history of the Universe
needs to be specified and we now describe in more details how this can
be achieved. Let us notice that equation~(\ref{paramoscillator}),
written with $N$ as time variable, takes the form
\begin{equation}
\label{eqmotionefold}
\frac{{\dd}^2\muST}{{\dd}N^2}+\frac{1}{{\cal H}}\frac{{\dd}{\cal
H}}{{\dd}N}\frac{{\dd}\muST}{{\dd}N}+\left[\left(\frac{k}{{\cal
H}}\right)^2 -V_{\mu}(N)\right]\muST=0 \, ,
\end{equation}
where $V_{\mu}(N)$ is the effective potential for the cosmological
perturbations. We see that, besides the background evolution, we need
to know the quantity $k/{\calH}$ during inflation as a function of
the time variable $N$. Let us assume that we are given a physical
scale today, say $\kstar/a_0$ (in $\mbox{Mpc}^{-1}$ for instance), where
$a_0$ is the present day scale factor. Then, one has
\begin{equation}
\frac{\kstar}{\cal H}=\frac{\kstar}{a_0H(N)}\times \frac{a_0}{a_{\uend}}
\times \frac{a_{\uend}}{a(N)}\equiv \frac{\Upsilon }{H(N)}{\rm
e}^{N_{_{\rm T}}-N}\, ,
\end{equation}
where $a_{\uend}$ is the scale factor at the end of inflation and
\begin{equation}
\label{eq:upsilon}
\Upsilon \equiv \frac{\kstar}{a_0} \frac{a_0}{a_{\uend}} \,,
\end{equation}
a constant which depends only on $\kstar$ and on the history of the
Universe through the ratio $a_0/a_{\uend}$. In order to evaluate this
constant we assume that, after inflation, there is a period of
reheating as described in section~\ref{sec:endinflation},
characterised by the two parameters $\wstate _{\ureh}$ and
$N_{\ureh}$, followed by a radiation dominated era that can be
described by $\Omega _{\urad}^0$, followed by the matter dominated
era. Let us be more precise about the reheating phase. It is
interesting to consider the quantity $R_{\urad}$ defined by
\begin{eqnarray}
\ln R_{\urad} &\equiv & \ln \left(\frac{a_{\uend}}{a_{\ureh}}\right)
-\frac{1}{4}\ln \left(\frac{\rho _{\ureh}}{\rho _{\uend}}\right)
=\frac14\left(-1+3\wstate _{\ureh}\right)N_{\ureh}\nonumber \\ & &
-\frac{1}{4} \ln
\left(\frac{3+3\wstate_{\ureh}}{5-3\wstate_{\ureh}}\right)\, ,
\end{eqnarray}
where $a_{\ureh}$ and $\rho _{\ureh}$ are respectively the scale
factor and the energy density in radiation at the end of the reheating
phase. This last quantity is obtained from equation~(\ref{rhoappro})
evaluated at $t_{\ureh}\equiv \Gamma ^{-1}$, the last term in the
squared bracket being neglected. From the previous definition, one
notices that $\ln R_{\urad}$ is exactly zero if $\wstate
_{\ureh}=1/3$. In this case, the reheating phase cannot be
distinguished from the radiation dominated era. If $-1/3<\wstate
_{\ureh}<1/3$ then $\ln R_{\urad}<0$ while if $1/3<\wstate _{\ureh}<1$
then $\ln R_{\urad}>0$. For reasons that will be explained below in
more details, it turns out to be convenient to define another quantity
$R$
\begin{equation}
\label{defR}
\ln R\equiv \ln R_{\urad}+\frac14 \ln\left(\kappa ^4\rho _{\uend}\right).
\end{equation}
The quantity $\rho _{\uend}$, the energy density stored in the scalar
field at the end of inflation ($\epsilon _1=1$), is also completely
determined once the background evolution has been solved. Therefore,
if $\ln R_{\urad}$ is known, one can deduce $\ln R$ and
vice-versa. Then, the ratio $a_0/a_{\uend}$ can be written
as~\cite{Liddle:2003as}
\begin{equation}
\label{a0aend}
%\frac{a_0}{a_{\uend}}=\left(\Omega _{\urad}^0\right)^{-1/4}
%\left(\kappa ^4\rho _{\ucri}^0\right)^{-1/4}\frac{R}{R_{\urad}^2}\, ,
\frac{a_0}{a_{\uend}}=\left(\Omega _{\urad}^0\right)^{-1/4}
\left(\kappa ^4\rho _{\ucri}^0\right)^{-1/4}\frac{\sqrt{\kappa^4
    \rho_\uend}}{R}\, ,
\end{equation}
where $\rho _{\ucri}^0$ is the critical energy density
today, \ie
\begin{equation}
\kappa^4 \rho_\ucri^0 = 3 \kappa^2 H_0^2\,.
\end{equation}
Therefore, given $\rho _{\ucri}^0$, $\Omega _{\rm rad}^0$ and $R$, \ie
a model of Universe between the end of inflation and the present time,
one can calculate the constant $\Upsilon $ and hence all the terms in
equation~(\ref{eqmotionefold}) are explicitly known.  In practice, we
implement $\ln R$ as the new inflationary parameter associated with
the reheating and that is sampled from the MCMC: an uniform prior has
been assumed on $\ln R$ in the range $[\ln R_{\min},\ln R_{\max}]$.

\par

Let us discuss how the limits $R_{\min}$ and $R_{\max}$ are
chosen. First of all there are limits on $\rho _{\uend}$. In order not
to spoil the success of the Big Bang Nucleosynthesis (BBN) it is
reasonable to require $\rho _{\uend}> \rho _{\unuc}$. Roughly
speaking, this means $\rho _{\uend}> 10^{-85}\mpl ^4$. On the other
hand, one must have $\rho _{\uend}< \mpl ^4$ in order for the whole
theoretical framework to be meaningful. In practice however, we do not
need to implement this upper bound because the constraint
$H_\uinf/\mpl <1.3 \times 10^{-5}$ derived in the slow-roll section (and
coming from $\epsilon _1< 0.022$) shows that the only viable
perturbations have to verify $\rho _{\uend}< 10^{-10}\mpl ^4$. In
other words, we have
\begin{equation}
-187 < \ln \left(\kappa ^4\rho_{\uend}\right) < -20\,,
\end{equation}
the upper bound being not required \emph{a priori} but being a
consistency check that we should recover for the viable inflationary
models. Now, if we use our reheating model, $R_{\urad}$ can be
rewritten as
\begin{equation}
\ln R_{\urad}=\frac{1-3\wstate _{\ureh}}{12(1+\wstate _{\ureh})} \ln
\left(\frac{\rho _{\ureh}}{\rho _{\uend}}\right)
-\frac{1}{3\left(1+\wstate_{\ureh}\right)}\ln
\left(\frac{3+3\wstate_{\ureh}} {5-3\wstate_{\ureh}}\right)\, ,
\end{equation}
where $\rho_{\unuc}<\rho _{\ureh}<\rho _{\uend}$ and $-1/3<\wstate
_{\ureh}<1$ in order to satisfy the strong and dominant energy
conditions. In this range, it is easy to see that the minimum is
obtain for $\wstate _{\ureh}=-1/3$ and $\rho _{\ureh}=\rho _{\unuc}$
whereas the maximum for $\wstate _{\ureh}=1$ and $\rho _{\ureh}=\rho
_{\unuc}$. Finally, if we use the link between $R_{\urad}$ and $R$,
one arrives at
\begin{equation}
\label{eq:lnRbounds}
\frac14 \ln \left(\kappa ^4 \rho _{\unuc}\right)
< \ln R < -\frac{1}{12} \ln \left(\kappa ^4 \rho _{\unuc}\right)
+\frac13 \ln \left(\kappa ^4 \rho _{\uend}\right),
\end{equation}
up to negligible factors depending on $\wstate _{\ureh}$ only. In the
above equation, the bounds are our definition of $\ln R_{\min}$ and
$\ln R_{\max}$. Notice that these bounds explicitly depends on
$\rho_\uend$ which varies with the inflaton potential parameters. As a
result, a ``hard prior'' has been coded in \COSMOMC to reject any
sample involving a $\ln R$ value that would not satisfy
equation~(\ref{eq:lnRbounds}). Note that from the knowledge of $\ln R$
and $\rho _{\uend}$, one can also derive $\ln R_{\urad}$ and
$a_0/a_{\uend}$.

\par

{}From a chosen set of these parameters, the numerical integration can
be straightforwardly performed to get the amplitude of the scalar and
tensor power spectra at the end of inflation for any comoving
wavenumber $k$ that can now be related to their corresponding physical
scales today.

\par

However, as mentioned at the beginning, this result has been obtained
with an arbitrarily chosen potential scale $M=1$. In fact, the MCMC
exploration is performed by directly sampling on the scalar power
spectra amplitude $P_*$ at a given scale $\kstar$ rather than on $M$
directly. An uniform prior has been chosen on $\ln(10^{10}P_*)$ in the
usual range $[2.7,4]$. However, one has to restore the consistency
between the value of $P_*$ and $M$ since the former is uniquely
determined by the latter. As we show in the following, a simple
rescaling of the relevant functions can be used for this purpose. This
rescaling has the advantage of being analytical and exact in the
framework of the linear perturbation theory.

\par

Let us consider the following rescaling for the scale $M$ or,
equivalently, for the potential
\begin{equation}
\label{rescaling} 
V(\phi )\rightarrow s V(\phi )\, ,
\end{equation}
where $s$ is a constant. What are the consequences of this rescaling
on the other quantities? From equation (\ref{Kleingordonefold}), one
sees that the field $\phi (N)$ and its derivative are unaffected
because only the logarithm of the potential appears in this
formula. On the other hand, if we write the Friedman equation as
\begin{equation}
H^2\left[1-\frac{\kappa^2}{3}\left(\frac{{\dd}\phi}{{\dd}N}\right)^2\right]
=\frac{\kappa^2 }{3}V\, ,
\end{equation}
then one notices that the Hubble parameter transforms as $H\rightarrow
s^{1/2}H$ which immediately implies that $\rho _{\uend}\rightarrow
s\rho_{\uend}$. This also means that ${\calH}=aH$ changes as
${\calH}\rightarrow s^{1/2}{\calH}$.

\par

The previous considerations are valid for the background. Let us now
see how the perturbed quantities are affected. The effective potential
for cosmological perturbations, $V_{\mu}$ in
equation~(\ref{eqmotionefold}), is invariant under the
rescaling~(\ref{rescaling}). This can be seen for instance on the
gravitational waves where $V_{\mu}(N)=a''/a=1-{\cal H}'/{\cal
  H}^2$. Therefore, requiring
\begin{equation}
k\rightarrow s^{1/2}k\, ,
\end{equation}
is sufficient to render equation~(\ref{eqmotionefold})
invariant. However, this does not mean that the amplitude itself $\mu
(k,\eta )$ is not changed. Indeed, although the equation of motion is
invariant, the initial conditions are modified since they read
\begin{equation}
\mu_\ini \propto (2k)^{-1/2},\qquad \left. \frac{\dd \mu}{\dd N}
\right \vert _\ini \propto {\calH}^{-1}_\ini i
\left(\frac{k}{2}\right)^{1/2}.
\end{equation}
Because we deal with a linear equation, this implies that
\begin{equation}
\mu \left(k,N\right)\rightarrow s^{-1/4}\mu \left(s^{1/2}k,N\right).
\end{equation} 
Therefore, the scalar power spectrum evaluated at the end of
inflation, \ie at $N=N_{_{\rm T}}$, transforms in a very simple way,
namely
\begin{equation}
\label{transPS}
k^3\left\vert \frac{\mu \left(k, N_{_{\rm T}}\right) }{a\sqrt{\epsilon
_1}} \right \vert ^2 \rightarrow s k^3\left\vert \frac{\mu
\left(s^{1/2}k,N_{_{\rm T}}\right)}{a\sqrt{\epsilon _1}} \right\vert
^2  .
\end{equation}
The last step is to determine the number $s$ required to restore the
consistency between our initial arbitrary normalisation $M=1$ and the
values of $P_*$ singled out during the MCMC exploration. From the
previous discussing, the required value of $s$ is simply given by the
ratio $P_*/\Pnum_\diamond$ where $\Pnum_\diamond$ is the amplitude of
the scalar power spectrum stemming from the mode by mode numerical
integration with $M=1$ and evaluated at $\kdiam = \kstar s^{-1/2}$.
Then, the final power spectra do not need to be recomputed from
scratch but can be directly deduced from the
transformation~(\ref{transPS}), and a similar one for the tensor
modes.

\par

In principle, the previous procedure is well-defined but the fact that
the wavenumber $k$ must be continuously rescaled makes it quite
difficult to implement in practice. In fact, it turns out to be more
convenient to use a last trick that we now describe. The quantity
$\rho _{\ureh}$ can be viewed as an independent model parameter and,
therefore, we also have the freedom to consider another model
corresponding to a rescaled $\rho _{\ureh}$. Let us consider the
following transformation
\begin{equation}
\rho _{\ureh}\rightarrow s^2\rho _{\ureh}\, ,
\end{equation}
where $s$ is precisely the scaling number used in the above discussion
on the rescaling of $M$. From these relations, one can immediately
check that $\ln R$ (or simply $R$) is invariant whereas $R_{\rm
rad}\rightarrow s^{-1/4}R_{\urad}$. From equation~(\ref{a0aend}), this
also implies that
\begin{equation}
\frac{a_0}{a_{\uend}}\rightarrow s^{1/2}\frac{a_0}{a_{\uend}}\, .
\end{equation}
If we now consider the quantity $\Upsilon $ we see that its
transformation under the previous rescaling is the same if we change
the scale $k$ and leave $\rho _{\ureh}$ fixed or fix the scale $k$ and
change $\rho _{\ureh}$. This is the main reason why it is more
convenient to perform a MCMC sampling on the parameter $R$ rather than
$R_{\urad}$ since at a given $R$, the rescaling required on $M$ does
not induce a rescaling of the wave numbers.

\par

In the following, we have used such an exact numerical integration of
the cosmological perturbations to constrain the models previously
discussed in section~\ref{sec:slowroll}. This leads to new insights
that we now describe and interpret.

\subsection{Large field models}
\label{sec:exactlf}

\begin{figure}
\begin{center}
\includegraphics[width=10cm]{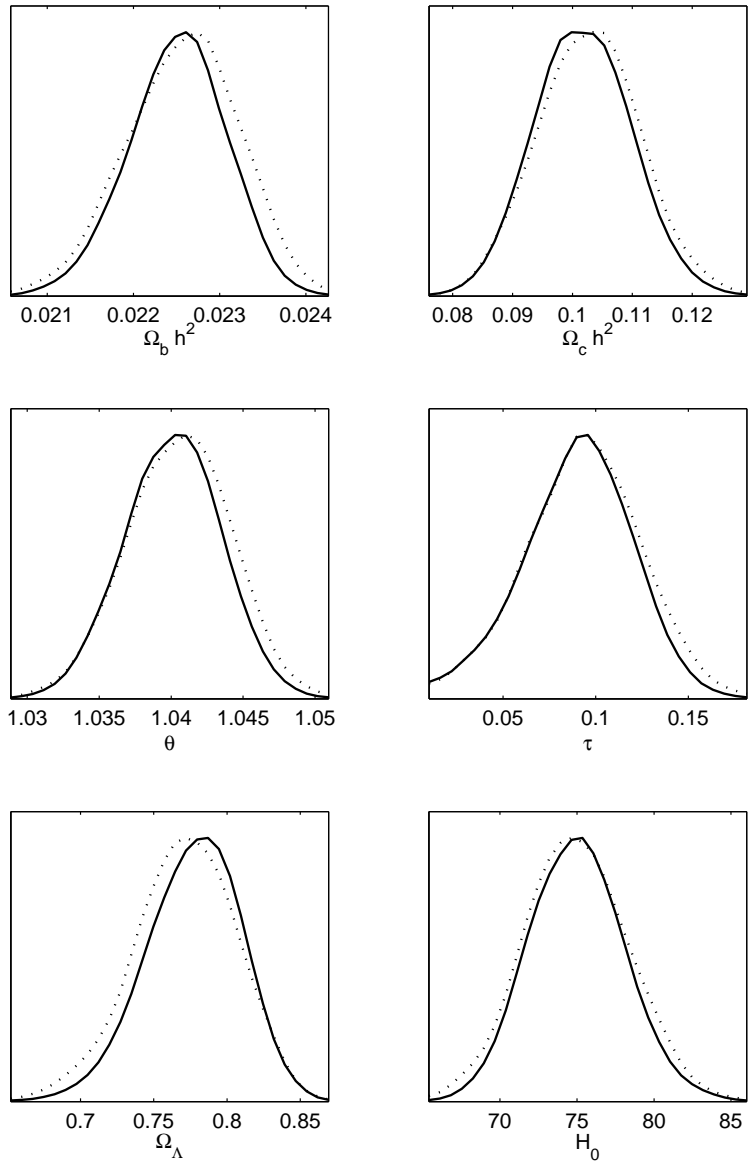}
\caption{Marginalised posterior probability distributions (solid
lines) and mean likelihood (dotted lines) for our $\Lambda$CDM base
cosmological parameters together with the cosmological constant and
the Hubble parameter obtained from the exact integration of large
field models power spectra.}
\label{large_cosmo_1D}
\end{center}
\end{figure}

In this section, we describe the results obtained for large field
models. The total number of inflationary parameters is now three: the
power of the potential $p$, the potential scale $M$, which is uniquely
determined from $P_*$, and the reheating parameter $R$. This accounts
for an overall number of parameters of seven given the four base
cosmological parameters. The best fit model has $\chi ^2 \simeq
11252.3$. In figure~\ref{large_cosmo_1D}, we have represented the
(one-dimensional) marginalised posterior probability distributions and
mean likelihoods for the base cosmological parameters. As one may
compare with the ones derived in the slow-roll models (see
figure~\ref{sr1st_cosmo_1D}), their determination is robust and their
values remain standard. This is just the consequence that, for large
field models, there exists values of the primordial parameters leading
to a power spectrum which allows a good fit of the data.

\begin{figure}
\begin{center}
\includegraphics[width=11cm]{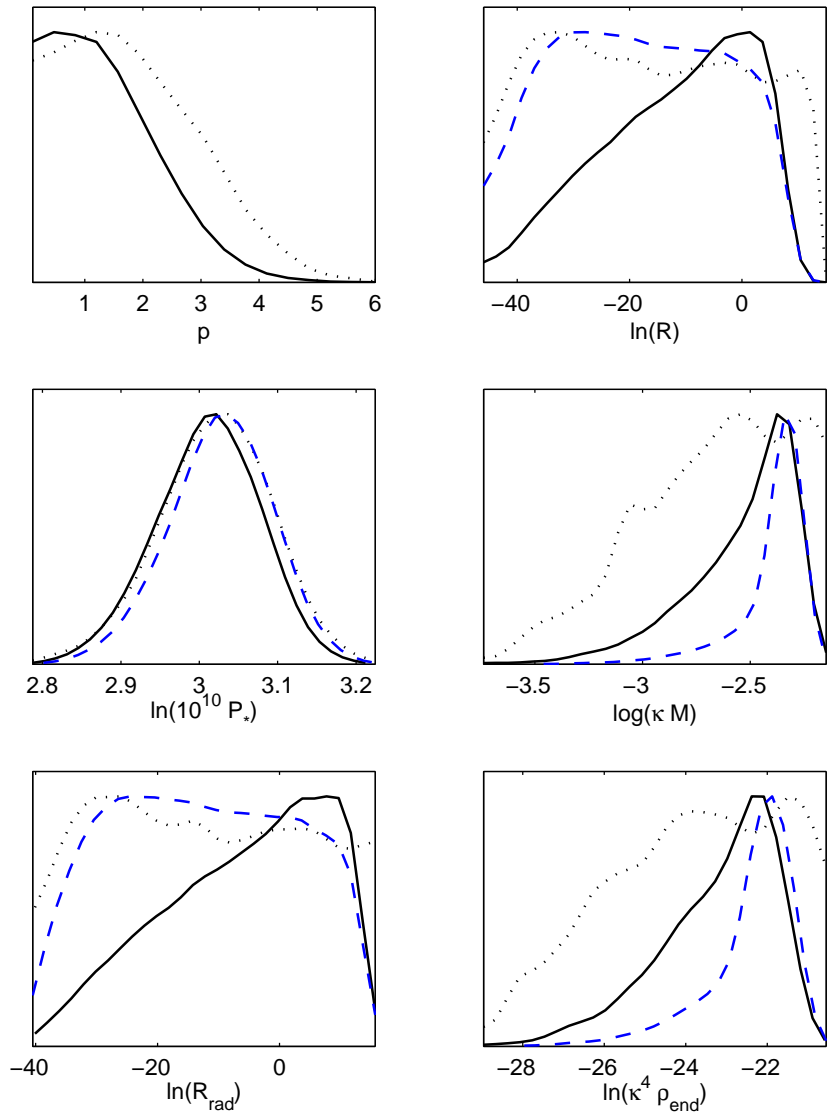}
\caption{Probability distributions for the inflationary parameters
characterising the large field models and the subsequent reheating
period. The dotted black lines represent the mean likelihood, the
solid black lines the marginalised posteriors associated with a flat
prior on the index $p$, and the blue dashed lines are the marginalised
posteriors coming from a flat prior on $\log p$.}
\label{large_inf_1D}
\end{center}
\end{figure}

\begin{figure}
\begin{center}
\includegraphics[width=12cm]{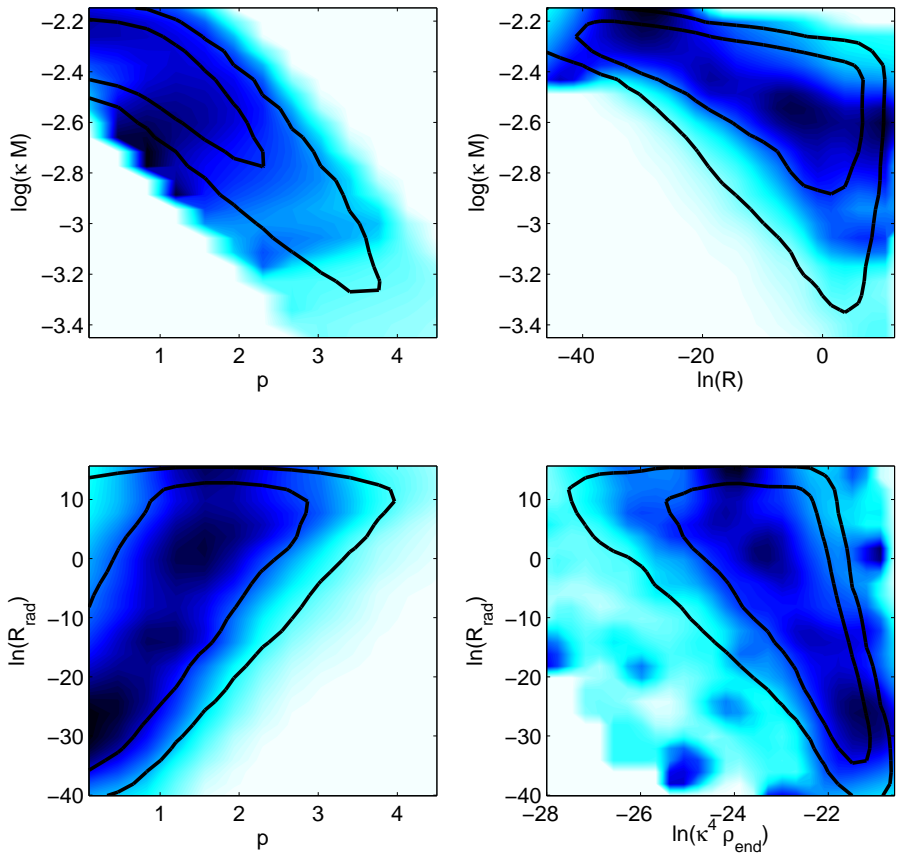}
\caption{Correlations between various pairs of inflationary parameters
characterising the large field models and the subsequent reheating
period. The coloured shaded regions represent the mean likelihoods
while the solid contours trace the one and two-sigma confidence
intervals of the two-dimensional marginalised posteriors.}
\label{large_inf_2D}
\end{center}
\end{figure}

Figure~\ref{large_inf_1D} shows the one-dimensional posterior
probabilities for the primordial parameters characterising the large
field models. The corresponding two-dimensional plots are presented in
figure~\ref{large_inf_2D}. Let us recall that the quantities $p$, $\ln
R$ and $P_*$ are directly sampled by the MCMC whereas $M$, $\ln
R_{\urad}$, $\rho _{\uend}$ and $a_0/a_\uend$ are derived
parameters. The curves represent the mean likelihood (dotted black
line) and the marginalised probability (solid black line) obtained
under a flat prior choice on $p$ in $[0.1,10]$, as well as an uniform
prior on $\log p$ in $[-1,1]$ (dashed line).

One sees that models with a potential power $p\gta 4$ are now strongly
incompatible with the data. For values of $p$ slightly greater than,
say, $3-4$, this is due a too high level of gravitational waves since
the corresponding value of the spectral index $\nS$ is still
compatible with the data. For values of $p$ much larger than $3-4$,
the power spectra are too red tilted and their associated $\nS$
becomes unacceptable. At two-sigma level, one gets the upper
marginalised bound
\begin{equation}
\label{eq:lfpbound}
p < 3.1 \,.
\end{equation}

The limiting case $p=1/10$ of our prior choice, \ie an almost
Harrison-Zeldovitch scale-invariant power spectrum, is still a
non-excluded model~\cite{Parkinson:2006ku, Pahud:2006kv}. Although the
maxima of the likelihood and of the marginalised probability are
located around $p\simeq 1.5$ and $p\simeq 0.5$ (respectively), the model
$p=1/10$ lies in the one-sigma confidence interval. This can also be
seen in figure~\ref{large_inf_2D}. The fact that values of $p<2$ are
favoured can be recovered in figure~\ref{lf} where the corresponding
slow-roll approximation predicts a line segment below the frontier
$\epsilon _2=2\epsilon _1$ (corresponding to $p=2$) in the
$(\epsilon_1,\epsilon_2)$ plane, \ie deeper into the $1\sigma$
confidence interval we have obtained in section~\ref{sec:wmapsr}.

\par

Let us now interpret the distribution obtained for the potential scale
$M$. The black solid line represents its marginalised probability when
a flat prior on $p$ is assumed while the blue dashed line is obtained
with a flat prior on $\log p$. The fact that the two curves are
significantly different signs that this parameter is poorly
constrained by the data. On the other hand, the general trend is
clearly the same. The value of the peak can be understood from
equation (\ref{scaleMlf}). In fact, the degeneracies between $\log
(\kappa M)$ and the index $p$ observed in figure~\ref{large_inf_2D}
can be reproduced almost exactly using this equation from which one
obtains
\begin{eqnarray}
\label{eq:lfpotscale}
\log \left(\kappa M\right) &\simeq & -1.57 +\frac14 \log p -\frac14
\log\left(p+4N_*\right) \nonumber \\ & + &
\frac{p}{8}\left[1.7-\log\left(p+4N_*\right)-\log p\right],
\end{eqnarray}
where $N_*\simeq 50$.

\par

The distribution of $\rho _{\uend}$ can be understood in the same
manner. Let us repeat that a theoretically motivated upper bound for
this quantity is only $\rho _{\uend}<\mpl ^4$. We observe a sharp drop
of the marginalised probability for values of $\ln (\kappa ^4 \rho
_{\uend})$ greater than $-20$. As expected and already mentioned, this
is nothing but literally the constraint on the energy scale of
inflation: in the slow-roll picture, $H/\mpl < 1.3 \times
10^{-5}$. With regards to the peak located at $\ln (\kappa ^4 \rho
_{\uend})\simeq -22$, it can be understood if one notices that $\rho
_{\uend}\simeq M^4(\phi _{\uend}/\mpl )^p$ and uses (\ref{phiendlf})
with the most probable $p\simeq 1$. At two-sigma level, one gets
\begin{equation}
\ln \left(\kappa^4 \rho_\uend\right) < -21.3 \,.
\end{equation}

\par

Let us turn to $\ln R$ and/or $\ln R_{\urad}$. As can be seen on the
one-dimensional or two-dimensional plots, these two quantities are not
constrained by the WMAP3 data. In particular, we see the strong
influence of the choice of the $p$ prior. With a flat prior on $p$
(solid black line), $\ln R$ peaks around zero while with a flat prior
on $\ln p$ (dashed blue line) the distribution is almost flat and cut
at the edges of its prior. Let us also remind that the prior on $\ln
R$ is not of top-hat shape but given by equation~(\ref{eq:lnRbounds})
which involves $\rho_\uend$. The data are simply not accurate enough
to constraint the reheating phase. Notice also that the long tails in
the $M$ and $\rho_\uend$ distributions are the result of their
correlation with $\ln R$ through $N_*$ as can be seen in
(\ref{eq:lfpotscale}).

\begin{figure}
\begin{center}
\includegraphics[width=16cm]{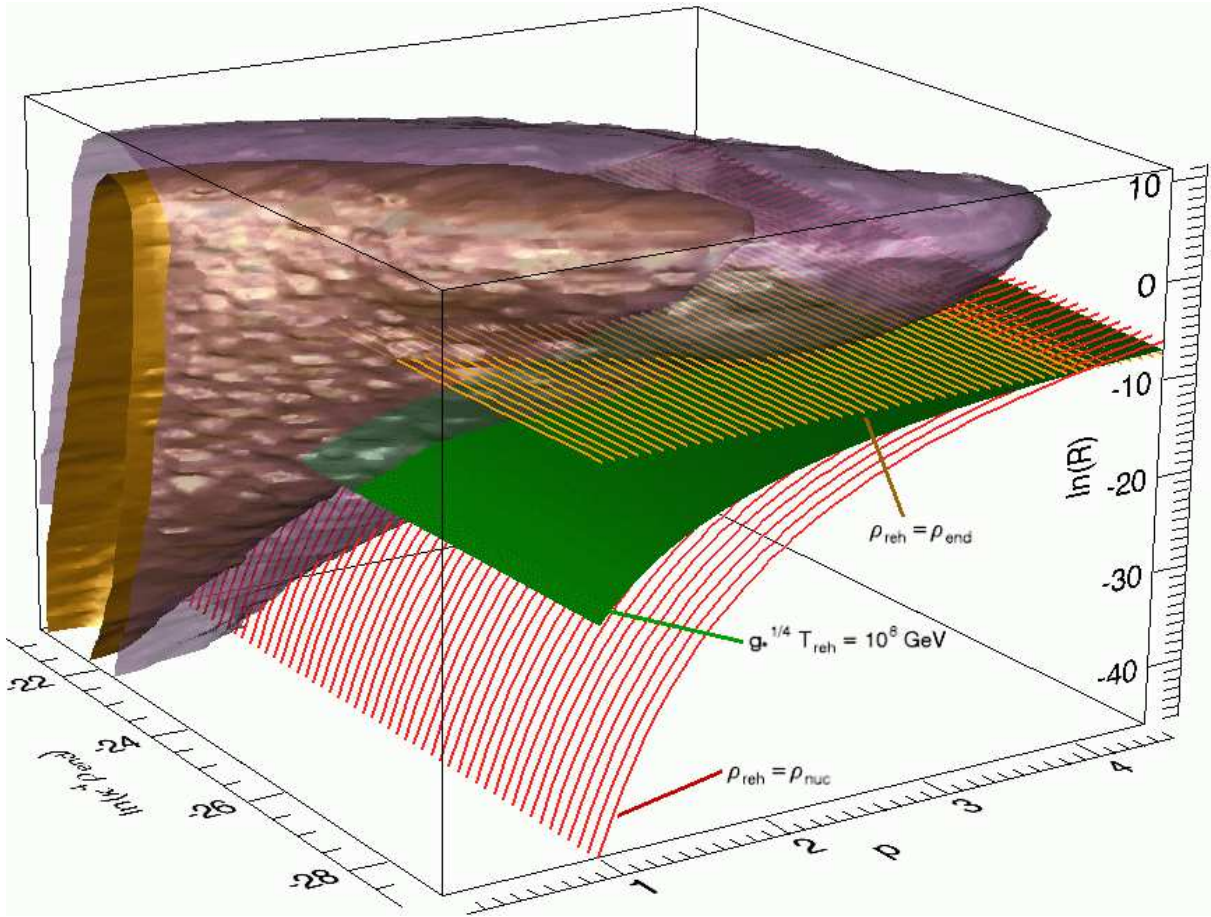}
\caption{One and two-sigma confidence intervals of the
  three-dimensional marginalised probability distribution in the space
  $[(\ln \kappa ^4\rho _{\uend}), p, \ln R]$. The three surfaces
  represent the locations of constant reheating temperature
  $g_*^{1/4}T_{\usssRH}$, the values of which are indicated in the
  figure. They intersect along a line for $p=4$ corresponding to a
  reheating period with an ultra-relativistic equation of state.}
\label{large_inf_4D}
\end{center}
\end{figure}

\begin{figure}
\begin{center}
\includegraphics[width=16cm]{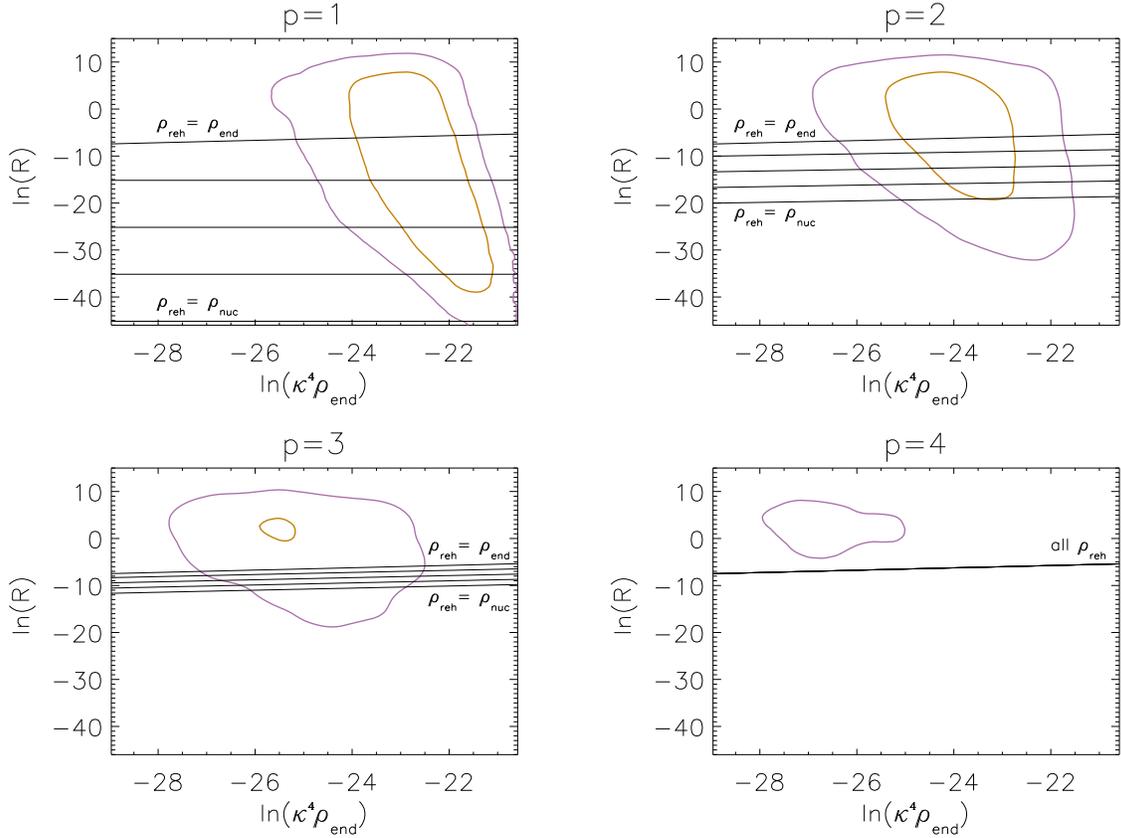}
\caption{Two-dimensional cuts of figure~\ref{large_inf_4D} along the
  planes of constant $p$. The various lines correspond to $(1/4)
  \ln(\kappa^4 \rho_{\ureh})$ equal to $-45$, $-35$, $-25$, $-15$ and
  $-(1/4)\ln(\kappa^4 \rho_{\uend})$ (bottom to top). For the case
  $p=4$, all the lines are merged because, in this case, $\wstate
  _{\ureh}=1/3$ and the reheating cannot be distinguished from the
  subsequent radiation dominated era.}
\label{large_inf_4D_cont}
\end{center}
\end{figure}

Our modelling of the reheating phase by means of $\ln R$ only can be
slightly improved in the special case of the large field
models. Indeed, the approach presented in
section~\ref{sec:endinflation} turns out to be rigourous for those
models with the equation of state parameter $\wstate _{\ureh}=P/\rho$
given by the potential power dependency $p$ through
(\ref{omegareh}). If we express the reheating
temperature~(\ref{reheatingT}) in terms of the outputs of our
computations, one obtains
\begin{eqnarray}
\label{reheatTlf}
g_*^{1/4}T_\usssRH & \simeq &\frac{30^{1/4}}{\sqrt{\pi}}\rho
_{\uend}^{1/4} \left(\frac{3p}{p+8}\right)^{1/4} \nonumber \\ & \times
& \exp\left\{-\frac{3p}{p-4}\left[\ln R-\frac{1}{4}\ln \left(\kappa
^4\rho _{\uend}\right)+\frac{1}{4}\ln\left(\frac{3p}{p+8}\right)
\right]\right\}.
\end{eqnarray}
In the space $[\ln R, \ln(\kappa ^4\rho _{\uend}), p]$, a given value
of the reheating temperature defines a surface. In fact, the above
equation can be worked out explicitly and reads
\begin{eqnarray}
\label{lnRTrh}
\ln R & = &\frac{p-4}{3p}\left[\ln \left(\frac{30^{1/4}}{\sqrt{\pi
  }}\right) -\frac{p+2}{2\left(p-4\right)}\ln
  \left(\frac{3p}{p+8}\right)-\ln \left(g_*^{1/4}\kappa
  T_\usssRH\right)\right] \nonumber \\ & + & \frac{p-1}{3p}\ln
  \left(\kappa ^4\rho _\uend \right).
\end{eqnarray}
The corresponding surfaces are represented in
figure~\ref{large_inf_4D} for three values of the reheating
temperature, or equivalently, for three values of $\rho _{\ureh}$
indicated in the figure. The $1\sigma $ and $2\sigma $ confidence
intervals of the three-dimensional marginalised probability have also
been plotted. For convenience, we have also represented in
figure~\ref{large_inf_4D_cont} a few non-marginalised two-dimensional
sections associated with several values of $p$. As can be seen in the
three-dimensional figure, the greater $p$, the more squeezed are the
lines of constant $\rho _{\ureh}$.

At that point, several comments are in order. Firstly, one notices
that the value $p=4$ plays a particular role. In this case, equation
(\ref{lnRTrh}) becomes $\ln R=(1/4)\ln (\kappa ^4 \rho _{\rm end})$
and $\ln R$ does no longer depend on $T_{\usssRH}$: this is why all
the surfaces intersect along the corresponding line as can be seen in
figure~\ref{large_inf_4D}. This is also why there is only one line in
figure~\ref{large_inf_4D_cont} for the panel $p=4$. Secondly, in this
last figure, one also remarks that the lines are not horizontal.
According to (\ref{lnRTrh}), the slope is $(p-1)/(3p)$ (except for
$\rho _{\ureh}=\rho _{\uend}$) and for $p=4$ one recovers the factor
$1/4$. Thirdly, the fact that the case $p=4$ plays a special role is
physically justified. The equation of state during reheating reduces
to a state parameter $\wstate_\ureh=1/3$ and, as a matter of fact, the
reheating phase cannot be distinguished from the subsequent radiation
dominated phase. Fourthly, and as mentioned before, the slope of the
lines is different if $\rho _{\ureh}=\rho _{\uend}$. This is because,
in this case, we have an extra contribution coming from the term
$\ln\left(g_*^{1/4}\kappa T_{\usssRH}\right)=\ln \left(g_*^{1/4}\kappa
^4\rho _{\uend}\right)/4$. This is especially visible in the top left
panel of figure~\ref{large_inf_4D_cont} where all the lines but the
one with $\rho _{\ureh}=\rho _{\uend}$ are horizontal.

\par

Let us now discuss the properties of such a reheating phase. We see in
the two top panels in figure~\ref{large_inf_4D_cont} that for $p=1$
and $p=2$, there are no constraint on the reheating temperature since
all the lines intersect the $1\sigma $ confidence interval. For $p=1$,
the central value is $\ln R\simeq -15$ and corresponds to
$\wstate_{\ureh}=-1/3$, $g_*^{1/4}\kappa T_{\usssRH}\simeq 3\times
10^{-8}$ and $N_{\ureh}\simeq 20$. For $p=1$, $\rho _{\ureh}$ is
unconstrained in the allowed range between $\rho _{\unuc}$ and
$\rho_\uend$. For $p=2$ the central value corresponds to
$\rho_{\ureh}=\rho _{\uend}$ and, therefore, to $N_{\ureh}\simeq 0$,
while $\rho _{\ureh}=\rho _{\unuc}$ is on the lower edge of the
one-sigma confidence interval. Although no constraints can be put on
$T_{\usssRH}$, these plots show that large field inflationary models
are compatible with the data in the sense that the accelerated phase
and the reheating period, which share the same field potential, can
indeed coexist without leading to contradictions when confronted to
the CMB observations. The cases $p=3$ and $p=4$ are slightly different
since the lines of constant reheating temperature are outside the
$1\sigma $ contour for $p=3$ and outside the $2\sigma $ for
$p=4$. However, these models were already strongly disfavoured because
of their ``high'' value of $p$.

\subsection{Small field models}
\label{sec:exactsf}

In this subsection, we study the exact numerical integration of small
field models. The number of inflationary parameters is now four
leading to an overall number of eight model parameters [see
equation~(\ref{potentialsf})]. The best fit model corresponds to $\chi
^2\simeq 11252.2$, a value very similar to the one obtained in the
large field case albeit involving the additional parameter $\mu$.

\begin{figure}
\begin{center}
\includegraphics[width=10cm]{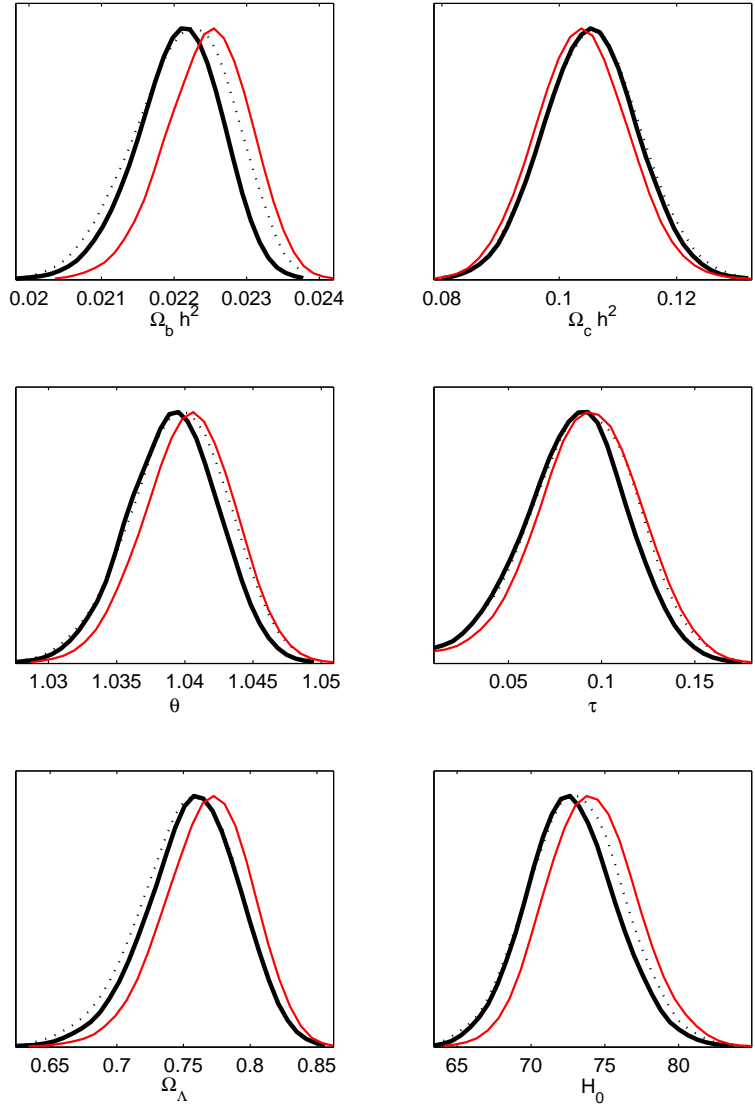}
\caption{Mean likelihoods (dotted black lines), marginalised posterior
  probability distributions from a flat prior on $\kappa \mu$ in
  $[0.1,10]$ (thick solid black lines) and marginalised posterior
  probability distributions from a flat prior on $\kappa \mu$ in
  $[0.1,100]$ (thin solid red lines) for the $\Lambda$CDM cosmological
  base parameters. The cosmological constant and the Hubble parameter
  posteriors are also represented. They have been obtained from a
  direct numerical integration of the small field models power
  spectra.}
\label{small_cosmo_1D}
\end{center}
\end{figure}

In figure~\ref{small_cosmo_1D}, we have plotted the posteriors
obtained for the base cosmological parameters. As previously, the
dotted black curves are the mean likelihoods, the thick solid black
curves represent the marginalised probabilities obtained under a flat
prior on the parameter $\mu /\mpl$ in the range $[0.1,10]$ and,
finally, the thin solid red lines refer to the marginalised
probability stemming from a flat prior on $\mu /\mpl $ but now in the
range $[0.1,100]$.  The most probable values of these parameters are
compatible with what was found in the slow-roll models and the exact
integration of the large field models. It is nevertheless interesting
to notice that the determination of these distributions is not
completely insensitive to the choice on the prior. This is
particularly the case for the base parameters $\OmegaB h^2$ and
$\tau$, which is not really surprising since both of them have
significant correlations with the spectral index $\nS$.

\begin{figure}
\begin{center}
\includegraphics[width=11cm]{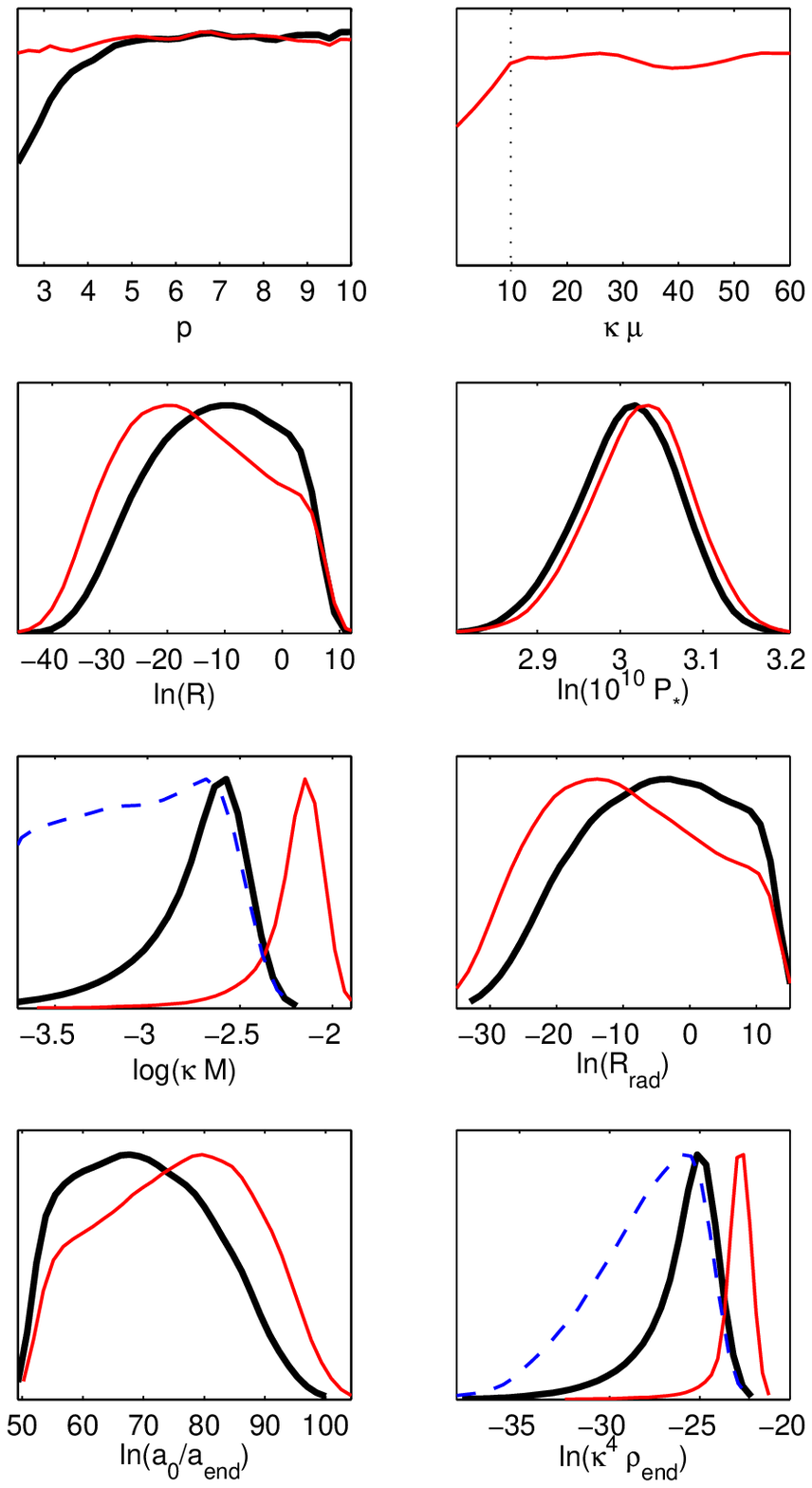}
\caption{Marginalised posterior probability distributions associated
  with a flat prior on $\kappa \mu$ in $[0.1,10]$ (thick solid black
  lines) and a flat prior on $\kappa \mu$ in $[0.1,100]$ (thin solid
  red lines) for the primordial parameters characterising small field
  models and the subsequent reheating phase. The blue dashed line
  represents the marginalised probabilities with a flat prior on $\log
  (\kappa \mu)$ in $[-1,1]$ and is drawn only when it is significantly
  different from the others. The dotted vertical line in the top right
  panel indicates the relevant range of values for the $\kappa \mu \in
  [0.1,10]$ prior.}
\label{small_inf_1D}
\end{center}
\end{figure}

In figure~\ref{small_inf_1D}, we have represented the posterior
distributions obtained for the inflationary parameters. The
conventions are the same as in figure~\ref{small_cosmo_1D}. The blue
dashed line is the marginalised probability with a flat prior on $\log
(\mu/\mpl )$ and is represented only when it leads to different
results. On the top left panel, one sees that the parameter $p$ is not
constrained since the corresponding distribution is basically
flat. Let us notice that a flat prior has been chosen for $p$ in the
range $[2.4,10]$, the case $p\rightarrow 2$ being very difficult to
handle numerically due to computational accuracy limitation. This
comes from the exponential behaviour of the $\epsilon_1$ parameter as
we discussed in section~\ref{sec:sfmodel}. In fact the case $p\simeq
2$ is interesting because one notices in figure~\ref{small_inf_1D}
that the value of the marginalised probability depends on the prior in
this regime. When the flat prior on $\mu/\mpl$ lies in $[0.1,10]$,
then the values $p\simeq 2$ are slightly disfavoured but this tendency
completely disappears when the prior is extended to $[0.1,100]$. This
effect can be physically interpreted by means of the slow-roll
approximation presented in section~\ref{sec:sfmodel}.  In the limit
$\mu/\mpl \ll 1$, the first slow-roll parameter is exponentially
suppressed while the second one is given by
equation~(\ref{srapproxsf}), \ie
\begin{equation}
\epsilon _2 \simeq \frac{1}{2\pi} \left(\frac{\mpl}{\mu} \right)^2 .
\end{equation}
In this limit, there is no means to comply with the observational
constraints because $\epsilon _2$ becomes large which is strongly
disfavoured by the data. However, as shown before, the above formula is
incorrect if $\mu /\mpl $ is not small. When the slow-roll parameters
are correctly evaluated, we have already pointed in
section~\ref{sec:sfmodel} that a compatible value of $\epsilon _2$ can
be obtained provided a large enough value of $\mu/\mpl $ is
chosen. Therefore, if the prior is too narrow, then such a large value
of $\mu/\mpl $ cannot be reached and the case $p\simeq 2$ is
disfavoured. On the other hand, if the $\mu/\mpl$ prior allows for
large enough value, then the tilt of the power spectra can be made
perfectly compatible with the data by singling out high $\mu/\mpl$
values. The case $p\simeq 2$ becomes perfectly compatible and, hence,
the posterior distribution on $p$ flattens. Therefore, from a data
analysis point of view, we conclude that the model with $p\simeq 2$ is
not excluded at all. From a theoretical point of view, the situation
is less clear since one could argue that the small field models
require $\mu < \mpl$. Note that, for $p=2$, it is necessary to have
$\mu/\mpl > 0.25$ in order for the energy scale $M$ to be above the
TeV (but $\mu/\mpl > 0.03$ only for $p=2.1$). Finally, let us
stress that we always have $\phi/\mu \ll 1$ and this is mandatory
because the small field potential~(\ref{potentialsf}) should be viewed
as the first leading terms of a Taylor expansion. However, this does
not require $\mu /\mpl \ll 1$. This condition appears when one
requires the vev of the inflaton field to be explicitly smaller than
the Planck mass: a theoretical prior choice but of some interest since
we have just shown that it modifies the $p$ posterior distribution
close to $p=2$. The question of knowing whether very large inflaton
vev in comparison with the Planck mass makes sense is
controversial~\cite{Linde:2004kg, Lyth:1998xn}. 

\par

The top right panel in figure~\ref{small_inf_1D} represents the
posterior associated with the parameter $\mu/\mpl $. As can be seen,
this parameter is basically unconstrained. The vertical dotted line is
the upper limit of the posterior derived under the flat prior in the
range $[0.1,10]$. In this case the respective posteriors obtained from
both the prior choices match. We notice that, if one focuses on the
range $[0.1,10]$ only, then there is a tendency for a ``large''
$\kappa \mu$, \ie close to the upper bound $\kappa \mu \simeq 10$. As
explained before, in the slow-roll language, very small values of
$\kappa \mu $ are indeed disfavoured since they would lead too large
values of the parameter $\epsilon _2$. In the full range $[0.1,100]$
the $\kappa \mu$ posterior is essentially flat and there is no
significant constraint on the values of $\kappa \mu$.

\begin{figure}
\begin{center}
\includegraphics[width=15cm]{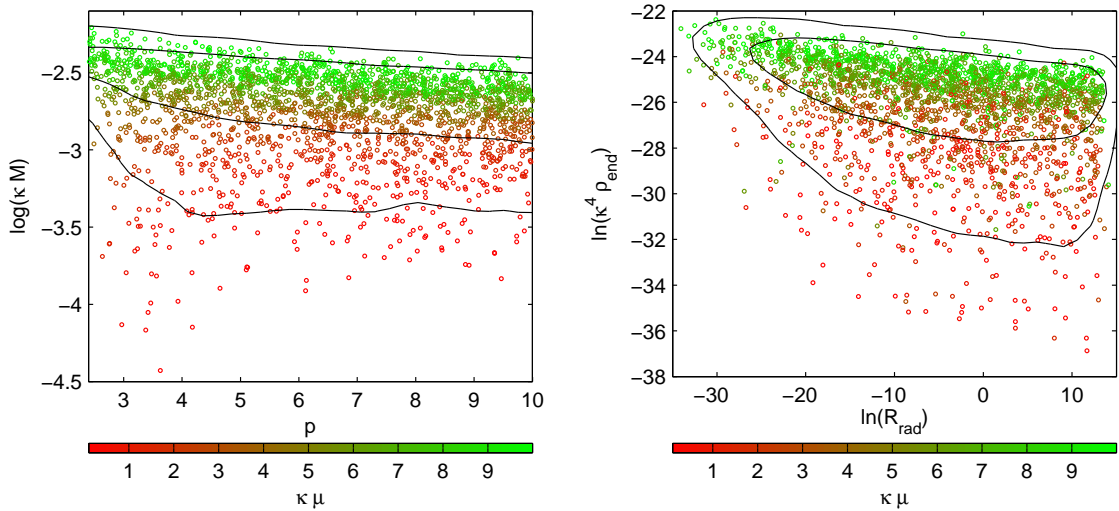}
\caption{One and two-sigma confidence contours of the two-dimensional
  marginalised posterior probability distributions (point density) in
  the plane $[\log (\kappa M),p]$ (left panel) and $[\ln (\kappa ^4
    \rho _{\uend}),\ln R_{\urad}]$ (right panel). The correlations
  with $\kappa \mu $ are indicated by the color bar.}
\label{small_inf_3D}
\end{center}
\end{figure}

Let us now turn to the energy scale $M$ and the energy density $\rho
_{\uend}$ at the end of inflation. A first remark is that these
parameters are poorly constrained and their corresponding posterior
distributions are strongly dependent on the prior choice. Let us try
to interpret these curves in more details making use our slow-roll
understanding. If $\mu /\mpl \ll 1$, then, in the slow-roll
approximation, the potential scale $M$ reads
\begin{equation}
\left(\frac{M}{\mpl}\right)^4\simeq \frac{45p^2}{32\pi }
\frac{Q_\mathrm{rms-PS}^2}{T^2}\left[N_*\frac{p(p-2)}{8\pi
}\right]^{-2(p-1)/(p-2)}\left(\frac{\mu }{\mpl }\right)^{2p/(p-2)}\, .
\end{equation}
As already discussed in section~\ref{sec:sfmodel}, this energy scale
can be very small. For $\mu /\mpl > 1$, following the discussion
after equation (\ref{phistarlfmusmall}), one may use an approximation
of equation~(\ref{normasf}) by using the limit $\phi _*\rightarrow
1$. One gets $\epsilon _1\simeq {\calO}(1)$ which gives a very rough
order of magnitude estimate (see figure~\ref{compeps}). Finally
\begin{equation}
\label{approxMsflargemu}
\left(\frac{M}{\mpl}\right)^4\simeq {\cal O}\left(
\frac{Q_\mathrm{rms-PS}^2}{T^2}\right)\, ,
\end{equation}
that is to say $\log (\kappa M)\simeq -1.9$. This is in good agreement
with what is observed in figure~\ref{small_inf_1D}. If we use the
prior $\kappa \mu$ in $[0.1,100]$ (thin red solid curve) then we
allows for larger value of $\kappa \mu $ than with the prior $\kappa
\mu$ in $[0.1,10]$ (thick solid black curve) and, consequently, the
marginalised probability is shifted towards the large energy scales by
volume effects in the parameter space. If, on the contrary, we use an
uniform prior on $\log(\mu /\mpl )$ (dashed blue line), then we favour
the small values of $\mu/\mpl $ and the marginalised probability opens
up in the direction of small $M/\mpl$, in accordance with the previous
considerations. These ones are also confirmed by inspecting the left
panel of figure~\ref{small_inf_3D}. For a fixed value of $p$, the
larger $\kappa \mu $ is, the larger $\log(\kappa M)$
becomes. Moreover, when $\kappa \mu > 1$, $\log(\kappa M)$ has a very
weak dependence on $p$ as one may guess from
(\ref{approxMsflargemu}). Finally, let us stress that the same
interpretation also applies to $\rho _{\uend}$. In addition we recover
the same cutoff at $\ln (\kappa ^4\rho_\uend)\simeq -20$ as for large
field models, which is once again simply the manifestation of the
constraint on the level of gravitational waves, \ie in the slow-roll
language the result of $\epsilon _1<0.022$.

\begin{figure}
\begin{center}
\includegraphics[width=15cm]{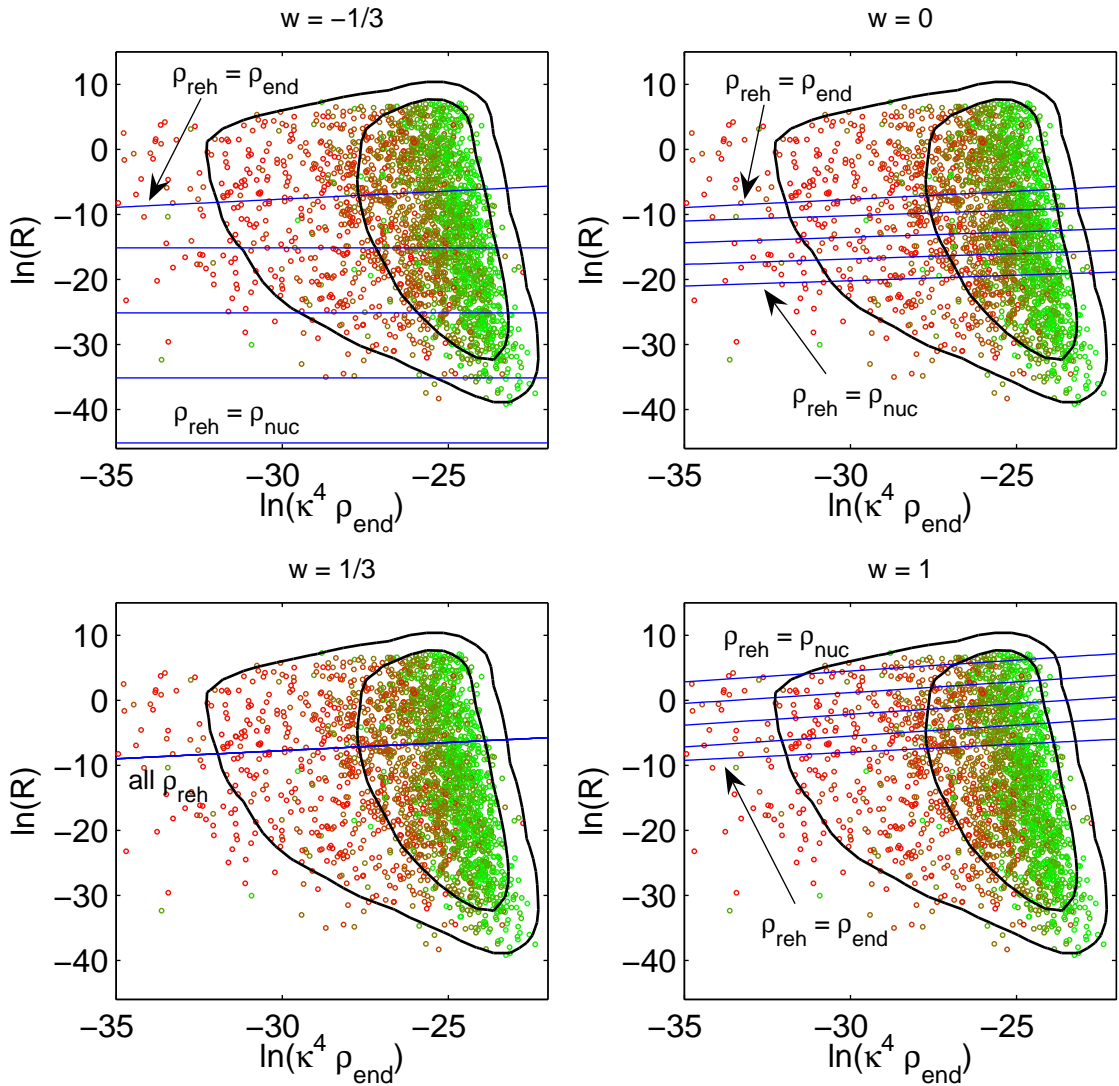}
\caption{One and two-sigma confidence intervals (solid contours) of
  the two-dimensional marginalised posteriors (point density) in the
  plane $[\ln R, \ln (\kappa ^4 \rho _{\uend})]$. The colour code is
  the same as in figure~\ref{small_inf_3D}. The reheating predictions
  are represented in the particular cases where the state parameter
  $\wstate _{\ureh}=-1/3$ (top left panel), $\wstate _{\ureh}=0$
  (top right panel), $\wstate _{\ureh}=1/3$ (bottom left panel) and
  $\wstate _{\ureh}=1$ (bottom right panel). The various solid lines
  correspond to different values of the reheating temperature $(1/4)
  \ln(\kappa^4 \rho_{\ureh})$ ranging from $-45$, $-35$, $-25$, $-15$
  to $-(1/4)\ln(\kappa^4 \rho_{\uend})$.}
\label{small_inf_4D}
\end{center}
\end{figure}

Finally, let us discuss the quantities related to the reheating phase,
in particular $\ln R$. To start with, it is important to remark that
contrary to the large field models, the quantities characterising the
reheating phase cannot be related to the free parameters of the small
field potential. In particular, the formula (\ref{omegareh}) is
specific to monomial potentials and can no longer be applied to the
present case. Now, from figure~\ref{small_inf_1D}, we see that the
quantity $\ln R$ has a lower bound slightly above the lower prior
bound and seems to be constrained. On the contrary, the upper limit
only comes from the upper bound of the prior. Let us recall that the
lower limit of the prior comes from the requirement $\rho _{\ureh} >
\rho _{\unuc}$ which approximately means $\ln R > -46$. Here, we
obtain the marginalised lower bound
\begin{equation}
\ln R > -34\,,
\end{equation}
at two-sigma level of confidence (and $\ln R > -22$ at
one-sigma). These values are obtained under the prior $\kappa\mu$ in
$[0.1,100]$ which weakens the lower bound of $\ln R$ (see
figure~\ref{small_inf_1D}). The corresponding bounds for the derived
parameter $R_\urad$ reads $\ln R_{\urad} > -31$ at $2\sigma$. Note
that these constraints are recovered on the shape of the $\ln
(a_0/a_{\uend})$ posterior at large values. It is important to notice
that these limits are modified when the prior is changed, reducing the
upper bound on the $\kappa \mu$ prior would induce tighter constraints
on $\ln R$. The small field models are the only class of models for
which we find some non trivial, albeit weak, constraints on the
reheating parameter.

\par

Let us now try to understand the physical origin of this bound in more
details. At first glance it may be surprising to get a constraint on a
parameter which is not directly involved in the generation of the
primordial cosmological perturbations. In fact, as we show below, this
is a side-effect coming from the constraint existing on the scalar
power spectrum tilt, that is to say the on the spectral index $\nS-1$,
or $\epsilon _2$ in the small field slow-roll approximation.

\par

In the case of small field models, the first slow-roll parameter is
extremely small. Using equation (\ref{a0aend}), one gets
\begin{equation}
\label{eq:lnRsf}
\ln R\simeq -58 +\ln \left[\frac{k}{a_0}\left(\mbox{Mpc}^{-1}\right)\right]
+N_k-\frac12\ln \frac{V_k}{V_{\uend}}\, ,
\end{equation}
where $k/a_0$ is a given physical scale, $N_k$ the number of e-folds
between the time at which the scale $k$ leaves the Hubble radius
during inflation and the end of inflation, and $V_k$ denotes the value
of the inflaton potential at Hubble exit. This equation means that
$\ln R$ is directly related to the time of Hubble exit. For different
values of $\ln R$, the modes of cosmological interest today probe
different ranges of field value along the inflationary potential. On
the other hand, not all the field values are compatible with the data,
simply because in a region where the potential is too steep (or even
too flat), the spectral index may take unacceptable values when
compared to the observations.

\par

One can quantitatively test the previous discussing. In practice, we
choose $k/a_0=0.05 \, \mbox{Mpc}^{-1}$, and one may approximate $V_k$
by $M^4$. Under these assumptions, the effect described before only
comes from the values taken by $N_k$. The slow-roll value of $\phi
_{\uend}$ is known from $\epsilon_1(\phi_\uend)=1$ whereas the value
of $\phi _k$ is obtained by solving the equation $\epsilon_2(\phi
_k)=\epsilon _2\vert _{\uobs}$. Recall that the function $\epsilon
_2(\phi)$ is explicitly given in (\ref{srsf}) and $\epsilon _2\vert
_{\uobs}$ is supposed to be a fiducial value of the second slow-roll
parameter compatible with the data. Using (\ref{trajecsf}) and
(\ref{trajecsfp2}) for $\phi_\uend$ and $\phi_k$ uniquely determines
$N_k$ and thus $\ln R$ by means of (\ref{eq:lnRsf}). Since only a
limited range of $\epsilon _2\vert _{\uobs}$ values is compatible with
the data, one indeed expect some constraints on the reheating
parameter $\ln R$. For instance, all $\ln R$ values associated with
$\epsilon _2\vert _{\uobs}= 0.07$ should be on the edge of the $\ln R$
marginalised posterior.

\begin{table}
\begin{center}
\begin{tabular}{|c|r|r|r|}
\hline \multicolumn{4}{|c|}{$\kappa \mu = 1$} \\
\hline $\epsilon_2$ & 0.03 & 0.05 & 0.07 \\
\hline$N_k$ & 100 & 60 & 42 \\
\hline$\ln R$ & 38 & -2 & -19 \\ \hline
\end{tabular}
\begin{tabular}{|c|r|r|r|}
\hline \multicolumn{4}{|c|}{$\kappa \mu = 15$} \\
\hline $\epsilon_2$ & 0.03 & 0.05 & 0.07 \\
\hline $N_k$ & 66 & 34 & 22 \\
\hline $\ln R$ & 3 & -28 & -40 \\ \hline
\end{tabular}
\begin{tabular}{|c|r|r|r|}
\hline \multicolumn{4}{|c|}{$\kappa \mu = 50$} \\
\hline $\epsilon_2$ & 0.03 & 0.05 & 0.07 \\
\hline$N_k$ & 40 & 23 & 16 \\
\hline$\ln R$ & -22 & -39 & -46 \\ \hline
\end{tabular}
\caption{Theoretical predicted slow-roll values of $\ln R$ for small
  field models with $p=4$ as a function of $\epsilon_2$ and for the
  mode $k/a_0 = 0.05 \, \mathrm{Mpc^{-1}}$. Some of these values are
  not compatible with the data (see text).}
\label{tab:rehsf}
\end{center}
\end{table}

In table~\ref{tab:rehsf}, we have derived the expected values of $\ln
R$ with respect to $\epsilon _2\vert _{\uobs}$ in the pure slow-roll
approximation for $p=4$ and for different values of $\kappa \mu$. For
$\kappa \mu=15$, the value $\epsilon _2=0.07$ is on the edge of the
confidence interval derived in section~\ref{sec:wmapsr} and
corresponds to $\ln R \simeq -40$: this is compatible with the drop of
the $\ln R$ posterior distribution around the same value seen in
figure~\ref{small_inf_1D}. It is interesting to note that for smaller
values of $\kappa \mu$, the relevant $N_k$ and $\ln R$ are pushed
toward higher values (see table~\ref{tab:rehsf}). This confirms that,
if one has the theoretical prejudice that $\mu/\mpl <1$, then the
bound on $\ln R$ is much tighter. Finally, to conclude this
discussion, one can estimate how much the change of the $\kappa \mu$
prior propagates to the other derived parameters. From
equation~(\ref{a0aend})
\begin{equation}
\Delta \left(\ln \frac{a_0}{a_{\uend}}\right)= -\Delta \left(\ln
R\right)+\frac12 \Delta \left[\ln \left(\kappa ^4 \rho _{\rm
end}\right)\right],
\end{equation}
and from table~\ref{tab:rehsf} and figure~\ref{small_inf_1D}, we see
that when one changes the prior from $\kappa \mu$ in $[0.1,10]$ to
$\kappa \mu$ in $[0.1,100]$, one has $\Delta \left(\ln R\right)\simeq
-10$ and $\Delta \left[\ln \left(\kappa ^4 \rho _{\uend}\right)
\right] \simeq +4$. Therefore, one gets $\Delta \left[\ln
(a_0/a_{\uend})\right]\simeq +12$ which is also the variation observed
in the bottom left panel in figure~\ref{small_inf_1D}.

\par

\begin{figure}
\begin{center}
\includegraphics[width=14cm]{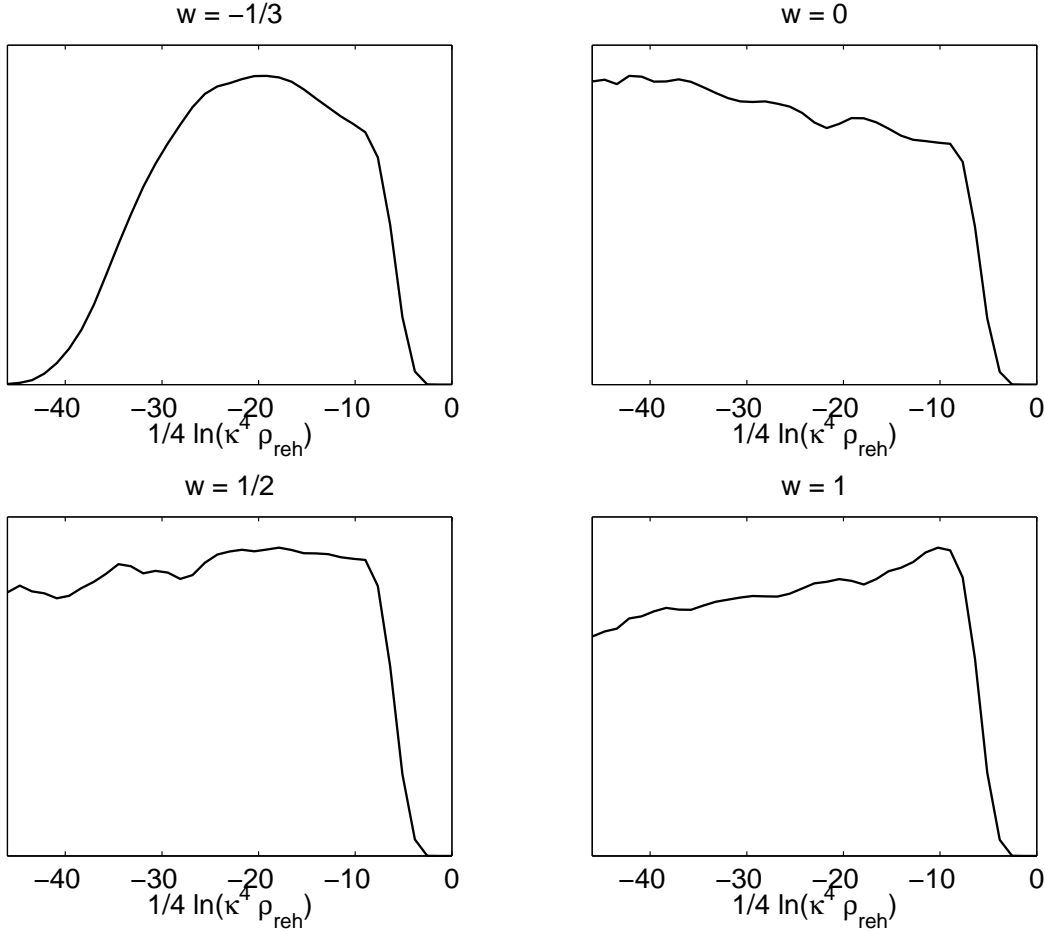}
\caption{Marginalised posterior probability distribution for the
  reheating energy scale $\ln(\kappa \rho_\ureh^{1/4})$ assuming a
  constant state parameter $\wstate_\ureh$. For the extreme case
  $\wstate_\ureh=\simeq-1/3$, the small field models prefer a reheating
  temperature $T_\ureh > 2\,\mbox{TeV}$ at $95\%$ confidence level.}
\label{small_inf_reh}
\end{center}
\end{figure}

Let us now try to propagate the weak constraint on $\ln R$ to the
phase of reheating itself. The reheating temperature can be evaluated
as in (\ref{reheatTlf}) except that, as already mentioned, there is no
longer a link between the state parameter $\wstate_\ureh$ and
$p$. However, for a constant $\wstate_\ureh$, one still has
\begin{eqnarray}
\label{eq:sftreh}
\ln \left(g_*^{1/4}\kappa T_{\usssRH}\right) &=&  \frac{3+ 3\wstate
  _{\ureh}}{1-3\wstate_\ureh} \ln R - \frac{1+3
  \wstate_\ureh}{2-6\wstate_\ureh} \ln (\kappa ^4\rho _{\uend})
\nonumber \\ & + & \frac{1}{1- 3\wstate_\ureh} \ln
\left(\frac{3+ 3\omega
    _{\ureh}}{5-3\wstate _\ureh} \right) + \ln
\dfrac{30^{1/4}}{\sqrt{\pi }}.
%\ln \left(g_*^{1/4}\kappa T_{\usssRH}\right) &=& \frac14 \ln (\kappa
%^4\rho _{\uend}) + \frac{3+ 3\wstate _{\ureh}}{1-3\wstate_\ureh} \ln R
%  \nonumber \\ & + & \ln \left[\frac{30^{1/4}}{\sqrt{\pi
%  }}\left(\frac{3+ 3\omega _{\ureh}}{5-3\wstate _\ureh}
%      \right)^{1/4}\right].
\end{eqnarray}
 For a fixed value of $T_{\usssRH}$ (and of $\wstate_{\ureh}$),
 equation~(\ref{eq:sftreh}) represents a line in the plane $[\ln R,
   \ln (\kappa ^4 \rho _\uend)]$. These lines are plotted in
 figure~\ref{small_inf_4D} for four values of $\wstate _{\ureh}$ and
 for the reheating temperature $(1/4) \ln(\kappa^4 \rho_{\ureh})$ equal
 to $-45$, $-35$, $-25$, $-15$, and $-(1/4)\ln(\kappa^4
 \rho_{\uend})$. As can be seen, the bound $\ln R > -34$ does not
 tighten significantly the allowed $T_{\usssRH}$ values. The only
 lower limit that can be extracted concerns the rather extreme
 reheating limit $\wstate _{\ureh}\simeq -1/3$ for which the lines
 corresponding to $g_*^{1/4} T_{\usssRH} \lesssim 1\,\mbox{MeV}$ lie
 outside the two-sigma level contour over $\ln R$ and
 $\rho_\uend$. The marginalised posteriors on $T_{\usssRH}$ can be
 derived by importance sampling~\cite{Lewis:2002ah} from the previous
 analysis and are plotted in figure~\ref{small_inf_reh} for four
 values of the parameter $\wstate_\ureh$. Notice that we have avoided
 the pure radiation-like case $\wstate_\ureh=1/3$ which is problematic
 in the inversion formula (\ref{eq:sftreh}), as previously
 discussed. In the extreme case $\wstate_\ureh \simeq -1/3$, the
 reheating temperature satisfies
\begin{equation}
g_*^{1/4} T_\ureh > 2 \, \mbox{TeV},
\end{equation}
at $95\%$ of confidence. Of course, such a bound close to the TeV
scale is already disfavored by BBN and may therefore seem
irrelevant. However, we would like to point out that, in the present
context, it was obtained from the CMB data only and to our knowledge,
this is the first time that this can be done.

\par

Let us conclude by noticing that, with better data (for instance
coming from the Planck satellite~\cite{Lamarre:2003zh}), it is likely
to obtain much better constraint on the reheating temperature in the
framework of small field models. As shown above, this is the result of
the influence of $\ln R$ on the \emph{observed} value of the the
spectral index. Therefore, much tighter constraints on $\nS$ would
probably lead to a narrower range of allowed reheating temperatures.

\subsection{Hybrid Models}
\label{sec:exacthyb}

\begin{figure}
\begin{center}
\includegraphics[width=10cm]{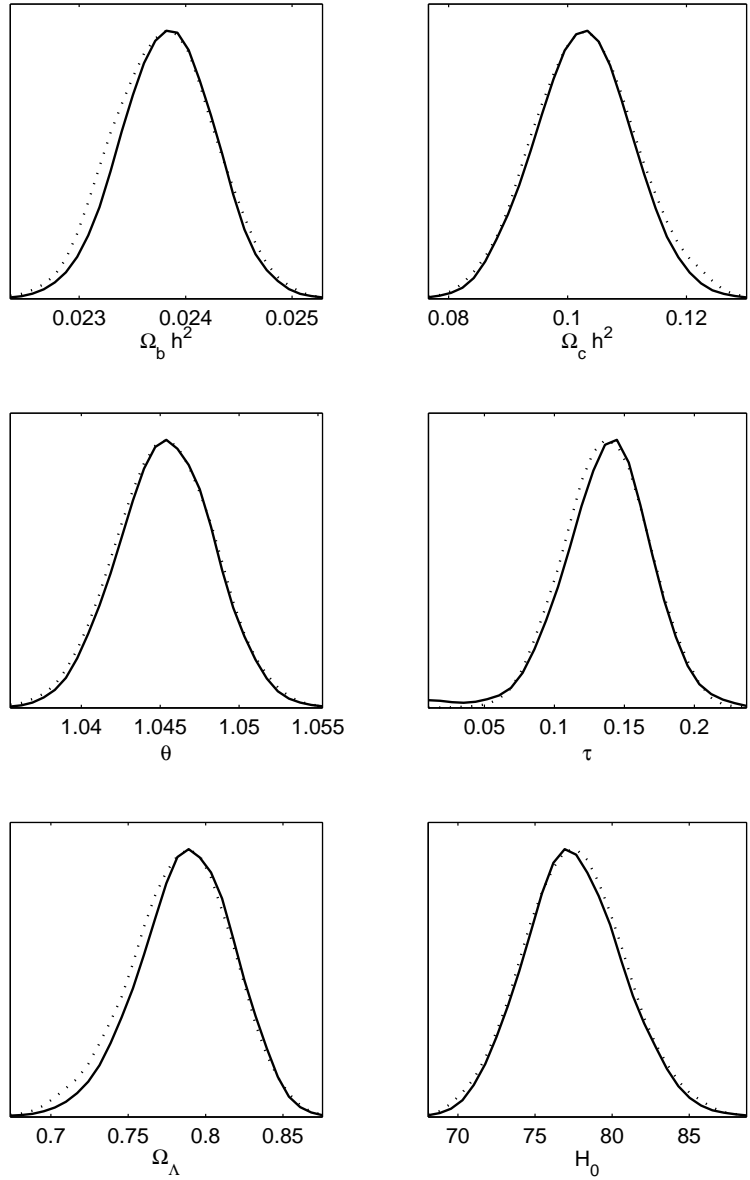}
\caption{Marginalised posterior probability distributions (solid black
  lines) and mean likelihoods (dotted lines) for the $\Lambda$CDM base
  cosmological parameters obtained for the hybrid models. The
  cosmological constant and the Hubble parameter derived posteriors
  are also shown.}
\label{hybrid_cosmo_1D}
\end{center}
\end{figure}

Let us now turn to the exact numerical integration of hybrid
models. This class of models involves nine parameters in total, the
four base cosmological parameters plus five inflationary
parameters. The best fit model has a $\chi ^2 \simeq 11257.2$, a value
which is larger than the one obtained for large and small field
models, the difference being $\Delta \chi^2\simeq +5$ in spite of one
extra parameter with respect to small field models and two with
respect to large field models [see equation~(\ref{potentialhyb})]. It
is therefore fair to say that this class of models does not fit the
data as well as the two previous ones.

\par

In Fig.~\ref{hybrid_cosmo_1D}, we have represented the posterior
distributions for the base cosmological parameters. The most striking
features of these plots are the values of the baryon number density
$\OmegaB h^2\simeq 0.024$, and of the optical depth, $\tau \simeq
0.14$. They should be compared to what was obtained before, say for
large or small fields model where $\OmegaB h^2 \simeq 0.0225$ and
$\tau \simeq 0.1$. In particular, such a large value of $\OmegaB$
exacerbates the tension between the results obtained with the CMB and
the BBN. Indeed, one has~\cite{Coc:2003ce} $0.020<\OmegaB h^2<0.024$
from the deuterium measurements and $0.007<\OmegaB h^2<0.014$ from
${}^7\mbox{Li}$. Even if one considers only the Deuterium
measurements, then the previous result pushes $\OmegaB h^2$ towards
the upper limit of the BBN allowed range.

\par

As discussed in section~\ref{sec:hybmodel} in the slow-roll context,
the hybrid models are associated with a blue tilted scalar power
spectrum, \ie typically $\nS>1$. Although not totally excluded by the
WMAP data, this situation is nevertheless disfavoured. Since the
parameter $\nS$ is degenerated with the two parameters $\OmegaB h^2$
and $\tau$, the more likely way to accommodate such primordial power
spectra with the observed CMB anisotropies is to compensate the
$\nS>1$ values of the spectral index by modifying these two
quantities. In particular, in order to compensate an excess of power
at small scales (blue spectrum), one has to increase $\OmegaB h^2$ and
$\tau$ as observed.

\begin{figure}
\begin{center}
\includegraphics[width=11.5cm]{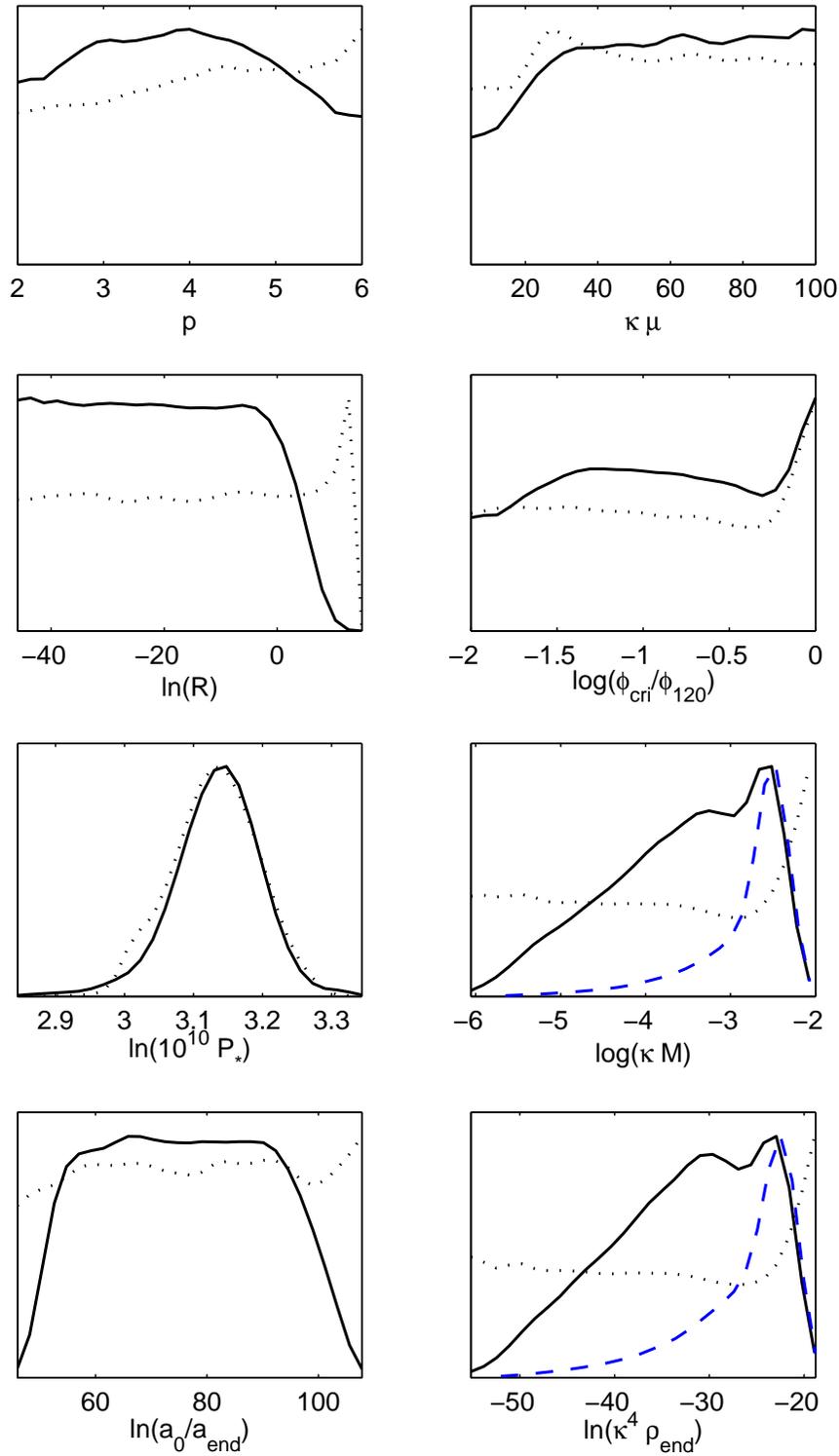}
\caption{Marginalised posterior probability distributions (solid black
  lines) and mean likelihoods (dotted curves) of the hybrid model
  inflationary parameters drawn from a flat prior on $\log \phi
  _{\ucri}$ in the range $[-2,0]$. The blue dashed lines represents
  the marginalised posteriors obtained from an uniform prior on $\phi
  _{\ucri}$ whenever they do not match with the former ones.}
\label{hybrid_inf_1D}
\end{center}
\end{figure}

In figure~\ref{hybrid_inf_1D}, we have plotted the posteriors obtained
for the inflationary parameters. As already discussed, the hybrid
models involve an extra parameter, namely the value of the inflaton
field at which the instability occurs and inflation stops. Also, we
have to pay attention to probe only the vacuum dominated regime which
is, by definition, what we mean by hybrid inflation. Otherwise, the
situation would be very similar to the large field models studied
before.

\par

{}From the slow-roll analysis of section~\ref{sec:hybmodel}, when
$\mu/\mpl <p/(8\sqrt{\pi})$, the field $\phi $ has to satisfy
$\phi<\phi _{\epsilon}^-$ (see also figure~\ref{pothyb}). However, for
a chosen value of $\phi_{\ucri}$ too close to $\phi _{\epsilon }^-$,
it is not guaranteed to obtains a total number of e-folds larger than
the required $60$. In the following, we define $\phi _{120}$ to be the
maximum value of $\phi _{\ucri}$ leading to at least $120$ e-folds of
inflation ($\phi$ is decreasing as hybrid inflation proceeds). In
other words, $\phi _{\ucri}$ should be smaller than $\phi _{120}$.

\par

If $\mu/\mpl $ is large enough, then $\phi _{\epsilon }^-$ does no
longer exist. In this case, $\phi _{120}$ is defined from the value of
$\phi $ corresponding to the inflexion point of the potential, \ie to
$\epsilon _2=0$ in the slow-roll framework, which is also the field
value maximising $\epsilon _1$. This is a convenient way to identify
the vacuum dominated regime for several reasons. Firstly, $\epsilon
_2=0$ corresponds to $\phi /\mu =1$. Therefore, the regime $\epsilon
_2<0$ means that $\phi /\mu <1$ and, hence, the term $M^4$ dominates
in the hybrid potential. Secondly, $\epsilon _2<0$ means that the
kinetic energy is decreasing absolutely, and relatively to the total
energy which, again, is exactly what we expect in a vacuum dominated
regime. In the following, $\phi _{\ucri}$ will be always measured in
units of $\phi _{120}$.

\begin{figure}
\begin{center}
\includegraphics[width=12.5cm]{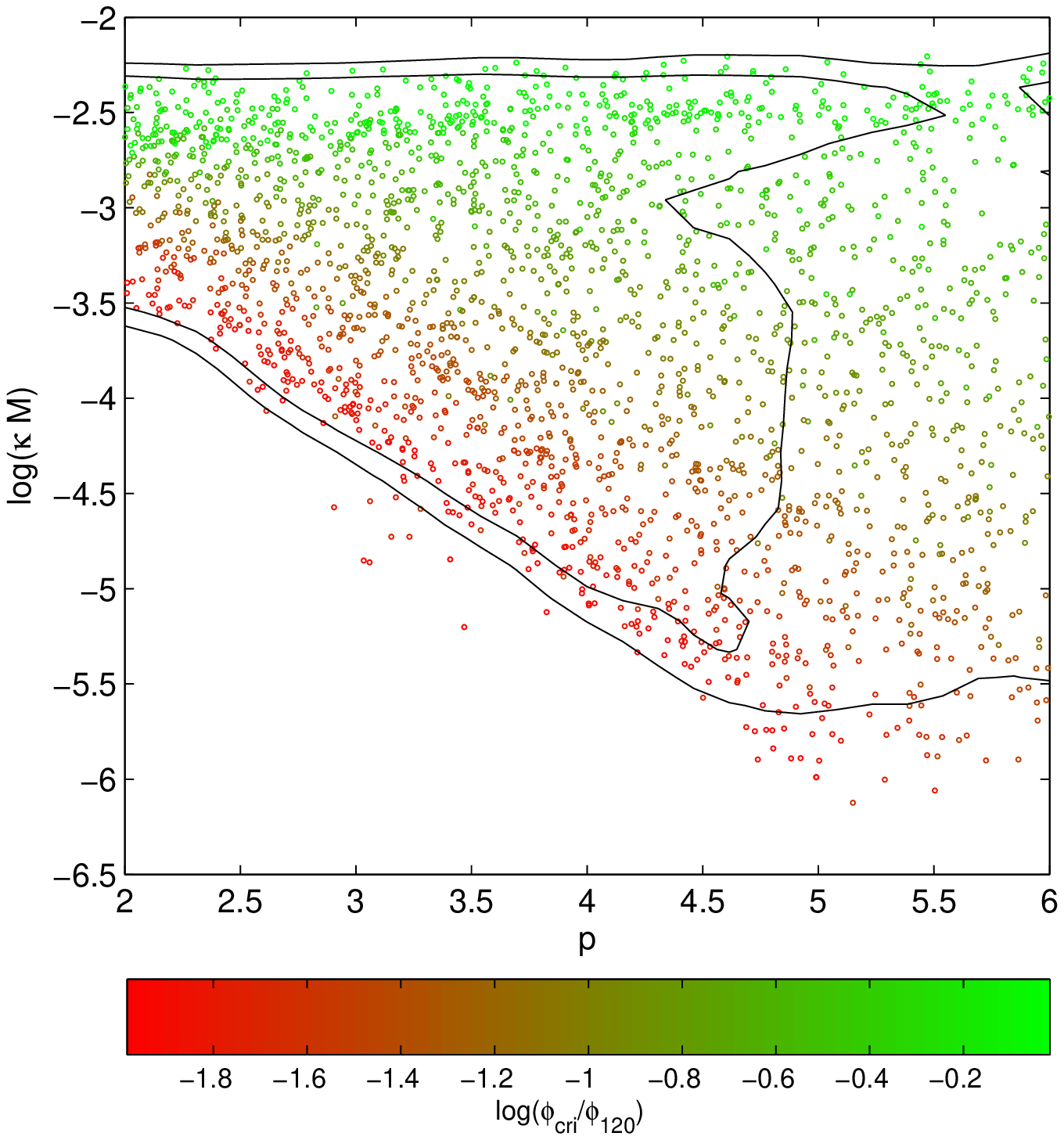}
\caption{Two-dimensional marginalised probability (point density) in
  the plane $[\log(\kappa M),p]$ for various values of $\phi _{\ucri}$
  measured in units of $\phi _{120}$ and indicated by the colour
  bar. The one-sigma and two-sigma confidence interval are represented
  by the solid contours.}
\label{hybrid_inf_3D}
\end{center}
\end{figure}

As seen in figure~\ref{hybrid_inf_1D}, the values of the hybrid
inflationary parameters are not really constrained. The posterior
distributions for $p$ or for the $\kappa \mu$ are basically flat
although small values of $\kappa \mu$ are slightly disfavoured. In the
same manner, the critical value $\phi _{\ucri}$ is not constrained but
values close to $\phi _{\ucri}\simeq \phi _{120}$ seem to be slightly
favoured. With regards to the energy scale $M$ and the energy density
at the end of inflation, $\rho _{\uend}$, we observe the usual cutoff
for the large values of these quantities, as it was the case for large
and small field models. This is still due to the constraint on the
amount of gravitational waves and in terms of slow-roll, on $\epsilon
_1$. These quantities are poorly determined as they strongly depend on
the choice of the prior on $\log \phi _{\ucri}$. In these plots, the
solid black lines represent the marginalised posterior probabilities
obtained with a flat prior on $\log(\phi _{\ucri}/\phi _{120})$,
whereas the dashed blue lines correspond to a flat prior on $\phi
_{\ucri}$. The order of magnitude of the results can be very simply
estimated. For instance, for the scale $M$, since we are in the vacuum
dominated regime, the term $(\phi/\mu)^p$ can be neglected and
$V\simeq M^4$. Then, the WMAP normalisation leads to the same equation
as for small field models, namely (\ref{approxMsflargemu}), which
implies $M/\mpl \simeq 10^{-4}-10^{-2}$, in agreement with
figure~\ref{hybrid_inf_1D}. Concerning the reheating parameter, there
is no constraint at all. The distribution of $\ln R$ is flat and is
only cut by the prior. This is of course the same for the posterior
distribution of the derived parameter $\ln
(a_0/a_{\uend})$. Therefore, for the hybrid models, nothing can be
said on the reheating with the WMAP3 data.

\par

Finally, figure~\ref{hybrid_inf_3D} represents the two-dimensional
marginalised probability in the plane $[\log(\kappa M),p]$ for various
values of $\phi _{\ucri}$ measured in units of $\phi _{120}$. We see
that the allowed range for $M$ is tighter for smaller values of $p$
and that models with large values of $\phi _{\ucri}/\phi _{120}$ tend
to have a scale $M$ larger than models with smaller $\phi
_{\ucri}/\phi _{120}$. Such an effect is expected since the potential
scale $M$, in the vacuum dominated regime, is roughly given by
\begin{equation}
\left(\frac{M}{\mpl}\right)^4\simeq \frac{45\epsilon
_1}{2}\frac{Q_\mathrm{rms-PS}^2}{T^2}\, .
\end{equation}
Now, if we decrease $\phi _{\ucri}$ then we decrease $\phi _*$ and
hence the observed value of $\epsilon _1$ according to
figure~\ref{pothyb} (bottom left panel).  As a result, decreasing
$\phi _{\ucri}$ implies decreasing the energy scale $M$, as observed
in figure~\ref{hybrid_inf_3D}.

\subsection{Running mass models}
\label{sec:exactrm}

\begin{figure}
\begin{center}
\includegraphics[width=10cm]{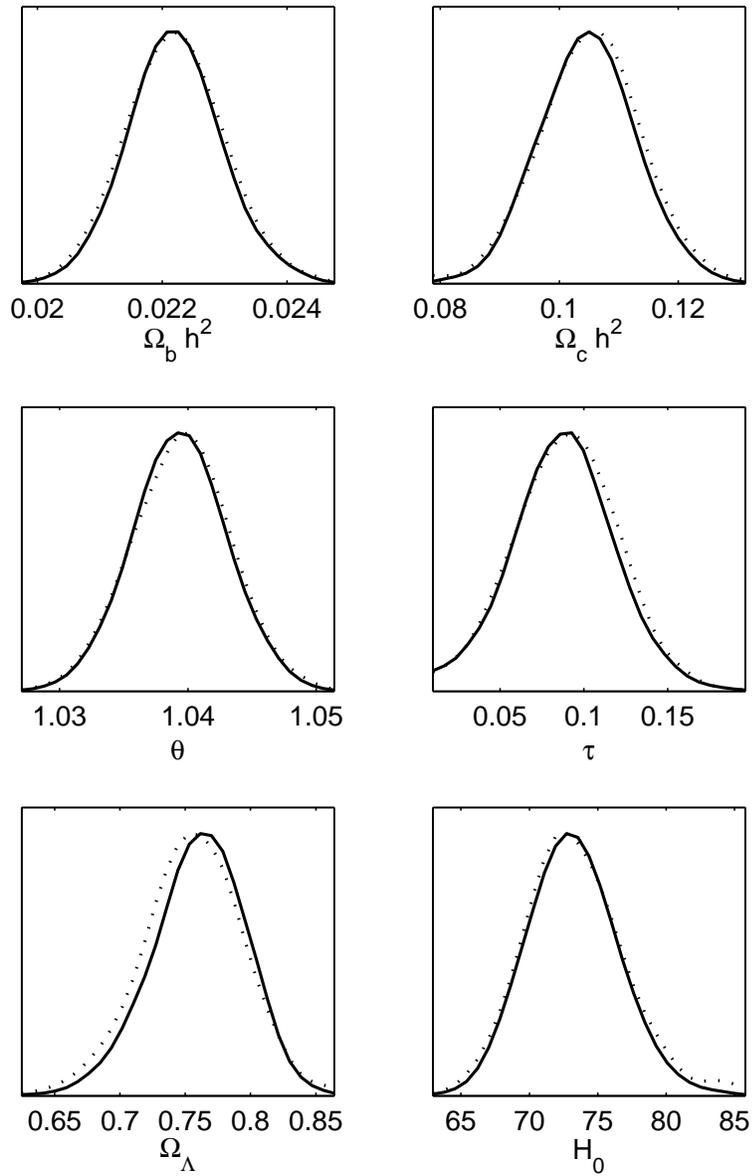}
\caption{Marginalised posteriors (solid curves) and mean likelihoods
  (dotted) for the cosmological parameters in the running mass
  inflation model.}
\label{runlogmass_cosmo}
\end{center}
\end{figure}

Finally, we now turn to the numerical integration of the running mass
models for which the best fit parameters lead to $\chi ^2 \simeq
11252.3$. As discussed in section~\ref{sec:rmmodel}, these models are
characterised by three additional parameters, $c$, $\phi _0$ and $\phi
_{\ucri}$ in comparison with the large field models. The total number
of inflationary parameters is therefore five, as for the hybrid
models, accounting for an overall number of nine parameters. Recall
however that $\phi_\ucri$ comes from the choice to stop inflation by
instability and according to the different versions of this
inflationary scenario, this is not always necessary.

\par

Our numerical integration of these models appears to be limited in the
parameter space due to finite accuracy issues. As pointed in
section~\ref{sec:rmmodel}, the slow-roll approximation gives a field
evolution $\phi (N)$ involving a double exponential behaviour with
respect to the number of e-folds [see
equation~(\ref{trajecrunning})]. As a result, integrating the
equations of motion along $60$ e-folds may require a computing
accuracy much smaller than the quadruple precision computing bound
$10^{-32}$. In order to satisfy this computing requirement, we have
considered a rather limited range for the MCMC running mass
parameters: an uniform prior has been chosen for $\log(\kappa \phi_0)$
in $[-3,0]$, for $c/2$ in $[-0.01,0.1]$ and for $\log(\kappa
\phi_\ucri)$ in $[-1.8,0]$.

\begin{figure}
\begin{center}
\includegraphics[width=14.5cm]{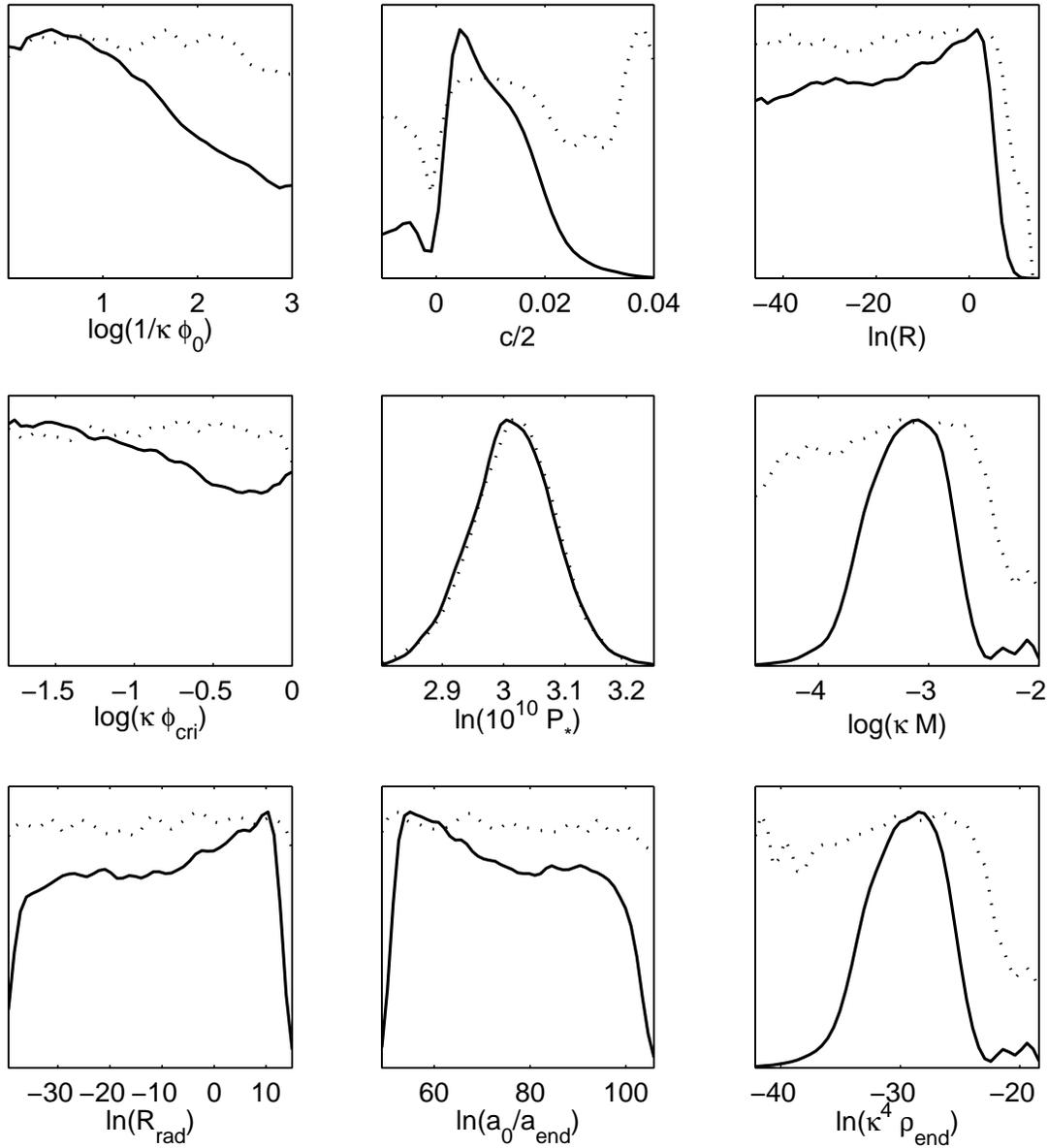}
\caption{Marginalised posteriors (solid curves) and mean likelihoods
  (dotted) for the running mass inflationary parameters. Apart from
  the power spectrum amplitude $\Pstar$ and its derived parameters,
  these posteriors are poorly constrained by the data and mainly
  result from the prior choice (see text).}
\label{runlogmass_1D}
\end{center}
\end{figure}

It may be convenient to compare this approach to the one used in the
literature~\cite{Covi:2004tp}. In that reference, when concerns with
the leading order only, the slow-roll equations $\nS-1\simeq 2(s-c)$
and $\alpha _\usssS\simeq 2sc$ are assumed. Motivated by theoretical
priors, an uniform sampling on $s$ and $c$ in $[-0.2,0.2]$ is
performed and used to compare the model predictions with the data,
presumably by assuming a power-law power spectrum whose tilt and
running are uniquely determined from the $s$ and $c$ values. An
advantage of this method is that one can consider a wide range of
values for the parameters $c$ and $s$. Our method differs from the
fact that no assumption is made on the shape of the primordial power
spectrum. Therefore, the exact numerical integration requires a
sampling on the fundamental model parameters $c$, $\phi_0$,
$\phi_\ucri$. The price to pay, as mentioned before, is that the
resulting domains probed by the derived parameters $s$ and $c$ are
reduced to maintain the required computational accuracy.

\par

The one-dimensional marginalised posteriors for the base cosmological
parameters are plotted in figure~\ref{runlogmass_cosmo} while the
inflationary parameters are represented in
figure~\ref{runlogmass_1D}. As for the other inflationary models which
reasonably fit well the data, the cosmological parameters are centred
to their fiducial values. Concerning the primordial parameters, we
recover the standard bounds on the energy scale of inflation and the
power spectrum amplitude, while the reheating parameters are not
constrained. On the other hand, the running mass parameters exhibit
distorted distributions and large values of $\phi_0$ seem to be
favoured. A priori, the posterior on $c/2$ could be associated with
some confidence intervals. However, as we show in the following, the
patterns observed in figure~\ref{runlogmass_1D} for the running mass
parameters are mainly dominated by the prior choices and the multiple
degeneracies that arise between the model parameters and their
observable effects. This means that these parameters cannot be
robustly constrained in a prior independent way with the current data,
at least in our prior range.

\begin{figure}
\begin{center}
\includegraphics[width=12cm]{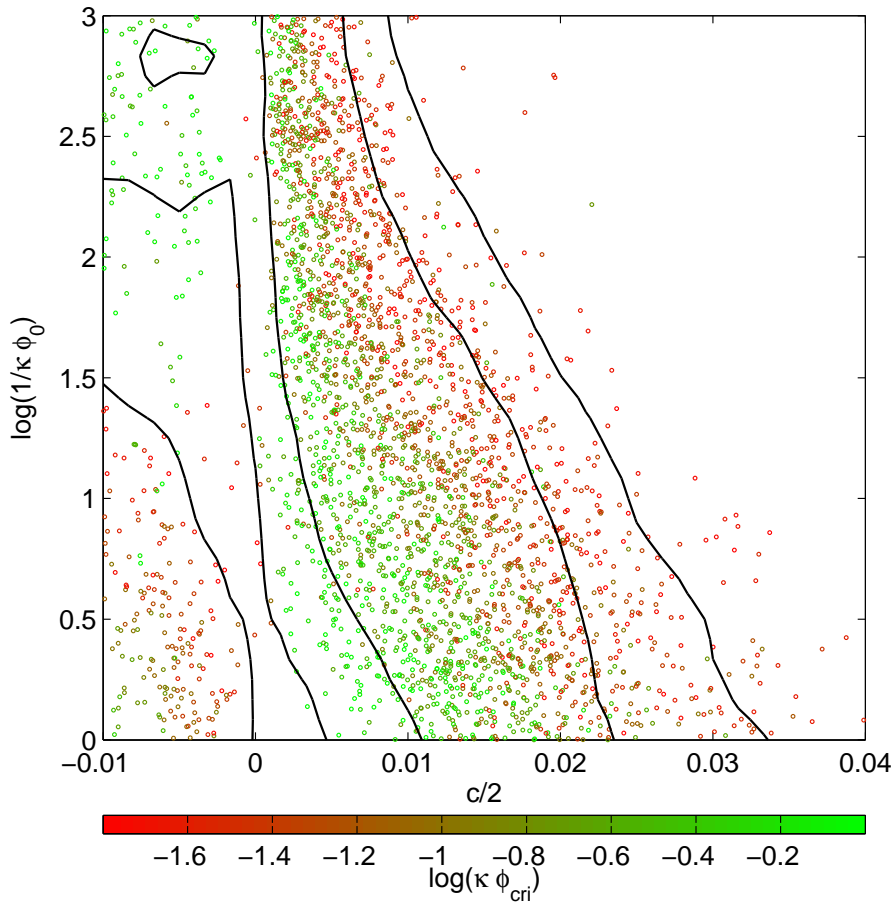}
\caption{Two-dimensional marginalised probability (dot density) in the
  plane $[\log(1/\phi _0),c/2]$ for various values of $\phi_\ucri$
  (colour bar). The one and two-sigma confidence intervals are the
  solid contours.}
\label{runlogmass_3D}
\end{center}
\end{figure}

The degeneracies between the parameters are clearly drawn in the
two-dimensional probability distribution (point density) plotted in
figure~\ref{runlogmass_3D}. The correlations between $c/2$,
$\log(\phi_\ucri)$ and $\log(\phi_0)$ appear as the preferred
``islands'' and ``directions'' explored by the MCMC. The topology of
these correlations can be understood from the slow-roll approximation
detailed in section~\ref{sec:rmmodel}. Indeed, since $\epsilon_1$ is
generically small for this class of models,
equation~(\ref{eps1eps2rm}) implies $1-\nS\simeq \epsilon _2$. Since
the spectral index $\nS$ is constrained by the data, one can recover
the range of model parameters that lead to such an observed value of
$\nS$, or $\epsilon_2$ in the present case.

\par

\begin{figure}
\begin{center}
\includegraphics[width=16cm]{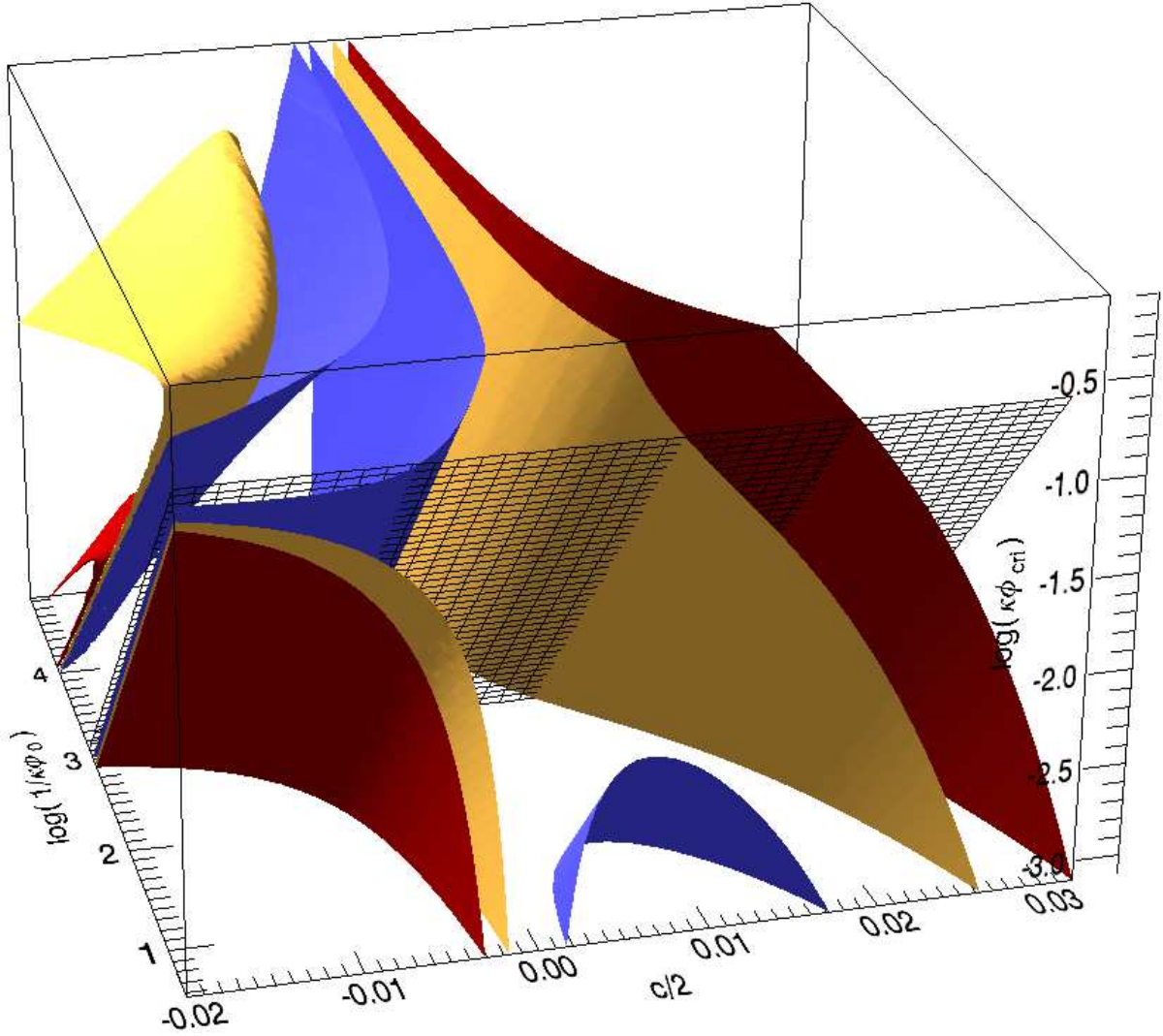}
\caption{Isosurfaces of constant $\epsilon _2$ in the
  three-dimensional parameter space of the running mass inflation
  model. The blue surface is associated with $\epsilon _2=-0.029$ and
  the dark red one with $\epsilon _2=0.074$. These two values
  correspond to the two-sigma confidence intervals at next-to-leading
  order in slow-roll expansion and for the Jeffreys' prior on
  $\epsilon_1$. The yellow surface corresponds to $\epsilon
  _2=0.034$. The black mesh corresponds to $\phi =\phi_0$ and marks
  the border between RM1, RM3 on one hand and RM2, RM4 on the other
  hand. Finally, the region $c>0$ corresponds to RM1 and RM2 while
  $c<0$ contains RM3 and RM4. The value $N_*=40$ has been used.}
\label{rmepsilon2}
\end{center}
\end{figure}

{}From the expression of $s$ derived in (\ref{eqs}), a given value of
$\epsilon_2$ defines a surface in the three-dimensional volume $[\log
(\kappa \phi _{\ucri}), c/2,\log(1/\kappa \phi _0)]$ given by
\begin{equation}
\label{eqphicrieps2}
\log (\kappa \phi _{\ucri})=\ue^{-cN_*}\left[1-\frac{\epsilon
    _2}{4} \left(\frac{c}{2}\right)^{-1}-\log\left(\frac{1}{\kappa
    \phi_0}\right)\right]\,,
\end{equation}
for a given $N_*$. This equation explicitly gives the degeneracies
between the running mass parameters leading to the same $\epsilon_2$
values. However, it is only an approximate solution of the slow-roll
equations valid as long as $\kappa \phi \ll 1$ and some precautions
should be in order when dealing with the RM4 and RM2 models. In fact,
one may alternatively invert (\ref{eq:rmefold}) numerically for a
given $N_*$ and use the equation (\ref{eps2running}) to recover the
surfaces of constant $\epsilon_2$. These surfaces are represented in
figure~\ref{rmepsilon2} for three values of $\epsilon _2$
corresponding to the mean value ($\epsilon_2=0.034$) and to the
two-sigma confidence bounds ($\epsilon _2=-0.029$ and $\epsilon
_2=0.074$) obtained for the second order slow-roll expansion and under
the Jeffreys' prior on $\epsilon_1$. If we compare
figure~\ref{runlogmass_3D} with figure~\ref{rmepsilon2}, we recover
that the MCMC spread out along the surfaces associated with $\epsilon
_2 \simeq 0.034$.

\par

Firstly, let us notice that the folded yellow surface in the region
$-\log(\kappa \phi _0)\simeq 4$, $c<0$ corresponds to models with small
values of $\kappa \phi _\ucri$ and $\kappa \phi _0$ but (and this
cannot be seen in the figure) with large values of $\kappa \phi _\ini$
(this is the RM4 model). These models are compatible with the data but
they may suffer a theoretical inconsistency since they are associated
with vev larger than the Planck mass as inflation proceeds. Note that
these parts of the yellow surface cannot be found by using
(\ref{eqphicrieps2}) since its domain of validity breaks down in that
particular case.

\par

Figure~\ref{rmepsilon2} demonstrates that the constraints appearing on
the one and two-dimensional marginalised probabilities on the running
mass parameters are mainly due to a preferential volume selection in
the parameter space by the prior choice. Indeed, if one would extend
the range of $\log(\kappa \phi _\ucri)$ below $-3$, then the dark red
surface would penetrate the region of higher $c$ values and would
render these $c$ values compatible with the data. The conclusion is
similar for the blue inverted half-pipe around the value
$c=0$. Extending the prior of $\kappa \phi _\ucri$ and $\kappa \phi
_0$ would render the region around $c=0$ compatible with the
data. However, the strong variations of $\kappa \phi _\ucri$ and
$\kappa \phi _0$ required to get a reasonable value of $\epsilon_2$
for an arbitrarily chosen $c$ suggest that some amount of fine-tuning
is required between these parameters.

\par

Nevertheless, for the running mass models, the main conclusion is that
the dependence on the priors prevents us to establish reliable
constraints on the free parameters. To avoid this difficulty, it would
be necessary to integrate numerically the spectrum for very small
values of $\kappa \phi _\ucri$ and $\kappa \phi _0$ but, as noticed at
the beginning of the section, this is a non-trivial technical issue
and some theoretical lower bound should also be set. On the other
hand, one can say that these models remain compatible with WMAP3 data.

\section{Are the power spectra really featureless?}
\label{sec:tpl}

\subsection{Basic equations}
\label{sec:basictpl}

Recently, the possibility that the power spectra could contain
non-expected features has been widely discussed. This question arose
because of the presence of the so-called cosmic variance outliers in
the first year WMAP data. Since these litigious points have
disappeared (at least some of them) in the new data, it seems at first
sight that this is no longer an interesting issue. This conclusion is
also consistent with the analysis of the WMAP
team~\cite{Spergel:2006hy} which quotes a $\chi^2$ drop of $\Delta
\chi ^2\simeq -4.5$ or $\Delta \chi ^2\simeq -9.5$ according to the
type of features considered. Notice that, although $\Delta \chi
^2\simeq -4.5$ seems indeed not of much interest, $\Delta \chi
^2\simeq -9.5$ deserves at least some attention. In this article, we
would like to consider this question further and derive the
marginalised bound the parameters associated with these features
satisfy. A tool that has been used in order to address this issue is
to confront the trans-Planckian power spectra to the CMB data. The
trans-Planckian problem originates from the fact that, due to the
exponential expansion during inflation, the scales of astrophysical
interest today were below the Planck length at the beginning of
inflation~\cite{Martin:2000xs, Brandenberger:2000wr, Niemeyer:2000eh,
  Kempf:2000ac, Kempf:2001fa, Easther:2001fi, Lemoine:2001ar,
  Easther:2001fz, Brandenberger:2002nq, Hassan:2002qk,
  Danielsson:2002kx, Easther:2002xe, Lizzi:2002ib, Alberghi:2003am,
  Niemeyer:2001qe, Niemeyer:2002kh, Armendariz-Picon:2003gd,
  Martin:2003kp, Greene:2004np, Brandenberger:2004kx, Kaloper:2002cs,
  Goldstein:2003ut, Collins:2003zv, Collins:2003mj,
  Kaloper:2003nv,deBoer:2004nd, Danielsson:2004xw}. In this regime,
the framework utilised to derive the inflationary predictions (namely
quantum field theory in curved space-time) breaks down. This is
similar to what happens in the context of black hole physics and the
derivation of the Hawking radiation~\cite{Unruh:1994je,
  Corley:1996ar,Corley:1997pr}. If we denote $\Mc$ the physical scale
at which new physical effects are supposed to become relevant, one
finds that superimposed oscillations appear in the power spectra the
amplitude of which is controlled by the $H$ to $\Mc$ ratio, the only
two scales available in the theory\footnote{Let us notice that
  superimposed oscillations could also originate from other physical
  mechanisms~\cite{Wang:2002hf,Burgess:2002ub,Martin:2003sf,Martin:2003bp,
    Hunt:2004vt}}. These superimposed oscillations are then tested
against the data and used as a tool to detect non-trivial features in
the primordial power spectrum~\cite{Bergstrom:2002yd, Elgaroy:2003gq,
  Okamoto:2003wk, Martin:2003sg, Martin:2004iv, Martin:2004yi,
  Barriga:2000nk, Kogo:2003yb, Huang:2003fw, Shafieloo:2003gf}.

\par

It was shown in Ref.~\cite{Armendariz-Picon:2003gd,
Martin:2003sg,Martin:2004yi,Martin:2004iv} that a crucial point is the
fact that the amplitude and the frequency of the oscillations are
independent quantities. Since this question was sometimes not fairly
appreciated in the recent literature, we would like to briefly review
where this comes from. In particular, we argue that considering
dependent amplitude and frequency is theoretically not justified in
the framework of the so-called ``minimal approach''. Moreover,
postulating that the amplitude is inversely proportional to the
frequency is phenomenologically restrictive since this has the
consequence that the region of interest in the parameter space is
missed.

\par

In the minimal approach, only the initial conditions for the
primordial perturbations are modified while their equation of motion
is left unchanged (this would no longer be the case if, for instance,
we had modelled the new physical effects by a modified dispersion
relation). The initial conditions are not the standard ones because
they are fixed when the wavelength of a given Fourier mode becomes
equal to a new fundamental characteristic scale $\ell_\uc$. The time
$\eta _k$ of mode ``appearance'' associated with a comoving wavenumber
$k$ stems from the condition
\begin{equation}
\lambda (\eta _k)=\frac{2\pi }{k}a(\eta _k)=\ell _{\mathrm{c}}
\equiv \frac{2\pi }{\Mc} \,,
\end{equation}
which implies that $\eta _k$ is a function of $k$. Therefore, this is
different from the standard inflationary calculations where the
initial time is taken to be $\eta _k=-\infty $ for any Fourier mode
$k$ and where, in a certain sense, the initial time does not depend on
$k$. This additional $k$-dependence is at the origin of the appearance
of the superimposed oscillations. Another crucial question is in which
state the Fourier mode is placed at the time $\eta _k$. At this point,
one would like to be as general as possible and we take
\begin{eqnarray}
\label{ci1}
\muST(\eta _k) &=& \mp
\frac{c_k+d_k}{\sqrt{2\omegaST (\eta _k)}}
\frac{4\sqrt{\pi }}{\mpl}\, , 
\\
\label{ci2}
\muST'(\eta _k) &=&
\pm i\sqrt{\frac{\omegaST (\eta _k)}{2}}
\frac{4\sqrt{\pi }(c_k-d_k)}{\mpl} \, .
\end{eqnarray}
The coefficients $c_k$ and $d_k$ are {\it a priori} two arbitrary
complex numbers satisfying the condition $\vert c_k\vert ^2-\vert
d_k\vert ^2=1$. Without restricting the physical content of the
problem, and given the fact that, in the limit $\Mc\rightarrow +\infty
$, one must recover the standard limit, \ie the Bunch-Davies vacuum,
one typically expects that
\begin{eqnarray}
c_k &=& 1+y_k\sigma_0+ \dots \, , \qquad d_k=x_k\sigma_0+ \dots
\, ,
\end{eqnarray}
where $\sigma_0\equiv H/\Mc$. Any other particular choice would be a
strong assumption to be justified. The only simplification that one
may consider is to assume that $x_k\simeq x$ and $y_k\sim y$, that is
to say that these two coefficients are not strongly scale dependent in
the range of scales under consideration. In the following, the
parameters $x$ and $y$ are considered as free parameters that are not
fixed by any existing well-established theories except that, of
course, they should satisfy $\vert c_k\vert ^2-\vert d_k\vert
^2=1$. This implies that the amplitude and the frequencies of the
superimposed oscillations are independent quantities as
announced. Finally, the form of the power spectra stemming from the
previous considerations read, for the scalar
modes~\cite{Martin:2003kp,Martin:2003sg}
\begin{eqnarray}
\label{pssrs2}
k^3P_{\zeta } &=&\frac{H^2}{\pi \epsilon_1 \mpl^2}
\left\{1-2(C+1)\epsilon_1 -C\epsilon _2 -\left(2\epsilon_1 +\epsilon
_2\right) \ln \frac{k}{\kstar} \right. \nonumber \\ & - &
\left. 2\vert x\vert \sigma _0 \left[ 1-2(C+1)\epsilon _1 -C\epsilon_2
  - \left(2\epsilon _1+\epsilon _2 \right) \ln \frac{k}{\kstar}\right]
\right.  \nonumber \\ & \times & \left. \cos \left[\frac{2}{\sigma _0}
  \left(1+\epsilon _1+\epsilon _1\ln \frac{k}{a_0 \Mc}\right)
  +\varphi \right] \right. \nonumber \\ & - & \left. \vert x\vert
\sigma _0\pi \left(2\epsilon _1+\epsilon _2 \right ) \sin
\left[\frac{2}{\sigma _0} \left(1+\epsilon _1+\epsilon _1\ln
  \frac{k}{a_0 \Mc}\right) +\varphi \right] \right\} ,
\end{eqnarray}
and for the gravitational waves
\begin{eqnarray}
k^3P_h &=& \frac{16 H^2}{\pi \mpl^2} \left\{1-2(C+1)\epsilon
_1-2\epsilon _1\ln \frac{k}{\kstar} \right. \nonumber \\ & - &
\left. 2\vert x\vert \sigma _0\left[1-2(C+1)\epsilon_1
  -2\epsilon_1 \ln \frac{k}{\kstar}\right] \right. \nonumber \\ &
\times & \left. \cos \left[\frac{2}{\sigma _0} \left(1+\epsilon_1
  +\epsilon_1 \ln \frac{k}{a_0 \Mc}\right)+\varphi \right] \right.
\nonumber \\ & - & \left.  2\vert x\vert \sigma _0\pi\epsilon_1 \sin
\left[\frac{2}{\sigma _0} \left(1+\epsilon_1 +\epsilon_1 \ln
  \frac{k}{a_0 \Mc}\right) +\varphi \right] \right\},
\label{pst}
\end{eqnarray}
where $\varphi $ is the argument of the complex number $x$, \ie
$x\equiv \vert x\vert {\rm e}^{i\varphi }$. We see that the new power
spectra depend, at most, on three new independent parameters, namely
the amplitude, the frequency and the phase of the superimposed
oscillations. In particular, their wavelength can be expressed as
\begin{equation}
\label{wl}
\frac{\Delta k}{k}=\frac{\sigma _0\pi }{\epsilon_1}\, .
\end{equation}

The derivation of the trans-Planckian corrections in the power spectra
(\ref{pssrs2}) and (\ref{pst}) assumes that the back-reaction effects
are not too important. For consistency, the energy density of the
perturbations must be smaller or equal than that of the inflationary
background. This leads to the condition $\vert x\vert \le \sqrt{3\pi}
\mpl/\Mc$, an estimate which is in agreement with the one derived
in~\cite{Armendariz-Picon:2003gd}. In order to put numbers on the
above constraint, we can use equations (\ref{HQ}) and (\ref{eq:Qrms})
together with $\Mc=H/\sigma _0$ to arrive at
\begin{equation}
\label{eq:brvalid}
\vert x\vert \sigma _0\le 10^4 \times \frac{\sigma
_0^2}{\sqrt{\epsilon _1}} \,,
\end{equation}
where, in order to derive an order of magnitude estimate, the
unimportant factors of order one have been neglected. It is important
to emphasise that the above constraint is only a sufficient condition,
but by no means, unless proven otherwise, a necessary condition for
the validity of the power spectra
calculations~\cite{Brandenberger:2004kx}.

\subsection{WMAP constraints on the oscillatory parameters}
\label{sec:wmaptpl}

In order to test the viability of superimposed oscillations in the
primordial power spectra, we perform an exploration of the
\emph{primordial} parameter space by using MCMC methods implemented in
\COSMOMC, given the third year WMAP data. Our analysis proceeds in two
steps. In a first part, the trans-Planckian power spectra in
(\ref{pssrs2}) and (\ref{pst}) are used to seed the CMB anisotropies
in the framework of the first order slow-roll expansion (see
section~\ref{sec:wmap1order}). In a second part, we reiterate the
analysis by using a power law primordial power spectrum supporting a
wider class of superimposed oscillations, namely oscillating with a
$k/\kstar$ power-law dependence. As detailed in~\cite{Martin:2003sg,
  Martin:2004iv, Martin:2004yi}, we use a modified version of \CAMB to
compute the CMB anisotropies whose required accuracy and computational
time renders impossible an exploration of the full parameter
space. Along the lines drawn in those references, only the primordial
parameter space is probed while the base cosmological parameters
remain fixed to their best fit values obtained from an MCMC analysis
without superimposed oscillations.

\subsubsection{Trans-Planckian power spectra}

\begin{figure}
\begin{center}
\includegraphics[width=12cm]{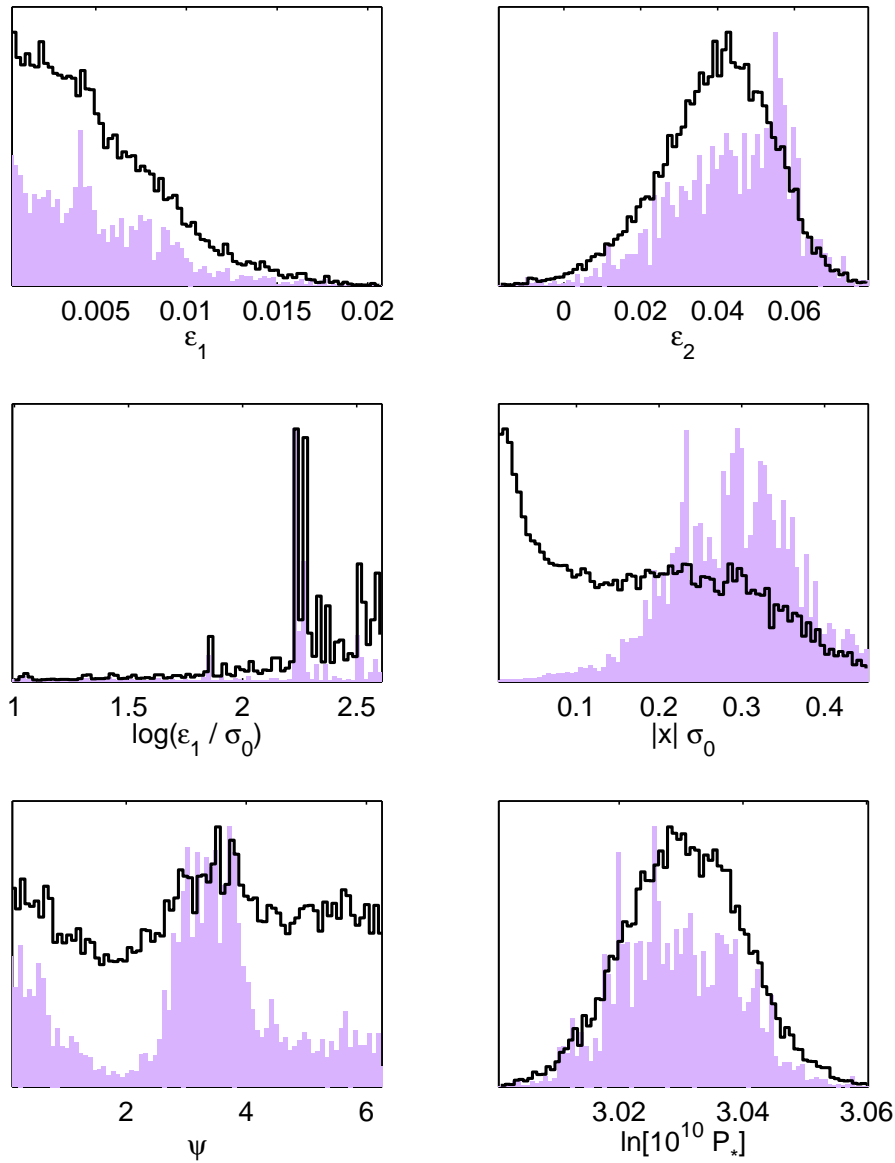}
\caption{Marginalised posterior probability distributions (solid right
stairs) and mean likelihood (shaded bars) for the trans-Planckian
primordial parameters. Note that these posteriors are derived under a
fix set of cosmological parameters.}
\label{fig:tpl_1D}
\end{center}
\end{figure}

As described in the previous section, in addition to the first order
Hubble-flow parameters $\epsilon_1$ and $\epsilon_2$, the scalar power
spectrum amplitude at the pivot scale $\Pstar$, considering the
trans-Planckian power spectra (\ref{pssrs2}) and (\ref{pst}) accounts
for three more primordial parameters. We have chosen uniform priors on
the overall oscillatory phase $\oscphase$ in $[0,2\pi]$, defined as
\begin{equation}
\label{eq:tpleffphase}
\oscphase \equiv \frac{2}{\sigma_0}\left(1+\epsilon_1 \right) +
\varphi,
\end{equation}
as well as on the parameter $|x| \sigma_0$ in $[0,0.45]$. Moreover, in
order to sample directly from the oscillation frequency, an uniform
prior has been chosen on the parameter $\log(\epsilon_1/\sigma_0)$ in
$[1,2.6]$. The other primordial parameters are sampled according to
the uniform prior choice on $\epsilon_1$ described in
section~\ref{sec:wmap1order}.

\par

The converged posteriors for the primordial parameters have been
plotted in figure~\ref{fig:tpl_1D}. As previously mentioned, they have
been obtained for a set of fixed cosmological parameters $\OmegaB h^2
= 0.021$, $\OmegaCDM h^2 = 0.0159$, $\theta = 1.0393$ and
$\tau=0.0942$ (implying $\Hzero \simeq 72\,\textrm{km/s/Mpc}$) and the
MCMC exploration has been stopped after approximately $200000$
elements for which the generalised Gelman and Rubin $R$--statistics
implemented in \COSMOMC~\cite{Lewis:2002ah,Brooks:1998} is less than
$10\%$. The constraints obtained on the standard primordial parameters
$\epsilon_1$, $\epsilon_2$ and $\Pstar$ are stronger than those
derived in section~\ref{sec:wmap1order}, as expected since the
cosmological parameters have been fixed to their best fit values.

\par

\begin{figure}
\begin{center}
\includegraphics[height=6.5cm,width=7.5cm]{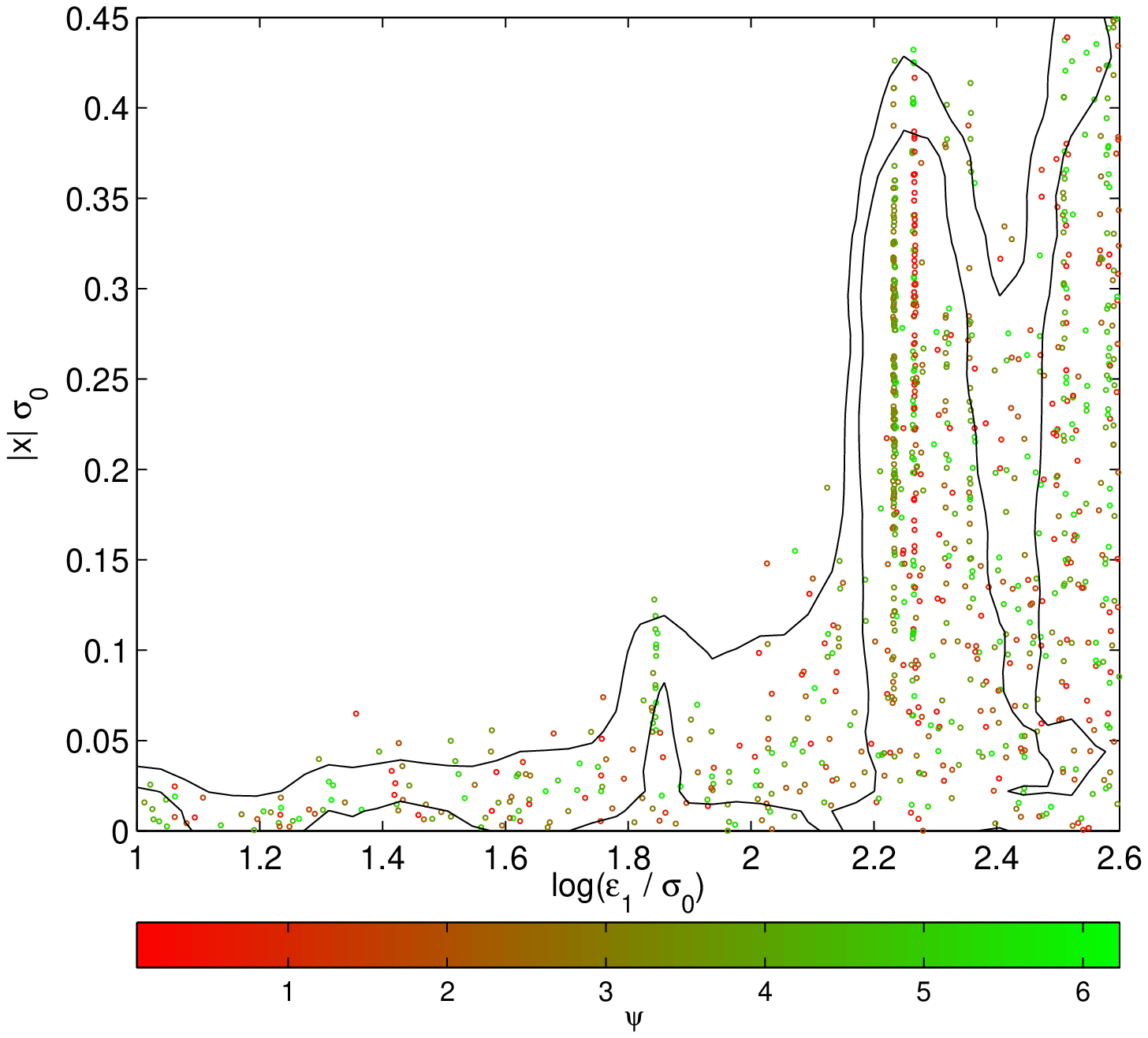}
\includegraphics[height=6.35cm,width=7.5cm]{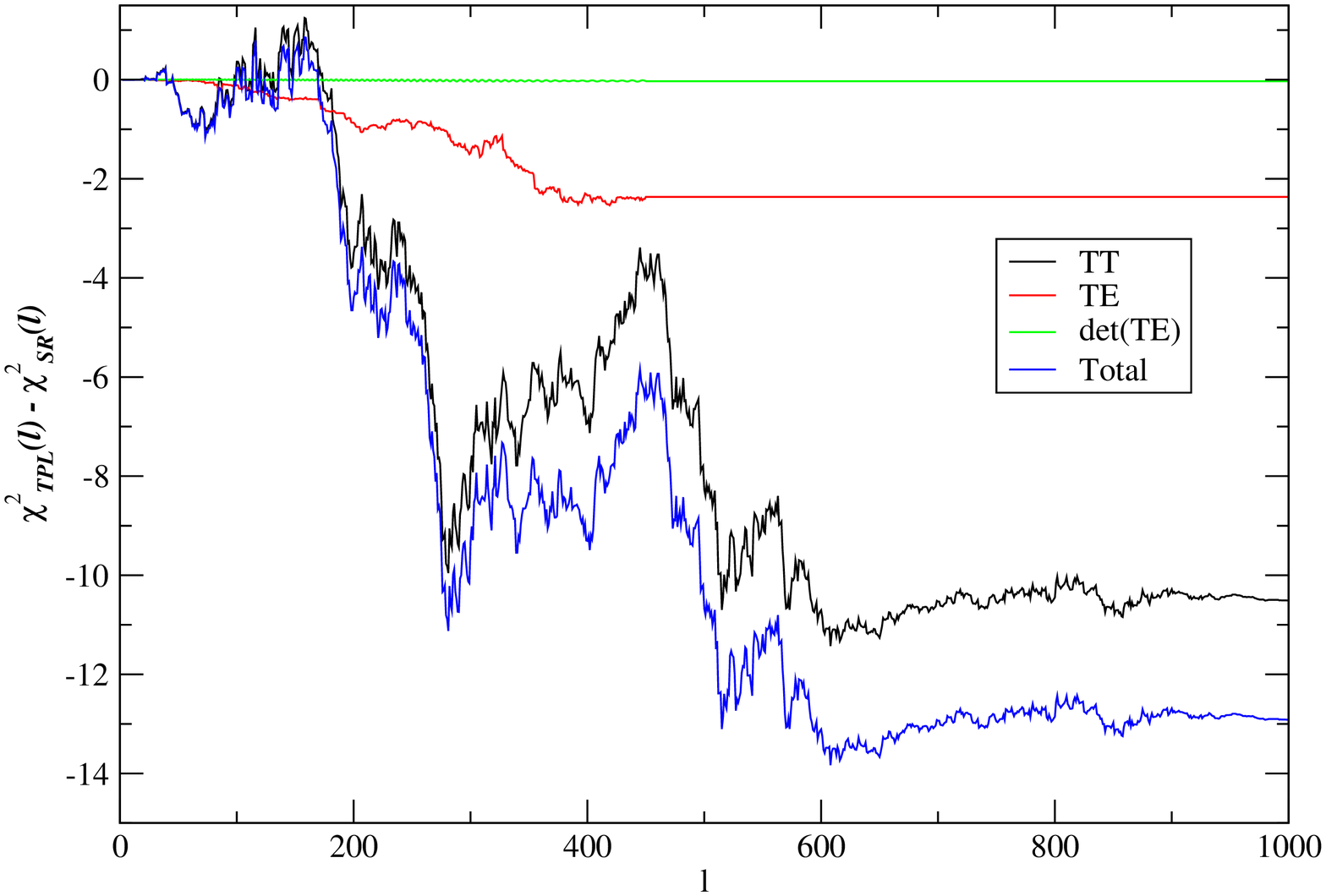}
\caption{In the left panel, smoothed one and two-sigma contours (solid
line) of the two-dimensional marginalised posterior probability
distributions (coloured dots) in the plane
$[\log(\epsilon_1/\sigma_0),|x| \sigma_0]$. The highest probable
frequencies clearly appear as the vertical coloured dot alignments. The
right panel represents the cumulative residual $\chi^2$ between
the trans-Planckian best fit model and the vanilla first order
slow-roll model.}
\label{fig:tpl_3D}
\end{center}
\end{figure}

The overall constraint on trans-Planckian superimposed oscillations is
given by the $|x|\sigma_0$ marginalised posterior. The vanishing value
of this quantity corresponds to the standard first order slow-roll
primordial power spectra without oscillation and is still the favoured
model given the third year WMAP data. However, as it was the case with
the first year data~\cite{Martin:2004yi}, the mean likelihood is
peaked over non-vanishing values of $|x| \sigma_0$ showing that
superimposed oscillations provide a better fit to the data. As
detailed in~\cite{Martin:2004yi}, the marginalised posterior remains
peaked around vanishing values due to volume effects in the parameter
space: the best fit region occupies a rather limited volume that does
not take over the accessible volume associated with the
non-oscillatory models in spite of their lower
likelihoods. Nevertheless, the statistical weight coming from the
highest likelihood values broadens out the posterior distribution on
$|x| \sigma_0$ and at two-sigma level one has
\begin{equation}
|x| \sigma_0 < 0.38\, .
\end{equation}
This number has to be compared with the limit derived
in~\cite{Martin:2004yi}, namely $|x| \sigma_0 < 0.11$. The best fit
volume confinement in the parameter space may be understood on the
marginali\-sed posterior associated with the frequency parameter
$\log(\epsilon_1/\sigma_0)$. As can be seen in
figure~\ref{fig:tpl_1D}, both the mean likelihood and marginalised
probability exhibits narrow peaks on particular frequencies only
thereby leading to a discrete set of best fit sub-manifolds in the
parameter space. From the WMAP team likelihood
code~\cite{Jarosik:2006ib, Spergel:2006hy, Hinshaw:2006ia,
Page:2006hz}, the best fit associated with the highest resonance peak
leads to an overall $\chi^2 =11239.9$, which is a fit
improvement of $\Delta \chi^2 \simeq -12$ with respect to the
first order vanilla slow-roll model, for three additional parameters
(see figure~\ref{fig:tpl_3D}). Note that this value is larger that the
one reported by the WMAP team~\cite{Spergel:2006hy}, possibly due to
the localisation of the highest likelihood resonance in a rather high
frequency region $\log(\epsilon_1/\sigma_0) \simeq 170$.

Eventually, compared to the first year WMAP data, we still find no
evidence for superimposed oscillations in the WMAP third year
data. However, considering trans-Planckian-like superimposed
oscillations still significantly improves the fit to the data, and
more importantly the overall statistical weight associated with these
best fit regions has increased.

\par

Let us now study what the above results imply for trans-Planckian
physics. As already mentioned, the best fit is obtained for
$\log(\epsilon_1/\sigma_0)\simeq 2.23$, $\epsilon _1\simeq 2.1 \times
10^{-3}$ and $\vert x\vert \sigma _0 \simeq 0.268$. This implies
$\sigma _0\simeq 1.2\times 10^{-5}$. Moreover, if one uses the value
of $\epsilon _1$ for the best fit (this does not mean, of course, that
we have detected a non-vanishing $\epsilon _1$ since this would imply
a detection of primordial gravitational waves), then one can estimate
$\Mc$ which reads $\Mc\simeq 0.3 \mpl$. Some serious problems show up
when one tries to see whether the best fit suffers from a back-reaction
problem. Using equation~(\ref{eq:brvalid}), one finds that this is not
the case provided $\vert x\vert \sigma _0\leq 3.3 \times
10^{-5}$. This limit is thus largely violated by the best fit. At this
point, several remarks are in order. Firstly, strictly speaking, this
results clearly invalidates the perturbative framework used in order
the derived the power spectra with the superimposed
oscillations. Secondly, as discussed in~\cite{Brandenberger:2004kx,
  Danielsson:2004xw}, the back-reaction is not necessarily a problem as
its effect might just ``renormalise'' the vacuum energy during
inflation. In other words, it may not necessarily prevent inflation to
proceed. Thirdly, the presence of superimposed oscillations is not
necessarily linked to trans-Planckian effects. In that case, the limit
given by equation~(\ref{eq:brvalid}) simply does not apply. But, then,
it becomes more difficult to physically motivate the logarithmic shape
of the oscillations which turns out to be favoured by the data (see
below).

\par

To conclude this section we would like to describe a few intriguing
features, although not statistically significant, that appears in the
previous analysis. As shown in~\cite{Easther:2004vq}, it is worth
stressing that multiple resonances in the likelihood are precisely
expected in presence of an oscillatory signal due to ``frequency
beating'' between the data and the model tested. These degeneracies
produce a multi-valued function in the recovery of a primordial
oscillation frequency, but also open a window on the \emph{a priori}
unobservable high frequency signals through their lower frequency
resonances. Compared to the first year data, we find the appearance of
new favoured frequencies that can be seen as the peaks in the
$\log(\epsilon_1/\sigma_0)$ posterior of figure~\ref{fig:tpl_1D}, or
through the dot alignments in figure~\ref{fig:tpl_3D}. Another
intriguing feature concerns the marginalised posterior and the mean
likelihood of the phase parameter $\oscphase$. As can be seen in
figure~\ref{fig:tpl_1D}, values around $\oscphase \simeq 0$ and
$\oscphase \simeq 3$ are slightly favoured by the data (once again,
not in a statistically significant way). As shown
in~\cite{Martin:2003sg}, in presence of high frequency oscillations,
$\oscphase = \pi$ modulo $\pi$ are the phase values expected to
maximise the oscillation amplitude in the multipole moments. As
already mentioned, even in presence of an oscillatory signal in the
data, one cannot conclude that it comes from a primordial origin and
it may be the result of some foreground contamination. However, one
would have to explain how such a pattern in the inter-multipole
correlations arises.
 
\subsubsection{Power law power spectrum}

In this section, we reiterate the previous analysis by using a
phenomenological primordial power spectrum. Only the scalar modes have
been considered with a power spectrum of the
form~\cite{Okamoto:2003wk}
\begin{equation}
\label{pspowosc}
k^3 P_\zeta = \Pstar \left(\frac{k}{\kstar}\right)^{\nS-1} \left( 1 -
\amposc \cos\left\{\frac{\freqosc}{\powosc} \left[\left(\frac{k}
  {\kstar}\right)^\powosc - 1\right] + \oscphase\right\} \right).
\end{equation}
In the limit $\powosc \rightarrow 0$, one recovers logarithmic-like
dependence in the oscillation frequency. The interest of this kind of
power spectrum resides in the comparison of models having superimposed
oscillations with a different frequency dependence in the wave number
$k$. In other words, one can test whether the logarithmic oscillations
are special or if any oscillatory signal can significantly improve the
best fit. For the sake of simplicity, we have not considered a running
spectral index and, as in the previous section, only the primordial
parameter space has been explored once the cosmological parameters
have been fixed to their best fit value obtained from the fiducial
power law scalar power spectrum: $\OmegaB h^2 = 0.0223$, $\OmegaCDM =
0.1064$, $\theta=1.040$ and $\tau=0.0885$. An uniform prior has been
chosen for $\log(\powosc)$ in the range $[-5,0.48]$ (the upper bound
corresponds to $p\simeq3$) as well as for $\log(\freqosc)$ in
$[1,2.5]$. The phase $\oscphase$ is sampled from a flat prior in
$[0,\pi]$ and $\nS$ from the range $[0.5,1.5]$, also with an uniform
prior.

\begin{figure}
\begin{center}
\includegraphics[width=12cm]{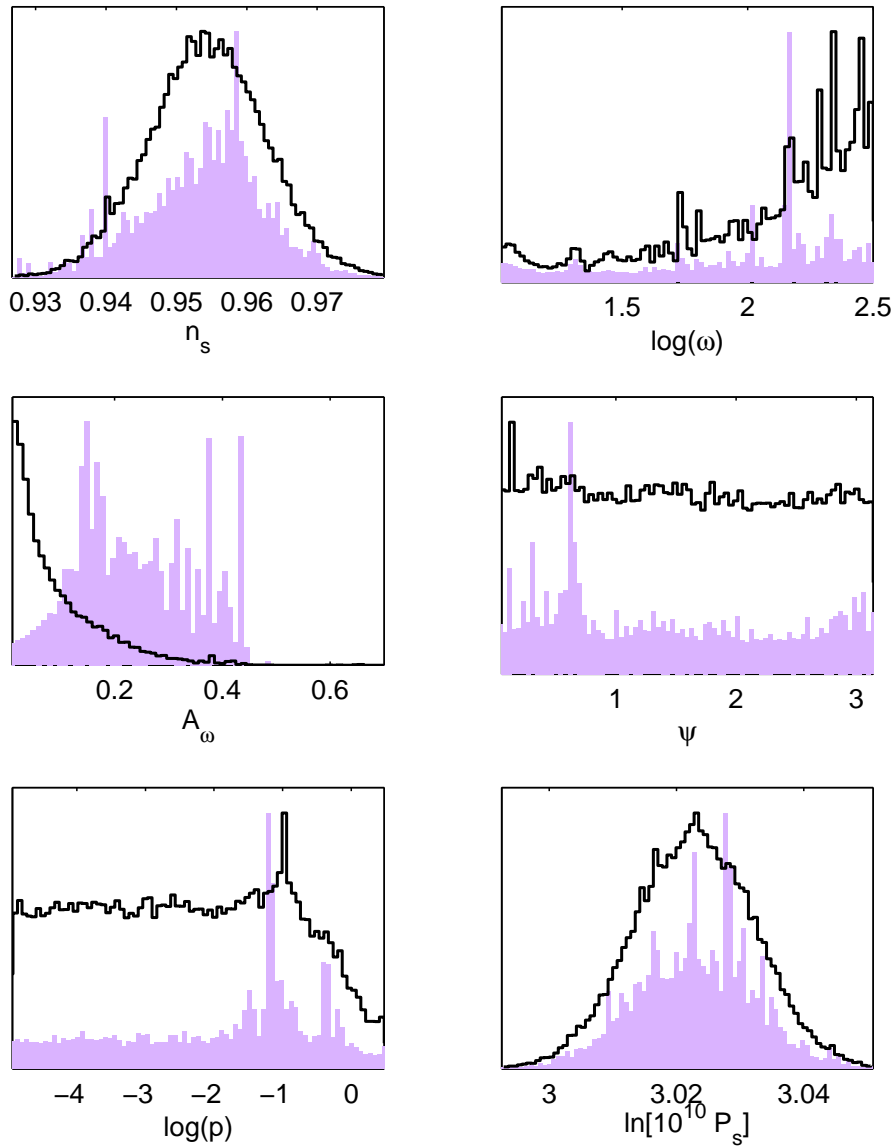}
\caption{Marginalised posterior probability distributions (solid right
  stairs) and mean likelihood (shaded bars) for the oscillatory power
  law parameters. Note that this posteriors are derived under a fix
  set of cosmological parameters.}
\label{fig:powosc_1D}
\end{center}
\end{figure}

The posterior marginalised distributions given the third year WMAP
data are plotted in figure~\ref{fig:powosc_1D}. One still observes
some particular frequencies improving the fit to the data while their
weight on the marginalised probability remains negligible (see the
$\amposc$ posterior). The associated best fit value corresponds to
$\chi^2 = 11242.7$, hence a $\Delta \chi^2 \simeq -10$ with
respect to a fiducial power law model, with however four additional
parameters. This result is in agreement with the one found by the WMAP
team~\cite{Spergel:2006hy}. However, it is important to stress that as
soon as one consider non-logarithmic oscillations, the effect coming
from modifying the pivot scale $\kstar$ can not longer being viewed as
a phase redefinition. In the present approach, we have not considered
this effect and $\kstar$ has been kept to its standard value $0.05
\,\mathrm{Mpc}^{-1}$. In this respect, the posterior on
$\log(\powosc)$ leads to a two-sigma level upper bound
\begin{equation}
\powosc < 0.68 \,,
\end{equation}
in favour of small $p$ value, \ie logarithmic-like superimposed
oscillations.

\section{Discussion and Conclusion}
\label{sec:end}

In this last section, we would like to briefly recap the results
obtained in this article. In a first step, we have studied the
compatibility of inflation with the WMAP3 CMB data using the slow-roll
approximation. We have found that, at leading order, the first
slow-roll parameter satisfies at $95\%$ of confidence
\begin{equation}
\epsilon _1<0.022\,.
\end{equation} 
These constraints implies an upper bound on the contribution of
primordial gravitational waves, namely $r_{10}<0.21$ at $2\sigma
$. This also leads to an upper bound on the energy scale of inflation
$H/\mpl < 1.3\times 10^{-5}$.

\par

The WMAP3 data also constraint the second slow-roll parameter (at the
$2\sigma $ level)
\begin{equation}
-0.02<\epsilon _2<0.09\, ,\qquad -0.07<\epsilon _2<0.07\, ,
\end{equation}
the first result being derived with a Jeffrey's prior on $\epsilon _1$
while the second is obtained with a uniform prior on $\epsilon _1$. We
see that positive $\epsilon _2$, hence red spectral index, are
slightly favoured although a scale-invariant power spectrum remains
compatible with the data. At the second order in the slow-roll
parameter a tendency for $\epsilon _3>0$ is observed but this is not
statistically significant. Together with $\epsilon _2>0$, this would
imply a negative scalar running.

\par

Our second step has been to exactly integrate, using numerical
methods, the inflationary power spectra for four fiducial models. The
four models considered were the large field, small field, hybrid and
running mass scenarios. For large field models $V(\phi) \propto
\phi^p$, we have found the $2 \sigma$ upper limit
\begin{equation}
p < 3.1 \,.
\end{equation}
With regards to the subsequent reheating period, constraints on the
reheating temperature can be found but only for the models that have
an index $p$ such that they are already excluded. These constraints
are therefore not of so much interest and are maybe just another
indication that these models are now incompatible with the CMB data.

\par

For small field models, the situation is slightly more
complicated. {\it A priori} no constraint on $p$ and/or $\mu $ can be
found given the WMAP3 data. But this statement is not completely prior
independent. Indeed, if one assumes that $\mu /\mpl< 100$, then the
marginalised probability over $\mu$ is flat. But if one considers that
$\mu/\mpl < 10$, then the case $p=2$ is disfavoured. The situation is
even more complicated because, if small values of $\mu/\mpl $ may
appear to be appealing from a theoretical point of view, the scale
$\mu $ should not be too small in order for the inflation energy scale
$M$ to be larger than, say, the MeV. Another interesting issue related
to the small field models is the reheating phase. These models are the
only ones for which it is possible to say something on the reheat
temperature given the WMAP3 data. The bounds are relatively weak since
we find after marginalisation
\begin{equation}
 T_\usssRH > 2\, \mbox{TeV} \,,
\end{equation}
at $95\%$ of confidence. Moreover, they are valid only for quite an
extreme equation of state during reheating, namely
$\wstate_\ureh\simeq -1/3$.

\par

We have also studied the hybrid and running mass models. For those
models, it is even more difficult to say something. Basically, hybrid
models are disfavoured ($\Delta \chi ^2=+5$) because of their blue
spectrum while no prior independent constraint can be put on the
running mass models because of the strong degeneracy among the
parameters. But this class of models remain compatible with the CMB
data.

\par

The last section of the paper was devoted to the possible presence of
superimposed oscillations in the power spectra. The marginalised
probability is still centred on $\vert x\vert \sigma _0 =0$, \ie still
compatible with no oscillations. At the $2\sigma $ level, one has
obtained
\begin{equation}
\vert x\vert \sigma _0<0.38\, .
\end{equation}
However, the likelihood is peaked at a non-vanishing value of $\vert
x\vert \sigma _0$ corresponding to $\Delta \chi^2=-12$ for $3$ extra
parameters. This apparent discrepancy is explained by the fact that
the best fit occupies a small volume in the parameters space. One can
nevertheless made the following two remarks. First, the overall
statistical weight of the superimposed oscillations has increased in
comparison with WMAP1 data and despite the disappearance of some of
the cosmic variance outliers. One could have expected exactly the
opposite. Secondly, we have also tested another functional shape for
the superimposed oscillations and have shown that it does not improve
the fit in the same manner, $\Delta \chi ^2=-10$ for $4$ extra
parameters. This suggests that the precise shape of the superimposed
oscillations is relevant and that logarithmic oscillations are
favoured. Notice, however, that the best fit solution, if interpreted
in the trans-Planckian framework, suffers from a severe backreaction
problem.

\par

Let us conclude this article by a few words about the future. The flow
of high accuracy data has not yet dried up and the forthcoming CMB
experiments, as Planck~\cite{Lamarre:2003zh}, will provide us with
even more accurate data which will help us to improve our constraints
on the various inflationary scenarios. Using the slow-roll language,
an exciting prospect would be to close the contour of the $\epsilon
_1$ parameter. This would imply a detection of primordial
gravitational waves and would open the possibility to test the
consistency check $r_{10}\sim -5\nT$, a smoking gun for slow-roll
inflation. But even if this cannot be done in a close future, the
example of small field models has taught us that this could also
improve our knowledge of the reheating period. Indeed, we have seen
that the constraint on $\ln R$ was directly linked to the constraint
on the shape of the primordial power spectra. Shrinking the error bars
on the tilt could therefore help us to put relevant limits on
$T_\usssRH$, at least for the small field models but, maybe, also for
the other class of inflationary scenarios. On the exact integration
side, the method we have presented could be applied to more
complicated models of inflation, especially to those transiently
violating the slow-roll conditions or to the ones involving several
interacting fields. Assuming only linear perturbation theory, such an
approach directly leads to marginalised constraints on the fundamental
parameters, as the ones involved in the inflaton potential and
reheating. A natural extension would be the determination of the
Bayesian evidence associated with each of the model
tested~\cite{Kunz:2006mc, Parkinson:2006ku}. This would allow a
statistical meaningful measure to prefer one inflationary model over
the others.

\acknowledgments

The computations have been performed thanks to super-computing
facilities made available by the Centre Informatique National de
l'Enseignement Sup\'erieur\footnote{\texttt{http://www.cines.fr}}, the
Institut du D\'eveloppement des Ressources en Informatique
Scientifique\footnote{\texttt{http://www.idris.fr}}, the French Data
Processing Center for
Planck-HFI\footnote{\texttt{http://www.planck.fr}} and by the
U.~K. Computational Cosmology
Consortium\footnote{\texttt{http://www.damtp.cam.ac.uk/cosmos}}.

\section*{References}

\bibliography{inflation}

\providecommand{\href}[2]{#2}\begingroup\raggedright\begin{thebibliography}{10%
0}

\bibitem{Jarosik:2006ib}
N.~Jarosik {\em et.~al.}, {\it Three-year wilkinson microwave anisotropy probe
  (wmap) observations: Beam profiles, data processing, radiometer
  characterization and systematic error limits},
  \href{http://xxx.lanl.gov/abs/astro-ph/0603452}{{\tt astro-ph/0603452}}.

\bibitem{Spergel:2006hy}
D.~N. Spergel {\em et.~al.}, {\it Wilkinson microwave anisotropy probe (wmap)
  three year results: Implications for cosmology},
  \href{http://xxx.lanl.gov/abs/astro-ph/0603449}{{\tt astro-ph/0603449}}.

\bibitem{Hinshaw:2006ia}
G.~Hinshaw {\em et.~al.}, {\it Three-year wilkinson microwave anisotropy probe
  (wmap) observations: Temperature analysis},
  \href{http://xxx.lanl.gov/abs/astro-ph/0603451}{{\tt astro-ph/0603451}}.

\bibitem{Page:2006hz}
L.~Page {\em et.~al.}, {\it Three year wilkinson microwave anisotropy probe
  (wmap) observations: Polarization analysis},
  \href{http://xxx.lanl.gov/abs/astro-ph/0603450}{{\tt astro-ph/0603450}}.

\bibitem{Alabidi:2006qa}
L.~Alabidi and D.~H. Lyth, {\it Inflation models after wmap year three},
  \href{http://xxx.lanl.gov/abs/astro-ph/0603539}{{\tt astro-ph/0603539}}.

\bibitem{Peiris:2006ug}
H.~Peiris and R.~Easther, {\it Recovering the inflationary potential and
  primordial power spectrum with a slow roll prior},
  \href{http://xxx.lanl.gov/abs/astro-ph/0603587}{{\tt astro-ph/0603587}}.

\bibitem{Easther:2006tv}
R.~Easther and H.~Peiris, {\it Implications of a running spectral index for
  slow roll inflation},  \href{http://xxx.lanl.gov/abs/astro-ph/0604214}{{\tt
  astro-ph/0604214}}.

\bibitem{Shafi:2006cs}
Q.~Shafi and V.~N. Senoguz, {\it Coleman-weinberg potential in good agreement
  with wmap},  \href{http://xxx.lanl.gov/abs/astro-ph/0603830}{{\tt
  astro-ph/0603830}}.

\bibitem{deVega:2006hb}
H.~J. de~Vega and N.~G. Sanchez, {\it Single field inflation models allowed and
  ruled out by the three years wmap data},
  \href{http://xxx.lanl.gov/abs/astro-ph/0604136}{{\tt astro-ph/0604136}}.

\bibitem{Lewis:1999bs}
A.~Lewis, A.~Challinor, and A.~Lasenby, {\it Efficient computation of cmb
  anisotropies in closed frw models},  {\em Astrophys. J.} {\bf 538} (2000)
  473--476, [\href{http://xxx.lanl.gov/abs/astro-ph/9911177}{{\tt
  astro-ph/9911177}}].

\bibitem{Lewis:2002ah}
A.~Lewis and S.~Bridle, {\it Cosmological parameters from cmb and other data: a
  monte- carlo approach},  {\em Phys. Rev.} {\bf D66} (2002) 103511,
  [\href{http://xxx.lanl.gov/abs/astro-ph/0205436}{{\tt astro-ph/0205436}}].

\bibitem{Guth:1980zm}
A.~H. Guth, {\it The inflationary universe: A possible solution to the horizon
  and flatness problems},  {\em Phys. Rev.} {\bf D23} (1981) 347--356.

\bibitem{Linde:2005ht}
A.~D. Linde, {\it Particle physics and inflationary cosmology},  {\em Contemp.
  Concepts Phys.} {\bf 5} (2005) 1--362,
  [\href{http://xxx.lanl.gov/abs/hep-th/0503203}{{\tt hep-th/0503203}}].

\bibitem{Martin:2004um}
J.~Martin, {\it Inflationary cosmological perturbations of quantum- mechanical
  origin},  {\em Lect. Notes Phys.} {\bf 669} (2005) 199--244,
  [\href{http://xxx.lanl.gov/abs/hep-th/0406011}{{\tt hep-th/0406011}}].

\bibitem{Martin:2003bt}
J.~Martin, {\it Inflation and precision cosmology},  {\em Braz. J. Phys.} {\bf
  34} (2004) 1307--1321, [\href{http://xxx.lanl.gov/abs/astro-ph/0312492}{{\tt
  astro-ph/0312492}}].

\bibitem{Lyth:1998xn}
D.~H. Lyth and A.~Riotto, {\it Particle physics models of inflation and the
  cosmological density perturbation},  {\em Phys. Rept.} {\bf 314} (1999)
  1--146, [\href{http://xxx.lanl.gov/abs/hep-ph/9807278}{{\tt
  hep-ph/9807278}}].

\bibitem{Ringeval:2005yn}
C.~Ringeval, P.~Brax, v.~de~Bruck, Carsten, and A.-C. Davis, {\it Boundary
  inflation and the wmap data},  {\em Phys. Rev.} {\bf D73} (2006) 064035,
  [\href{http://xxx.lanl.gov/abs/astro-ph/0509727}{{\tt astro-ph/0509727}}].

\bibitem{Lucchin:1984yf}
F.~Lucchin and S.~Matarrese, {\it Power law inflation},  {\em Phys. Rev.} {\bf
  D32} (1985) 1316.

\bibitem{Turner:1983he}
M.~S. Turner, {\it Coherent scalar field oscillations in an expanding
  universe},  {\em Phys. Rev.} {\bf D28} (1983) 1243.

\bibitem{Kofman:1997yn}
L.~Kofman, A.~D. Linde, and A.~A. Starobinsky, {\it Towards the theory of
  reheating after inflation},  {\em Phys. Rev.} {\bf D56} (1997) 3258--3295,
  [\href{http://xxx.lanl.gov/abs/hep-ph/9704452}{{\tt hep-ph/9704452}}].

\bibitem{Garcia-Bellido:1997wm}
J.~Garcia-Bellido and A.~D. Linde, {\it Preheating in hybrid inflation},  {\em
  Phys. Rev.} {\bf D57} (1998) 6075--6088,
  [\href{http://xxx.lanl.gov/abs/hep-ph/9711360}{{\tt hep-ph/9711360}}].

\bibitem{Felder:2000hj}
G.~N. Felder {\em et.~al.}, {\it Dynamics of symmetry breaking and tachyonic
  preheating},  {\em Phys. Rev. Lett.} {\bf 87} (2001) 011601,
  [\href{http://xxx.lanl.gov/abs/hep-ph/0012142}{{\tt hep-ph/0012142}}].

\bibitem{Micha:2004bv}
R.~Micha and I.~I. Tkachev, {\it Turbulent thermalization},  {\em Phys. Rev.}
  {\bf D70} (2004) 043538, [\href{http://xxx.lanl.gov/abs/hep-ph/0403101}{{\tt
  hep-ph/0403101}}].

\bibitem{Senoguz:2004vu}
V.~N. Senoguz and Q.~Shafi, {\it Reheat temperature in supersymmetric hybrid
  inflation models},  {\em Phys. Rev.} {\bf D71} (2005) 043514,
  [\href{http://xxx.lanl.gov/abs/hep-ph/0412102}{{\tt hep-ph/0412102}}].

\bibitem{Bassett:2005xm}
B.~A. Bassett, S.~Tsujikawa, and D.~Wands, {\it Inflation dynamics and
  reheating},  {\em Rev. Mod. Phys.} {\bf 78} (2006) 537--589,
  [\href{http://xxx.lanl.gov/abs/astro-ph/0507632}{{\tt astro-ph/0507632}}].

\bibitem{Podolsky:2005bw}
D.~I. Podolsky, G.~N. Felder, L.~Kofman, and M.~Peloso, {\it Equation of state
  and beginning of thermalization after preheating},  {\em Phys. Rev.} {\bf
  D73} (2006) 023501, [\href{http://xxx.lanl.gov/abs/hep-ph/0507096}{{\tt
  hep-ph/0507096}}].

\bibitem{Desroche:2005yt}
M.~Desroche, G.~N. Felder, J.~M. Kratochvil, and A.~Linde, {\it Preheating in
  new inflation},  {\em Phys. Rev.} {\bf D71} (2005) 103516,
  [\href{http://xxx.lanl.gov/abs/hep-th/0501080}{{\tt hep-th/0501080}}].

\bibitem{Allahverdi:2005mz}
R.~Allahverdi and A.~Mazumdar, {\it Supersymmetric thermalization and
  quasi-thermal universe: Consequences for gravitinos and leptogenesis},
  \href{http://xxx.lanl.gov/abs/hep-ph/0512227}{{\tt hep-ph/0512227}}.

\bibitem{Allahverdi:2006wh}
R.~Allahverdi and A.~Mazumdar, {\it Towards a successful reheating within
  supersymmetry},  \href{http://xxx.lanl.gov/abs/hep-ph/0603244}{{\tt
  hep-ph/0603244}}.

\bibitem{Allahverdi:2006iq}
R.~Allahverdi, K.~Enqvist, J.~Garcia-Bellido, and A.~Mazumdar, {\it Gauge
  invariant mssm inflaton},  \href{http://xxx.lanl.gov/abs/hep-ph/0605035}{{\tt
  hep-ph/0605035}}.

\bibitem{Abramovitz:1970aa}
M.~Abramowitz and I.~A. Stegun, {\em Handbook of mathematical functions with
  formulas, graphs, and mathematical tables}.
\newblock National Bureau of Standards, Washington, US, ninth~ed., 1970.

\bibitem{Gradshteyn:1965aa}
I.~S. Gradshteyn and I.~M. Ryzhik, {\em Table of Integrals, Series, and
  Products}.
\newblock Academic Press, New York and London, 1965.

\bibitem{Bardeen:1980kt}
J.~M. Bardeen, {\it Gauge invariant cosmological perturbations},  {\em Phys.
  Rev.} {\bf D22} (1980) 1882--1905.

\bibitem{Mukhanov:1990me}
V.~F. Mukhanov, H.~A. Feldman, and R.~H. Brandenberger, {\it Theory of
  cosmological perturbations. part 1. classical perturbations. part 2. quantum
  theory of perturbations. part 3. extensions},  {\em Phys. Rept.} {\bf 215}
  (1992) 203--333.

\bibitem{Mukhanov:1981xt}
V.~F. Mukhanov and G.~V. Chibisov, {\it Quantum fluctuation and 'nonsingular'
  universe. (in russian)},  {\em JETP Lett.} {\bf 33} (1981) 532--535.

\bibitem{Grishchuk:1974ny}
L.~P. Grishchuk, {\it Amplification of gravitational waves in an isotropic
  universe},  {\em Sov. Phys. JETP} {\bf 40} (1975) 409--415.

\bibitem{Grishchuk:1975uf}
L.~P. Grishchuk, {\it The amplification of gravitational waves and creation of
  gravitons in the isotropic universes. (erratum)},  {\em Nuovo Cim. Lett.}
  {\bf 12} (1975) 60--64.

\bibitem{Martin:1997zd}
J.~Martin and D.~J. Schwarz, {\it The influence of cosmological transitions on
  the evolution of density perturbations},  {\em Phys. Rev.} {\bf D57} (1998)
  3302--3316, [\href{http://xxx.lanl.gov/abs/gr-qc/9704049}{{\tt
  gr-qc/9704049}}].

\bibitem{Stewart:1993bc}
E.~D. Stewart and D.~H. Lyth, {\it A more accurate analytic calculation of the
  spectrum of cosmological perturbations produced during inflation},  {\em
  Phys. Lett.} {\bf B302} (1993) 171--175,
  [\href{http://xxx.lanl.gov/abs/gr-qc/9302019}{{\tt gr-qc/9302019}}].

\bibitem{Martin:1999wa}
J.~Martin and D.~J. Schwarz, {\it The precision of slow-roll predictions for
  the cmbr anisotropies},  {\em Phys. Rev.} {\bf D62} (2000) 103520,
  [\href{http://xxx.lanl.gov/abs/astro-ph/9911225}{{\tt astro-ph/9911225}}].

\bibitem{Martin:2000ak}
J.~Martin, A.~Riazuelo, and D.~J. Schwarz, {\it Slow-roll inflation and cmb
  anisotropy data},  {\em Astrophys. J.} {\bf 543} (2000) L99--L102,
  [\href{http://xxx.lanl.gov/abs/astro-ph/0006392}{{\tt astro-ph/0006392}}].

\bibitem{Schwarz:2001vv}
D.~J. Schwarz, C.~A. Terrero-Escalante, and A.~A. Garcia, {\it Higher order
  corrections to primordial spectra from cosmological inflation},  {\em Phys.
  Lett.} {\bf B517} (2001) 243--249,
  [\href{http://xxx.lanl.gov/abs/astro-ph/0106020}{{\tt astro-ph/0106020}}].

\bibitem{Schwarz:2004tz}
D.~J. Schwarz and C.~A. Terrero-Escalante, {\it Primordial fluctuations and
  cosmological inflation after wmap 1.0},  {\em JCAP} {\bf 0408} (2004) 003,
  [\href{http://xxx.lanl.gov/abs/hep-ph/0403129}{{\tt hep-ph/0403129}}].

\bibitem{Leach:2002ar}
S.~M. Leach, A.~R. Liddle, J.~Martin, and D.~J. Schwarz, {\it Cosmological
  parameter estimation and the inflationary cosmology},  {\em Phys. Rev.} {\bf
  D66} (2002) 023515, [\href{http://xxx.lanl.gov/abs/astro-ph/0202094}{{\tt
  astro-ph/0202094}}].

\bibitem{Liddle:1994dx}
A.~R. Liddle, P.~Parsons, and J.~D. Barrow, {\it Formalizing the slow roll
  approximation in inflation},  {\em Phys. Rev.} {\bf D50} (1994) 7222--7232,
  [\href{http://xxx.lanl.gov/abs/astro-ph/9408015}{{\tt astro-ph/9408015}}].

\bibitem{Martin:2002vn}
J.~Martin and D.~J. Schwarz, {\it Wkb approximation for inflationary
  cosmological perturbations},  {\em Phys. Rev.} {\bf D67} (2003) 083512,
  [\href{http://xxx.lanl.gov/abs/astro-ph/0210090}{{\tt astro-ph/0210090}}].

\bibitem{Casadio:2004ru}
R.~Casadio, F.~Finelli, M.~Luzzi, and G.~Venturi, {\it Improved wkb analysis of
  cosmological perturbations},  {\em Phys. Rev.} {\bf D71} (2005) 043517,
  [\href{http://xxx.lanl.gov/abs/gr-qc/0410092}{{\tt gr-qc/0410092}}].

\bibitem{Casadio:2005xv}
R.~Casadio, F.~Finelli, M.~Luzzi, and G.~Venturi, {\it Higher order slow-roll
  predictions for inflation},  {\em Phys. Lett.} {\bf B625} (2005) 1--6,
  [\href{http://xxx.lanl.gov/abs/gr-qc/0506043}{{\tt gr-qc/0506043}}].

\bibitem{Casadio:2005em}
R.~Casadio, F.~Finelli, M.~Luzzi, and G.~Venturi, {\it Improved wkb analysis of
  slow-roll inflation},  {\em Phys. Rev.} {\bf D72} (2005) 103516,
  [\href{http://xxx.lanl.gov/abs/gr-qc/0510103}{{\tt gr-qc/0510103}}].

\bibitem{Gong:2001he}
J.-O. Gong and E.~D. Stewart, {\it The density perturbation power spectrum to
  second-order corrections in the slow-roll expansion},  {\em Phys. Lett.} {\bf
  B510} (2001) 1--9, [\href{http://xxx.lanl.gov/abs/astro-ph/0101225}{{\tt
  astro-ph/0101225}}].

\bibitem{Choe:2004zg}
J.~Choe, J.-O. Gong, and E.~D. Stewart, {\it Second order general slow-roll
  power spectrum},  {\em JCAP} {\bf 0407} (2004) 012,
  [\href{http://xxx.lanl.gov/abs/hep-ph/0405155}{{\tt hep-ph/0405155}}].

\bibitem{Leach:2003us}
S.~M. Leach and A.~R. Liddle, {\it Constraining slow-roll inflation with wmap
  and 2df},  {\em Phys. Rev.} {\bf D68} (2003) 123508,
  [\href{http://xxx.lanl.gov/abs/astro-ph/0306305}{{\tt astro-ph/0306305}}].

\bibitem{Lewis:2006ma}
A.~Lewis, {\it Observational constraints and cosmological parameters},
  \href{http://xxx.lanl.gov/abs/astro-ph/0603753}{{\tt astro-ph/0603753}}.

\bibitem{Freedman:2000cf}
W.~L. Freedman {\em et.~al.}, {\it Final results from the hubble space
  telescope key project to measure the hubble constant},  {\em Astrophys. J.}
  {\bf 553} (2001) 47--72,
  [\href{http://xxx.lanl.gov/abs/astro-ph/0012376}{{\tt astro-ph/0012376}}].

\bibitem{Gelman:1992}
A.~Gelman and D.~Rubin, {\it Inference from iterative simulations using
  multiple sequences},  {\em Statistical Science} {\bf 7} (1992) 457--511.

\bibitem{Parkinson:2006ku}
D.~Parkinson, P.~Mukherjee, and A.~R. Liddle, {\it A bayesian model selection
  analysis of wmap3},  \href{http://xxx.lanl.gov/abs/astro-ph/0605003}{{\tt
  astro-ph/0605003}}.

\bibitem{Pahud:2006kv}
C.~Pahud, A.~R. Liddle, P.~Mukherjee, and D.~Parkinson, {\it Model selection
  forecasts for the spectral index from the planck satellite},
  \href{http://xxx.lanl.gov/abs/astro-ph/0605004}{{\tt astro-ph/0605004}}.

\bibitem{Huang:2006um}
Q.-G. Huang and M.~Li, {\it Running spectral index in noncommutative inflation
  and wmap three year results},
  \href{http://xxx.lanl.gov/abs/astro-ph/0603782}{{\tt astro-ph/0603782}}.

\bibitem{Vilenkin:2004vx}
A.~Vilenkin, {\it Eternal inflation and chaotic terminology},
  \href{http://xxx.lanl.gov/abs/gr-qc/0409055}{{\tt gr-qc/0409055}}.

\bibitem{Linde:1984st}
A.~D. Linde, {\it Chaotic inflating universe},  {\em JETP Lett.} {\bf 38}
  (1983) 176--179.

\bibitem{Vilenkin:1983xp}
A.~Vilenkin, {\it Quantum fluctuations in the new inflationary universe},  {\em
  Nucl. Phys.} {\bf B226} (1983) 527.

\bibitem{Vilenkin:1983xq}
A.~Vilenkin, {\it The birth of inflationary universes},  {\em Phys. Rev.} {\bf
  D27} (1983) 2848.

\bibitem{Goncharov:1987ir}
A.~S. Goncharov, A.~D. Linde, and V.~F. Mukhanov, {\it The global structure of
  the inflationary universe},  {\em Int. J. Mod. Phys.} {\bf A2} (1987)
  561--591.

\bibitem{Linde:1993xx}
A.~D. Linde, D.~A. Linde, and A.~Mezhlumian, {\it From the big bang theory to
  the theory of a stationary universe},  {\em Phys. Rev.} {\bf D49} (1994)
  1783--1826, [\href{http://xxx.lanl.gov/abs/gr-qc/9306035}{{\tt
  gr-qc/9306035}}].

\bibitem{Starobinsky:1986fx}
A.~A. Starobinsky, {\it Stochastic de sitter (inflationary) stage in the early
  universe},  in {\em Lecture Notes in Physics} (H.~J. de~Vega and N.~Sanchez,
  eds.), vol.~246, (Berlin), Springer-Verlag, 1986.

\bibitem{Liguori:2004fa}
M.~Liguori, S.~Matarrese, M.~Musso, and A.~Riotto, {\it Stochastic inflation
  and the lower multipoles in the cmb anisotropies},  {\em JCAP} {\bf 0408}
  (2004) 011, [\href{http://xxx.lanl.gov/abs/astro-ph/0405544}{{\tt
  astro-ph/0405544}}].

\bibitem{Martin:2005ir}
J.~Martin and M.~Musso, {\it Solving stochastic inflation for arbitrary
  potentials},  {\em Phys. Rev.} {\bf D73} (2006) 043516,
  [\href{http://xxx.lanl.gov/abs/hep-th/0511214}{{\tt hep-th/0511214}}].

\bibitem{Martin:2005hb}
J.~Martin and M.~Musso, {\it On the reliability of the langevin pertubative
  solution in stochastic inflation},  {\em Phys. Rev.} {\bf D73} (2006) 043517,
  [\href{http://xxx.lanl.gov/abs/hep-th/0511292}{{\tt hep-th/0511292}}].

\bibitem{Liddle:2003as}
A.~R. Liddle and S.~M. Leach, {\it How long before the end of inflation were
  observable perturbations produced?},  {\em Phys. Rev.} {\bf D68} (2003)
  103503, [\href{http://xxx.lanl.gov/abs/astro-ph/0305263}{{\tt
  astro-ph/0305263}}].

\bibitem{Linde:1981mu}
A.~D. Linde, {\it A new inflationary universe scenario: A possible solution of
  the horizon, flatness, homogeneity, isotropy and primordial monopole
  problems},  {\em Phys. Lett.} {\bf B108} (1982) 389--393.

\bibitem{Albrecht:1982wi}
A.~Albrecht and P.~J. Steinhardt, {\it Cosmology for grand unified theories
  with radiatively induced symmetry breaking},  {\em Phys. Rev. Lett.} {\bf 48}
  (1982) 1220--1223.

\bibitem{Kinney:1995cc}
W.~H. Kinney and K.~T. Mahanthappa, {\it Inflation at low scales: General
  analysis and a detailed model},  {\em Phys. Rev.} {\bf D53} (1996)
  5455--5467, [\href{http://xxx.lanl.gov/abs/hep-ph/9512241}{{\tt
  hep-ph/9512241}}].

\bibitem{German:2001tz}
G.~German, G.~G. Ross, and S.~Sarkar, {\it Low-scale inflation},  {\em Nucl.
  Phys.} {\bf B608} (2001) 423--450,
  [\href{http://xxx.lanl.gov/abs/hep-ph/0103243}{{\tt hep-ph/0103243}}].

\bibitem{Valluri:2000aa}
S.~Vallury, D.~Jeffrey, and R.~Corless, {\it Some applications of the lambert w
  function to physics},  {\em Can. J. Phys.} {\bf 78} (2000) 823.

\bibitem{Linde:1991km}
A.~D. Linde, {\it Axions in inflationary cosmology},  {\em Phys. Lett.} {\bf
  B259} (1991) 38--47.

\bibitem{Copeland:1994vg}
E.~J. Copeland, A.~R. Liddle, D.~H. Lyth, E.~D. Stewart, and D.~Wands, {\it
  False vacuum inflation with einstein gravity},  {\em Phys. Rev.} {\bf D49}
  (1994) 6410--6433, [\href{http://xxx.lanl.gov/abs/astro-ph/9401011}{{\tt
  astro-ph/9401011}}].

\bibitem{Ringeval:2001xd}
C.~Ringeval, {\it Fermionic massive modes along cosmic strings},  {\em Phys.
  Rev.} {\bf D64} (2001) 123505,
  [\href{http://xxx.lanl.gov/abs/hep-ph/0106179}{{\tt hep-ph/0106179}}].

\bibitem{Rocher:2004et}
J.~Rocher and M.~Sakellariadou, {\it Supersymmetric grand unified theories and
  cosmology},  {\em JCAP} {\bf 0503} (2005) 004,
  [\href{http://xxx.lanl.gov/abs/hep-ph/0406120}{{\tt hep-ph/0406120}}].

\bibitem{Bastero-Gil:2006cm}
M.~Bastero-Gil, S.~F. King, and Q.~Shafi, {\it Supersymmetric hybrid inflation
  with non-minimal kaehler potential},
  \href{http://xxx.lanl.gov/abs/hep-ph/0604198}{{\tt hep-ph/0604198}}.

\bibitem{Covi:1998mb}
L.~Covi and D.~H. Lyth, {\it Running-mass models of inflation, and their
  observational constraints},  {\em Phys. Rev.} {\bf D59} (1999) 063515,
  [\href{http://xxx.lanl.gov/abs/hep-ph/9809562}{{\tt hep-ph/9809562}}].

\bibitem{Covi:2002th}
L.~Covi, D.~H. Lyth, and A.~Melchiorri, {\it New constraints on the
  running-mass inflation model},  {\em Phys. Rev.} {\bf D67} (2003) 043507,
  [\href{http://xxx.lanl.gov/abs/hep-ph/0210395}{{\tt hep-ph/0210395}}].

\bibitem{Covi:2004tp}
L.~Covi, D.~H. Lyth, A.~Melchiorri, and C.~J. Odman, {\it The running-mass
  inflation model and wmap},  {\em Phys. Rev.} {\bf D70} (2004) 123521,
  [\href{http://xxx.lanl.gov/abs/astro-ph/0408129}{{\tt astro-ph/0408129}}].

\bibitem{Salopek:1988qh}
D.~S. Salopek, J.~R. Bond, and J.~M. Bardeen, {\it Designing density
  fluctuation spectra in inflation},  {\em Phys. Rev.} {\bf D40} (1989) 1753.

\bibitem{Grivell:1999wc}
I.~J. Grivell and A.~R. Liddle, {\it Inflaton potential reconstruction without
  slow-roll},  {\em Phys. Rev.} {\bf D61} (2000) 081301,
  [\href{http://xxx.lanl.gov/abs/astro-ph/9906327}{{\tt astro-ph/9906327}}].

\bibitem{Adams:2001vc}
J.~A. Adams, B.~Cresswell, and R.~Easther, {\it Inflationary perturbations from
  a potential with a step},  {\em Phys. Rev.} {\bf D64} (2001) 123514,
  [\href{http://xxx.lanl.gov/abs/astro-ph/0102236}{{\tt astro-ph/0102236}}].

\bibitem{Makarov:2005uh}
A.~Makarov, {\it On the accuracy of slow-roll inflation given current
  observational constraints},  {\em Phys. Rev.} {\bf D72} (2005) 083517,
  [\href{http://xxx.lanl.gov/abs/astro-ph/0506326}{{\tt astro-ph/0506326}}].

\bibitem{Sanchez:2005pi}
A.~G. Sanchez {\em et.~al.}, {\it Cosmological parameters from cmb measurements
  and the final 2dfgrs power spectrum},  {\em Mon. Not. Roy. Astron. Soc.} {\bf
  366} (2006) 189--207, [\href{http://xxx.lanl.gov/abs/astro-ph/0507583}{{\tt
  astro-ph/0507583}}].

\bibitem{Linde:2004kg}
A.~Linde, {\it Prospects of inflation},  {\em Phys. Scripta} {\bf T117} (2005)
  40--48, [\href{http://xxx.lanl.gov/abs/hep-th/0402051}{{\tt
  hep-th/0402051}}].

\bibitem{Lamarre:2003zh}
J.-M. Lamarre {\em et.~al.}, {\it The planck high frequency instrument, a 3rd
  generation cmb experiment, and a full sky submillimeter survey},
  \href{http://xxx.lanl.gov/abs/astro-ph/0308075}{{\tt astro-ph/0308075}}.

\bibitem{Coc:2003ce}
A.~Coc, E.~Vangioni-Flam, P.~Descouvemont, A.~Adahchour, and C.~Angulo, {\it
  Updated big bang nucleosynthesis confronted to wmap observations and to the
  abundance of light elements},  {\em Astrophys. J.} {\bf 600} (2004) 544--552,
  [\href{http://xxx.lanl.gov/abs/astro-ph/0309480}{{\tt astro-ph/0309480}}].

\bibitem{Martin:2000xs}
J.~Martin and R.~H. Brandenberger, {\it The trans-planckian problem of
  inflationary cosmology},  {\em Phys. Rev.} {\bf D63} (2001) 123501,
  [\href{http://xxx.lanl.gov/abs/hep-th/0005209}{{\tt hep-th/0005209}}].

\bibitem{Brandenberger:2000wr}
R.~H. Brandenberger and J.~Martin, {\it The robustness of inflation to changes
  in super-planck- scale physics},  {\em Mod. Phys. Lett.} {\bf A16} (2001)
  999--1006, [\href{http://xxx.lanl.gov/abs/astro-ph/0005432}{{\tt
  astro-ph/0005432}}].

\bibitem{Niemeyer:2000eh}
J.~C. Niemeyer, {\it Inflation with a high frequency cutoff},  {\em Phys. Rev.}
  {\bf D63} (2001) 123502,
  [\href{http://xxx.lanl.gov/abs/astro-ph/0005533}{{\tt astro-ph/0005533}}].

\bibitem{Kempf:2000ac}
A.~Kempf, {\it Mode generating mechanism in inflation with cutoff},  {\em Phys.
  Rev.} {\bf D63} (2001) 083514,
  [\href{http://xxx.lanl.gov/abs/astro-ph/0009209}{{\tt astro-ph/0009209}}].

\bibitem{Kempf:2001fa}
A.~Kempf and J.~C. Niemeyer, {\it Perturbation spectrum in inflation with
  cutoff},  {\em Phys. Rev.} {\bf D64} (2001) 103501,
  [\href{http://xxx.lanl.gov/abs/astro-ph/0103225}{{\tt astro-ph/0103225}}].

\bibitem{Easther:2001fi}
R.~Easther, B.~R. Greene, W.~H. Kinney, and G.~Shiu, {\it Inflation as a probe
  of short distance physics},  {\em Phys. Rev.} {\bf D64} (2001) 103502,
  [\href{http://xxx.lanl.gov/abs/hep-th/0104102}{{\tt hep-th/0104102}}].

\bibitem{Lemoine:2001ar}
M.~Lemoine, M.~Lubo, J.~Martin, and J.-P. Uzan, {\it The stress-energy tensor
  for trans-planckian cosmology},  {\em Phys. Rev.} {\bf D65} (2002) 023510,
  [\href{http://xxx.lanl.gov/abs/hep-th/0109128}{{\tt hep-th/0109128}}].

\bibitem{Easther:2001fz}
R.~Easther, B.~R. Greene, W.~H. Kinney, and G.~Shiu, {\it Imprints of short
  distance physics on inflationary cosmology},  {\em Phys. Rev.} {\bf D67}
  (2003) 063508, [\href{http://xxx.lanl.gov/abs/hep-th/0110226}{{\tt
  hep-th/0110226}}].

\bibitem{Brandenberger:2002nq}
R.~Brandenberger and P.-M. Ho, {\it Noncommutative spacetime, stringy spacetime
  uncertainty principle, and density fluctuations},  {\em Phys. Rev.} {\bf D66}
  (2002) 023517, [\href{http://xxx.lanl.gov/abs/hep-th/0203119}{{\tt
  hep-th/0203119}}].

\bibitem{Hassan:2002qk}
S.~F. Hassan and M.~S. Sloth, {\it Trans-planckian effects in inflationary
  cosmology and the modified uncertainty principle},  {\em Nucl. Phys.} {\bf
  B674} (2003) 434--458, [\href{http://xxx.lanl.gov/abs/hep-th/0204110}{{\tt
  hep-th/0204110}}].

\bibitem{Danielsson:2002kx}
U.~H. Danielsson, {\it A note on inflation and transplanckian physics},  {\em
  Phys. Rev.} {\bf D66} (2002) 023511,
  [\href{http://xxx.lanl.gov/abs/hep-th/0203198}{{\tt hep-th/0203198}}].

\bibitem{Easther:2002xe}
R.~Easther, B.~R. Greene, W.~H. Kinney, and G.~Shiu, {\it A generic estimate of
  trans-planckian modifications to the primordial power spectrum in inflation},
   {\em Phys. Rev.} {\bf D66} (2002) 023518,
  [\href{http://xxx.lanl.gov/abs/hep-th/0204129}{{\tt hep-th/0204129}}].

\bibitem{Lizzi:2002ib}
F.~Lizzi, G.~Mangano, G.~Miele, and M.~Peloso, {\it Cosmological perturbations
  and short distance physics from noncommutative geometry},  {\em JHEP} {\bf
  06} (2002) 049, [\href{http://xxx.lanl.gov/abs/hep-th/0203099}{{\tt
  hep-th/0203099}}].

\bibitem{Alberghi:2003am}
G.~L. Alberghi, R.~Casadio, and A.~Tronconi, {\it Trans-planckian footprints in
  inflationary cosmology},  {\em Phys. Lett.} {\bf B579} (2004) 1--5,
  [\href{http://xxx.lanl.gov/abs/gr-qc/0303035}{{\tt gr-qc/0303035}}].

\bibitem{Niemeyer:2001qe}
J.~C. Niemeyer and R.~Parentani, {\it Trans-planckian dispersion and
  scale-invariance of inflationary perturbations},  {\em Phys. Rev.} {\bf D64}
  (2001) 101301, [\href{http://xxx.lanl.gov/abs/astro-ph/0101451}{{\tt
  astro-ph/0101451}}].

\bibitem{Niemeyer:2002kh}
J.~C. Niemeyer, R.~Parentani, and D.~Campo, {\it Minimal modifications of the
  primordial power spectrum from an adiabatic short distance cutoff},  {\em
  Phys. Rev.} {\bf D66} (2002) 083510,
  [\href{http://xxx.lanl.gov/abs/hep-th/0206149}{{\tt hep-th/0206149}}].

\bibitem{Armendariz-Picon:2003gd}
C.~Armendariz-Picon and E.~A. Lim, {\it Vacuum choices and the predictions of
  inflation},  {\em JCAP} {\bf 0312} (2003) 006,
  [\href{http://xxx.lanl.gov/abs/hep-th/0303103}{{\tt hep-th/0303103}}].

\bibitem{Martin:2003kp}
J.~Martin and R.~Brandenberger, {\it On the dependence of the spectra of
  fluctuations in inflationary cosmology on trans-planckian physics},  {\em
  Phys. Rev.} {\bf D68} (2003) 063513,
  [\href{http://xxx.lanl.gov/abs/hep-th/0305161}{{\tt hep-th/0305161}}].

\bibitem{Greene:2004np}
B.~R. Greene, K.~Schalm, G.~Shiu, and J.~P. van~der Schaar, {\it Decoupling in
  an expanding universe: Backreaction barely constrains short distance effects
  in the cmb},  {\em JCAP} {\bf 0502} (2005) 001,
  [\href{http://xxx.lanl.gov/abs/hep-th/0411217}{{\tt hep-th/0411217}}].

\bibitem{Brandenberger:2004kx}
R.~H. Brandenberger and J.~Martin, {\it Back-reaction and the trans-planckian
  problem of inflation revisited},  {\em Phys. Rev.} {\bf D71} (2005) 023504,
  [\href{http://xxx.lanl.gov/abs/hep-th/0410223}{{\tt hep-th/0410223}}].

\bibitem{Kaloper:2002cs}
N.~Kaloper, M.~Kleban, A.~Lawrence, S.~Shenker, and L.~Susskind, {\it Initial
  conditions for inflation},  {\em JHEP} {\bf 11} (2002) 037,
  [\href{http://xxx.lanl.gov/abs/hep-th/0209231}{{\tt hep-th/0209231}}].

\bibitem{Goldstein:2003ut}
K.~Goldstein and D.~A. Lowe, {\it A note on alpha-vacua and interacting field
  theory in de sitter space},  {\em Nucl. Phys.} {\bf B669} (2003) 325--340,
  [\href{http://xxx.lanl.gov/abs/hep-th/0302050}{{\tt hep-th/0302050}}].

\bibitem{Collins:2003zv}
H.~Collins, R.~Holman, and M.~R. Martin, {\it The fate of the alpha-vacuum},
  {\em Phys. Rev.} {\bf D68} (2003) 124012,
  [\href{http://xxx.lanl.gov/abs/hep-th/0306028}{{\tt hep-th/0306028}}].

\bibitem{Collins:2003mj}
H.~Collins and R.~Holman, {\it Taming the alpha vacuum},  {\em Phys. Rev.} {\bf
  D70} (2004) 084019, [\href{http://xxx.lanl.gov/abs/hep-th/0312143}{{\tt
  hep-th/0312143}}].

\bibitem{Kaloper:2003nv}
N.~Kaloper and M.~Kaplinghat, {\it Primeval corrections to the cmb
  anisotropies},  {\em Phys. Rev.} {\bf D68} (2003) 123522,
  [\href{http://xxx.lanl.gov/abs/hep-th/0307016}{{\tt hep-th/0307016}}].

\bibitem{deBoer:2004nd}
J.~de~Boer, V.~Jejjala, and D.~Minic, {\it Alpha-states in de sitter space},
  {\em Phys. Rev.} {\bf D71} (2005) 044013,
  [\href{http://xxx.lanl.gov/abs/hep-th/0406217}{{\tt hep-th/0406217}}].

\bibitem{Danielsson:2004xw}
U.~H. Danielsson, {\it Transplanckian energy production and slow roll
  inflation},  {\em Phys. Rev.} {\bf D71} (2005) 023516,
  [\href{http://xxx.lanl.gov/abs/hep-th/0411172}{{\tt hep-th/0411172}}].

\bibitem{Unruh:1994je}
W.~G. Unruh, {\it Sonic analog of black holes and the effects of high
  frequencies on black hole evaporation},  {\em Phys. Rev.} {\bf D51} (1995)
  2827--2838.

\bibitem{Corley:1996ar}
S.~Corley and T.~Jacobson, {\it Hawking spectrum and high frequency
  dispersion},  {\em Phys. Rev.} {\bf D54} (1996) 1568--1586,
  [\href{http://xxx.lanl.gov/abs/hep-th/9601073}{{\tt hep-th/9601073}}].

\bibitem{Corley:1997pr}
S.~Corley, {\it Computing the spectrum of black hole radiation in the presence
  of high frequency dispersion: An analytical approach},  {\em Phys. Rev.} {\bf
  D57} (1998) 6280--6291, [\href{http://xxx.lanl.gov/abs/hep-th/9710075}{{\tt
  hep-th/9710075}}].

\bibitem{Wang:2002hf}
X.~Wang, B.~Feng, M.~Li, X.-L. Chen, and X.~Zhang, {\it Natural inflation,
  planck scale physics and oscillating primordial spectrum},  {\em Int. J. Mod.
  Phys.} {\bf D14} (2005) 1347,
  [\href{http://xxx.lanl.gov/abs/astro-ph/0209242}{{\tt astro-ph/0209242}}].

\bibitem{Burgess:2002ub}
C.~P. Burgess, J.~M. Cline, F.~Lemieux, and R.~Holman, {\it Are inflationary
  predictions sensitive to very high energy physics?},  {\em JHEP} {\bf 02}
  (2003) 048, [\href{http://xxx.lanl.gov/abs/hep-th/0210233}{{\tt
  hep-th/0210233}}].

\bibitem{Martin:2003sf}
J.~Martin and P.~Peter, {\it Parametric amplification of metric fluctuations
  through a bouncing phase},  {\em Phys. Rev.} {\bf D68} (2003) 103517,
  [\href{http://xxx.lanl.gov/abs/hep-th/0307077}{{\tt hep-th/0307077}}].

\bibitem{Martin:2003bp}
J.~Martin and P.~Peter, {\it On the 'causality argument' in bouncing
  cosmologies},  {\em Phys. Rev. Lett.} {\bf 92} (2004) 061301,
  [\href{http://xxx.lanl.gov/abs/astro-ph/0312488}{{\tt astro-ph/0312488}}].

\bibitem{Hunt:2004vt}
P.~Hunt and S.~Sarkar, {\it Multiple inflation and the wmap 'glitches'},  {\em
  Phys. Rev.} {\bf D70} (2004) 103518,
  [\href{http://xxx.lanl.gov/abs/astro-ph/0408138}{{\tt astro-ph/0408138}}].

\bibitem{Bergstrom:2002yd}
L.~Bergstrom and U.~H. Danielsson, {\it Can map and planck map planck
  physics?},  {\em JHEP} {\bf 12} (2002) 038,
  [\href{http://xxx.lanl.gov/abs/hep-th/0211006}{{\tt hep-th/0211006}}].

\bibitem{Elgaroy:2003gq}
O.~Elgaroy and S.~Hannestad, {\it Can planck-scale physics be seen in the
  cosmic microwave background?},  {\em Phys. Rev.} {\bf D68} (2003) 123513,
  [\href{http://xxx.lanl.gov/abs/astro-ph/0307011}{{\tt astro-ph/0307011}}].

\bibitem{Okamoto:2003wk}
T.~Okamoto and E.~A. Lim, {\it Constraining cut-off physics in the cosmic
  microwave background},  {\em Phys. Rev.} {\bf D69} (2004) 083519,
  [\href{http://xxx.lanl.gov/abs/astro-ph/0312284}{{\tt astro-ph/0312284}}].

\bibitem{Martin:2003sg}
J.~Martin and C.~Ringeval, {\it Superimposed oscillations in the wmap data?},
  {\em Phys. Rev.} {\bf D69} (2004) 083515,
  [\href{http://xxx.lanl.gov/abs/astro-ph/0310382}{{\tt astro-ph/0310382}}].

\bibitem{Martin:2004iv}
J.~Martin and C.~Ringeval, {\it Addendum to ``superimposed oscillations in the
  wmap data?''},  {\em Phys. Rev.} {\bf D69} (2004) 127303,
  [\href{http://xxx.lanl.gov/abs/astro-ph/0402609}{{\tt astro-ph/0402609}}].

\bibitem{Martin:2004yi}
J.~Martin and C.~Ringeval, {\it Exploring the superimposed oscillations
  parameter space},  {\em JCAP} {\bf 0501} (2005) 007,
  [\href{http://xxx.lanl.gov/abs/hep-ph/0405249}{{\tt hep-ph/0405249}}].

\bibitem{Barriga:2000nk}
J.~Barriga, E.~Gaztanaga, M.~G. Santos, and S.~Sarkar, {\it On the apm power
  spectrum and the cmb anisotropy: Evidence for a phase transition during
  inflation?},  {\em Mon. Not. Roy. Astron. Soc.} {\bf 324} (2001) 977,
  [\href{http://xxx.lanl.gov/abs/astro-ph/0011398}{{\tt astro-ph/0011398}}].

\bibitem{Kogo:2003yb}
N.~Kogo, M.~Matsumiya, M.~Sasaki, and J.~Yokoyama, {\it Reconstructing the
  primordial spectrum from wmap data by the cosmic inversion method},  {\em
  Astrophys. J.} {\bf 607} (2004) 32--39,
  [\href{http://xxx.lanl.gov/abs/astro-ph/0309662}{{\tt astro-ph/0309662}}].

\bibitem{Huang:2003fw}
Q.-G. Huang and M.~Li, {\it Power spectra in spacetime noncommutative
  inflation},  {\em Nucl. Phys.} {\bf B713} (2005) 219--234,
  [\href{http://xxx.lanl.gov/abs/astro-ph/0311378}{{\tt astro-ph/0311378}}].

\bibitem{Shafieloo:2003gf}
A.~Shafieloo and T.~Souradeep, {\it Primordial power spectrum from wmap},  {\em
  Phys. Rev.} {\bf D70} (2004) 043523,
  [\href{http://xxx.lanl.gov/abs/astro-ph/0312174}{{\tt astro-ph/0312174}}].

\bibitem{Brooks:1998}
S.~P. Brooks and A.~Gelman {\em J. Comp. Graph. Stat.} {\bf 7} (1998) 434.

\bibitem{Easther:2004vq}
R.~Easther, W.~H. Kinney, and H.~Peiris, {\it Observing trans-planckian
  signatures in the cosmic microwave background},  {\em JCAP} {\bf 0505} (2005)
  009, [\href{http://xxx.lanl.gov/abs/astro-ph/0412613}{{\tt
  astro-ph/0412613}}].

\bibitem{Kunz:2006mc}
M.~Kunz, R.~Trotta, and D.~Parkinson, {\it Measuring the effective complexity
  of cosmological models},  {\em Phys. Rev.} {\bf D74} (2006) 023503,
  [\href{http://xxx.lanl.gov/abs/astro-ph/0602378}{{\tt astro-ph/0602378}}].

\end{thebibliography}\endgroup

\end{document}